\documentclass[12pt,a4paper]{article}
\usepackage{cite,mathtools,amsmath,amssymb}
\usepackage[dvipdfmx]{graphicx}
\usepackage[usenames]{color}
\usepackage[bookmarksopen,colorlinks=true,linkcolor=dark_green,citecolor=dark_red,urlcolor=dark_red,linktocpage=false]{hyperref}

\definecolor{dark_blue}{rgb}{0,0,0.6}
\definecolor{dark_green}{rgb}{0,0.4,0}
\definecolor{dark_red}{rgb}{0.6,0,0}

\usepackage[height=22.5cm,width=16.5cm,centering]{geometry}

\begin{document}
%%%%%%%%%%%%%%%%%%%%%%%%%%%%%%%%%%%%%%%%%%%%%%%%%%

%%%%%%%%%%%%%%%%%%%%%%%%%%%%%%%%%%%%%%%%%%%%%%%%%%
\begin{titlepage}

\begin{center}

\hfill CTPU-17-26 \\
\hfill KEK-TH-1986 \\

\vskip .75in

{\LARGE \bf 
Gravitational waves from bubble dynamics: 
\\ \vspace{5mm} 
Beyond the Envelope
}

\vskip .75in

{\large Ryusuke Jinno$^{a,b}$ and Masahiro Takimoto$^{b,c}$}

\vskip 0.25in

\begin{tabular}{ll}
$^{a}$ &\!\! {\em Center for Theoretical Physics of the Universe, Institute for Basic Science (IBS),}\\
&{\em Daejeon 34051, Korea}\\[.3em]
$^{b}$ &\!\! {\em Theory Center, High Energy Accelerator Research Organization (KEK),}\\
&{\em Oho, Tsukuba, Ibaraki 305-0801, Japan}\\[.3em]
$^{c}$ &\!\! {\em Department of Particle Physics and Astrophysics, Weizmann Institute of Science,}\\
&{\em Rehovot 7610001, Israel}\\[.3em]
\end{tabular}

\end{center}
\vskip .5in

\begin{abstract}
We study gravitational-wave production from bubble dynamics 
(bubble collisions and sound waves) 
during a cosmic first-order phase transition with an analytic approach.
We first propose modeling the system with the thin-wall approximation
but without the envelope approximation often adopted in the literature,
in order to take bubble propagation after collisions into account.
The bubble walls in our setup are considered as modeling 
the scalar field configuration and/or the bulk motion of the fluid.
We next write down analytic expressions for the gravitational-wave spectrum, 
and evaluate them with numerical methods.
It is found that, in the long-lasting limit of the collided bubble walls,
the spectrum grows from $\propto f^3$ to $\propto f^1$ in low frequencies,
showing a significant enhancement compared to the one with the envelope approximation.
It is also found that the spectrum saturates in the same limit,
indicating a decrease in the correlation of the energy-momentum tensor at late times.
We also discuss the implications of our results to gravitational-wave production 
both from bubble collisions (scalar dynamics) and sound waves (fluid dynamics).
\end{abstract}

\end{titlepage}
%%%%%%%%%%%%%%%%%%%%%%%%%%%%%%%%%%%%%%%%%%%%%%%%%%

%%%%%%%%%%%%%%%%%%%%%%%%%%%%%%%%%%%%%%%%%%%%%%%%%%
\thispagestyle{empty}
\tableofcontents
\thispagestyle{empty}
\newpage
\setcounter{page}{1}
%%%%%%%%%%%%%%%%%%%%%%%%%%%%%%%%%%%%%%%%%%%%%%%%%%

%%%%%%%%%%%%%%%%%%%%%%%%%%%%%%%%%%%%%%%%%%%%%%%%%%
\section{Introduction}
\label{sec:Intro}
\setcounter{equation}{0}
%%%%%%%%%%%%%%%%%%%%%%%%%%%%%%%%%%%%%%%%%%%%%%%%%%

Gravitational waves (GWs) provide us with a unique probe to the early universe.
Various cosmological dynamics are known to produce GWs:
inflationary quantum fluctuations~\cite{Starobinsky:1979ty}, 
preheating~\cite{Khlebnikov:1997di},
topological defects such as domain walls and cosmic strings~\cite{Vilenkin:2000jqa},
first-order phase transitions~\cite{Witten:1984rs,Hogan:1986qda}, and so on.
Gravitational waves from these cosmological sources, if detected,
will give us an important clue to the high energy physics we are seeking for.
On the observational side there has recently been a remarkable progress of
the detection of GWs from black hole binaries reported by the LIGO collaboration~\cite{Abbott:2016blz,Abbott:2016nmj,Abbott:2017vtc},
and gravitational-wave astronomy has now been established.
In the future, space interferometers such as LISA~\cite{Seoane:2013qna}, 
BBO~\cite{Harry:2006fi} and DECIGO~\cite{Seto:2001qf}
are expected to open up a new era of gravitational-wave cosmology.

In this paper, we study GW production from first-order phase transitions.
Though it has been shown that first-order phase transitions do not occur 
within the standard model~\cite{Kajantie:1996mn,Gurtler:1997hr,Csikor:1998eu},
various types of motivated particle physics models 
still predict first-order phase transitions in the early Universe
(see Refs.~\cite{Apreda:2001tj,Apreda:2001us,Grojean:2006bp,Huber:2007vva,Espinosa:2008kw,
Ashoorioon:2009nf,Kang:2009rd,Jarvinen:2009mh,Konstandin:2010cd,
No:2011fi,Wainwright:2011qy,Barger:2011vm,Leitao:2012tx,
Dorsch:2014qpa,Kozaczuk:2014kva,Schwaller:2015tja,Kakizaki:2015wua,Jinno:2015doa,Huber:2015znp,Leitao:2015fmj,
Huang:2016odd,Garcia-Pepin:2016hvs,Jaeckel:2016jlh,Dev:2016feu,Hashino:2016rvx,
Jinno:2016knw,Barenboim:2016mjm,Kobakhidze:2016mch,Hashino:2016xoj,Artymowski:2016tme,
Kubo:2016kpb,Balazs:2016tbi,Vaskonen:2016yiu,Dorsch:2016nrg,
Huang:2017laj,Baldes:2017rcu,Chao:2017vrq,Beniwal:2017eik,Addazi:2017gpt,Kobakhidze:2017mru,
Tsumura:2017knk,Marzola:2017jzl,Bian:2017wfv,Huang:2017rzf,Iso:2017uuu,Addazi:2017oge,Kang:2017mkl,
Cai:2017tmh,Hashino:2018wee},
and also Refs.~\cite{Caprini:2015zlo,Cai:2017cbj} and references therein for reviews).
Excitingly, planned GW detectors are sensitive to phase transitions around TeV-PeV scales,
and thus such GWs offer one of the promising tools to probe new physics beyond the standard model.

In a thermal first-order phase transition, 
true-vacuum bubbles start to nucleate at some temperature,
and then they expand because of the pressure difference between the false and true vacua. 
These bubbles eventually collide with each other and the phase transition completes.
Though uncollided bubbles do not radiate GWs because of the spherical symmetry of each bubble,
the collision process breaks the symmetry and as a result GWs are produced.
The analysis of GW production from such processes was initiated 
in Refs.~\cite{Kosowsky:1991ua,Kosowsky:1992rz,Kosowsky:1992vn,Kamionkowski:1993fg}.
In the first numerical simulation carried out in Ref.~\cite{Kosowsky:1991ua} 
in a vacuum transition, 
it was noticed that the main GW production comes from the uncollided regions of the bubble walls.
This observation made the basis for the ``envelope approximation," 
in which only the uncollided regions of the bubble walls are taken into account in calculating GW production
(see Fig.~\ref{fig:Envelope}).
This approximation has been widely used in the subsequent literature 
together with the ``thin-wall approximation," 
in which the released energy is assumed to be concentrated in infinitely thin bubble walls.\footnote{
Though in the early literature the envelope approximation includes 
the thin-wall approximation~\cite{Kosowsky:1991ua,Kosowsky:1992rz,Kosowsky:1992vn,Kamionkowski:1993fg},
we distinguish them in this paper.
}
Later a numerical simulation with the same approximations  
has been performed in Ref.~\cite{Huber:2008hg} with an increased number of bubbles,
and a more precise form of the GW spectrum has been obtained.

In the literature mentioned above, the bubble walls are thought to represent the energy concentration 
by the profile of the scalar field that drives the transition 
or by the bulk motion of the fluid coupled to the scalar field.
It has recently been noticed in a series of 
numerical simulations~\cite{Hindmarsh:2013xza,Hindmarsh:2015qta,Hindmarsh:2017gnf}\footnote{
See also Refs.~\cite{Child:2012qg,Giblin:2013kea,Giblin:2014qia} for other numerical simulations.
}
that the latter bulk motion propagates even after bubble collisions, 
and works as a long-lasting source of GWs.
It has been found that the GWs from such sound waves typically dominate 
the other sources of GWs,\footnote{
In addition to bubble collisions and sound waves, turbulence is another important source for 
GWs~\cite{Kamionkowski:1993fg,Kosowsky:2001xp,Nicolis:2003tg,
Caprini:2006jb,Caprini:2009yp,Kahniashvili:2009mf}.
Note that the sound-wave regime as we study in this paper can develop into turbulent regime at late times.
}
and that the resulting spectrum cannot be modeled by the envelope approximation 
because of the long-lasting nature of 
the source~\cite{Hindmarsh:2013xza,Hindmarsh:2015qta,Hindmarsh:2017gnf,Weir:2016tov}.

These numerical simulations have brought significant developments in our understanding on GW sourcing
both from bubble collisions and from sound waves.
In this paper, however, we stress the importance of the cooperation between
\begin{itemize}
\item[]
\begin{center}
Analytic understanding
\;\;\;\;\;\;\;\; 
\&
\;\;\;\;\;\;\;\; 
Numerical understanding,
\end{center}
\end{itemize}
and aim to develop the former.
This is partly because the former approach sometimes goes beyond 
the barrier of computational resources,
and also because it often gives a clearer understanding of the system.\footnote{
Note that another analytic modeling (sound-shell model~\cite{Hindmarsh:2016lnk}) 
has also been proposed
to explain the enhancement of the GW spectrum around the scale of sound shell thickness.
}
For this purpose we adopt the method of relating the GW spectrum 
with the two-point correlator of the energy-momentum tensor $\left< T(x)T(y) \right>$,
which is pioneered in Ref.~\cite{Caprini:2007xq}
in the context of GW production from bubble dynamics.
In Ref.~\cite{Jinno:2016vai} it has been pointed out that, under the thin-wall approximation, 
various contributions to the correlator $\left< T(x)T(y) \right>$
reduce to finite number of classes.
This observation made it possible to write down the GW spectrum analytically 
in the same setup as the numerical study in Ref.~\cite{Huber:2008hg},
i.e. the GW spectrum with the thin-wall and envelope approximations.\footnote{
By using the same formalism, it is also possible to investigate 
the effect of the bubble nucleation rate on the GW spectrum analytically~\cite{Jinno:2017ixd}.
}
In this paper we further develop this method, and write down
\begin{itemize}
\item[]
\begin{center}
Gravitational-wave spectrum with the thin-wall approximation 
\;\;\;\;\;\;\;\; 
\\
but {\it without the envelope approximation}.\footnote{
There are other assumptions and approximations 
such as constant wall velocity, free propagation after collision
and the absence of cosmic expansion: see Sec.~\ref{sec:Setup}.
}
\;\;\;\;\;\;\;\;
\end{center}
\end{itemize}
As explained in Sec.~\ref{sec:Setup}, 
the bubble walls in our setup can be regarded as modeling the energy concentration 
in the scalar field and/or in the bulk motion of the fluid, 
and therefore the resulting spectrum is considered to be relevant to GW production 
both from bubble collisions (scalar field contribution) and sound waves (fluid contribution).
We discuss the implications and also limitations of our modeling there.
The analytic expressions for the GW spectrum, Eqs.~(\ref{eq:MainDeltaS}) and (\ref{eq:MainDeltaD}), 
have multi-dimensional integrations, and therefore we evaluate them with numerical methods.\footnote{
Note that this is essentially different from numerical simulations:
the GW spectrum obtained this paper is the spectrum taking infinitely many bubbles into account.
}
As a result, the growth and saturation of the spectrum are clearly observed
as a function of the duration time of the collided walls.

The organization of the paper is as follows.
In Sec.~\ref{sec:Setup} we present our setup and summarize basic ingredients to estimate 
the GW spectrum. We also discuss the implications and limitations of our setup in this section.
In Sec.~\ref{sec:Analytic} we write down the analytic expressions for the GW spectrum,
and we evaluate them in Sec.~\ref{sec:Numerical} with Monte-Carlo integration.
Sec.~\ref{sec:DiscussionConclusion} is devoted to discussion and conclusions.

%%%%%%%%%%%%%%%%%%%%%%%%%%%%%%%%%%%%%%%%%%%%%%%%%%
\section{Setup and basic ingredients}
\label{sec:Setup}
\setcounter{equation}{0}
%%%%%%%%%%%%%%%%%%%%%%%%%%%%%%%%%%%%%%%%%%%%%%%%%%

In this section, we present our setup and basic ingredients to estimate the GW spectrum.
In order to estimate the spectrum, 
we have to clarify the energy-momentum tensor of the system.
It is determined by
\begin{itemize}
\item[(1)]
Spacetime distribution of bubbles (i.e. nucleation rate of bubbles),
\item[(2)]
Energy-momentum tensor profile around a bubble wall,
\item[(3)]
Dynamics after bubble collisions.
\end{itemize}
Since it is generically hard to solve the full dynamics of the system,
reasonable approximations are necessary for practical calculations.
The aim of the following subsections is to clarify our approximations and their validity.

In Sec.~\ref{subsec:GWV}, 
we give a brief overview of bubble dynamics in a cosmological phase transition,
and explain our approximations for (2) and (3).
In this subsection we do not show the explicit expression 
for the energy-momentum profile in order to avoid possible complications.
In Sec.~\ref{subsec:AA}, we present the explicit form of the profile.
In Sec.~\ref{subsec:rate}, 
we give our assumption about the nucleation rate which determines (1).
In Sec.~\ref{subsec:gwfrom}, 
we present a formalism to calculate the GW spectrum
from the correlation function of the energy-momentum tensor.
In Sec.~\ref{subsec:SetupSum} we summarize our setup and discuss its physical implications.

Before moving on, we comment on the meaning of ``wall" in the present paper.
This word usually refers to the energy localization in the scalar field gradient.
We use the word ``scalar wall" for such energy localization throughout the paper.
This is because, as mentioned in Sec.~\ref{sec:Intro}, 
the main energy carrier around bubbles can be not only the scalar field 
but also the bulk motion of the fluid.
Since our modeling aims to represent (at least some aspects of) 
the scalar field and sound waves,
we refer to the energy localization from both as ``wall"
after modeling the energy-momentum profile.

%%%%%%%%%%%%%%%%%%%%%%%%%%%%%%%%%%%%%%%%%%%%%%%%%%
\subsection{Bubble dynamics in cosmological phase transitions}
\label{subsec:GWV}
%%%%%%%%%%%%%%%%%%%%%%%%%%%%%%%%%%%%%%%%%%%%%%%%%%

In this subsection we give a brief overview of GW production 
in a cosmological phase transition paying particular attention to the bubble dynamics.
We also introduce our approximations for (2) and (3) and discuss their validity.

%%%%%%%%%%%%%%%%%%%%%%%%%%%%%%%%%%%%%%%%%%%%%%%%%%
\subsubsection*{Classification}
%%%%%%%%%%%%%%%%%%%%%%%%%%%%%%%%%%%%%%%%%%%%%%%%%%

In cosmological first-order phase transitions, 
bubbles of the true vacuum nucleate, expand and then collide with each other.
The released free energy accumulates around the bubble surfaces during their expansion.
Though uncollided bubbles do not radiate GWs because of the spherical symmetry of each bubble, 
their collisions break it and produce GWs.
In order to estimate the resulting GW spectrum,
we have to know the energy-momentum tensor of the system determined by the bubble dynamics.
The behavior of the expanding bubbles can be categorized into three classes:
\begin{itemize}
\item[(a)] 
Runaway,
\item[(b)]
Low terminal velocity,
\item[(c)] 
High terminal velocity.
\end{itemize}
Roughly speaking, the balance between the released energy and 
the friction coming from the background thermal plasma determines which case is realized 
(see e.g. Refs.~\cite{Espinosa:2010hh,Leitao:2015ola}).
Schematically, the acceleration of the scalar wall is given by 
\begin{align}
         \dot{v}_w \propto \rho_0 - F_{\rm fric}(v_w),
\end{align}
where $v_w$ is the scalar wall velocity, 
$\rho_0$ denotes the released energy, 
and $F_{\rm fric}(v_w)$ denotes the friction term.

In the non-relativistic regime, 
the friction term is proportional to the velocity: $F_{\rm fric} \propto v_w$.
If the ratio $\alpha$ between the released energy $\rho_0$ and 
that of the surrounding plasma $\rho_{\rm rad}$
\begin{align}
\alpha 
&\equiv 
\frac{\rho_0}{\rho_{\rm rad}},
\label{eq:alpha}
\end{align}
is suppressed $\alpha \lesssim {\mathcal O}(0.1)$, 
the acceleration tends to cease within non-relativistic regime.\footnote{
In order to determine the terminal velocity,
we have to specify couplings between the scalar field and the plasma.
Here we do not consider details of friction, 
assuming that the couplings are not extremely suppressed.
}
We refer to such cases as (b) low terminal velocity.
On the other hand, if the released energy is large enough $\alpha \gtrsim {\mathcal O}(0.1)$,
the scalar wall velocity enters the relativistic regime $v_w\simeq 1$.
In order to deal with this regime, we have to know the behavior of the friction term 
in $\gamma_w \equiv 1/\sqrt{1 - v_w^2} \rightarrow \infty$ limit.
Though it had long been considered that the friction saturates in the relativistic limit 
and cannot stop the acceleration of the scalar wall~\cite{Bodeker:2009qy},
the authors of Ref.~\cite{Bodeker:2017cim} have recently pointed out that
particle splitting processes around the wall generate a friction term proportional to $\gamma_w$.
This term becomes larger and larger in $\gamma_w \rightarrow \infty$ limit, 
and stops the acceleration before bubble collisions in some cases.
In fact, as we discuss in Sec.~\ref{subsec:GWV} (c),
the acceleration may stop due to this splitting process 
for most cases of our interest.
We refer to such cases as (c) high terminal velocity.
We also denote by (a) runaway
those cases in which the acceleration continues all the way until bubble collisions.

\begin{figure}
\begin{center}
\includegraphics[scale=0.3]{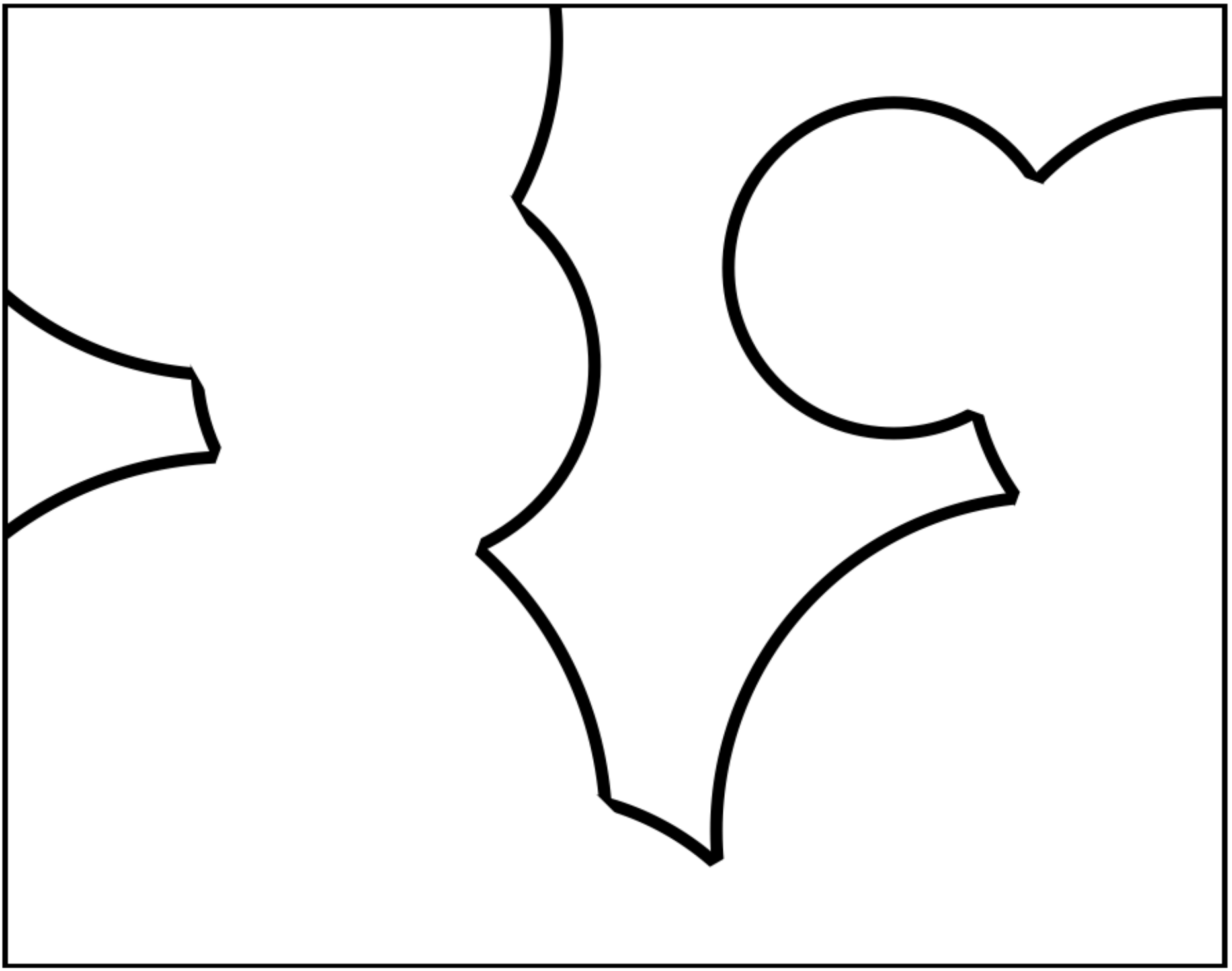}
\caption{\small
Rough sketch of the bubble dynamics with the envelope approximation.
The bubble walls, denoted by the black lines, 
accumulate energy and momentum as they expand and then lose them instantly 
when they collide with each other.
Compare this figure with Fig.~\ref{fig:BeyondEnv}.
This figure is the same as in Ref.~\cite{Jinno:2016vai}.
}
\label{fig:Envelope}
\end{center}
\begin{center}
\includegraphics[scale=0.3]{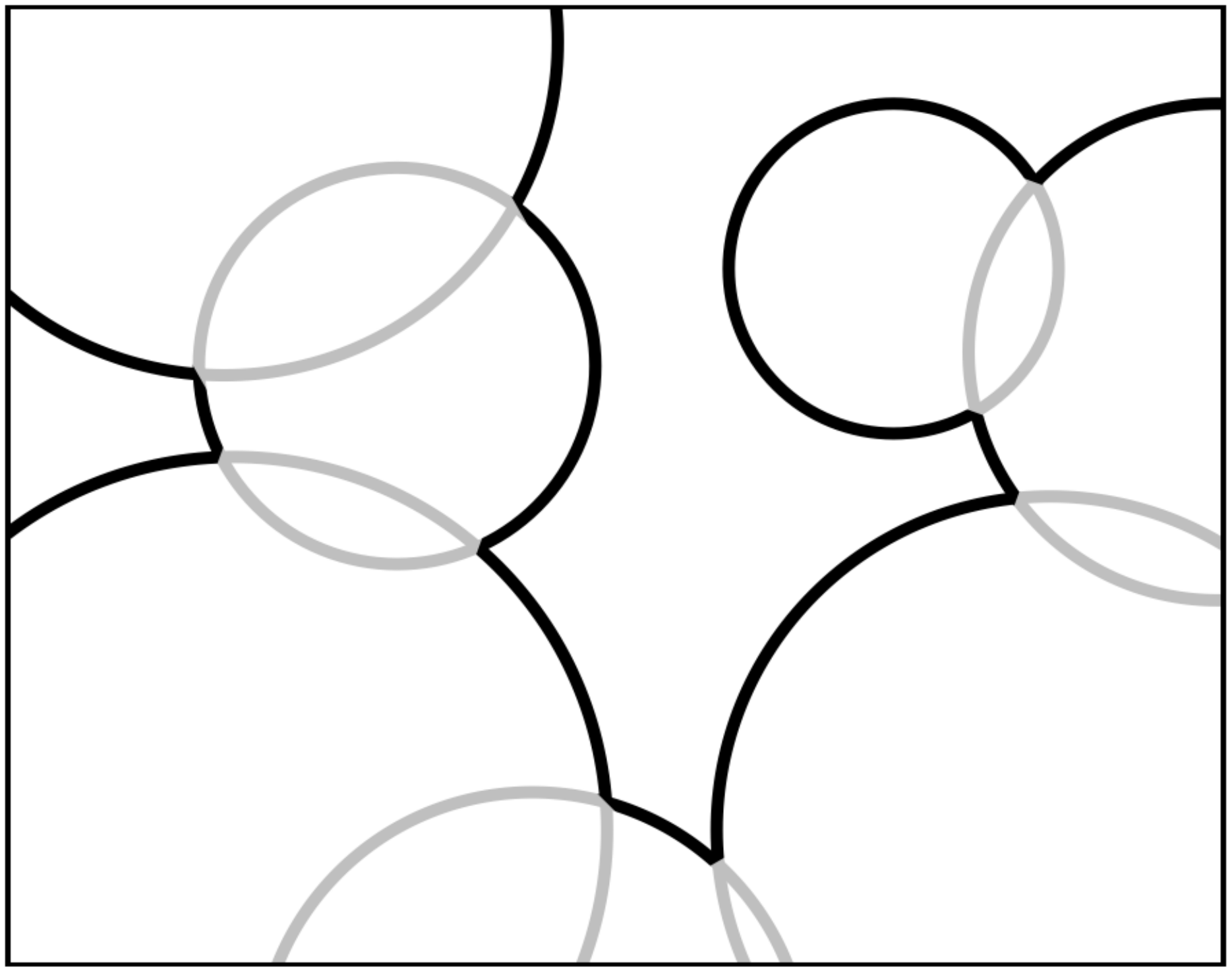}
\caption{\small
Rough sketch of the bubble dynamics without the envelope approximation.
The collided bubble walls, denoted by the gray lines, 
gradually lose their energy and momentum densities after collisions.
}
\label{fig:BeyondEnv}
\end{center}
\end{figure}

As mentioned in the beginning of this section,
in order to have the energy-momentum tensor of the system, we have to clarify
\begin{itemize}
\item[(2)]
Energy-momentum tensor profile around a bubble wall,
\item[(3)]
Dynamics after bubble collisions,
\end{itemize}
for each of (a), (b) and (c).
Especially, the importance of (3) has recently been pointed out in 
Refs.~\cite{Hindmarsh:2013xza,Hindmarsh:2015qta,Hindmarsh:2017gnf}, 
in which the authors observed a sizable GW production from sound waves after bubble collisions.
In this paper, we adopt 
\begin{itemize}
\item[(2)]
Thin-wall approximation,
\item[(3)]
Free propagation,
\end{itemize}
for these two.
The former assumes that the released energy is localized in an infinitesimally thin 
surface of a bubble,
while the latter assumes that the energy and momentum accumulated 
around the bubble surfaces until the first collision just pass through after that. 
See also Eq.~(\ref{eq:rhoBBeyond_2}) for the explicit form of the energy-momentum tensor.
Fig.~\ref{fig:BeyondEnv} is a schematic picture of the system with these approximations.
The black lines denote the thin walls, 
which model the energy and momentum localization in the scalar wall and/or in the bulk motion of the fluid.
The released energy accumulates until their first collisions.
On the other hand, the gray lines denote the evolution of such localized energy and momentum after collisions.
They gradually lose the accumulated energy and momentum densities after collisions, 
as a result of free propagation.
In addition, we assume that the propagation velocity of such localized energy 
is constant both before and after collisions, and denote it by $v$.
Below we discuss the validity of such approximations 
(thin-wall and free propagation with a constant velocity) for both (a) and (b).
Unfortunately, we do not know much about the dynamics realized in case (c).
Thus, we only mention some features and possible procedures to determine the GW spectrum in this case.
We summarize the properties of each case in Table~\ref{tab:type}.

Before moving on, we comment on turbulence after bubble collisions.
In all of the three cases plasma turbulence can be a sizable source for GWs 
at late times~\cite{Kamionkowski:1993fg,Kosowsky:2001xp,Nicolis:2003tg,
Caprini:2006jb,Caprini:2009yp,Kahniashvili:2009mf}.
However, since the turbulence dynamics is highly nonlinear,
we restrict ourselves to the bubble dynamics before the onset of turbulence in this paper.

%%%%%%%%%%%%%%%%
%%%%%%%%%%%%%%%%

%%%%%%%%%%%%%%%%
\begin{table}[htb]
\begin{center}
\caption{Classification of bubble dynamics}
\vskip 0.05in 
\begin{footnotesize}
\begin{tabular}{|c|c|c|c|} \hline 
Type & Width of source & Scalar wall velocity & Dynamics after collision \\ \hline\hline
Runaway & thin & $\gamma_w \gg 1$ & free propagation with speed of light \\ \hline
Low terminal velocity&thick&constant $v_w$&free propagation with speed of sound \\ \hline
High terminal velocity & ? & constant $\gamma_w \gg 1$& ? \\ \hline
\end{tabular}
\label{tab:type}
\end{footnotesize}
\end{center}
\end{table}
%%%%%%%%%%%%%%%%

%%%%%%%%%%%%%%%%%%%%%%%%%%%%%%%%%%%%%%%%%%%%%%%%%%
\subsubsection*{(a) Runaway}
\label{subsubsec:a}
%%%%%%%%%%%%%%%%%%%%%%%%%%%%%%%%%%%%%%%%%%%%%%%%%%

In the runaway case the released energy is relatively large ($\alpha \gtrsim {\mathcal O}(0.1)$) 
and the friction term cannot stop the acceleration of the scalar wall by the time of collisions.
Below we assume that the friction from the surrounding plasma, 
especially the splitting effect explained in (c) high terminal velocity, is negligible.
We also assume that the energy density of the Universe is dominated by the vacuum energy density.\footnote{
This is indeed a good assumption, since extremely large values of $\alpha$ are required for runaway walls: 
see the estimate in (c) High terminal velocity.
}

In the runaway case, the released energy accumulates 
around a thin surface of a bubble in the form of scalar gradient.
Let us first consider the behavior of the scalar field before collisions.
Suppose that the scalar field difference between the true and false vacua is $\Delta \phi$,
the released energy density is $\rho_0$, the bubble radius is $R$, 
and the width of energy localization is $l_b$.
We may estimate the width of the surface as
\begin{align}
\left(\frac{\Delta\phi}{l_b}\right)^2 l_b R^2
&\sim \rho_0R^3
\;\;\;\;
\to 
\;\;\;\;
l_b
\sim \frac{\Delta\phi^2}{\rho_0R}.
\end{align}
We see that the scalar wall becomes thiner and thiner as the bubble expands.
As we see in Sec.~\ref{subsec:rate}, 
the typical bubble size just before collision is sub-horizon, 
which we denote $R_{\rm coll} = \epsilon / H$ 
with $\epsilon \lesssim {\mathcal O}(0.1)$ with $H$ denoting the Hubble parameter.
For clarity,  we parametrize $\rho_0$ as $\rho_0 \sim m_{\rm typ}^2\Delta\phi^2$
where $m_{\rm typ}$ denotes the typical mass scale of the potential.
Usually $m_{\rm typ} \sim T$ holds with $T$ being the temperature of the plasma
around the time of transition.
Also $m_{\rm typ} \sim T \lesssim \Delta\phi$ holds for most of the runaway cases
since the released energy dominates the radiation energy.
The typical momentum $k_b$ of the scalar field configuration just before collision is given by
\begin{align}
k_b
&\sim 
\left.
\frac{1}{l_b} 
\right|_{R=R_{\rm coll}} 
\sim \epsilon M_P
\left(
\frac{m_{\rm typ}}{\Delta\phi}
\right),
\end{align}
where $M_P$ denotes the reduced Planck mass and 
we used the Friedmann equation $M_P^2H^2 \sim \rho_0$ satisfied in the runaway case.
Note that the typical momentum is much larger than other physical parameters such as $m_{\rm typ}$
as long as the phase transition occurs well below the Planck scale.
In particle analogy, 
the number density $n_b$ of such high momentum modes just before collision is given by
\begin{align}
k_b n_b l_b R_{\rm coll}^2 
&\sim \rho_0 R_{\rm coll}^3 
\;\;\;\;
\rightarrow 
\;\;\;\;
n_b
\sim \epsilon M_P m_{\rm typ} \Delta \phi.
\end{align}

Now let us consider the effect of bubble collisions, assuming that 
the particle analogy is applicable.
We focus only on those high momentum modes where most of the released energy is accumulated.
Though the scalar field profile is deformed to some extent during the collision process,
deformation in such high momentum modes is expected to be small.
To see this, let us consider a scattering process caused by $\lambda \phi^4$ interaction for example.
Denoting the change in the number density of such high momentum modes by $\Delta n_b$, we have the following relation:
\begin{align}
\frac{\Delta n_b}{n_b}
&\sim 
\frac{\lambda^2}{k_b^2} n_b \Delta t_{\rm coll}, 
\;\;\;\;
\Delta t_{\rm coll}
\sim 
\frac{1}{l_b}
\;\;\;\;
\rightarrow 
\;\;\;\;
\frac{\Delta n_b}{n_b}
\sim 
\frac{\lambda^2}{\epsilon^2}
\left(
\frac{\Delta \phi}{m_{\rm typ}}
\right)
\left(
\frac{H}{M_P}
\right)
\ll 1,
\end{align}
with $\Delta t_{\rm coll}$ indicating the duration time of the collision process.
We see that the effect of collision is typically negligible for the high momentum modes
because of the last factor.

Next let us consider the effect of decay processes after bubble collisions.
Denoting the mass and decay rate of the scalar field at the true vacuum by
$\sim m_{\rm typ}$ and $\sim y^2 m_{\rm typ}$, respectively,
we may estimate the lifetime of high momentum modes $\Delta t_{\rm decay}$ as
\begin{align}
H\Delta t_{\rm decay}
&\sim 
\frac{m_{\rm typ} \Delta \phi}{M_P}
\frac{1}{y^2 m_{\rm typ}}
\frac{k_b}{m_{\rm typ}}
\sim
\frac{\epsilon}{y^2}.
\end{align}
Therefore the lifetime can be comparable to the Hubble time 
though it depends on the model parameters.

Finally let us consider the validity of the thin-wall approximation after collisions.
After bubbles collide with each other, 
the energy injection into the scalar motion ceases and the scalar field start to evolve with free propagation. 
Since the scalar motion has a finite momentum width $\Delta k_b$, 
the width of the energy concentration becomes thicker and thicker.
On the other hand,
the relevant scale in the GW spectrum is 
the typical bubble size around the time of collisions.
Therefore, the thin-wall approximation is expected to hold 
until the width of the energy concentration becomes comparable to this length scale.
Let us denote by $\Delta t_{\rm thin}$ 
the timescale with which the scalar wall width grows to this length scale.
Approximating the momentum width by $\Delta k_b\sim k_b$, 
we may estimate $\Delta t_{\rm thin}$ as
\begin{align}
\Delta t_{\rm thin} 
\Delta v_w
&\sim R_{\rm coll}
\;\;\;\;
\rightarrow 
\;\;\;\;
H \Delta t_{\rm thin}
\sim 
\epsilon
\left( \frac{k_b}{m_{\rm typ}} \right)^2
\gg 1,
\end{align}
with $\Delta v_w$ being the velocity dispersion corresponding to $\Delta k_b$.\footnote{
This can be estimated from the relation between the velocity dispersion and the momentum dispersion:
\begin{align}
\Delta \gamma_w
&\sim 
\frac{v_w\Delta v_w}{(1 - v_w^2)^{3/2}}
\sim \frac{\Delta k_b}{m_{\rm typ}}
\sim \frac{k_b}{m_{\rm typ}}
~~
\to
~~
\Delta v_w
\sim
\frac{1}{\gamma_w^3}
\frac{k_b}{m_{\rm typ}}
\sim 
\frac{m_{\rm typ}^2}{k_b^2}.
\end{align}
}
Hence, the thin-wall approximation can be valid until a Hubble time
as long as the phase transition occurs much below the Planck scale.

Let us summarize the runaway case.
During bubble expansion, the released energy accumulates within extremely thin regions 
in the form of scalar gradient, and the scalar field becomes ultra-relativistic.
As long as particle analogy is applicable,
the effect of bubble collisions is typically negligible for such ultra-relativistic modes,
and the scalar walls are almost luminal both before and after collisions.\footnote{
Note that, if nontrivial trapping of the scalar field at the false vacuum occurs after collisions,
the dynamics can be quite different from the one described here~\cite{Konstandin:2011ds}.
}
Though it should be confirmed in future studies that the above particle analogy is applicable 
in a system with extremely relativistic scalar configurations $\gamma_w \gg 1$,\footnote{
Note that numerical simulations with $\gamma_w \gg1$, which is relevant in the runaway case, are generically difficult.
}
it is at least expected that the scalar field configurations are still energetic for some time after collisions.

%%%%%%%%%%%%%%%%%%%%%%%%%%%%%%%%%%%%%%%%%%%%%%%%%%
\subsubsection*{(b) Low terminal velocity}
\label{subsubsec:b}
%%%%%%%%%%%%%%%%%%%%%%%%%%%%%%%%%%%%%%%%%%%%%%%%%%

Next let us consider those cases where the released energy is subdominant compared to that of radiation 
($\alpha \lesssim {\mathcal O}(0.1)$).
In such cases, 
the dynamics of bubble expansion is determined by coupled equations between the scalar field and the plasma.
Soon after bubbles nucleate,
the pressure difference between the true and false vacua gets balanced with 
the friction from the thermal plasma,
and as a result the scalar wall velocity $v_w$ approaches a constant value~\cite{Espinosa:2010hh}.

During bubble expansion,
the released energy is converted into the bulk motion of the plasma surrounding the scalar wall.
Since there is no distance scale in the fluid equations,
the fluid profile (enthalpy $\omega$, fluid velocity $\vec{u}$, and so on) depends only on the variable $\xi \equiv r/t$,
where $r$ is the distance from the bubble nucleation point and $t$ is the time after nucleation.
Generally the fluid bulk motion is localized around the position of the scalar wall $\xi \sim v_w$,\footnote{
Note that this does not necessarily mean that the fluid velocity $u$ is close to the scalar wall velocity $v_w$.
}
and the width of this energy localization is smaller than but comparable to the bubble size.
The fraction $\kappa$, called the efficiency factor,
is defined as the fraction of the released energy $\rho_0$ which goes into the plasma motion.
It is obtained from the velocity and enthalpy profile as~\cite{Espinosa:2010hh}
\begin{align}
\kappa
&= \frac{3}{\rho_0v_w^3}\int _0^{\infty}d\xi~\xi^2\omega \gamma^2 u^2,
\label{eq:kappa}
\end{align}
with $\gamma \equiv 1 / \sqrt{1 - u^2}$ and $u \equiv |\vec{u}|$. 

After the bubbles collide, the energy injection into plasma motion ceases.
However, as pointed out in Refs.~\cite{Hindmarsh:2013xza,Hindmarsh:2015qta,Hindmarsh:2017gnf}, 
the plasma motion remains after bubble collisions and produce a sizable amount of GWs.
In the present case the released energy is typically subdominant compared to the plasma energy density,
and thus $\delta \rho/\rho_{\rm rad} \ll 1$ and $u \ll1$ hold in most cases
with $\delta \rho \sim \kappa \rho_0$ being the energy density in the plasma motion localized around the bubbles.
When these two conditions hold,
the dynamics of the plasma motion is well described by linear theory (sound wave dynamics).
For example, the fluid velocity obeys the ordinary wave equation
\begin{align}
\left(
\frac{\partial^2}{\partial t^2} - c_s^2 \Delta
\right) \vec{u}
&= 0,
\end{align}
with $c_s\simeq 1/\sqrt{3}$ being the speed of sound.
Therefore the plasma dynamics after bubble collisions
is described just by the free propagation of plasma motions with the speed of sound.
The width of the energy localization is fixed at the first collision,
and the plasma velocity $u$ starts to decrease after that
because of the increase in the volume of energy localization and because of the energy conservation.\footnote{
On this point our modeling of the system differs from the one in Ref.~\cite{Caprini:2007xq}.
}
Though sound waves might be damped by viscosity at late times,
the timescale of such an effect can be larger than the Hubble time~\cite{Hindmarsh:2017gnf}.

In short, the bubble dynamics in the low terminal velocity case is as follows:
during bubble expansion the released energy is converted into the plasma bulk motion 
around bubbles within relatively thick regions (compared to (a) runaway case),
while after bubble collisions the fluid motions obey simple wave equations 
and they freely propagate with the speed of sound $c_s$.

Now let us discuss the validity of our assumptions, 
i.e. thin-wall and free propagation with a constant velocity $v$.
For the thin-wall approximation, 
as long as we are interested in frequencies corresponding to length scales
around or larger than the typical bubble size,
we expect that the thin-wall approximation works as a reasonable approximation
because such modes cannot see the width of the energy concentration.\footnote{
One might worry that the thin-wall approximation may fail to describe the present system
after the region of energy localization fills the whole Universe.
(Note that the volume of energy localization continues to increase after bubble collisions.)
See the discussion in the latter part of Sec.~\ref{sec:DiscussionConclusion} on this point.
}
However, for length scales around or smaller than the thickness of energy localization,
our modeling misses an important contribution from superpositions of fluid velocity.\footnote{
In fact, in the sound-shell model~\cite{Hindmarsh:2016lnk}, 
it is the overlap of sound shells that contributes to the continuous GW sourcing.
}
Therefore, our modeling should be regarded as capturing possible infrared structure in the GW spectrum.
Regarding free propagation after collisions, it is justified as long as the fluid obeys the ordinary wave equation.
On the other hand, the assumption of a constant velocity may fail in some cases,
because the velocity in the present case is not unique:
the region of energy concentration expands with a velocity around $\xi \sim v_w$ before collisions,
while it propagates with the speed of sound $c_s$ after collisions.
However, as long as $v_w \simeq c_s \sim \mathcal{O}(0.1)$, 
our modeling is expected to work as a reasonable estimate for the infrared structure
by setting the velocity $v$ to be $c_s$.\footnote{
Note that phase transitions with $v_w \gtrsim \mathcal{O}(0.1)$ is most relevant from the viewpoint of detection,
because otherwise GW production is suppressed.
}
To summarize, our setup will work as a reasonable estimate on the GW spectrum 
for low frequencies (around or lower than the inverse of the typical bubble size at the collision time)
as long as $v_w\sim\mathcal{O}(0.1)$.

%%%%%%%%%%%%%%%%%%%%%%%%%%%%%%%%%%%%%%%%%%%%%%%%%%
\subsubsection*{(c) High terminal velocity}
\label{subsubsec:c}
%%%%%%%%%%%%%%%%%%%%%%%%%%%%%%%%%%%%%%%%%%%%%%%%%%

Finally let us discuss the high terminal velocity case.
Recently, the authors of Ref.~\cite{Bodeker:2017cim} have pointed out that 
particle splitting processes generate a friction term proportional to $\gamma_w$.
We denote by (c) high terminal velocity 
those cases in which such a friction term stops the acceleration of the bubble walls before collisions.
Below we discuss when this is realized instead of (a) runaway.
We also mention the behavior of the energy-momentum tensor.

First let us consider when (c) is realized by estimating the terminal velocity in (a) and (c).
The friction term from the particle splitting processes is given by~\cite{Bodeker:2017cim}
\begin{align}
F_{\rm fric}(v_w)
&\sim 
\gamma_w g_{\rm typ}^2 \Delta{m} T^3,
\end{align}
where $g_{\rm typ}$ denotes a typical value of the coupling of such species to the scalar field,
and $\Delta m$ denotes the mass of some particle species 
gained by the transition from the false to the true vacuum.
This friction term stops the acceleration of the wall when $F_{\rm fric}$ becomes comparable to $\rho_0$, 
which gives
\begin{align}
\gamma_w^{\rm high-terminal}
&\sim 
\frac{\alpha}{g_{\rm typ}^2}
\left( \frac{T}{\Delta m} \right).
\end{align}
On the other hand, if we assume runaway,
the wall continues to be accelerated until collision, 
and $\gamma_w$ becomes 
\begin{align}
\gamma_w^{\rm runaway}
&\sim 
\epsilon \left(\frac{M_P}{\Delta \phi}\right).
\end{align}
The condition for runaway is then written as 
$\gamma_w^{\rm runaway} \lesssim \gamma_w^{\rm high-terminal}$, which gives
\begin{align}
\alpha
&\gtrsim 
\epsilon g_{\rm typ}^2
\left( \frac{\Delta m}{\Delta \phi} \right)
\left(
\frac{M_P}{T}
\right).
\end{align}
Therefore, runaway seems unlikely unless a huge amount of latent heat is released in the transition.

Now let us consider the behavior of the energy-momentum tensor.
Before bubble collisions, the released energy is converted into the plasma bulk motion 
localized around the scalar walls.\footnote{
The situation near the scalar walls may be far from thermal equilibrium because of the production of energetic particles.
However, the fluid description is still expected to be valid
for the length scale relevant to GW production $R_{\rm coll} \sim \epsilon / H$,
which is much larger than the typical length scale of particle scattering.
}
Note that in the present case the energy localization is much thinner than 
(b) low terminal velocity case~\cite{Espinosa:2010hh}.
Therefore, the thin-wall approximation is valid at least until the time of bubble collisions.
However, because of the huge energy release, 
the energy density around the bubble surfaces is much larger than that of background 
$\delta \rho / \rho_{\rm rad} \gg 1$, and the velocity field $u$ is no more nonrelativistic.
Therefore, the fluid dynamics enters the nonlinear regime, 
and full numerical simulations are necessary in order to obtain the behavior of the energy-momentum tensor
during and after bubble collisions.
Such a study is beyond the scope of this paper.

%%%%%%%%%%%%%%%%%%%%%%%%%%%%%%%%%%%%%%%%%%%%%%%%%%
\subsection{Explicit form of the energy-momentum tensor}
\label{subsec:AA}
%%%%%%%%%%%%%%%%%%%%%%%%%%%%%%%%%%%%%%%%%%%%%%%%%%

So far we have discussed the validity of our assumptions 
(thin-wall approximation and free propagation with a constant velocity).
In this subsection we present the explicit form of the energy-momentum tensor we use in the following.
The final expression is Eq.~(\ref{eq:TB}) with $\rho_B$ given by Eq.~(\ref{eq:rhoBBeyond_2}).

%%%%%%%%%%%%%%%%%%%%%%%%%%%%%%%%%%%%%%%%%%%%%%%%%%
\subsubsection*{Thin-wall approximation}
%%%%%%%%%%%%%%%%%%%%%%%%%%%%%%%%%%%%%%%%%%%%%%%%%%

We first give the explicit form of the energy-momentum tensor for uncollided bubble walls
to illustrate the thin-wall approximation.
Here note that ``walls" refer to the energy concentration by the scalar field gradient and/or
the bulk motion of the fluid, as mentioned just before Sec.~\ref{subsec:GWV}.
Let us consider a setup where a single bubble nucleates at a spacetime point $x_n \equiv (t_{xn}, \vec{x}_n)$
and expands with a constant velocity $v$,
and denote the infinitesimal width of bubble walls by $l_B$.
Here the subscript ``$n$" denotes ``nucleation."
In this setup, the $ij$-part of the energy-momentum tensor $T_B$ of this bubble is given by\footnote{
Though in general there are isotropic contributions $\propto \delta_{ij}$ from 
the false vacuum energy, we can neglect them because it does not contribute to GW production.
}
\begin{align}
T_{Bij}(x)
&= \rho_B(x)\widehat{(x - x_n)}_i\widehat{(x - x_n)}_j,
\label{eq:TB}
\end{align}
with $\rho_B$ being $\rho_B^{\rm (uncollided)}$ defined as
\begin{align}
\rho_B^{\rm (uncollided)}(x)
&\equiv
\left\{
\begin{array}{cc}
\displaystyle 
\frac{4\pi}{3} r_B(t_x,t_{xn})^3 \kappa\rho_0
\Big/ 4\pi r_B(t_x,t_{xn})^2 l_B
&
r_B(t_x,t_{xn}) < |\vec{x} - \vec{x}_n| < r'_B(t_x,t_{xn}) \\
0
& 
{\rm otherwise}
\end{array}
\right. ,
\label{eq:rhoBNoColl}
\end{align}
and $r_B, r'_B$ denoting the bubble radius
\begin{align}
r_B(t_x,t_{xn})
&= v(t_x - t_{xn}), 
\;\;\;
r'_B(t_x,t_{xn})
= r_B(t_x,t_{xn}) + l_B.
\end{align}
Here $x$ denotes the spacetime point $x = (t_x,\vec{x})$, 
the Latin indices run over $1, 2, 3$ throughout the paper,
and the hat on the vector $\hat{\bullet}$ indicates the unit vector in $\vec{\bullet}$ direction.
Also, as mentioned before, the efficiency factor $\kappa$ 
is the fraction of the released energy density $\rho_0$ 
localized around the walls\footnote{
This corresponds to the energy of the bulk fluid when the scalar wall reaches a terminal velocity, 
while it corresponds to the energy of the scalar wall itself when the scalar field carries most of the released energy. 
}~\cite{Kamionkowski:1993fg}.
In the runaway case we have $\kappa \simeq 1$, while in the terminal velocity cases it depends on the setup 
(see Eq.~(\ref{eq:kappa}) and Refs.~\cite{Espinosa:2010hh,Leitao:2015ola}).

%%%%%%%%%%%%%%%%%%%%%%%%%%%%%%%%%%%%%%%%%%%%%%%%%%
\subsubsection*{Free propagation with arbitrary damping}
%%%%%%%%%%%%%%%%%%%%%%%%%%%%%%%%%%%%%%%%%%%%%%%%%%

Now we give the explicit form of the energy-momentum tensor after collisions.
Let us consider a bubble wall fragment which experiences the first collision at $x_i = (t_{xi}, \vec{x}_i)$.
In the following we often call this first collision ``interception," and label it by the subscript ``$i$."
Note that the interception point differs among each fragment.
Assuming free propagation of the walls after the first collisions,
we write $\rho_B(x)$ for this particular fragment after the first collision as
\begin{align}
\rho_B^{({\rm collided})}(x)
&=
\displaystyle 
\left[
\frac{4\pi}{3} r_B(t_{xi},t_{xn})^3 \kappa\rho_0
\Big/ 4\pi r_B(t_{xi},t_{xn})^2 l_B
\right]
\times 
\frac{r_B(t_{xi},t_{xn})^2}{r_B(t_x,t_{xn})^2}
\times 
D(t_x,t_{xi})
\label{eq:rhoBBeyond_2}
\\
&=
\rho_B^{({\rm uncollided})}(x)
\times 
\frac{r_B(t_{xi} ,t_{xn})^3}{r_B(t_x ,t_{xn})^3}
\times
D(t_x,t_{xi}),
\label{eq:rhoBBeyond_3}
\end{align}
for $r_B(t_x,t_{xn}) < |\vec{x} - \vec{x}_n| < r'_B(t_x,t_{xn})$,
while it vanishes otherwise.
Note that we take only the first collisions into account 
and neglect the effect of subsequent collisions on the energy-momentum tensor. 
The second factor in the R.H.S. of Eq.~(\ref{eq:rhoBBeyond_2}) takes into account
the increase in the wall area and the total energy conservation.
Here we have introduced a ``damping function" $D$, which satisfies $D(t_x, t_{xi} = t_x) = 1$.
This function accounts for how the collided walls lose their energy and momentum densities from $t_{xi}$ to $t_x$
in addition to the second factor in Eq.~(\ref{eq:rhoBBeyond_2}).
For free propagation we have $D = 1$ for an arbitrary combination of $t_x$ and $t_{xi}$.
Though our analytic expressions for the GW spectrum are applicable to any form of $D$,\footnote{
The dumping function $D$ depends on the underlying particle model 
which describes the strength of the coupling of the scalar field with light particles.
As discussed in~Sec.~\ref{subsec:GWV}, in some cases we expect $D\simeq 1$ during one Hubble time.
Though it may also depend on the nucleation time $t_n$,
our final expressions for the GW spectrum can be applied to such cases as well.
}
we adopt the following form for practical calculations:
\begin{align}
D(t,t_i)
&= e^{-(t - t_i)/\tau}.
\label{eq:D}
\end{align}
Here $\tau$ denotes a typical damping timescale of the walls,
which generally depends on the underlying particle model.
The instant damping $\tau = 0$ corresponds to the envelope approximation (see Fig.~\ref{fig:Envelope}), 
while $\tau = \infty$ corresponds to free propagation, i.e. no damping.
The introduction of the damping function makes it possible to clarify
when GWs are sourced for each wavenumber, as we see in Sec.~\ref{sec:Numerical}.

Note that Eq.~(\ref{eq:rhoBBeyond_2}) reduces to Eq.~(\ref{eq:rhoBNoColl}) for $t_{xi} = t_x$.
Therefore, Eq.~(\ref{eq:TB}) with $\rho_B$ given by Eq.~(\ref{eq:rhoBBeyond_2}) 
(for uncollided walls we take $t_{xi} = t_x$) is our assumption for the energy-momentum tensor.

%%%%%%%%%%%%%%%%%%%%%%%%%%%%%%%%%%%%%%%%%%%%%%%%%%
\subsection{Nucleation rate of bubbles}
\label{subsec:rate}
%%%%%%%%%%%%%%%%%%%%%%%%%%%%%%%%%%%%%%%%%%%%%%%%%%

As mentioned in the beginning of this section, we need 
\begin{itemize}
\item[(1)]
Spacetime distribution of bubbles (i.e., nucleation rate of bubbles),
\end{itemize}
in order to estimate the GW spectrum.
In this paper we assume the following form for the bubble nucleation rate per unit time and volume:
\begin{align}
\Gamma(t)
&= \Gamma_*e^{\beta(t - t_*)},
\label{eq:Gamma}
\end{align}
where $t_*$ denotes some typical time for bubble nucleation,
$\Gamma_*$ indicates the nucleation rate at $t = t_*$, 
and $\beta$ is assumed to be constant.
The typical nucleation time $t_*$ is calculated from the condition $H_*^4 \Gamma_* \sim 1$
with $H_*$ being the Hubble parameter at $t = t_*$,
and this is equivalent to specifying the temperature $T_*$ just before the nucleation time.
The expression (\ref{eq:Gamma}) applies to thermal phase transitions,\footnote{
Those cases in which this expression does not hold have recently been 
studied by the authors of Ref.~\cite{Megevand:2016lpr}.
}
and the parameter $\beta$ is calculated with the instanton method
from the underlying particle model~\cite{Linde:1977mm,Linde:1981zj}.
The inverse of $\beta$ gives a typical timescale from nucleation to collision,
and the typical bubble size when bubbles collide
(or the typical distance between two neighboring bubbles)
is given by $\sim v/\beta$ correspondingly.\footnote{
This is understood as follows.
The nucleation rate grows with a typical timescale $\sim 1/\beta$.
This means that, after bubbles start to fill the Universe, 
they can expand only for a timescale $\sim 1/\beta$ before they collide with others.
Therefore this gives the typical timescale from nucleation to collision,
while the typical bubble size at collisions is given by $\sim v/\beta$.
}

Before moving on, we comment on the cosmic expansion.
In this paper we neglect the cosmic expansion during the phase transition,
because the transition typically completes 
in a short period compared to the Hubble time:
$\beta/H_* \sim \mathcal{O}(10^{1-5})\gg1$~\cite{Kamionkowski:1993fg}.

%%%%%%%%%%%%%%%%%%%%%%%%%%%%%%%%%%%%%%%%%%%%%%%%%%
\subsection{GW power spectrum from the energy-momentum tensor}
\label{subsec:gwfrom}
%%%%%%%%%%%%%%%%%%%%%%%%%%%%%%%%%%%%%%%%%%%%%%%%%%

In this subsection we summarize a formalism to estimate the GW spectrum
from the energy-momentum tensor of the system when the produced GWs are stochastic.
The point is that the GW spectrum is determined by the two-point correlator of the energy-momentum tensor, 
which we symbolically denote by $\left< T(x)T(y) \right>$.
Here the angular bracket denotes an ensemble average:
the nucleation rate gives the probability of bubble nucleation,
and in this respect the energy-momentum tensor can be regarded as a stochastic variable.
This subsection closely follows Ref.~\cite{Caprini:2007xq}.

%%%%%%%%%%%%%%%%%%%%%%%%%%%%%%%%%%%%%%%%%%%%%%%%%%
\subsubsection{GW power spectrum at the transition time}
\label{subsec:GWAtTransition}
%%%%%%%%%%%%%%%%%%%%%%%%%%%%%%%%%%%%%%%%%%%%%%%%%%

%%%%%%%%%%%%%%%%%%%%%%%%%%%%%%%%%%%%%%%%%%%%%%%%%%
\subsubsection*{Equation of motion and its solution}
%%%%%%%%%%%%%%%%%%%%%%%%%%%%%%%%%%%%%%%%%%%%%%%%%%

As mentioned in Sec.~\ref{subsec:rate}, we neglect the effect of cosmic expansion during the transition.
With this assumption the metric is well described by the Minkowski background with tensor perturbations:
\begin{align}
ds^2
&= - dt^2 + (\delta_{ij}+2h_{ij})dx^idx^j.
\end{align}
The tensor perturbations, satisfying the transverse and traceless conditions $h_{ii} = \partial_i h_{ij} = 0$, 
obey the following evolution equation
\begin{align}
\ddot{h}_{ij}(t,\vec{k})+k^2h_{ij} (t,\vec{k}) 
&= {8\pi G } \Pi_{ij}(t,\vec{k}).
\label{eq:hEOM}
\end{align}
Here the dot denotes the time derivative,
and $\bullet (t, \vec{k})$ indicates the Fourier mode of the corresponding quantity
with $\vec{k}$ being the wave vector.\footnote{
The convention for Fourier transformation is taken to be 
$\int d^3x \; e^{i\vec{k}\cdot\vec{x}}$ and $\int d^3k/(2\pi)^3 \; e^{-i\vec{k}\cdot\vec{x}}$.
}
The source term $\Pi_{ij}$  denotes the projected energy-momentum tensor:
\begin{align}
\Pi_{ij}(t,\vec{k})
&= K_{ijkl}(\hat{k}) T_{kl}(t,\vec{k}),
\label{eq:PiKT}
\\
K_{ijkl}(\hat{k})
&\equiv P_{ik}(\hat{k})P_{jl}(\hat{k}) - \frac{1}{2}P_{ij}(\hat{k})P_{kl}(\hat{k}), 
\;\;\;\;
P_{ij}(\hat{k})
\equiv \delta_{ij}-\hat{k}_i\hat{k}_j.
\label{eq:K}
\end{align}
We assume that the source is switched on from $t_{\rm start}$ to $t_{\rm end}$,
which we take $t_{\rm start/end} \rightarrow \mp \infty$ in the following calculation.

Eq.~(\ref{eq:hEOM}) is formally solved by using the Green function $G_k$ 
satisfying $G_k(t,t) = 0$ and $\partial G_k(t,t')/\partial t |_{t = t'} = 1$ as
\begin{align}
h_{ij}(t,\vec{k})
&= 8\pi G
\int_{t_{\rm start}}^t dt' \;
G_k(t,t') \Pi_{ij}(t',\vec{k}),
\;\;\;\;\;\;
t < t_{\rm end},
\label{eq:hsol_G}
\end{align}
where $G_k(t,t') = \sin (k(t - t'))/k$.
Matching conditions at $t = t_{\rm end}$ give 
\begin{align}
h_{ij}(t,\vec{k})
&= A_{ij}(\vec{k}) \sin(k(t - t_{\rm end})) + B_{ij}(\vec{k}) \cos(k(t - t_{\rm end})),
\label{eq:hsol}
\end{align}
for $t > t_{\rm end}$. 
Here the coefficients are given by
\begin{align}
A_{ij}(\vec{k})
&= \frac{8\pi G}{k}\int_{t_{\rm start}}^{t_{\rm end}}dt \;
\cos(k(t_{\rm end} - t)) \Pi_{ij}(t,\vec{k}), \\
B_{ij}(\vec{k})
&= \frac{8\pi G}{k}\int_{t_{\rm start}}^{t_{\rm end}}dt \;
\sin(k(t_{\rm end} - t)) \Pi_{ij}(t,\vec{k}).
\end{align}
%%

%%%%%%%%%%%%%%%%%%%%%%%%%%%%%%%%%%%%%%%%%%%%%%%%%%
\subsubsection*{Power spectrum}
%%%%%%%%%%%%%%%%%%%%%%%%%%%%%%%%%%%%%%%%%%%%%%%%%%

Now we give the expression for the GW spectrum using the formal solution (\ref{eq:hsol}).
The energy density of GWs is given by
\begin{align}
\rho_{\rm GW}(t)
&= \frac{\langle \dot{h}_{ij}(t,\vec{x})\dot{h}_{ij}(t,\vec{x}) \rangle}{8\pi G},
\end{align}
where the angular bracket denotes taking an oscillation average for several oscillation periods
and also an ensemble average.
The latter procedure is justified because of the stochasticity of GWs produced in phase transitions.\footnote{
The effect of cosmic variance is extremely suppressed 
because we have a huge number of samples at the time of observations.
}
The energy density of GWs per each logarithmic wavenumber
normalized by the total energy density of the universe is given by
\begin{align}
\Omega_{\rm GW}(t,k) 
&\equiv \frac{1}{\rho_{\rm tot}(t)} \frac{d\rho_{\rm GW}}{d\ln k}(t,k)
=
\frac{k^3}{16\pi^3 G}P_{\dot{h}}(t,k).
\label{eq:OmegaGW}
\end{align}
Here we have defined the power spectrum $P_{\dot{h}}$ as
\begin{align}
\langle \dot{h}_{ij}(t,\vec{k}) \dot{h}_{ij}^*(t,\vec{q}) \rangle
&= (2\pi)^3 \delta^{(3)}(\vec{k} - \vec{q}) P_{\dot{h}}(t,k).
\label{eq:dothdoth}
\end{align}
The power spectrum $P_{\dot{h}}$ is related to the source term $\Pi_{ij}$ in the following way. 
First we define the unequal-time correlator $\Pi$ of the source term as
\begin{align}
\langle \Pi_{ij}(t_x,\vec{k})\Pi^*_{ij}(t_y,\vec{q}) \rangle
&= (2\pi)^3 \delta^{(3)}(\vec{k} - \vec{q})\Pi(t_x,t_y,k).
\label{eq:PiPi}
\end{align}
This correlator has the following relation to the original energy-momentum tensor
\begin{align}
\Pi(t_x,t_y,k) 
&= K_{ijkl}(\hat{k})K_{ijmn}(\hat{k})
\int d^3r \; e^{i \vec{k} \cdot \vec{r}} \langle T_{kl} T_{mn} \rangle (t_x,t_y,\vec{r}),
\label{eq:Pi}
\end{align}
where 
\begin{align}
\langle T_{ij} T_{kl} \rangle (t_x,t_y,\vec{r})
&\equiv \langle T_{ij}(t_x,\vec{x}) T_{kl}(t_y,\vec{y}) \rangle,
\end{align}
with $\vec{r} \equiv \vec{x} - \vec{y}$.
This correlator depends on $\vec{x}$ and $\vec{y}$ only through the combination $\vec{r}$
because of the spacial homogeneity of the system.
Then, since $\dot{h}$ is related to the source term through Eq.~(\ref{eq:hsol}) for $t > t_{\rm end}$,
we obtain the following relation by using Eqs.~(\ref{eq:hsol}), (\ref{eq:dothdoth}) and (\ref{eq:PiPi}): 
\begin{align}
P_{\dot{h}}(t,k) 
&= 32\pi^2G^2
\int_{t_{\rm start}}^{t_{\rm end}} dt_x
\int_{t_{\rm start}}^{t_{\rm end}} dt_y \;
\cos(k(t_x - t_y))\Pi (t_x,t_y,k),
\;\;\;\;
t > t_{\rm end}.
\label{eq:Pdoth}
\end{align}
Though we put the argument $t$ in the L.H.S., 
the R.H.S. has no dependence on it
because the source term is switched off for $t > t_{\rm end}$ and 
because there is no dilution of GWs by the cosmic expansion in the present system.
Substituting Eq.~(\ref{eq:Pdoth}) into Eq.~(\ref{eq:OmegaGW}), one finds
\begin{align}
\Omega_{\rm GW} (t,k)
&= \frac{2Gk^3}{\pi \rho_{\rm tot}}
\int_{t_{\rm start}}^{t_{\rm end}} dt_x
\int_{t_{\rm start}}^{t_{\rm end}} dt_y \;
\cos(k(t_x - t_y))\Pi (t_x,t_y,k),
\;\;\;\;
t > t_{\rm end}.
\label{eq:OmegaPi}
\end{align}
Eq.~(\ref{eq:OmegaPi}) means that the GW spectrum is obtained straightforwardly 
once one finds expressions for $\Pi(t_x,t_y,k)$,
or equivalently the two-point correlator of the energy-momentum tensor $\langle T(x)T(y) \rangle$
with $x \equiv (t_x,\vec{x})$ and $y \equiv (t_y,\vec{y})$.

Here we rewrite the expression for the GW spectrum for later convenience.
As mentioned in Eq.~(\ref{eq:alpha}), 
we define the ratio of the released energy density 
to the background radiation energy density $\rho_{\rm rad}$ just before the transition:
\begin{align}
\alpha
&\equiv \frac{\rho_0}{\rho_{\rm rad}},
\;\;\;\;\;\;
\rho_{\rm tot}
= \rho_0 + \rho_{\rm rad},
\end{align}
Then we may factor out some parameter dependences from the GW spectrum:
\begin{align}
\Omega_{\rm GW}(t,k)
&\equiv
\kappa^2 \left(\frac{H_*}{\beta}\right)^2\left(\frac{\alpha}{1+\alpha}\right)^2
\Delta(k/\beta),
\label{eq:OmegaDelta}
\end{align}
where $\Delta$ is given by
\begin{align}
\Delta(k/\beta)
&=
\frac{3}{8\pi G}\frac{\beta^2 \rho_{\rm tot}}{\kappa^2 \rho_0^2}\Omega_{\rm GW}(t,k) \nonumber \\
&= 
\frac{3}{4\pi^2}\frac{\beta^2k^3}{\kappa^2\rho_0^2}
\int_{t_{\rm start}}^{t_{\rm end}} dt_x
\int_{t_{\rm start}}^{t_{\rm end}} dt_y \;
\cos(k(t_x - t_y))\Pi (t_x,t_y,k).
\label{eq:DeltaPi}
\end{align}
In deriving Eq.~(\ref{eq:DeltaPi}) we have used the Friedmann equation $H_*^2 = (8\pi G / 3)\rho_{\rm tot}$.
The definition (\ref{eq:OmegaDelta}) factors out $G$, $\kappa$, $\rho_0$ and $\rho_{\rm tot}$ dependences.
The dimensionless GW spectrum $\Delta$ depends on dimensionless quantities
such as $v$, $\tau$ and $k/\beta$,
though we have kept only $k/\beta$ dependence explicitly in Eqs.~(\ref{eq:OmegaDelta})--(\ref{eq:DeltaPi}).

%%%%%%%%%%%%%%%%%%%%%%%%%%%%%%%%%%%%%%%%%%%%%%%%%%
\subsubsection{GW power spectrum at present}
\label{subsec:GWAtPresent}
%%%%%%%%%%%%%%%%%%%%%%%%%%%%%%%%%%%%%%%%%%%%%%%%%%

Gravitational waves are redshifted after production until the present time.
The relation between the scale factor just after the transition $a_*$ 
and the one at present $a_0$ is given by
\begin{align}
\frac{a_*}{a_0}
&= 8.0\times10^{-16}
\left( \frac{g_*}{100} \right)^{-1} \left(\frac{T_*}{100~{\rm GeV}}\right)^{-1},
\end{align}
where $g_*$ is the total number of relativistic degrees of freedom in the thermal bath at temperature $T_*$.
The present frequency is obtained by redshifting:
\begin{align}
f_0
&=
\left( \frac{a_*}{a_0} \right) f_*
= 
1.65 \times 10^{-5} {\rm Hz}
\left( \frac{f_*}{\beta} \right) \left( \frac{\beta}{H_*} \right)
\left( \frac{T_*}{100~{\rm GeV}} \right)
\left( \frac{g_*}{100} \right)^{\frac{1}{6}}.
\label{eq:EnvCalSPresent}
\end{align}
The present GW amplitude is obtained by 
taking into account that GWs are non-interacting radiation:
\begin{align}
\left.
\Omega_{\rm GW}h^2 
\right|_{t = t_0}
&=1.67\times 10^{-5} \left( \frac{g_*}{100}\right)^{-\frac{1}{3}}
\left.
\Omega_{\rm GW}h^2 
\right|_{t = t_{\rm end}}\nonumber \\
&=1.67\times 10^{-5}\kappa^2 \Delta \left( \frac{\beta}{H_*} \right)^{-2}
\left( \frac{\alpha}{1 + \alpha} \right)^2
\left( \frac{g_*}{100} \right)^{-\frac{1}{3}}.
\label{eq:OmegaPresent}
\end{align}
Note that in Eq.~(\ref{eq:OmegaPresent}) the argument in $\Delta$ is redshifted:
the present frequency $f_0$ in the argument of $\Omega_{\rm GW}$ is related to 
the argument $k/\beta$ in $\Delta$ by $f_0 = (a_*/a_0) f_* = (a_*/a_0) (k/2\pi)$.\footnote{
In the present paper we regard $k$ as the physical wavenumber,
not the comoving one.
}
Also note that the shape of the GW spectrum is encoded in $\Delta$, 
which is obtained from the two-point correlator of the energy-momentum tensor $\left< T(x)T(y) \right>$
by using Eq.~(\ref{eq:DeltaPi}).

%%%%%%%%%%%%%%%%%%%%%%%%%%%%%%%%%%%%%%%%%%%%%%%%%%
\subsection{Summary of the setup and its implications}
\label{subsec:SetupSum}
%%%%%%%%%%%%%%%%%%%%%%%%%%%%%%%%%%%%%%%%%%%%%%%%%%

Let us summarize this section.
In order to estimate the GW spectrum from bubble dynamics in a cosmological first-order phase transition,
we need to specify the behavior of the energy-momentum tensor of the system.
More specifically, we need to know the three ingredients (1)--(3) listed in the beginning of this section.
In this paper we approximate the system 
with the thin-wall approximation and free propagation after first collisions 
with a constant velocity (Eqs.~(\ref{eq:TB}) and (\ref{eq:rhoBBeyond_2})) for (2) and (3), 
while we assume an exponential form for the nucleation rate (Eq.~(\ref{eq:Gamma})) for (1).
With these ingredients, the GW spectrum is obtained from the two-point correlator 
of the energy-momentum tensor by using the stochastic property of GWs produced 
in first-order phase transitions (Eq.~(\ref{eq:DeltaPi})).

The resulting GW spectrum is determined by six parameters:
\begin{align}
&\alpha, 
\;\;
\beta,
\;\;
T_*, 
\;\;
\kappa, 
\;\;
\tau, 
\;\;
v.
\end{align} 
The first three parameters $\alpha$ (Eq.~(\ref{eq:alpha})), 
$\beta$ (Eq.~(\ref{eq:Gamma}))
and $T_*$ (below Eq.~(\ref{eq:Gamma}))
can be estimated by thermal field theory.
The efficiency factor $\kappa$ (Eqs.~(\ref{eq:rhoBNoColl}) or (\ref{eq:rhoBBeyond_2})),
which parameterizes the fraction of the released energy localized around the bubble walls,
is determined by the energy-momentum profile of a single bubble.
For the runaway case with sufficiently large $\alpha$ we expect $\kappa \simeq 1$,
while for the terminal velocity case it depends on the setup.
The typical damping timescale $\tau$ of collided walls (Eq.~(\ref{eq:D})) 
depends on the underlying model and is expected to be much larger than the duration time of the phase transition 
$\tau \gg 1/\beta$ in most cases of our interest.

Regarding the validity of our modeling,
the GW spectrum for the runaway case can be estimated by setting $v = 1$ 
and the result is expected to be rather precise
as long as nontrivial scalar field trapping does not occur, as discussed in Sec.~\ref{subsec:GWV} (a).
For the low terminal velocity case, 
we expect that our modeling captures the low-frequency structure (frequencies around or lower than the peak)
of the GW spectrum by setting $v = c_s$, as discussed in Sec.~\ref{subsec:GWV} (b).
However, note that our modeling does not take into account the thickness of the energy localizations
and hence the effect of their overlapping.
Also, if $v_w$ is much different from the speed of sound, e.g. $v_w\ll c_s$,
the approximation $v = c_s$ may fail to capture the system.
One possible remedy for this can be as follows.
As we see in Sec.~\ref{sec:Numerical}, 
GW production is dominated by the dynamics after collisions.
Then, the main role of the bubble dynamics before collisions can be regarded as determining the typical bubble size, 
which gives one of the initial conditions for GW production at late times.\footnote{
For $v_w \ll c_s$, the bubble shapes after collisions deviate from spherical ones.
This can also affect the GW spectrum to some extent.
}
The typical bubble size in our setup ($v = c_s$) 
is adjusted to the one realized in such a system ($v_w \ll c_s$)
by the following replacement:
\begin{align}
\beta
&\rightarrow 
\beta_{\rm eff} 
\equiv \frac{c_s}{v_w}\beta.
\end{align}
Therefore, we expect that setting $v = c_s$ and replacing $\beta$ with $\beta_{\rm eff}$ 
give at least an order-of-magnitude estimate for the GW spectrum in the low terminal velocity case.

%%%%%%%%%%%%%%%%%%%%%%%%%%%%%%%%%%%%%%%%%%%%%%%%%%
\section{Analytic expressions}
\label{sec:Analytic}
\setcounter{equation}{0}
%%%%%%%%%%%%%%%%%%%%%%%%%%%%%%%%%%%%%%%%%%%%%%%%%%

Now that we have defined the properties of the energy-momentum tensor of the system 
in Sec.~\ref{subsec:AA} and \ref{subsec:rate},
we can calculate $\left< T(x)T(y) \right>$ and 
relate it with the GW spectrum by using the method explained in Sec.~\ref{subsec:gwfrom}.
In this section we summarize the basic strategy to calculate it and 
present the resulting analytic expressions for the GW spectrum.
The derivation is explained in detail in Appendix~\ref{app:Envelope}--\ref{app:Simple}.

%%%%%%%%%%%%%%%%%%%%%%%%%%%%%%%%%%%%%%%%%%%%%%%%%%
\subsection{Basic strategy}
%%%%%%%%%%%%%%%%%%%%%%%%%%%%%%%%%%%%%%%%%%%%%%%%%%

First we summarize the basic strategy to calculate 
the correlator of the energy-momentum tensor and the resulting GW spectrum.
From Eq.~(\ref{eq:DeltaPi}) we see that the spectrum is essentially the unequal-time correlator
of the energy-momentum tensor $\Pi(t_x,t_y,k)$, which is the Fourier transform of 
$\Pi(t_x,t_y,\vec{r}) \sim \left< T(t_x,\vec{x}) T(t_y,\vec{y}) \right>$ 
with $T$ symbolically denoting the energy-momentum tensor.
Calculating $\left< T(t_x,\vec{x}) T(t_y,\vec{y}) \right>$ means
\begin{itemize}
\item
Fix the spacetime points $x = (t_x, \vec{x})$ and $y = (t_y, \vec{y})$.
\item
Find those bubble configurations that give nonzero $T(x)T(y)$, 
estimate the probability for such configurations to occur,
and calculate the value of $T(x)T(y)$ in each case.
\item
Sum over all such configurations.
\end{itemize}
As detailed in Appendix~\ref{app:Envelope},
we may classify the bubble configurations depending on whether the energy-momentum tensor
at $(t_x,\vec{x})$ and $(t_y,\vec{y})$ comes from the same bubble or different bubbles
(in other words, the same nucleation point or different nucleation points).
We refer to each contribution as
\begin{itemize}
\item
Single-bubble,
\item
Double-bubble.
\end{itemize}
Therefore, the resulting GW spectrum $\Delta$ becomes the sum of these two contributions:
\begin{align}
\Delta
&= 
\Delta^{(s)} + \Delta^{(d)},
\end{align}
where the superscripts denote ``single" and ``double," respectively.\footnote{
The spherical symmetry of a single bubble does not mean that
the single-bubble contribution vanishes.
See Appendix~\ref{app:saru} on this point.
}

%%%%%%%%%%%%%%%%%%%%%%%%%%%%%%%%%%%%%%%%%%%%%%%%%%
\subsection{Analytic expressions}
%%%%%%%%%%%%%%%%%%%%%%%%%%%%%%%%%%%%%%%%%%%%%%%%%%

Now we present the analytic expressions for the GW spectrum.
After a short calculation (see Appendix~\ref{app:Envelope}--\ref{app:Simple}),
we obtain the single-bubble spectrum
\begin{align}
\Delta^{(s)}
= 
&
\int_{-\infty}^\infty dt_x
\int_{-\infty}^\infty dt_y
\int_{v|t_{x,y}|}^\infty dr 
\int_{-\infty}^{t_{\rm max}} dt_n 
\int_{t_n}^{t_x} dt_{xi}
\int_{t_n}^{t_y} dt_{yi}
\nonumber \\[1ex]
\frac{k^3}{3}
&\left[
\begin{matrix*}[l]
\displaystyle
\;
e^{-I(x_i,y_i)} \;
\Gamma(t_n) \;
\frac{r}{r_{xn}^{(s)}r_{yn}^{(s)}}
\\
\displaystyle
\times 
\left[
j_0(kr){\mathcal K}_0(n_{xn\times},n_{yn\times})
+ \frac{j_1(kr)}{kr}{\mathcal K}_1(n_{xn\times},n_{yn\times})
+ \frac{j_2(kr)}{(kr)^2}{\mathcal K}_2(n_{xn\times},n_{yn\times})
\right]
\\[2.5ex]
\displaystyle
\times \;
\partial_{txi}
\left[
r_B(t_{xi},t_n)^3
D(t_x,t_{xi})
\right]
\partial_{tyi}
\left[
r_B(t_{yi},t_n)^3
D(t_y,t_{yi})
\right]
\cos(kt_{x,y})
\end{matrix*}
\right],
\label{eq:MainDeltaS}
\end{align}
and the double-bubble spectrum
\begin{align}
\Delta^{(d)}
=
&
\int_{-\infty}^\infty dt_x 
\int_{-\infty}^\infty dt_y 
\nonumber \\
&
\int_0^\infty dr 
\int_{-\infty}^{t_x} dt_{xn} 
\int_{-\infty}^{t_y} dt_{yn} 
\int_{t_{xn}}^{t_x} dt_{xi} 
\int_{t_{yn}}^{t_y} dt_{yi} 
\int_{-1}^1 dc_{xn}
\int_{-1}^1 dc_{yn}
\int_0^{2\pi} d\phi_{xn,yn}
\nonumber \\[1ex]
\frac{k^3}{3}
&\left[
\begin{matrix*}[l]
\displaystyle
\;
\Theta_{\rm sp}(x_i,y_n)
\Theta_{\rm sp}(x_n,y_i)
e^{-I(x_i,y_i)} 
\Gamma(t_{xn})\Gamma(t_{yn})
\\[1ex]
\displaystyle
\times \;
r^2
\left[
j_0(kr){\mathcal K}_0(n_{xn},n_{yn})
+ \frac{j_1(kr)}{kr}{\mathcal K}_1(n_{xn},n_{yn})
+ \frac{j_2(kr)}{(kr)^2}{\mathcal K}_2(n_{xn},n_{yn})
\right]
\\[2ex]
\displaystyle
\times \;
\partial_{txi}
\left[
r_B(t_{xi},t_{xn})^3
D(t_x,t_{xi})
\right]
\partial_{tyi}
\left[
r_B(t_{yi},t_{yn})^3
D(t_y,t_{yi})
\right]
\cos(kt_{x,y})
\end{matrix*}
\right].
\label{eq:MainDeltaD}
\end{align}
Here $j_n$ are the spherical Bessel functions, and 
${\mathcal K}_n$ functions, expressed by elementary functions, 
are defined in Appendix~\ref{app:useful}.
See Appendix~\ref{app:Envelope}--\ref{app:Simple} for the definition of 
the other quantities.
These expressions apply to a general damping function $D$ and a general nucleation rate $\Gamma$.\footnote{
In Appendix~\ref{app:Envelope}--\ref{app:Simple}
we keep the nucleation-time dependence of $D$
in order to make the discussion as general as possible.
This dependence is omitted in Eqs.~(\ref{eq:MainDeltaS})--(\ref{eq:MainDeltaD}).
}
In addition, though we have derived these expressions by assuming 
Eqs.~(\ref{eq:TB}) and (\ref{eq:rhoBBeyond_2}) for the energy-momentum tensor, 
the derivation is basically the same for other forms.

Given the specific forms for the damping function (\ref{eq:D}) and the nucleation rate (\ref{eq:Gamma}),
we can simplify the expressions and obtain Eq.~(\ref{eq:BeyondDeltaSConcFinal})
for the single-bubble and Eq.~(\ref{eq:BeyondDeltaDConcConc}) for the double-bubble,
which have now been reduced to five- and nine-dimensional integrations, respectively.
For the double-bubble spectrum,
we further arrange the expression using a technique detailed in Appendix~\ref{app:tech}
to obtain the seven-dimensional integration (\ref{eq:AppBeyondDeltaDConcConcFinalSubtracted}).

In Sec.~\ref{sec:Numerical} we evaluate the spectrum numerically.
However, it is possible to show directly that the spectrum behaves $\propto k$ for small $k$
in the long-lasting limit $\tau \to \infty$. 
See the discussion in the latter part of Sec.~\ref{sec:DiscussionConclusion} and 
Appendix~\ref{app:Analytic} on this point.

%%%%%%%%%%%%%%%%%%%%%%%%%%%%%%%%%%%%%%%%%%%%%%%%%%
\section{Numerical results}
\label{sec:Numerical}
\setcounter{equation}{0}
%%%%%%%%%%%%%%%%%%%%%%%%%%%%%%%%%%%%%%%%%%%%%%%%%%

In this section we show the results for numerical evaluation of the spectrum
(\ref{eq:MainDeltaS}) and (\ref{eq:MainDeltaD}).
We show the total spectrum $\Delta = \Delta^{(s)} + \Delta^{(d)}$,
changing the duration time of the collided walls $\tau$.
This will tell us how GWs are sourced in time by the bubble dynamics after collisions.

For practical evaluations we use 
(\ref{eq:BeyondDeltaSConcFinal}) and (\ref{eq:AppBeyondDeltaDConcConcFinalSubtracted})
for the single- and double-bubble spectrum, respectively.
In this section we show the total spectrum $\Delta = \Delta^{(s)} + \Delta^{(d)}$ only,
and the numerical results for each contribution are shown in Appendix~\ref{app:Other}.
Also, all the dimensionful quantities are normalized by $\beta$ in the following.
In evaluating the spectrum we use a multi-dimensional integration algorithm VEGAS 
in the CUBA library~\cite{Hahn:2004fe},
and cut the calculation at a relative error of $5\%$.

In the present paper we report only for $k \lesssim 1$.
This is because for larger $k$
the spectrum becomes highly oscillatory and numerical difficulties arise.
We leave numerical evaluations for higher wavenumbers as future work.

%%%%%%%%%%%%%%%%%%%%%%%%%%%%%%%%%%%%%%%%%%%%%%%%%%
\subsection{Spectral shape}
%%%%%%%%%%%%%%%%%%%%%%%%%%%%%%%%%%%%%%%%%%%%%%%%%%

\begin{figure}[h]
\begin{center}
\includegraphics[width=0.7\columnwidth]{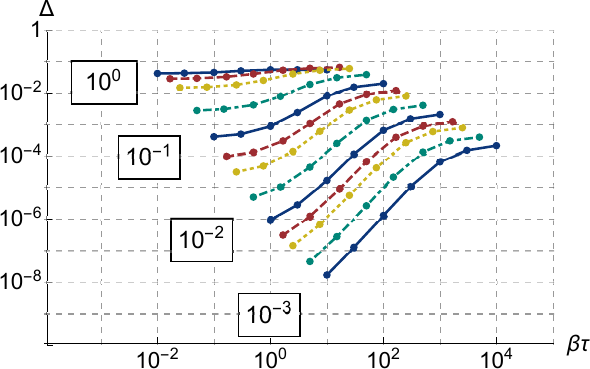}
\caption{\small
The total spectrum $\Delta$ as a function of the duration time $\tau$ for $v = 1$.
The blue, red, yellow and green lines correspond to 
$k = (1, 0.6, 0.4, 0.2) \times 10^{-n}$ ($n \in \mathbb{Z}$), respectively.
}
\label{fig:tauDeltaT_v=1}
\end{center}
\end{figure}

\begin{figure}[h]
\begin{center}
\includegraphics[width=0.7\columnwidth]{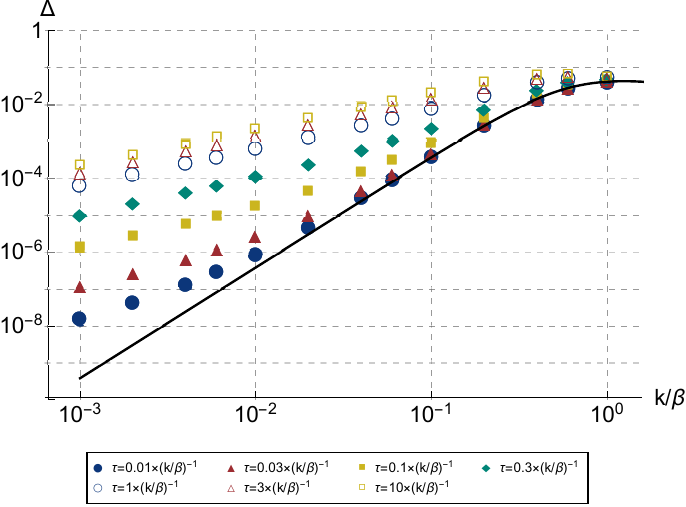}
\caption{\small
The total spectrum $\Delta$ as a function of wavenumber $k$ for $v = 1$.
Each data point corresponds to $\tau = (0.01, 0.03, 0.1, 0.3, 1, 3, 10) \times (1/k)$,
while the black line shows the spectrum with the envelope approximation
obtained in Ref.~\cite{Jinno:2016vai}.
}
\label{fig:kDeltaT_v=1}
\end{center}
\end{figure}

\begin{figure}[h]
\begin{center}
\includegraphics[width=0.7\columnwidth]{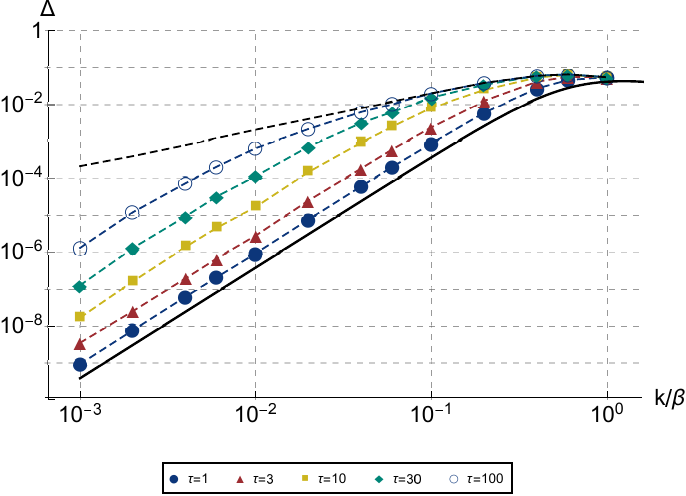}
\caption{\small
The total spectrum $\Delta$ as a function of wavenumber $k$ for $v = 1$.
Each colored line corresponds to $\tau = 1, 3, 10, 30, 100$ from bottom to top,
while the black-dashed line corresponds to the data points for $\tau = 10 \times (1/k)$.
The black-solid line is the same as Fig.~\ref{fig:kDeltaT_v=1}.
}
\label{fig:kDeltaTSlice_v=1}
\end{center}
\end{figure}

First, we show the spectrum for $v = 1$ in Figs.~\ref{fig:tauDeltaT_v=1}--\ref{fig:kDeltaTSlice_v=1}.
Fig.~\ref{fig:tauDeltaT_v=1} shows the GW spectrum as a function of the duration time $\tau$
for various wavenumbers from $k = 0.001$ to $1$.
The blue, red, yellow and green lines denote 
$k = (1,0.6,0.4,0.2) \times 10^{-n}$ $(n \in \mathbb{Z})$, respectively,
and these wavenumbers are comparable or smaller than 
the inverse of the typical bubble size around the time of collisions (which corresponds to $k \sim 1$).
The sampling points for the duration time are 
$\tau = (0.01,0.03,0.1,0.3,1,3,10) \times (1/k)$ for each wavenumber.
The $\tau$ dependence of the spectrum can be interpreted as denoting 
the typical sourcing time:
for example, if the spectrum grows around $\tau \sim 10$ for some wavenumber, 
it means that in the long-lasting limit ($\tau \to \infty$) the growth typically occurs 
around time $10$ after the typical collision time.
Important features in the spectrum are
\begin{itemize}
\item
For a wide range of wavenumbers 
(smaller than the inverse of the typical bubble size around the time of collisions),
the spectrum grows significantly as $\tau$ increases. 
\item
The growth occurs at $\tau \sim 1/vk$, and it stops after that.
\end{itemize}
There is a physical interpretation for the latter: 
for a fixed wavenumber, GW sourcing occurs when the typical bubble size grows to $\sim 1/k$.
In addition, there is a reason for the termination of the sourcing at late times:
see Sec.~\ref{sec:DiscussionConclusion}.

Fig.~\ref{fig:kDeltaT_v=1} is essentially the same as Fig.~\ref{fig:tauDeltaT_v=1},
except that the horizontal axis is the wavenumber $k$.
Different markers correspond to different values of $\tau$ mentioned above,
and the black line is the spectrum with the envelope approximation (instant damping $\tau = 0$)
reported in Ref.~\cite{Jinno:2016vai}.
It is seen that the spectrum approaches to the black line for small $\tau$, as expected.\footnote{
For wavenumbers $0.001 \lesssim k \lesssim 0.01$, 
we have checked that smaller values for $\tau$ than shown in this plot reproduce the black line.
These data are also used in making constant-$\tau$ slices in Fig.~\ref{fig:kDeltaTSlice_v=1}.
}
An important feature is that
\begin{itemize}
\item
The spectrum for low frequencies grows from $\propto k^3$ to $\propto k$,
\end{itemize}
in the long-lasting limit.
There is an explanation for this behavior (see the latter part of Sec.~\ref{sec:DiscussionConclusion}) 
and also an analytic proof on this linear behavior (see Appendix~\ref{app:Analytic}).

Fig.~\ref{fig:kDeltaTSlice_v=1} shows the spectrum at fixed $\tau$.
The colored lines correspond to $\tau = 1, 3, 10, 30, 100$ from bottom to top,
while the black-dashed line corresponds to the maximal value of $\tau$ in the data, 
i.e. $\tau = 10 \times (1/k)$.
In making this figure we have interpolated the data points shown 
in Figs.~\ref{fig:tauDeltaT_v=1} and \ref{fig:kDeltaT_v=1} to make constant-$\tau$ slices.
Also, for large wavenumbers $k > 0.1$
we have extrapolated the value at $\tau = 10 \times (1/k)$ to larger $\tau$
by assuming that the spectrum is constant for $\tau > 10 \times (1/k)$. 
It is seen that the saturation of the GW spectrum starts from higher wavenumbers as $\tau$ increases.

\begin{figure}[h]
\begin{center}
\includegraphics[width=0.7\columnwidth]{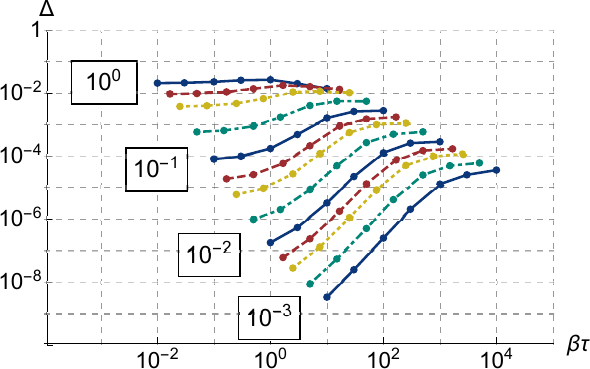}
\caption{\small
Same as Fig.~\ref{fig:tauDeltaT_v=1} except that $v = c_s$.
}
\label{fig:tauDeltaT_v=cs}
\end{center}
\end{figure}

\begin{figure}[h]
\begin{center}
\includegraphics[width=0.7\columnwidth]{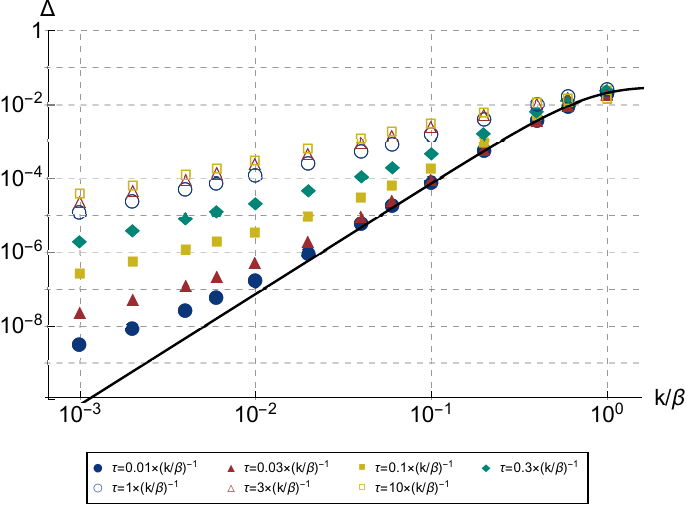}
\caption{\small
Same as Fig.~\ref{fig:kDeltaT_v=1} except that $v = c_s$.
}
\label{fig:kDeltaT_v=cs}
\end{center}
\end{figure}

\begin{figure}[h]
\begin{center}
\includegraphics[width=0.7\columnwidth]{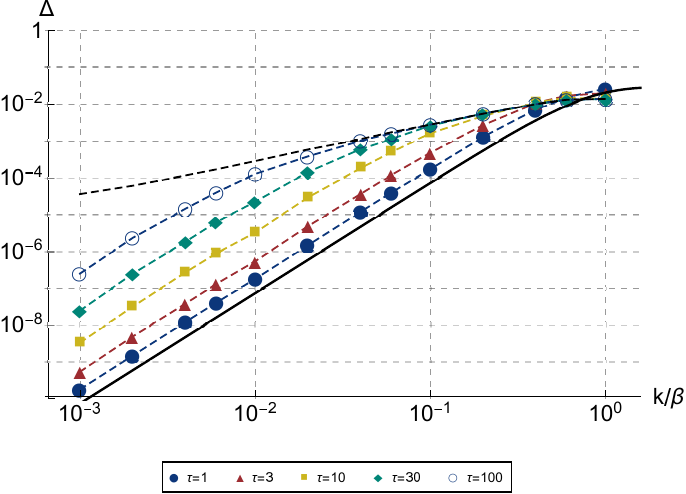}
\caption{\small
Same as Fig.~\ref{fig:kDeltaTSlice_v=1} except that $v = c_s$.
}
\label{fig:kDeltaTSlice_v=cs}
\end{center}
\end{figure}

In Figs.~\ref{fig:tauDeltaT_v=cs}--\ref{fig:kDeltaTSlice_v=cs} we show the spectrum for $v = c_s$.
As discussed in Sec.~\ref{sec:Setup}, these figures are considered to be relevant to 
(b) low terminal velocity.
The basic features are the same as Figs.~\ref{fig:tauDeltaT_v=1}--\ref{fig:kDeltaTSlice_v=1}.

In Figs.~\ref{fig:tauDeltaT_v=1} and \ref{fig:tauDeltaT_v=cs},
the low-frequency behavior in the long-lasting limit (black-dashed lines) 
is approximately given by
\begin{align}
\Delta
&\simeq
\left\{
\begin{matrix}
0.202 
\times 
\displaystyle \left( \frac{k}{\beta} \right)
& (v = 1),
\\[3ex]
0.0292
\times 
\displaystyle \left( \frac{k}{\beta} \right)
& (v = c_s),
\end{matrix}
\right. 
\end{align}
for $k/\beta \lesssim 1$.\footnote{
For $k/\beta \lesssim H/\beta$ the cosmic expansion will no longer be negligible,
and the spectrum is expected to be suppressed.
}

%%%%%%%%%%%%%%%%%%%%%%%%%%%%%%%%%%%%%%%%%%%%%%%%%%
\subsection{Peak position}
%%%%%%%%%%%%%%%%%%%%%%%%%%%%%%%%%%%%%%%%%%%%%%%%%%

\begin{figure}[h]
\begin{center}
\includegraphics[width=0.5\columnwidth]{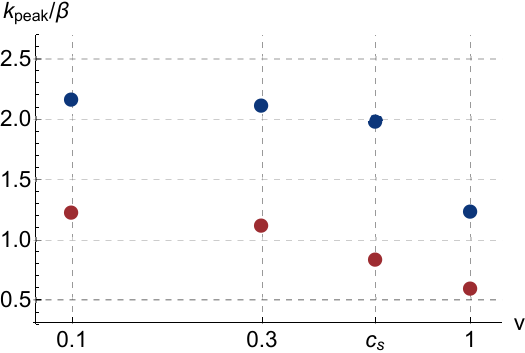}
\caption{\small
Peak position of the spectrum with the envelope approximation (blue) 
and without the envelope approximation (red).
}
\label{fig:Peak}
\end{center}
\end{figure}

In Fig.~\ref{fig:Peak} we show the peak wavenumber of the spectrum 
with the envelope approximation (blue) and without the envelope approximation (red). 
In calculating the latter we have used the data for $\tau = 10 \times (1/k)$,
assuming that the spectrum is constant for larger values of $\tau$.
It is seen that the peak position moves to lower $k$ in the long-lasting limit.

\vskip 0.5in

%%%%%%%%%%%%%%%%
%%%%%%%%%%%%%%%%

%%%%%%%%%%%%%%%%
%%%%%%%%%%%%%%%%

%%%%%%%%%%%%%%%%
%%%%%%%%%%%%%%%%

%%%%%%%%%%%%%%%%
%%%%%%%%%%%%%%%%

%%%%%%%%%%%%%%%%
%%%%%%%%%%%%%%%%

%%%%%%%%%%%%%%%%
%%%%%%%%%%%%%%%%

%%%%%%%%%%%%%%%%
%%%%%%%%%%%%%%%%

\clearpage

%%%%%%%%%%%%%%%%%%%%%%%%%%%%%%%%%%%%%%%%%%%%%%%%%%
\section{Discussion and conclusions}
\label{sec:DiscussionConclusion}
\setcounter{equation}{0}
%%%%%%%%%%%%%%%%%%%%%%%%%%%%%%%%%%%%%%%%%%%%%%%%%%

In this paper we have studied GW production from bubble dynamics in cosmic first-order phase transitions.
We have used the method of relating the GW spectrum with the two-point correlator
of the energy-momentum tensor $\left< T(x) T(y) \right>$
by using the stochasticity of produced GWs.
In calculating the correlator, the main approximations we have adopted in this paper are
\begin{itemize}
\item[(1)]
Exponential form for the nucleation rate (Eq.~(\ref{eq:Gamma})),
\item[(2)]
Thin-wall approximation (Eqs.~(\ref{eq:TB}) and (\ref{eq:rhoBBeyond_2})),
\item[(3)]
Free propagation after bubble collisions (Eqs.~(\ref{eq:TB}) and (\ref{eq:rhoBBeyond_2})),
\end{itemize}
and thus we have generalized our previous result \cite{Jinno:2016vai}
by removing the envelope approximation.
As explained in Sec.~\ref{sec:Setup}, 
the approximations we adopt in this paper are expected to give a reasonable modeling of the system
for frequencies around or lower than the inverse of the typical bubble size around the time of collisions,
as long as the phase transition proceeds with either (a) runaway or (b) low terminal velocity.\footnote{
In Ref.~\cite{Hindmarsh:2017gnf} 
it has been reported that the peak of the spectrum
is located at frequencies around the inverse of the bubble wall width.
In fact, bubble wall width seems to be one of the characteristic scales 
in cosmological first-order phase transitions.
However,
our current formalism cannot take such a finite wall width into account, 
which will affect the spectrum at frequencies higher than the inverse of 
the bubble radius around the time of collisions.
It would be one of the future directions to consider how to deal with the finite wall width in the present formalism.
}
The remaining case, (c) high terminal velocity, would require the analysis of nonlinear dynamics,
and beyond the scope of this paper.

Our main results are the analytic expressions in Sec.~\ref{sec:Analytic}
(Eqs.~(\ref{eq:MainDeltaS}) and (\ref{eq:MainDeltaD})),
and the numerical results presented in Sec.~\ref{sec:Numerical}:
\begin{itemize}
\item
For the analytic expressions, 
they apply to a general damping function $D$ (see Sec.~\ref{subsec:AA})
and a general nucleation rate $\Gamma$ (see Sec.~\ref{subsec:rate}).
Also, it can be shown that
the spectrum behaves as $\propto k$ for small $k$ 
in the long-lasting limit of bubble walls 
(see the discussion below and also Appendix~\ref{app:Analytic}).
\item
In Sec.~\ref{sec:Numerical}, we have performed numerical evaluation of the spectrum
obtained in Sec.~\ref{sec:Analytic}.
It is found that the spectrum shows a significant enhancement 
in the long-lasting limit of bubble walls,
compared to the one with the envelope approximation.
It is also found that the spectrum growth occurs as a transition 
from $\propto k^3$ to $\propto k$ for small wavenumbers.
For a fixed wavenumber $k$, such a transition typically occurs time $\sim 1/vk$ after collision,
which has a physical interpretation that the sourcing occurs when the (collided) bubbles
expand to size $\sim 1/k$.
Also, the sourcing terminates after this typical sourcing time.
\end{itemize}

\begin{figure}[h]
\begin{center}
\includegraphics[width=1\columnwidth]{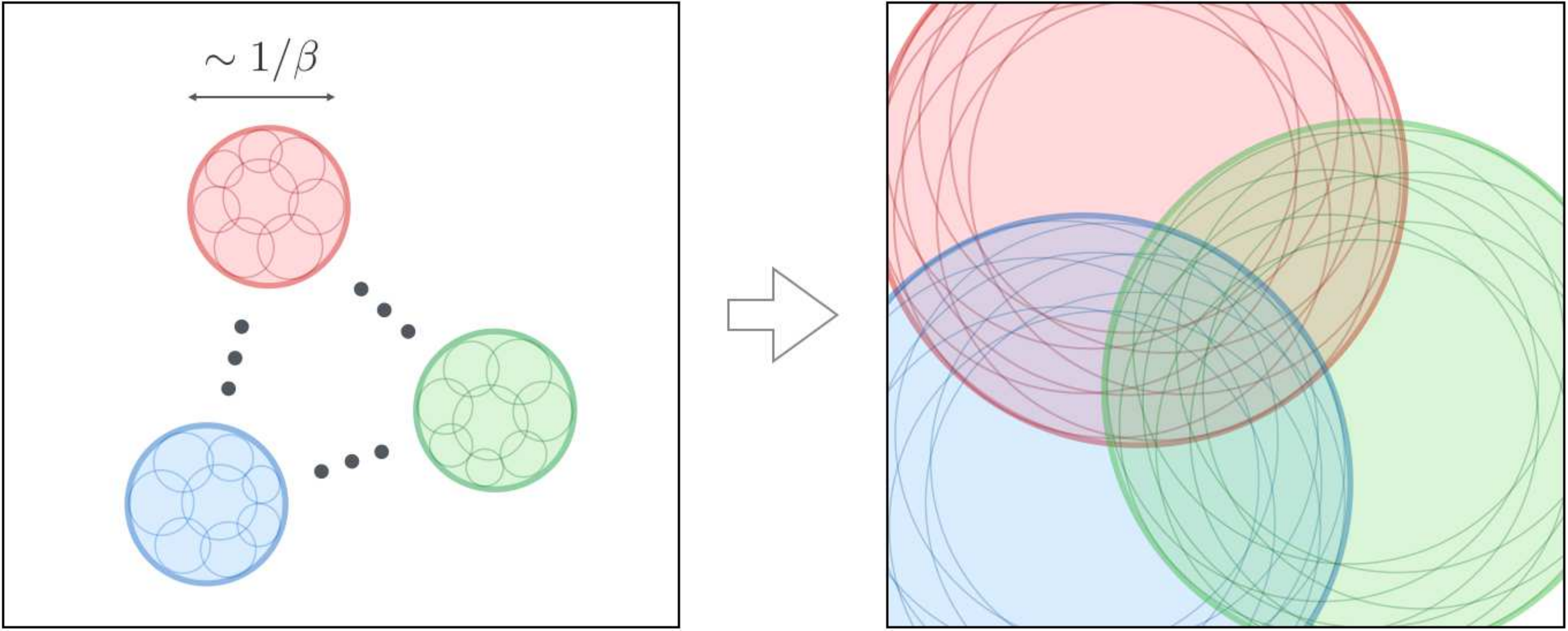}
\caption{\small
Schematic picture of the ``correlation groups."
Bubbles are divided into groups which typically have radius $\sim 1/\beta$ 
around the time of phase transition (left).
These bubbles have correlation with each other only within each group,
and correlation across different groups are (exponentially) suppressed.
As long as free propagation after bubble collisions makes a good approximation of the system,
these groups have no correlation even after they expand and overlap with each other,
and can be regarded as independent sources of GWs (right).
}
\label{fig:Correlation}
\end{center}
\end{figure}

At this point we compare our results with the literature.
The results we have obtained in Sec.~\ref{sec:Numerical} and 
the ones in the numerical simulation literature seems to have some discrepancies.
It is commonly considered that GW sourcing from sound waves continues 
all the way until the Hubble time after bubble collisions 
because of the long-lasting nature of the source.
This argument is based on the following ansatz on the correlator $\Pi$ at late times:
\begin{align}
\Pi(t_x,t_y,k)
&\overset{?}{=} 
\Pi(t_x - t_y,k).
\label{eq:PiIncorrect}
\end{align}
However, if this argument holds true,
the GW spectrum presented in this paper would show a linear enhancement 
in the duration time of the bubble walls $\tau$,
as long as our modeling of the system correctly captures the dynamics at least for low frequencies
(i.e. frequencies lower than the inverse of the typical bubble size around the time of collisions).
Instead, what we have observed is the termination of the sourcing and the resulting saturation 
of the spectrum in the long-lasting limit.
Within the modeling of the system we have presented in this paper,
there is a clear reason why Eq.~(\ref{eq:PiIncorrect}) does not hold at least for low frequencies:
\begin{itemize}
\item
GWs are sourced by the two-point correlator of 
the energy-momentum tensor $\left< T(x)T(y) \right>$, or more precisely, 
the projected correlator $KK \left< T(x)T(y) \right>$ (see Sec.~\ref{subsec:gwfrom}).
On the other hand, the projected one-point correlator $K \left< T(x) \right>$ vanishes
because of the spherical symmetry of the system makes $\left< T(x) \right> \propto \delta_{ij}$.

\item
Therefore, in order to produce nonzero $KK \left< T(x)T(y) \right>$,
the contribution to the energy-momentum tensor at the spacetime point $x$ 
must affect the energy-momentum tensor at $y$ in some way.
Taking into account the fact that bubble nucleation finishes within the timescale of $\sim 1/\beta$,
this means that the bubble which affects $T(x)$ and the one which affects $T(y)$
must nucleate within a distance of $\sim {\mathcal O}(1/\beta)$\footnote{
In the single-bubble case this is automatically satisfied because the two bubbles are identical,
while in the double-bubble case the two bubbles give a net contribution to
$KK \left< T(x)T(y) \right>$ only when the nucleation points are close with each other $\sim {\mathcal O}(1/\beta)$.
}
(Fig.~\ref{fig:Cherry} might be of some help).

\item
Let us see this from a different viewpoint.
In the system under consideration, we may divide the bubbles into some groups 
in which the nucleation points are within a radius of $\sim {\mathcal O}(1/\beta)$,
which we call ``correlation groups" (see Fig.~\ref{fig:Correlation}).
These correlation groups just expand as time goes without affecting each other.
Gravitational-wave production in this system can be modeled 
just by the sum of sourcing from these independent sources,
because they have only suppressed correlations with each other even after they overlap.
In this modeling, 
GW sourcing at wavenumber $k$ occurs 
only when the correlation groups expand to a size $\sim 1/k$, and there is no sourcing at later times.

\item
In addition, one can show that this modeling reproduces 
the observed behavior of the spectrum $\Delta \propto k$ for low frequencies.
First note that 
\begin{itemize}
\item
Relativistic objects with energy density $\rho_{\rm source}$ and size $\sim 1/k$ 
which last for a period $\Delta t$ produces GWs with a typical amplitude
$\Omega_{\rm GW} \sim (\rho_{\rm tot} / M_P^2) (\rho_{\rm source} / \rho_{\rm tot})^2 (\Delta t)^2$ 
at wavenumber $k$.\footnote{
This can be derived for example from the equation of motion as
\begin{align}
\Box h
&\sim
\frac{\rho_{\rm source}}{M_P^2}
\;\;\;\;
\to 
\;\;\;\;
\rho_{\rm GW}
\sim
\frac{\rho_{\rm source}^2 \Delta t^2}{M_P^2}
\;\;\;\;
\to 
\;\;\;\;
\Omega_{\rm GW}
\sim
\frac{\rho_{\rm tot}}{M_P^2}
\frac{\rho_{\rm source}^2}{\rho_{\rm tot}^2}
\Delta t^2.
\end{align}
Also, with $\rho_{\rm source} \sim \kappa \rho_0$, this reproduces the well-known behavior 
of GWs from bubble collisions
\begin{align}
\Omega_{\rm GW}
\sim 
\kappa^2 \left( \frac{H_*}{\beta} \right)^2 \left( \frac{\alpha}{1 + \alpha} \right)^2.
\end{align}
}
\end{itemize}
Then also note that in the present setup 
there are $\sim (k/\beta)^{-3}$ overlapping independent sources 
at time $\sim 1/k$ after collisions.
Each source has energy density $\rho_{\rm source} \propto k^3$ 
and lasts for $\Delta t \sim 1/k$.
Therefore one finds
$\Omega_{\rm GW} \propto k^{-3} \cdot (k^3)^2 \cdot (k^{-1})^2 \propto k$,\footnote{
More precisely, from $\rho_{\rm source} \sim (k/\beta)^3 \kappa \rho_0$, 
the GW spectrum becomes 
\begin{align}
\Omega_{\rm GW}
\sim 
\left( \frac{\beta}{k} \right)^3 
\cdot
\frac{\rho_{\rm tot}}{M_P^2}
\left( \frac{(k / \beta)^3 \kappa \rho_0}{\rho_{\rm tot}} \right)^2
\left( \frac{1}{k} \right)^2
\sim
\left( \frac{k}{\beta} \right)
\cdot
\kappa^2 \left( \frac{H_*}{\beta} \right)^2 \left( \frac{\alpha}{1 + \alpha} \right)^2,
\end{align}
at low frequencies $k \lesssim \beta$.
}
and hence our modeling of the system by independent expanding sources 
captures the late time GW sourcing quite well.

\end{itemize}
At the current stage we do not have any argument which reconciles above reasoning 
with the sound-wave enhancement of GWs in the literature.

Finally we discuss the effect of finite wall width in the low terminal velocity case.
As mentioned in Sec.~\ref{subsec:GWV} (b), 
the region of energy concentration increases in volume after bubble collisions.
Such wall regions eventually fill the whole Universe and start to overlap with each other.\footnote{
Note that, 
as mentioned in Sec.~\ref{subsec:GWV} (b), 
the width of such wall regions remains to be constant while their area increases as bubbles expand.
}
One might worry that the description of the present system by the thin-wall approximation 
would not be valid any longer after such overlaps start to develop.
However, these overlaps do not necessarily mean 
the breakdown of the thin-wall approximation.
This is because most overlaps are supposed to be irrelevant in GW production:
as we have just seen, 
two bubbles with nucleation points more than ${\mathcal O}(1/\beta)$ 
distant from each other have only suppressed correlation,
and their overlap is quite unlikely to affect GW production.
In this sense, we only have to focus on each correlation group in discussing GW production.
For each group the volume fraction of the wall regions does not increase even well after collisions (see Fig.~\ref{fig:Correlation}).
Therefore, the thin-wall approximation is still expected to be a good description of the system 
for GW frequencies lower than the inverse of the wall width.

Though much remains to be settled,
the analytic approach to the dynamics of GW sourcing in first-order phase transitions
as we have presented in this paper will work complementarily with numerical simulations.
We believe that such a direction is worth further investigation in the future.

%%%%%%%%%%%%%%%%
%%%%%%%%%%%%%%%%

%%%%%%%%%%%%%%%%%%%%%%%%%%%%%%%%%%%%%%%%%%%%%%%%%%%%%%%
\section*{Acknowledgments}
%%%%%%%%%%%%%%%%%%%%%%%%%%%%%%%%%%%%%%%%%%%%%%%%%%%%%%%

R.J. is grateful to M.~Hindmarsh, S.~Huber, T.~Konstandin and J.~M.~No for helpful discussions.
The work of R.J. was supported by IBS under the project code, IBS-R018-D1.
The work of R.J. and M.T. was supported by JSPS Research Fellowships for Young Scientists.

\clearpage

%%%%%%%%%%%%%%%%%%%%%%%%%%%%%%%%%%%%%%%%%%%%%%%%%%
\appendix
%%%%%%%%%%%%%%%%%%%%%%%%%%%%%%%%%%%%%%%%%%%%%%%%%%

%%%%%%%%%%%%%%%%%%%%%%%%%%%%%%%%%%%%%%%%%%%%%%%%%%
\section{GW spectrum with the envelope approximation}
\label{app:Envelope}
\setcounter{equation}{0}
%%%%%%%%%%%%%%%%%%%%%%%%%%%%%%%%%%%%%%%%%%%%%%%%%%

In this appendix we derive the GW spectrum by evaluating the unequal-time power spectrum 
$\Pi(t_x,t_y,k)$, or equivalently the correlator $\langle T(x)T(y) \rangle$.
Though our main goal is to derive it without the envelope approximation,
we first illustrate the calculation procedure with this approximation.
This is because the full derivation of the correlator in Sec.~\ref{app:BeyondEnv}
is somewhat complicated, and therefore it would be better to see a simpler example first.
We use Eq.~(\ref{eq:TB})--(\ref{eq:rhoBNoColl}) for the energy-momentum tensor of uncollided bubble walls,
while we assume that it vanishes instantly after collision.
This appendix basically follows Ref.~\cite{Jinno:2016vai}.

%%%%%%%%%%%%%%%%%%%%%%%%%%%%%%%%%%%%%%%%%%%%%%%%%%
\subsection{Basic strategy}
\label{subsec:EnvStrategy}
%%%%%%%%%%%%%%%%%%%%%%%%%%%%%%%%%%%%%%%%%%%%%%%%%%

We first explain the essence for the derivation of the GW spectrum.
From the definition of the ensemble average,
all we have to do to obtain $\langle T(x)T(y) \rangle$ is
\begin{itemize}
\item
Fix the spacetime points $x = (t_x, \vec{x})$ and $y = (t_y, \vec{y})$.
\item
Find bubble configurations giving nonzero $T(x)T(y)$,
estimate the probability for such configurations to occur, 
and calculate the value of $T(x)T(y)$ in each case.
\item
Sum over all possible configurations.
\end{itemize}
We call $x$ and $y$ in the arguments of $\langle T(x)T(y) \rangle$ 
``(spacetime) evaluation point" in the following.
Also, $t_x$ and $t_y$ are called ``evaluation time", 
while $\vec{x}$ and $\vec{y}$ are called ``(spacial) evaluation point".

\begin{figure}
\begin{center}
\includegraphics[scale=0.35]{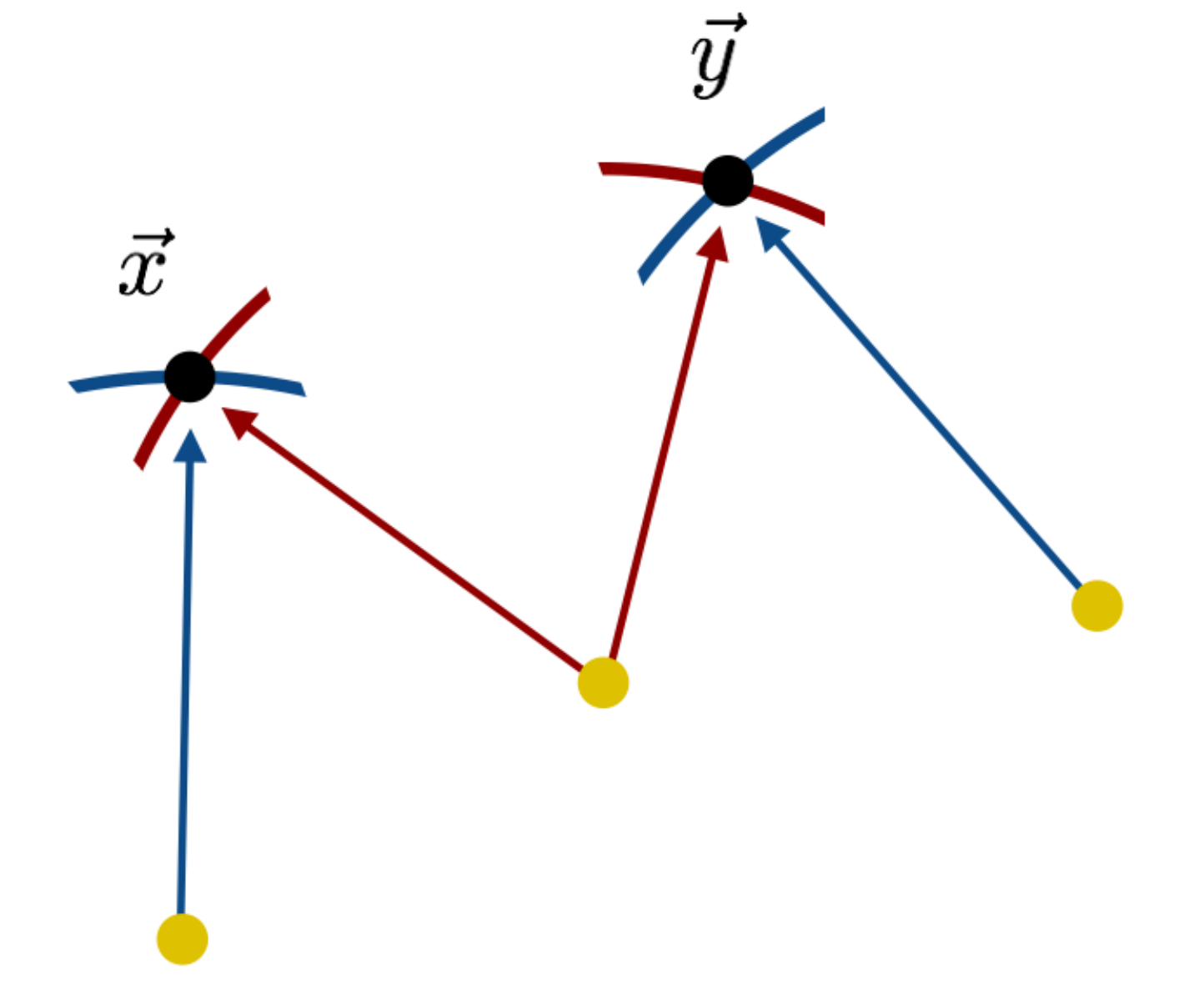}
\caption{\small
Rough sketch of the single- and double-bubble contributions.
In single-bubble contribution (red), the bubble wall fragments passing through $\vec{x}$ and $\vec{y}$
come from the single nucleation point,
while in double-bubble contribution (blue) they belong to different nucleation points.
In this figure the evaluation times are taken to be $t_x = t_y$ for illustrative purpose.
}
\label{fig:SingleDouble}
\end{center}
\end{figure}

Let us consider which bubble configurations give nonzero $T(x)T(y)$.
In order for this to occur, some bubble wall fragment(s) must be at $\vec{x}$ at time $t_x$,
and other(s) must be at $\vec{y}$ at time $t_y$\footnote{
If one takes radiation component and the false-vacuum energy into account, 
the energy-momentum tensor is nonzero even when there is no bubble wall at $x$ or $y$.
However, these contributions are isotropic and 
they vanish after multiplied to the projection operator (\ref{eq:K}).
}.
We refer to such bubbles whose wall pass through $\vec{x}$ at $t_x$
or $\vec{y}$ at $t_y$ as $x$-bubble or $y$-bubble, respectively,
and call the wall fragments which pass these spacetime evaluation points 
$x$-fragment or $y$-fragment, respectively.
See Fig.~\ref{fig:SingleDouble} for illustration.
In this figure, bubble nucleation points are denoted by the yellow circles.
The red bubble is $x$-bubble {\it and} $y$-bubble (which we call $xy$-bubble),
while the left and right blue ones are $x$-bubble and $y$-bubble,
respectively.

Next we take the thin-wall limit $l_B \to 0$ into account.
In this limit we do not have to consider 
those cases where two different fragments exist at a single spacetime evaluation point,
because such a probability is infinitely smaller than the probability for one fragment to exist at the point.
Therefore we consider one wall fragment at $x$, and another at $y$.
There are two possibilities for this: 
one is that these fragments originate from the same nucleation point
(red lines in Fig.~\ref{fig:SingleDouble}), 
while the other is that these come from different nucleation points
(blue lines in Fig.~\ref{fig:SingleDouble}).
This leads to the following classification:\footnote{
One may wonder why the single-bubble contribution has to be taken into account,
because it is well known that a spherical object does not radiate GWs.
The answer is that fixing the spacetime points $x$ and $y$ breaks the spherical symmetry
of a single bubble. See Appendix~\ref{app:saru} on this point.
}
\begin{itemize}
\item
Single-bubble: \\
The wall fragments passing through $x$ and $y$ originate from a single nucleation point.
\item
Double-bubble: \\
The wall fragments passing through $x$ and $y$ originate from two different nucleation points.
\end{itemize}
In the rest of this appendix we calculate these contributions in turn,
using the envelope approximation mentioned in 
Sec.~\ref{sec:Intro} and Sec.~\ref{subsec:AA}.
The final expressions are shown in 
Eqs.~(\ref{eq:EnvDeltaS}) and (\ref{eq:EnvDeltaD}) for single- and double-bubble contributions, 
respectively.

Before going into calculation of the spectrum, 
we first fix our notations and then introduce the ``false vacuum probability".
These are repeatedly used in the following calculations.
Also we take $\beta = 1$ unit without loss of generality.

%%%%%%%%%%%%%%%%
%%%%%%%%%%%%%%%%

%%%%%%%%%%%%%%%%%%%%%%%%%%%%%%%%%%%%%%%%%%%%%%%%%%
\subsection{Prerequisites}
\label{subsec:EnvPre}
%%%%%%%%%%%%%%%%%%%%%%%%%%%%%%%%%%%%%%%%%%%%%%%%%%

%%%%%%%%%%%%%%%%%%%%%%%%%%%%%%%%%%%%%%%%%%%%%%%%%%
\subsubsection*{Notations}
%%%%%%%%%%%%%%%%%%%%%%%%%%%%%%%%%%%%%%%%%%%%%%%%%%

\begin{figure}
\begin{center}
\includegraphics[width=0.8\columnwidth]{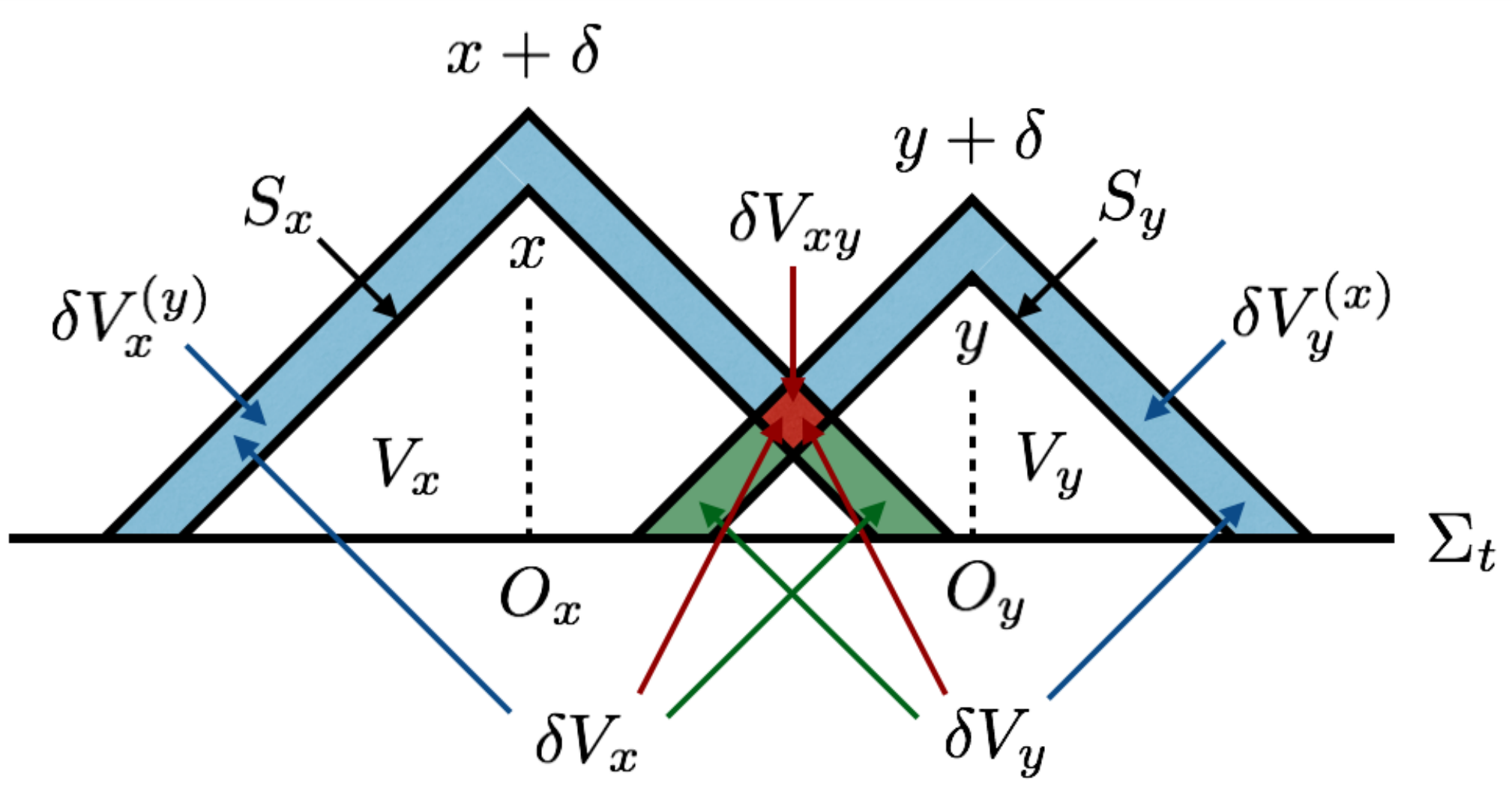}
\caption{\small
Definitions of various quantities.
The horizontal axis denotes the spacial direction,
while the vertical one denotes the time direction.
Note that the spacial direction has three dimensions in reality,
and therefore the region $\delta V_{xy}$ stretches in the time direction.
See also Fig.~\ref{fig:3DLightCone}.
In addition, see Fig.~\ref{fig:Circle} for the intersection of $v$-cones with $\Sigma_t$.
This figure is the same as in Ref.~\cite{Jinno:2016vai}.
}
\label{fig:LightCone}
\end{center}
\end{figure}

In this subsection we fix our notations and conventions.
We denote the two spacetime points in the correlator as (see Fig.~\ref{fig:LightCone})
\begin{align}
x 
&= (t_x,\vec{x}),
\;\;\;
y = (t_y,\vec{y}),
\end{align}
and write the average and difference of the time coordinates as
\begin{align}
t_{\langle x,y \rangle}
&\equiv \frac{t_x + t_y}{2}, 
\;\;\;
t_{x,y}
\equiv t_x - t_y.
\end{align}
We sometimes use these quantities in place of $t_x$ and $t_y$.
Also, the distance between $\vec{x}$ and $\vec{y}$ is denoted by
\begin{align}
\vec{r}
\equiv \vec{x} - \vec{y},
\;\;\;
r
&\equiv |\vec{r}|.
\end{align}
For later convenience, we define the spacial distance 
normalized by the wall velocity $v$:
\begin{align}
r_v
&\equiv \frac{r}{v}.
\end{align}

\begin{figure}
\begin{center}
\includegraphics[width=0.6\columnwidth]{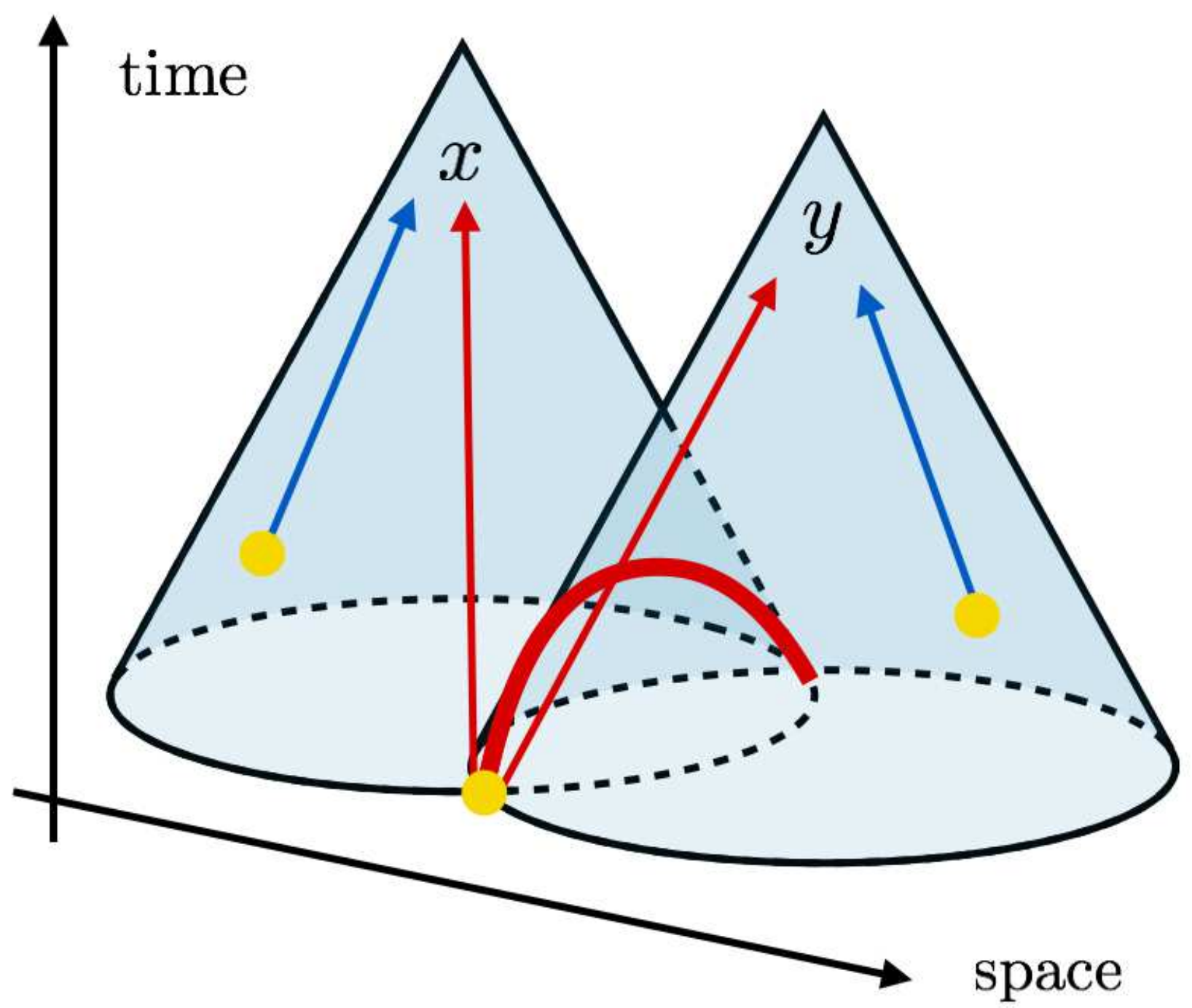}
\caption{\small
How Fig.~\ref{fig:LightCone} looks in $1 + 2$ dimensions.
The region $\delta V_{xy}$ is shown as the red region along the intersection of the two $v$-cones.
The single-bubble contribution refers to the wall fragments 
from bubbles which nucleate in the red region (denoted by the red arrows),
while the double-bubble contribution refers to 
those from different nucleation points (denoted by the blue arrows).
This figure is the same as in Ref.~\cite{Jinno:2016vai}.
}
\label{fig:3DLightCone}
\end{center}
\end{figure}

In what follows we often consider past cones with velocity $v$ originating from $x$ and $y$
(see Fig.~\ref{fig:LightCone} and \ref{fig:3DLightCone}),
and we refer to these as past $v$-cones.
These coincide with past light cones for luminal wall case $v = 1$.
We label these by $S_x$ and $S_y$.
The regions inside $S_x$ and $S_y$ are called $V_x$ and $V_y$, respectively,
and their union is written as $V_{xy} \equiv V_x \cup V_y$.
Also, we define the following spacetime points
\begin{align}
x + \delta 
&\equiv (t_x + l_B/v,\vec{x}),
\;\;\;
y + \delta \equiv (t_y + l_B/v,\vec{y}).
\end{align}
We denote their past $v$-cones by $S_{x + \delta}$ and $S_{y + \delta}$,
and their inner regions by $V_{x + \delta}$ and $V_{y + \delta}$.
The thin regions on the surface of $V_x$ and $V_y$ with width $l_B/v$ are written as
\begin{align}
\delta V_x
&\equiv V_{x+\delta} - V_x, 
\;\;\;
\delta V_y
\equiv V_{y+\delta} - V_y.
\end{align}
The intersection of these regions is denoted by
\begin{align}
\delta V_{xy}
&\equiv \delta V_x \cap \delta V_y.
\end{align}
We also define the following region for later use
\begin{align}
\delta V_x^{(y)}
&\equiv \delta V_x - V_{y+\delta},
\;\;\;
\delta V_y^{(x)}
\equiv \delta V_y - V_{x+\delta}.
\end{align}
Fig.~\ref{fig:LightCone} summarizes these notations.

\begin{figure}
\begin{center}
\includegraphics[width=0.65\columnwidth]{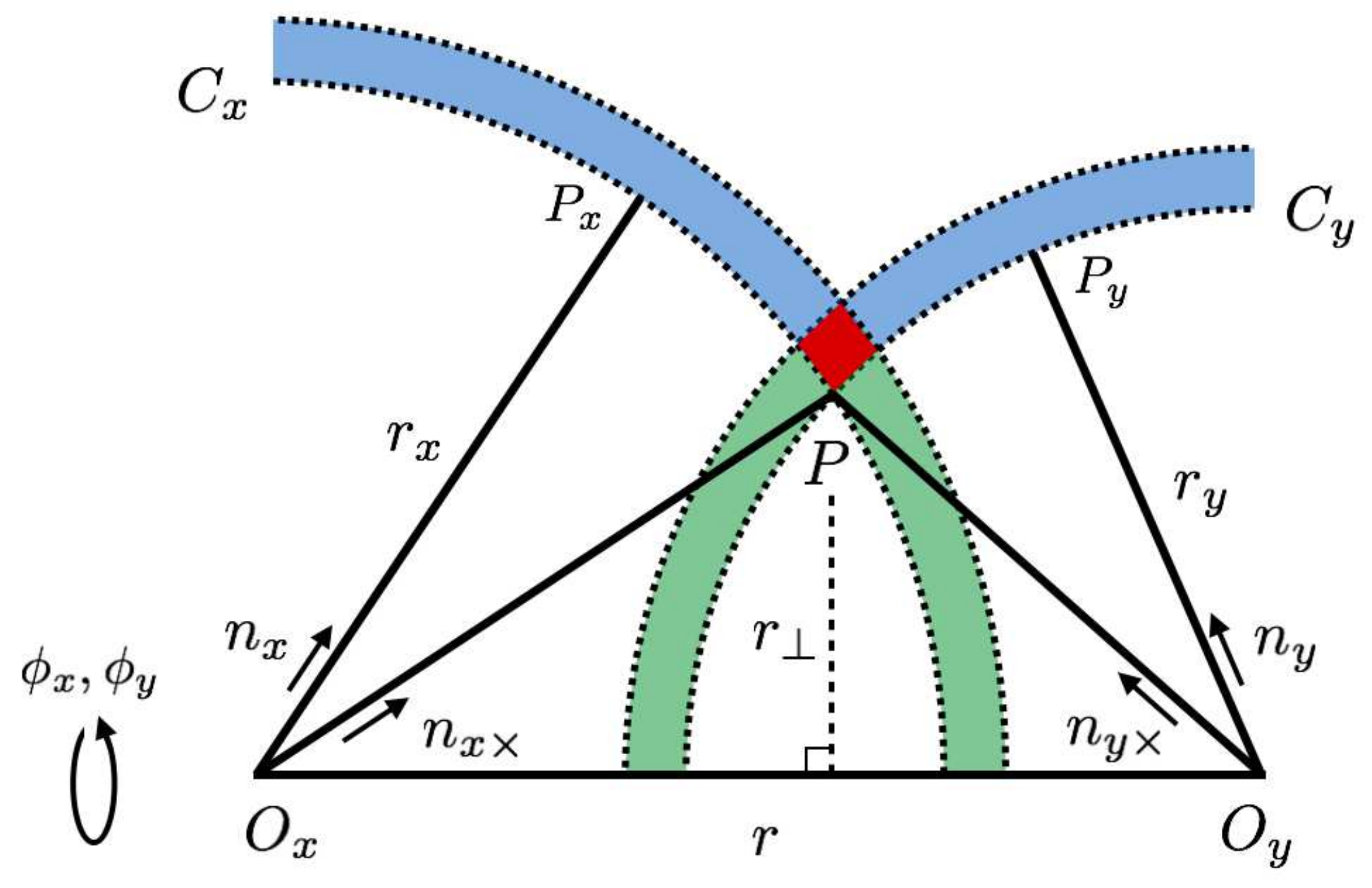}
\caption{\small
Intersection of the past light cones of $x$ and $y$ with the constant time hypersurface $\Sigma_t$.
In this figure, $t$ is taken to be before $t_{\rm max}$.
When $P_x = P_y$, 
i.e. when both points are on $C_x \cap C_y$ with $\phi_x = \phi_y$, we call them $P$.
The width of the circles denotes $l_B$, which is taken to be $+0$ in the thin-wall limit.
This figure is the same as in Ref.~\cite{Jinno:2016vai}.
}
\label{fig:Circle}
\end{center}
\end{figure}

Next, let us consider a constant-time hypersurface $\Sigma_t$ at time $t$.
The two past $v$-cones $S_x$ and $S_y$ form spheres on this hypersurface
as shown in Fig.~\ref{fig:Circle}.
We call these two spheres $C_x$ and $C_y$,
and call their centers $O_x$ and $O_y$.
The radii of $C_x$ and $C_y$ are given by
\begin{align}
r_x
& \equiv
r_B(t_x,t),
\;\;\;\;
r_y
\equiv
r_B(t_y,t),
\label{eq:rxry}
\end{align}
respectively.
These spheres $C_x$ and $C_y$ have intersection 
only for $t < t_{\rm max}$ with
\begin{align}
t_{\rm max}
&\equiv \frac{t_x + t_y - r_v}{2}.
\label{eq:ttimes}
\end{align}
Let $P_x$ and $P_y$ be some arbitrary points on $C_x$ and $C_y$.
These points are parametrized by
$n_x \equiv \overrightarrow{O_xP_x}/|\overrightarrow{O_xP_x}|$ 
and $n_y \equiv \overrightarrow{O_yP_y}/|\overrightarrow{O_yP_y}|$.
We parametrize these unit vectors so that 
$\theta_x$ and $\theta_y$ denote the polar angle
and $\phi_x$ and $\phi_y$ denote the azimuthal angle 
with respect to $\vec{r}$:
\begin{align}
n_x
&\equiv (s_x c_{\phi x},s_x s_{\phi x},c_x),
\;\;\;
n_y
\equiv (s_y c_{\phi y},s_y s_{\phi y},c_y),
\end{align}
and often write $\cos \theta_x (\theta_y)$ and $\sin \theta_x(\theta_y)$ as
$c_x(c_y)$ and $s_x(s_y)$, respectively, 
and $\cos \phi_x (\phi_y)$ and $\sin \phi_x (\phi_y)$
as $c_{\phi x} (c_{\phi y})$ and $s_{\phi x} (s_{\phi y})$, respectively.
We also define the following product $N$ of the unit vectors for later convenience:
\begin{align}
N_{ijkl}
&\equiv 
(n_x)_i(n_x)_j(n_y)_k(n_y)_l.
\end{align}
In addition, 
we often consider those cases where $P_x$ and $P_y$ are identical
(i.e. they are an identical point on the intersection of $C_x$ and $C_y$).
In such cases we write $P_x = P_y = P$, 
and label the quantities introduced above by ``$\times$":
$n_x$ and $n_y$ as $n_{x\times}$ and $n_{y\times}$,
$\theta_x$ and $\theta_y$ as $\theta_{x\times}$ and $\theta_{y\times}$,
and
$\phi_x$ and $\phi_y$ as $\phi_{\times}$, respectively.
In such cases, the cosines of the polar angles are related with 
the radii of $C_x$ and $C_y$ as
\begin{align}
\cos\theta_{x\times}
&=
- \frac{r^2 + r_x^2 - r_y^2}{2rr_x},
\;\;\;\;
\cos\theta_{y\times}
= \frac{r^2 + r_y^2 - r_x^2}{2rr_y}.
\label{eq:cos}
\end{align}
Likewise, the product of the unit vectors $N$ is written as $N_{\times}$:
\begin{align}
N_{\times,ijkl}
&=
(n_{x\times})_i(n_{x\times})_j(n_{y\times})_k(n_{y\times})_l.
\label{eq:Ntimes}
\end{align}
%%

%%%%%%%%%%%%%%%%
%%%%%%%%%%%%%%%%

%%%%%%%%%%%%%%%%
%%%%%%%%%%%%%%%%

%%%%%%%%%%%%%%%%
%%%%%%%%%%%%%%%%

%%%%%%%%%%%%%%%%%%%%%%%%%%%%%%%%%%%%%%%%%%%%%%%%%%
\subsubsection*{False vacuum probability}
%%%%%%%%%%%%%%%%%%%%%%%%%%%%%%%%%%%%%%%%%%%%%%%%%%

We define the false vacuum probability $P(x,y,\cdots)$ with $x,y,\cdots$ being arbitrary spacetime points
as the probability for these points to be in the false vacuum.
This is equivalent to the probability for no bubbles to nucleate in the union 
$V_{xy\cdots} \equiv V_x \cup V_y \cup \cdots$.
If we divide $V_{xy\cdots}$ into infinitesimal regions $\delta V_i$
with $i$ being the label for these regions,
the probability we need is the product of $1 - \Gamma_i \delta V_i$
over the whole region in $V_{xy\cdots}$.
Therefore~\cite{Turner:1992tz},
\begin{align}
P(x,y,\cdots)
&=
\prod_i \left( 1 - \Gamma_i dV_i \right)
=
e^{-I(x,y,\cdots)},
\;\;\;\;
I(x,y,\cdots)
= \int_{V_{xy\cdots}} d^4z \; \Gamma(t_z).
\label{eq:FalseP}
\end{align}
Here $z = (t_z,\vec{z})$.
Using Eq.~(\ref{eq:FalseP}), we can explicitly write down the probability 
for two spacetime points $x$ and $y$ to remain in the false vacuum.
As seen from Fig.~\ref{fig:LightCone}, the intersection of the union $V_{xy}$ with the hypersurface $\Sigma_t$
consists of a union of two spheres for $t < t_{\rm max}$, 
while it consists of separate spheres for $t > t_{\rm max}$.
Therefore, the $I$ function is written as
\begin{align}
I(x,y)
&= 
\int_{-\infty}^{t_{\rm max}}dt \; 
\Gamma(t) 
\left[ \frac{\pi}{3} r_x^3(2 + c_{x\times})(1 - c_{x\times})^2 
+ \frac{\pi}{3} r_y^3(2 - c_{y\times})(1 + c_{y\times})^2 \right]
\nonumber \\
&\;\;\;\;
+ 
\int_{t_{\rm max}}^{t_x} dt \; 
\Gamma(t) 
\left[ \frac{4\pi}{3} r_x^3 \right]
+
\int_{t_{\rm max}}^{t_y} dt \; 
\Gamma(t) 
\left[ \frac{4\pi}{3} r_y^3 \right].
\end{align}
Here the quantity in the square parenthesis in the first line is the three-dimensional volume 
of the intersection of $V_{xy}$ with $\Sigma_t$. 
Note that this expression does not assume any specific form for $\Gamma(t)$.

For the nucleation rate (\ref{eq:Gamma}), it is straightforward to evaluate the time integration by using 
Eqs.~(\ref{eq:rxry}), (\ref{eq:ttimes}) and (\ref{eq:cos}).
We get\footnote{
We have changed the definition of ${\mathcal I}$ from Ref.~\cite{Jinno:2016vai}.
}
\begin{align}
I(x,y)
&\equiv
v^3 \Gamma_* e^{t_{\left< x,y \right>}} \; {\mathcal I}(t_{x,y},r_v),
\;\;\;\;
{\mathcal I}(t_{x,y},r_v)
= 
8\pi
\left[
e^{t_{x,y}/2} 
+ e^{- t_{x,y}/2}
+ \frac{t_{x,y}^2 - (r_v^2 + 4r_v)}{4r_v} e^{-r_v/2}
\right].
\end{align}
Note that we adopt $\beta = 1$ unit without loss of generality.

%%%%%%%%%%%%%%%%%%%%%%%%%%%%%%%%%%%%%%%%%%%%%%%%%%
\subsection{Single-bubble spectrum}
\label{subsec:EnvSingle}
%%%%%%%%%%%%%%%%%%%%%%%%%%%%%%%%%%%%%%%%%%%%%%%%%%

%%%%%%%%%%%%%%%%%%%%%%%%%%%%%%%%%%%%%%%%%%%%%%%%%%
\subsubsection*{Conditions for bubble configurations}
%%%%%%%%%%%%%%%%%%%%%%%%%%%%%%%%%%%%%%%%%%%%%%%%%%

We now calculate the single-bubble contribution to the GW spectrum
(see the red arrows in Figs.~\ref{fig:SingleDouble} and \ref{fig:3DLightCone}).
First let us make clear necessary and sufficient conditions
for the wall of a single bubble to contribute to the energy-momentum tensor both at $x$ and $y$.
They are summarized as
\begin{itemize}
\item
No bubbles nucleate in $V_{xy}$.
\item
One bubble nucleates in $\delta V_{xy}$.
\end{itemize}
The former condition is understood as follows.
If this is not satisfied, some bubble(s) nucleate in $V_{xy}$.
Such bubbles expand, and the evaluation point $\vec{x}$ or $\vec{y}$ enters these bubbles 
before the evaluation time $t_x$ or $t_y$.
Here note that, in the envelope approximation,
every spacial point is passed through by bubble walls only once (see Fig.~\ref{fig:Envelope}).
Therefore there cannot be any wall fragment at spacetime points $x$ or $y$.
The latter condition is understood as follows.
As seen from Figs.~\ref{fig:LightCone} or \ref{fig:Circle},
those bubbles which contribute to the energy-momentum tensor at $x$ and $y$
must have their nucleation points in $\delta V_{xy}$.
Though more than one bubbles can nucleate in this region,
such a probability is infinitely smaller than the one for a single bubble to nucleate 
in the thin-wall limit $l_B \to 0$. 
Therefore we have only to take one bubble nucleation into account.

%%%%%%%%%%%%%%%%%%%%%%%%%%%%%%%%%%%%%%%%%%%%%%%%%%
\subsubsection*{Expression for GW spectrum}
%%%%%%%%%%%%%%%%%%%%%%%%%%%%%%%%%%%%%%%%%%%%%%%%%%

Now let us move on to the calculation of the correlator.
First we calculate the probability part.
Since the region $\delta V_{xy}$ exists only before $t_{\rm max}$,
we limit ourselves to the nucleation time $t_n < t_{\rm max}$. 
Let us take a constant-time hypersurface $\Sigma_{t_n}$ at time $t_n$.
We label ``$n$" to all the quantities introduced around Eqs.~(\ref{eq:rxry})--(\ref{eq:cos}) 
to denote that these quantities are on $\Sigma_{t_n}$,
like $n_{xn}$, $\theta_{xn}$, $\phi_{xn}$, $n_{yn}$, $\theta_{yn}$, $\phi_{yn}$ and $N_n$.
Denoting by $dP^{(s)}$ (``$s$" denotes ``single") 
the probability for no bubbles to nucleate in $V_{xy}$ 
and for one bubble to nucleate in $\delta V_{xy}$ 
in the time interval $[t_n, t_n + dt_n]$
and in the azimuthal angle interval $[\phi_n, \phi_n + d\phi_n]$,
we obtain 
\begin{align}
dP^{(s)}
&= 
P(x,y) 
\times \frac{r_\perp l_B^2}{\sin(\theta_{xn\times} - \theta_{yn\times})} d\phi_n
\times \Gamma(t_n) dt_n 
= 
P(x,y) \times 
\frac{l_B^2}{\displaystyle \frac{c_{yn\times}}{r_{xn}^{(s)}} - \frac{c_{xn\times}}{r_{yn}^{(s)}}} d\phi_n
\times \Gamma(t_n) dt_n,
\label{eq:EnvSingledP}
\end{align}
with $r_{xn}^{(s)} \equiv r_B(t_x,t_n)$ and $r_{yn}^{(s)} \equiv r_B(t_y,t_n)$, 
and $r_\perp = r_{xn}^{(s)} s_{xn\times} = r_{yn}^{(s)} s_{yn\times}$ denotes 
the distance from $P$ to the line $O_xO_y$.
See Fig.~\ref{fig:Circle}.
Here the factor $r_\perp l_B^2 d\phi_n / \sin(\theta_{xn\times} - \theta_{yn\times})$ 
comes from the infinitesimal three-dimensional volume 
of the intersection of $\delta V_{xy}$ with $\Sigma_{t_n}$,
i.e. the volume of the red diamond in Fig.~\ref{fig:Circle} rotated around $\vec{r}$
with an infinitesimal azimuthal angle $d\phi_n$.
The next step is to calculate the value of $T_{ij}(x)T_{kl}(y)$ 
when such an event with probability $dP^{(s)}$ occurs.
From the expression for the energy-momentum tensor (\ref{eq:TB})--(\ref{eq:rhoBNoColl}), it is given by
\begin{align}
\left[
T_{ij}(x)T_{kl}(y)
\right]^{(s)}
&= 
\left( \frac{4\pi}{3}r_{xn}^{(s)3} \kappa \rho_0\frac{1}{4\pi r_{xn}^{(s)2}l_B} \right)
\left( \frac{4\pi}{3}r_{yn}^{(s)3} \kappa \rho_0\frac{1}{4\pi r_{yn}^{(s)2}l_B} \right)
N_{n\times,ijkl},
\label{eq:EnvSingleTT}
\end{align}
with $N_{n\times,ijkl}$ being $N_{n,ijkl}$ when $P_x$ and $P_y$ are identical:
see Eq.~(\ref{eq:Ntimes}).
From these expressions, the correlator is obtained by the summation over $t_n$ as
\begin{align}
\left< T_{ij}T_{kl} \right>^{(s)}(t_x,t_y,r)
&= 
\int_{t_n = -\infty}^{t_n = t_{\rm max}} dP^{(s)} \; 
\left[ 
T_{ij}(x)T_{kl}(y) 
\right]^{(s)}.
\label{eq:EnvSingledPTT}
\end{align}
Now everything is straightforward.
Let us first calculate the unequal-time power spectrum $\Pi^{(s)}$ in Eq.~(\ref{eq:Pi}),
and then the GW spectrum $\Delta^{(s)}$ in Eq.~(\ref{eq:DeltaPi}).
We can perform the angular integration in Eq.~(\ref{eq:Pi}) with the formula (\ref{eq:AppKN}).
In the present case, we substitute the special value 
$n_{x\times}$ and $n_{y\times}$ into $n_x$ and $n_y$,
and therefore the polar angles are given by Eq.~(\ref{eq:cos}) 
and $\phi_x - \phi_y = \phi_\times - \phi_\times = 0$.
After integrating out the angular direction of $\vec{r}$, we obtain
\begin{align}
&\Pi^{(s)}(t_x,t_y,k)
\nonumber \\
&= 
\int_{v|t_{x,y}|}^\infty dr 
\int_{-\infty}^{t_{\rm max}}dt_n
\nonumber \\[1ex]
&\;\;\;\;
\frac{4\pi^2}{9}\kappa^2\rho_0^2 
\left[
\begin{matrix*}[l]
\displaystyle
\;
e^{-I(x,y)}
\Gamma(t_n) 
r r_{xn}^{(s)2}r_{yn}^{(s)2}
\\[1ex]
\displaystyle
\times 
\left[
j_0(kr){\mathcal K}_0(n_{xn\times},n_{yn\times})
+ \frac{j_1(kr)}{kr}{\mathcal K}_1(n_{xn\times},n_{yn\times})
+ \frac{j_2(kr)}{(kr)^2}{\mathcal K}_2(n_{xn\times},n_{yn\times})
\right]
\end{matrix*}
\right],
\label{eq:EnvPiSGeneral}
\end{align}
where $j_n$ are the spherical Bessel functions given in Eqs.~(\ref{eq:j})
and the explicit forms of $\mathcal{K}_n$ are given in Appendix~\ref{app:useful}.
The lower bound of the integration region for $r$
is because the intersection $\delta V_{xy}$ does not exist for $r_v \equiv r/v < |t_{x,y}|$.
Though at this stage we can perform the integration by $t_n$ for the nucleation rate (\ref{eq:Gamma}),
we leave it general for a while.
Substituting this into Eq.~(\ref{eq:DeltaPi}),
we obtain the single-bubble spectrum for a general nucleation rate
\begin{align}
\Delta^{(s)}
= 
&
\int_{-\infty}^\infty dt_x 
\int_{-\infty}^\infty dt_y
\int_{v|t_{x,y}|}^\infty dr 
\int_{-\infty}^{t_{\rm max}}dt_n 
\nonumber \\[1ex]
\frac{k^3}{3}
&\left[
\begin{matrix*}[l]
\displaystyle
\; 
e^{-I(x,y)}
\Gamma(t_n) 
r r_{xn}^{(s)2}r_{yn}^{(s)2}
\\[1ex]
\displaystyle
\times
\left[
j_0(kr){\mathcal K}_0(n_{xn\times},n_{yn\times})
+ \frac{j_1(kr)}{kr}{\mathcal K}_1(n_{xn\times},n_{yn\times})
+ \frac{j_2(kr)}{(kr)^2}{\mathcal K}_2(n_{xn\times},n_{yn\times})
\right]
\cos(kt_{x,y})
\end{matrix*}
\right].
\label{eq:EnvDeltaSGeneral}
\end{align}
Here we have substituted $t_{\rm start}, t_{\rm end} = \mp \infty$.

Finally let us consider the nucleation rate (\ref{eq:Gamma}).
Given this specific form, we can perform $t_{\left< x,y \right>}$ and $t_n$ integrations
to obtain the GW spectrum as
\begin{align}
\Delta^{(s)}
&= 
\int_{-\infty}^\infty dt_{x,y} \int_{|t_{x,y}|}^\infty dr_v \;
v^3k^3 
\frac{e^{-\frac{r_v}{2}}}{{\mathcal I}(t_{x,y},r_v)}
\left[
j_0(vkr_v){\mathcal S}_0 
+ \frac{j_1(vkr_v)}{vkr_v}{\mathcal S}_1 
+ \frac{j_2(vkr_v)}{(vkr_v)^2}{\mathcal S}_2
\right]
\cos(kt_{x,y}).
\label{eq:EnvDeltaS}
\end{align}
Here we have changed the integration variable from $r$ to $r_v \equiv r/v$.
Also, ${\mathcal S}_n$ functions are
\begin{align}
{\mathcal S}_0
&= 
\frac{2}{3}\frac{(t_{x,y}^2 - r_v^2)^2}{r_v^3}(r_v^2 + 6r_v + 12), 
\nonumber \\
{\mathcal S}_1
&= 
\frac{2}{3}\frac{t_{x,y}^2 - r_v^2}{r_v^3}
\left[
- t_{x,y}^2(r_v^3 + 12r_v^2 + 60r_v + 120)
+ r_v^2(r_v^3 + 4r_v^2 + 12r_v + 24)
\right],
\nonumber \\
{\mathcal S}_2
&= 
\frac{1}{6}
\frac{1}{r_v^3}
\left[
t_{x,y}^4(r_v^4 + 20r_v^3 + 180r_v^2 + 840r_v + 1680)
\right.
\nonumber \\
&\;\;\;\;\;\;\;\;\;\;\;\;\;\;\;\;\;\;\;\;\;
\left.
\frac{}{}
- 2t_{x,y}^2r_v^2(r_v^4 + 12r_v^3 + 84r_v^2 + 360r_v + 720)
\right.
\nonumber \\
&\;\;\;\;\;\;\;\;\;\;\;\;\;\;\;\;\;\;\;\;\;\;\;\;\;\;\;\;\;\;\;\;\;\;\;\;\;\;\;\;\;\;\;\;\;\;
\left.
+ r_v^4(r_v^4 + 4r_v^3 + 20r_v^2 + 72r_v + 144)
\right],
\label{eq:EnvCalS}
\end{align}
and we have used $\int_{-\infty}^\infty dY e^{-Xe^Y + nY} = (n - 1)!/X^n$ 
for $t_{\left< x,y \right>}$ integration.

The exponential factor $e^{-\frac{r_v}{2}}$ in Eq.~(\ref{eq:EnvDeltaS}) has a physical origin.
For the nucleation rate (\ref{eq:Gamma}), 
there is a typical time when spacial points experience the transition 
from the false to true vacuum.
Fig.~\ref{fig:Envelope} roughly corresponds to such a time.
This typical transition time is also a typical time for bubble collisions,
as seen from this figure.
Let us set the evaluation times $t_x$ and $t_y$ around this typical collision time,
because GWs are mainly sourced around this time in the envelope approximation.
If we take spacial evaluation points $\vec{x}$ and $\vec{y}$ far separated with each other,
the $xy$-bubble has to nucleate well before the typical collision time.
Probabilities for bubble nucleation at such early times are suppressed by $e^{-r_v/2}$,
and this is the origin for the exponential factor in Eq.~(\ref{eq:EnvDeltaS}).

For the numerical evaluation of Eq.~(\ref{eq:EnvDeltaS}), see Ref.~\cite{Jinno:2016vai}.
It is shown that $\Delta^{(s)}$ is proportional to $k^3$ for $k / \beta \lesssim 1 / v$,
while it behaves as $\propto k^{-1}$ for $k / \beta \gtrsim 1 / v$.

%%%%%%%%%%%%%%%%%%%%%%%%%%%%%%%%%%%%%%%%%%%%%%%%%%
\subsection{Double-bubble spectrum}
\label{subsec:EnvDouble}
%%%%%%%%%%%%%%%%%%%%%%%%%%%%%%%%%%%%%%%%%%%%%%%%%%

%%%%%%%%%%%%%%%%%%%%%%%%%%%%%%%%%%%%%%%%%%%%%%%%%%
\subsubsection*{Conditions for bubble configurations}
%%%%%%%%%%%%%%%%%%%%%%%%%%%%%%%%%%%%%%%%%%%%%%%%%%

For the double-bubble spectrum, the procedure is basically the same as the single-bubble case.
The necessary and sufficient conditions for the walls of two different bubbles to
contribute to the energy-momentum tensor at $x$ and $y$ are summarized as
\begin{itemize}
\item
No bubbles nucleate in $V_{xy}$.
\item
One bubble nucleates in $\delta V_x^{(y)}$,
and another nucleates in $\delta V_y^{(x)}$.
\end{itemize}
Note that the second condition is different from the single-bubble case.
It means that one bubble nucleates in the left blue-shaded region
in Fig.~\ref{fig:LightCone}, 
and another nucleates in the right blue-shaded region.
As mentioned in the beginning of this section,
we call the former and latter bubbles $x$- and $y$-bubbles, respectively.
Since they nucleate independently of each other, 
we denote the nucleation times for these bubbles by $t_{xn}$ and $t_{yn}$, respectively, 
and consider constant-time hypersurfaces $\Sigma_{t_{xn}}$ and $\Sigma_{t_{yn}}$ 
separately when we discuss the nucleation process.
We again label all the quantities introduced around Eqs.~(\ref{eq:rxry})--(\ref{eq:cos}) 
by ``$n$" to denote that these quantities are on these hypersurfaces.

%%%%%%%%%%%%%%%%%%%%%%%%%%%%%%%%%%%%%%%%%%%%%%%%%%
\subsubsection*{Expression for GW spectrum}
%%%%%%%%%%%%%%%%%%%%%%%%%%%%%%%%%%%%%%%%%%%%%%%%%%

Let us consider the probability $dP^{(d)}$ (``$d$" denotes ``double")
where no bubbles nucleate in $\delta V_{xy}$
and one bubble nucleates in each of $\delta V_x^{(y)}$ and $\delta V_y^{(x)}$
in time intervals $dt_{xn}$ and $dt_{yn}$ 
within infinitesimal angular intervals $dc_{xn}d\phi_{xn}$ and $dc_{yn}d\phi_{yn}$, respectively.
Such a probability is given by
\begin{align}
dP^{(d)}
= 
P(x,y) 
&\times 
r_{xn}^{(d)2} l_B \Gamma(t_{xn})dt_{xn} dc_{xn} d\phi_{xn}
\times
r_{yn}^{(d)2} l_B \Gamma(t_{yn})dt_{yn} dc_{yn} d\phi_{yn}.
\label{eq:EnvDobledP}
\end{align}
Here we have defined $r_{xn}^{(d)} \equiv r_B(t_x,t_{xn})$ and $r_{yn}^{(d)} \equiv r_B(t_y,t_{yn})$.
The expression (\ref{eq:EnvDobledP}) holds only for $t_{xn} < t_x$ and $t_{yn} < t_y$,
because otherwise there is no allowed region for bubble nucleation
(i.e. the hypersurface $\Sigma_{t_{xn}}$ or $\Sigma_{t_{yn}}$ 
does not intersect with $V_x$ or $V_y$).
Also, the integration regions for $c_{xn}$ or $c_{yn}$ depend on 
whether the nucleation time is before or after $t_{\rm max}$.
For the former case, the allowed region 
(the blue-shaded region in Figs.~\ref{fig:LightCone} and \ref{fig:Circle})
does not form a complete sphere,
and therefore the cosines must satisfy $c_{xn} > c_{x\times}$ or $c_{yn} < c_{y\times}$.
For the latter case, such region forms a complete sphere 
and $c_{xn}$ or $c_{yn}$ is allowed to take any value in $[-1,1]$.
When such an event with probability $dP^{(d)}$ occurs,
the expression for the product of the energy-momentum tensor at the evaluation points $x$ and $y$ becomes
\begin{align}
\left[
T_{ij}(x)T_{kl}(y)
\right]^{(d)}
&= 
\left( \frac{4\pi}{3}r_{xn}^{(d)3} \kappa \rho_0\frac{1}{4\pi r_{xn}^{(d)2}l_B} \right)
\left( \frac{4\pi}{3}r_{yn}^{(d)3} \kappa \rho_0\frac{1}{4\pi r_{yn}^{(d)2}l_B} \right)
N_{n,ijkl}.
\label{eq:EnvTTDouble}
\end{align}
Here $N_{n,ijkl} \equiv (n_{xn})_i(n_{xn})_j(n_{yn})_k(n_{yn})_l$ 
with $n_{xn}$ and $n_{yn}$ being $n_x$ and $n_y$ 
on the constant-time hypersurface $\Sigma_{t_{xn}}$ and $\Sigma_{t_{yn}}$, respectively.
Then, by summing up all the contributions from various nucleation time,
we get the following expression for the correlator of the energy-momentum tensor
\begin{align}
\left< T_{ij}T_{kl} \right>^{(d)}(t_x,t_y,r)
&= 
\int_{t_{xn} = -\infty}^{t_{xn} = t_x} 
\int_{t_{yn} = -\infty}^{t_{yn} = t_y}
dP^{(d)} \; 
\left[ 
T_{ij}(x)T_{kl}(y) 
\right]^{(d)}.
\label{eq:EnvDoubledPTT}
\end{align}

\begin{figure}
\begin{center}
\includegraphics[width=0.55\columnwidth]{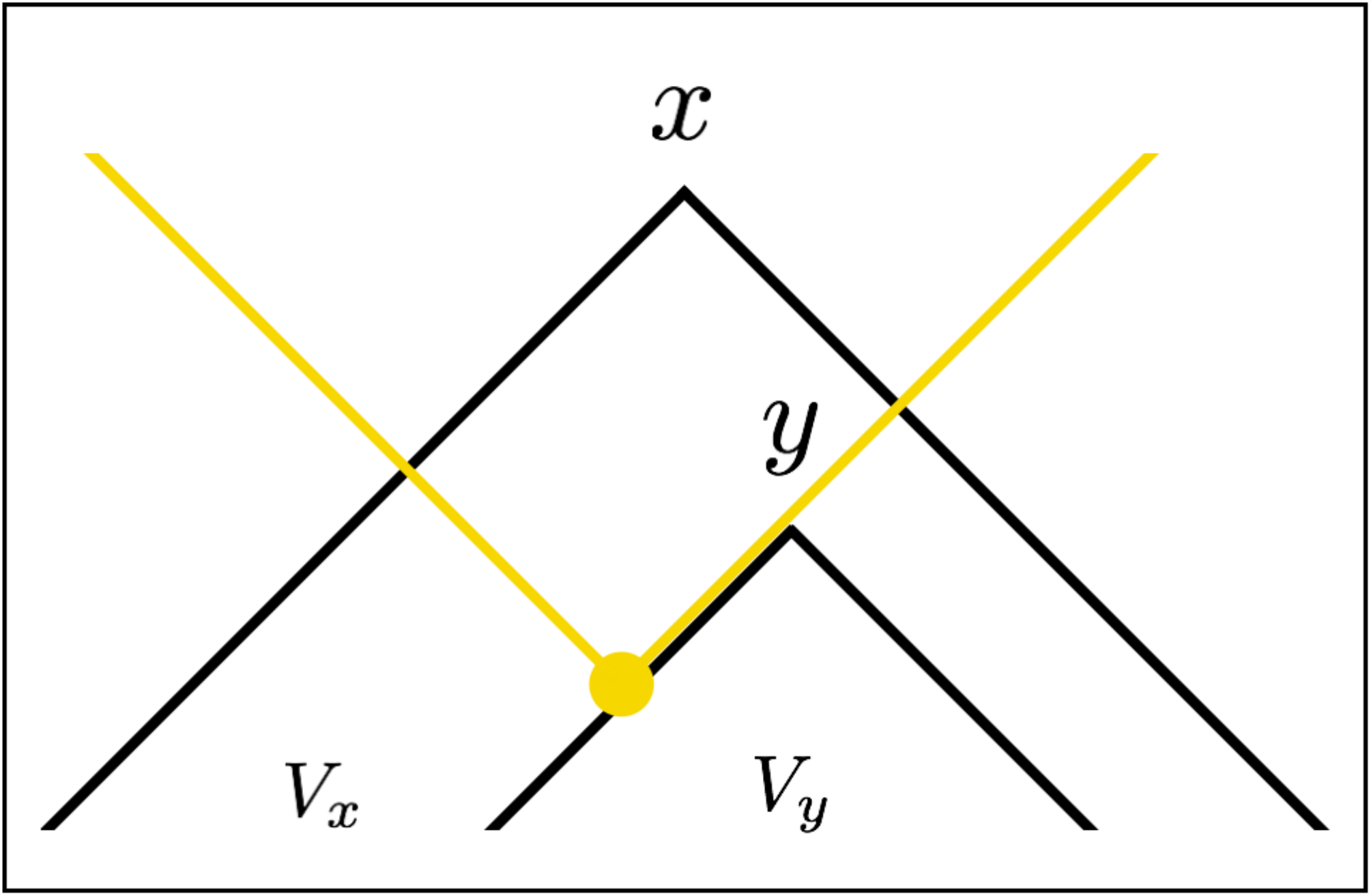}
\caption{\small
Illustration for why we do not have to take the integration region 
$r_v < t_{x,y}$ into account in the envelope approximation.
The horizontal and vertical axes denote the space and time direction, respectively.
For $r_v < t_{x,y}$, the wall of the bubble nucleated on the past $v$-cone of $y$
passes through $\vec{x}$ before the evaluation time $t_x$,
and hence there can be no bubble wall at $\vec{x}$ at time $t_x$.
}
\label{fig:Smallr}
\end{center}
\end{figure}

\begin{figure}
\begin{center}
\includegraphics[width=0.6\columnwidth]{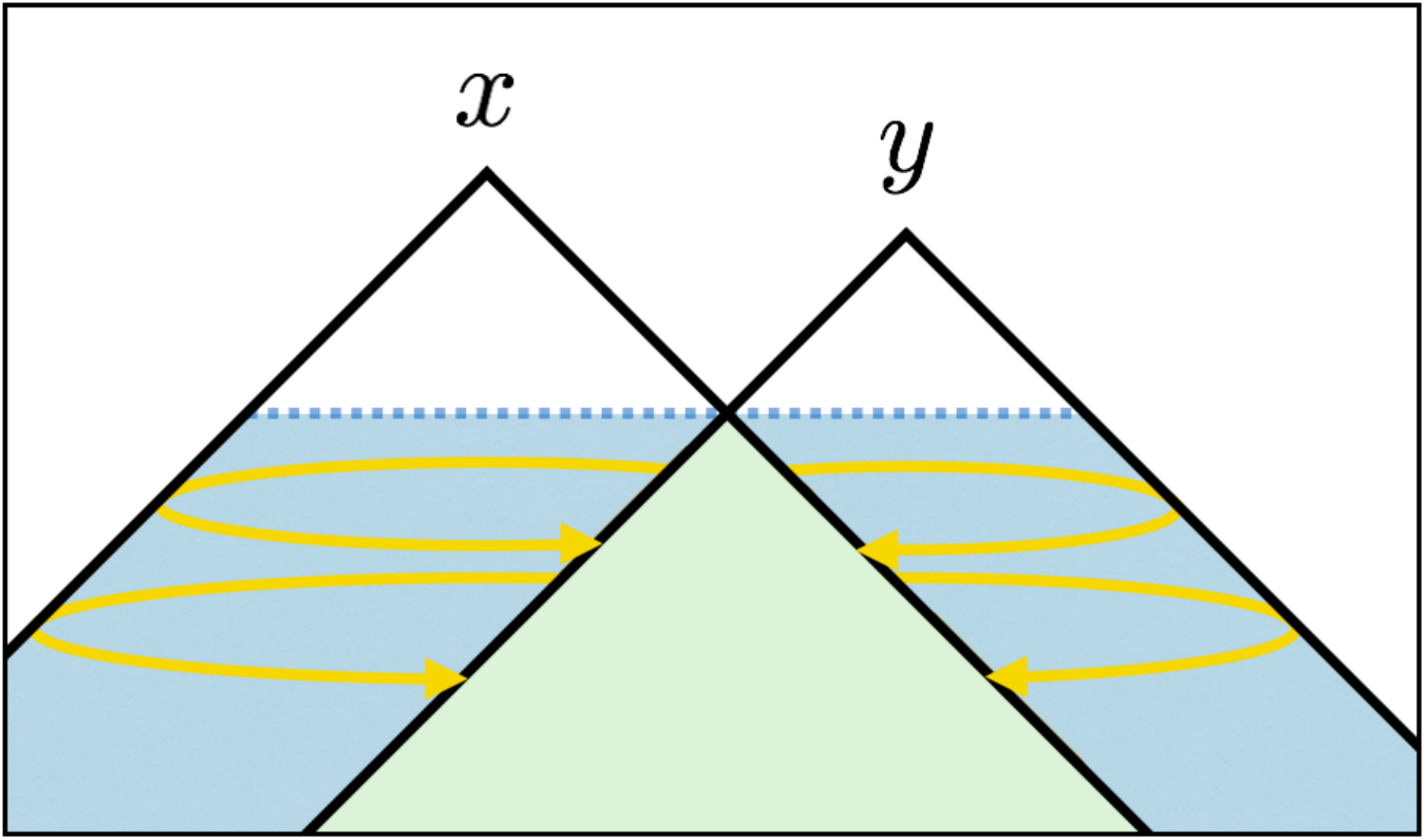}
\caption{\small
Illustration for the integration regions for $t_{xn}$ and $t_{yn}$,
and also for $c_{xn}$ and $c_{yn}$, in Eq.~(\ref{eq:EnvPiDGeneralOnTheWay}).
Integration regions for $t_{xn}$ and $t_{yn}$ are restricted below the blue-dotted line
because contributions from above this line vanish due to the spherical symmetry.
For $t_{xn}$ and $t_{yn}$ below this line,
the integration regions for $c_{xn}$ and $c_{yn}$ are restricted only on the yellow arrows.
}
\label{fig:EnvDouble}
\end{center}
\end{figure}

Taking Eqs.~(\ref{eq:EnvDobledP}) and (\ref{eq:EnvTTDouble}) into account, one sees that 
the integrations with respect to the $x$ bubble and $y$ bubble factorize
and therefore can be done separately.
Substituting Eq.~(\ref{eq:EnvDoubledPTT}) into Eq.~(\ref{eq:Pi}), and using Eq.~(\ref{eq:AppKN}),
we have
\begin{align}
&\Pi^{(d)}(t_x,t_y,k)
= 
\int_{v|t_{x,y}|}^\infty dr
\int_{-\infty}^{t_{\rm max}} dt_{xn}
\int_{-\infty}^{t_{\rm max}} dt_{yn}
\int_{c_{xn\times}}^1 dc_{xn}
\int_{-1}^{c_{yn\times}} dc_{yn}
\int_0^{2\pi} d\phi_{xn,yn}
\nonumber \\
&\;\;\;\;\;\;\;\;\;\;\;\;\;\;\;\;\;\;\;\;
\frac{4\pi^2}{9} \kappa^2\rho_0^2
\left[
\begin{matrix*}[l]
\;
\displaystyle
e^{-I(x,y)}
\Gamma(t_{xn})\Gamma(t_{yn}) \;
r^2r_{xn}^{(d)3}r_{yn}^{(d)3}
\\[1ex]
\displaystyle
\times
\left[
j_0(kr){\mathcal K}_0(n_{xn},n_{yn})
+ \frac{j_1(kr)}{kr}{\mathcal K}_1(n_{xn},n_{yn})
+ \frac{j_2(kr)}{(kr)^2}{\mathcal K}_2(n_{xn},n_{yn})
\right]
\end{matrix*}
\right].
\label{eq:EnvPiDGeneralOnTheWay}
\end{align}
Here $\phi_{xn,yn} \equiv \phi_{xn} - \phi_{yn}$,
and we have integrated out $\phi_{\left< xn,yn \right>} \equiv (\phi_{xn} + \phi_{yn})/2$ direction
by using Eq.~(\ref{eq:AppKN}).
Below we explain the integration ranges in this expression.
First, the lower bound for the integration region for $r$ comes from the following argument.
Under the envelope approximation,
there is no bubble configuration where 
bubble walls contribute to the energy-momentum tensor at both $x$ and $y$
if $r_v < t_{x,y}$ is satisfied.
We illustrate this point in Fig.~\ref{fig:Smallr}.
If $y$ is inside the past $v$-cone of $x$,
the evaluation point $\vec{x}$ is already inside the $y$-bubble
before the evaluation time $t_x$.
In other words,
some fragment of the $y$-bubble has already passed through $\vec{x}$ before $t_x$.
Since any spacial point is passed through by bubble walls only once in the envelope approximation,
there cannot be any wall fragment at $\vec{x}$ at time $t_x$.
The same argument holds when $x$ is inside the past $v$-cone of $y$,
and therefore we can safely restrict the integration region to $r_v > t_{x,y}$.
Second, we explain $t_{xn}$ and $t_{yn}$ integrations.
We have removed 
$[t_{\rm max},t_x]$ and $[t_{\rm max},t_y]$
from their integration regions
(note that $t_{\rm max}$ exists because
$x$ and $y$ are now taken to be spacelike from the first argument).
This is because the contributions from 
$[t_{\rm max},t_x]$ and $[t_{\rm max},t_y]$
vanish because of the spherical symmetry.
See Fig.~\ref{fig:EnvDouble} for illustration.
The removed regions correspond to above the blue-dotted line.
Bubbles which nucleate in these regions contribute to the energy-momentum tensor 
at the evaluation points from every direction, 
and such contributions vanish after summing over the whole solid angle.
In contrast, for $t_{xn}$ and $t_{yn}$ in $[-\infty,t_{\rm max}]$,
only limited directions are allowed for nucleation
for both $x$- and $y$-bubbles, as shown in the yellow lines. 
These allowed directions correspond to the blue-shaded regions in Fig.~\ref{fig:Circle}.
This explains the lower (upper) bound $c_{xn\times}$ ($c_{yn\times}$) 
for the $c_{xn}$ ($c_{yn}$) integration.

%%%%%%%%%%%%%%%%
%%%%%%%%%%%%%%%%

%%%%%%%%%%%%%%%%
%%%%%%%%%%%%%%%%

We can proceed further by integrating out the angular variables.
First, note that $\phi_{xn,yn}$ appears only in ${\mathcal K}$ functions.
After integrating them out, one sees that only the term proportional 
to the first term in Eq.~(\ref{eq:AppK2}) survives.
It has a form like $(3c_{xn}^2 - 1)(3c_{yn}^2 - 1)$, 
and the variables $c_{xn}$ and $c_{yn}$ appear only in this part.
Therefore we can complete the angular integration and obtain
\begin{align}
&\Pi^{(d)}(t_x,t_y,k)
\nonumber \\
&= 
\int_{v|t_{x,y}|}^\infty dr
\int_{-\infty}^{t_{\rm max}} dt_{xn}
\int_{-\infty}^{t_{\rm max}} dt_{yn} \;
\frac{16\pi^3}{9} \kappa^2\rho_0^2
\left[
\begin{matrix*}[l]
\displaystyle
\;
e^{-I(x,y)}
\Gamma(t_{xn})\Gamma(t_{yn}) \;
r^2r_x^{(d)3}r_y^{(d)3}
\\[1ex]
\displaystyle
\times
(c_{xn\times}^3 - c_{xn\times})
(c_{yn\times} - c_{yn\times}^3) \;
\frac{j_2(kr)}{(kr)^2}
\end{matrix*}
\right].
\label{eq:EnvPiDGeneral}
\end{align}
Given this expression, it is straightforward to construct the spectrum $\Delta^{(d)}$
by using Eq.~(\ref{eq:DeltaPi}):
\begin{align}
\Delta^{(d)}
= 
&
\int_{-\infty}^\infty dt_x 
\int_{-\infty}^\infty dt_y
\int_{v|t_{x,y}|}^\infty dr
\int_{-\infty}^{t_{\rm max}} dt_{xn}
\int_{-\infty}^{t_{\rm max}} dt_{yn}
\nonumber \\
&\frac{4\pi}{3}k^3 
\left[
\begin{matrix*}[l]
\displaystyle
\;
e^{-I(x,y)}
\Gamma(t_{xn})\Gamma(t_{yn}) \;
r^2r_x^{(d)3}r_y^{(d)3}
\\[1ex]
\displaystyle
\times
(c_{xn\times}^3 - c_{xn\times})
(c_{yn\times} - c_{yn\times}^3) \;
\frac{j_2(kr)}{(kr)^2}
\cos(kt_{x,y})
\end{matrix*}
\right].
\label{eq:EnvDeltaDGeneral}
\end{align}
These (\ref{eq:EnvPiDGeneral})--(\ref{eq:EnvDeltaDGeneral}) 
are general expressions for the correlator and the GW spectrum 
with an arbitrary nucleation rate.

Finally let us take the nucleation rate (\ref{eq:Gamma}) into account.
We can perform $t_{xn}$ and $t_{yn}$ integrations in $\Pi^{(d)}$,
and also $t_{\left< x,y \right>}$ integration in $\Delta^{(d)}$ by using 
$\int_{-\infty}^\infty dY e^{-Xe^Y + nY} = (n - 1)!/X^n$.
As a result, we have
\begin{align}
\Delta^{(d)}
&= 
\int_{-\infty}^\infty dt_{x,y} 
\int_{|t_{x,y}|}^\infty dr_v \;
v^3k^3\frac{e^{-r_v}}{{\mathcal I}(t_{x,y},r_v)^2}
\frac{j_2(vkr_v)}{(vkr_v)^2} \;
{\mathcal D}(t_{x,y},r_v){\mathcal D}(- t_{x,y},r_v)
\cos(kt_{x,y}).
\label{eq:EnvDeltaD}
\end{align}
Here ${\mathcal D}$ function is defined as
\begin{align}
{\mathcal D}(t_{x,y},r_v)
&= 
\sqrt{\frac{\pi}{3}}
\frac{t_{x,y}^2 - r_v^2}{r_v^2}
\left[
t_{x,y}(r_v^2 + 6r_v + 12)
+ (r_v^3 + 2r_v^2)
\right].
\end{align}
The exponential factor $e^{-r_v}$ in Eq.~(\ref{eq:EnvDeltaD}) has a physical origin.
To see this, let us consider the evaluation times $t_x$ and $t_y$ around the typical collision time.
In order for the $x$- and $y$-bubbles to contribute to the GW spectrum,
they must nucleate before $t_{\rm max}$.
See Fig.~\ref{fig:EnvDouble} and note that the past $v$-cones of $x$ and $y$ 
have an overlap only before $t_{\rm max}$.
For a large separation between the evaluation points $\vec{x}$ and $\vec{y}$,
these bubbles must nucleate well before the evaluation times,
and such probabilities are exponentially suppressed.
This is the origin of the exponential factor in Eq.~(\ref{eq:EnvDeltaD}).

For the numerical evaluation of Eq.~(\ref{eq:EnvDeltaD}), see Ref.~\cite{Jinno:2016vai}.
It is shown that $\Delta^{(d)}$ is smaller than $\Delta^{(s)}$ for all $k$.
In particular, 
$\Delta^{(d)}$ shows a rapid decrease $\propto k^{-2}$ for $k / \beta \gtrsim 1$
in the luminal case $v = c$,
while it decreases as $\propto k^{-1}$ for other cases.

\clearpage

%%%%%%%%%%%%%%%%%%%%%%%%%%%%%%%%%%%%%%%%%%%%%%%%%%
\section{GW spectrum beyond the envelope approximation}
\label{app:BeyondEnv}
\setcounter{equation}{0}
%%%%%%%%%%%%%%%%%%%%%%%%%%%%%%%%%%%%%%%%%%%%%%%%%%

In this appendix we derive the GW spectrum without the envelope approximation.
The difference from the envelope case is that the bubble walls now 
keep their energy and momentum after collision (see Fig.~\ref{fig:BeyondEnv}).
This collision, which we refer to as ``interception" in the following, makes the evaluation of 
$\left< T(x) T(y) \right>$ much more complicated.

%%%%%%%%%%%%%%%%%%%%%%%%%%%%%%%%%%%%%%%%%%%%%%%%%%
\subsection{Basic strategy}
\label{subapp:BeyondEnvStrategy}
%%%%%%%%%%%%%%%%%%%%%%%%%%%%%%%%%%%%%%%%%%%%%%%%%%

Basic strategy is the same as in the envelope case.
In the present case as well, 
there are two types of contributions: 
\begin{itemize}
\item
Single-bubble
\item
Double-bubble
\end{itemize}
The difference appears in
the assumption on the functional form of the energy-momentum tensor.
It  is given by Eq.~(\ref{eq:TB}),
but $\rho_B$ is now taken to be $\rho_B^{\rm (collided)}$, where
\begin{align}
\rho_B^{({\rm collided})}(x)
&=
\displaystyle 
\left[
\frac{4\pi}{3} r_B(t_{xi},t_{xn})^3 \kappa\rho_0
\Big/ 4\pi r_B(t_{xi},t_{xn})^2 l_B
\right]
\times 
\frac{r_B(t_{xi},t_{xn})^2}{r_B(t_x,t_{xn})^2}
\times 
D(t_x,t_{xi},t_{xn}),
\nonumber \\
&=
\rho_B^{({\rm uncollided})}(x)
\times 
\frac{r_B(t_{xi} ,t_{xn})^3}{r_B(t_x ,t_{xn})^3}
\times
D(t_x,t_{xi},t_{xn}),
\label{eq:rhoBBeyond}
\end{align}
for $r_B(t_x,t_{xn}) < |\vec{x} - \vec{x}_n| < r'_B(t_x,t_{xn})$,
and it vanishes otherwise.
Here $t_{xi}$ denotes the time when the bubble wall fragment is intercepted by 
some other walls for the first time.
Note that different wall fragments have different $t_{xi}$.
The second factor in the R.H.S. of the first line in Eq.~(\ref{eq:rhoBBeyond}) takes into account
the increase of the bubble-wall area and the resulting loss of 
the energy and momentum densities after interception.
Also, $D$ is a ``damping function",
which accounts for how fast collided walls lose their energy and momentum
on top of the loss coming from the increase in their area.
It depends on the underlying particle model
which describes the couplings between the scalar field and light species.
Therefore we take $D$ to be arbitrary in the following, except for the condition $D(t_x,t_{xi} = t_x,t_n)  = 1$.
This is because there is no time for damping if the interception occurs at the evaluation time.
Note that, in Eq.~(\ref{eq:rhoBBeyond}),
we have assumed that the bubble walls are affected only by the first collision
and not by the subsequent collisions.
This is because the energy and momentum in bubble walls are sourced 
by the released energy (latent heat) until their first collisions,
while there is no such energy sourcing after that.
In Fig.~\ref{fig:BeyondEnv}, the collided walls, denoted by gray lines, propagate inside other bubbles
without energy sourcing after their first collisions.

In the following, we first summarize our notation
and then proceed to the calculation of $\left< T(x) T(y) \right>$
from single- and double-bubble contributions in turn.

%%%%%%%%%%%%%%%%%%%%%%%%%%%%%%%%%%%%%%%%%%%%%%%%%%
\subsection{Prerequisites}
\label{subsec:Prerequisites}
%%%%%%%%%%%%%%%%%%%%%%%%%%%%%%%%%%%%%%%%%%%%%%%%%%

%%%%%%%%%%%%%%%%%%%%%%%%%%%%%%%%%%%%%%%%%%%%%%%%%%
\subsubsection*{Labels for spacetime points}
%%%%%%%%%%%%%%%%%%%%%%%%%%%%%%%%%%%%%%%%%%%%%%%%%%

As before, we use
$x = (t_x, \vec{x})$ and $y = (t_y, \vec{y})$ 
for the arguments in $\left< T(x) T(y) \right>$.
The bubble walls which pass through these evaluation points 
must nucleate somewhere before the evaluation times $t_x$ and $t_y$.
We denote these nucleation points as 
$x_n = (t_{xn},\vec{x}_n)$ and $y_n = (t_{yn},\vec{y}_n)$, 
and call these bubbles $x$- and $y$-bubbles, respectively.
The bubble wall fragments experience the interception (the first collision) before the evaluation time
in some cases.
In such cases we denote the interception points
by $x_i = (t_{xi},\vec{x}_i)$ and $y_i = (t_{yi},\vec{y}_i)$.
Note that, as mentioned before, the interception times of two wall fragments are
generally different even if they belong to the same bubble,
and therefore $x_i$ and $y_i$ depend on the bubble itself and 
the direction in which the fragment is propagating.
In summary, our notation is
\begin{itemize}
\item
$n$ : label for nucleation points
\item
$i$ : label for interception points
\end{itemize}
%%

%%%%%%%%%%%%%%%%%%%%%%%%%%%%%%%%%%%%%%%%%%%%%%%%%%
\subsubsection*{Spacelike theta function}
%%%%%%%%%%%%%%%%%%%%%%%%%%%%%%%%%%%%%%%%%%%%%%%%%%

In the following discussion,
we often require two spacetime points to be spacelike.
We impose this condition by inserting the spacelike step function $\Theta_{\rm sp}$ defined by
\begin{align}
\Theta_{\rm sp}(x,y)
&\equiv \Theta(|\vec{x} - \vec{y}|^2/v^2 - (t_x - t_y)^2),
\label{eq:Thetasp}
\end{align}
to the integrands,\footnote{
There are other ways to impose the spacelike condition,
e.g. $\Theta((|\vec{x} - \vec{y}|^2/v^2 - (t_x - t_y)^2)^2)$.
Our results do not depend on how we choose the functional form.
}
where $\Theta$ is the Heaviside theta function.

%%%%%%%%%%%%%%%%%%%%%%%%%%%%%%%%%%%%%%%%%%%%%%%%%%
\subsubsection*{Shorthand notations}
%%%%%%%%%%%%%%%%%%%%%%%%%%%%%%%%%%%%%%%%%%%%%%%%%%

In the calculation below, we often encounter 
time differences and spacial distances between two spacetime points.
Therefore, for brevity, we introduce 
\begin{align}
t_{x\bullet,y\circ}
&\equiv t_{x\bullet} - t_{y\circ},
\;\;\;\;
r_{x\bullet,y\circ}
\equiv |\vec{x}_\bullet - \vec{y}_\circ|.
\end{align}
Also, we define the average as
\begin{align}
t_{\left< x\bullet,y\circ \right>}
&\equiv 
\frac{t_{x\bullet} + t_{y\circ}}{2}.
\end{align}
In addition, we often take time derivatives along the propagation direction of the wall fragment.
We write such a derivative of a quantity $Q$ with respect to $t_\bullet$
and derivatives with respect to $t_\bullet$ and $t_\circ$ as
\begin{align}
\left[ Q \right]_{t\bullet},
\;\;\;\;
\left[ Q \right]_{t\bullet,t\circ},
\end{align}
and so on.
For example, the derivative with respect to $t_{xi}$
along the propagation of the $x$-fragment is written as $\left[ Q \right]_{txi}$.
The derivatives with respect to $t_{xi}$ (along the $x$-fragment) and $t_{yi}$ (along the $y$-fragment) 
is written as $\left[ Q \right]_{txi,tyi}$.
Note that the order of the derivatives with respect to $t_\bullet$ and $t_\circ$ is not important 
in all the following calculations.
We comment on these derivatives below Eqs.~(\ref{eq:BeyondDeltaSConc}) 
and (\ref{eq:BeyondDeltaDConc}) with concrete examples.

%%%%%%%%%%%%%%%%
\begin{figure}
\begin{center}
\includegraphics[width=0.45\columnwidth]{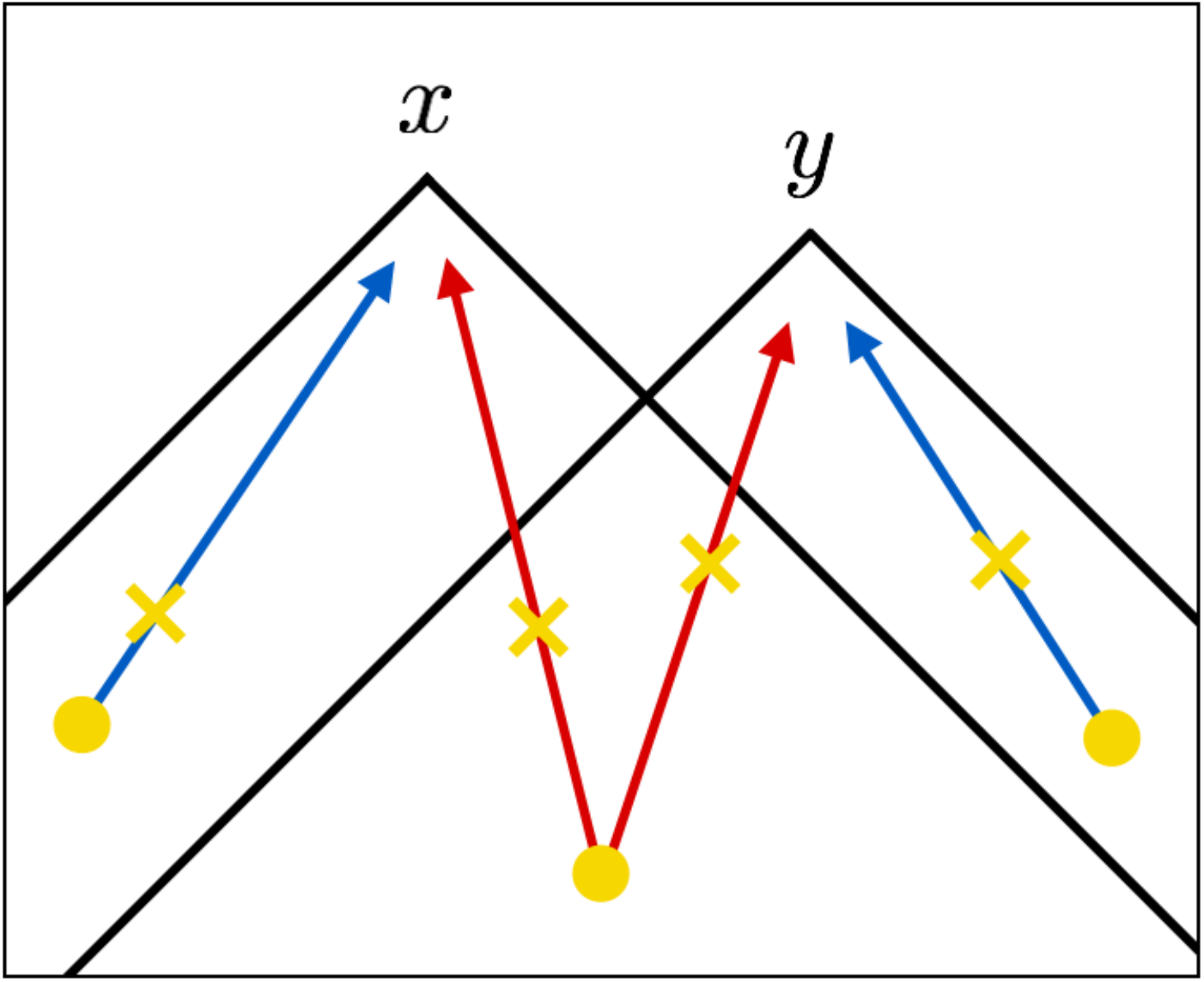}
\includegraphics[width=0.45\columnwidth]{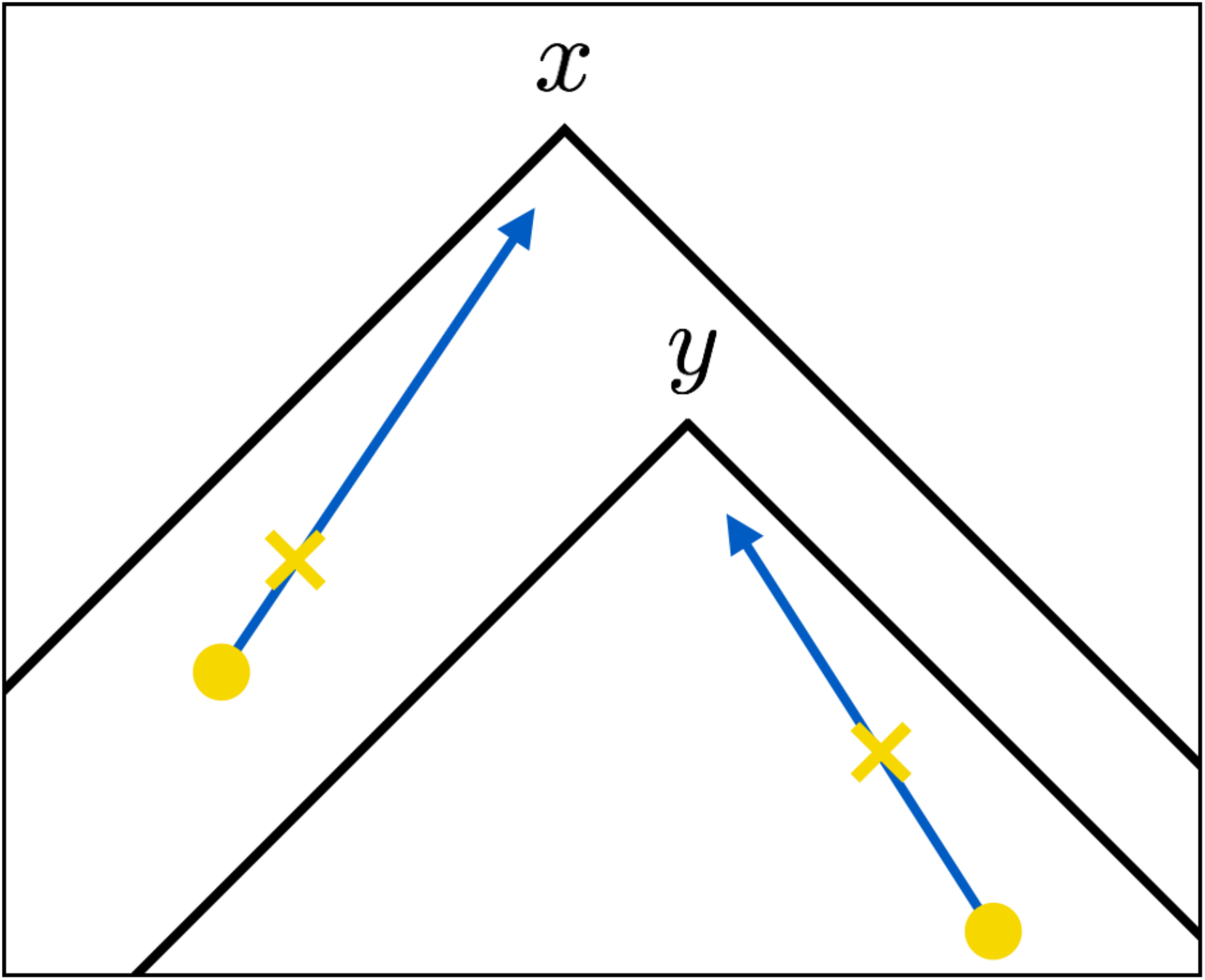}
\caption{\small
Past $v$-cones relevant in the calculation beyond the envelope.
Bubble wall fragments are now subject to the interception dynamics,
which occurs at the points denoted by the yellow crosses.
For the double-bubble contribution (blue)
$x$ and $y$ are not necessarily spacelike,
in contrast to the envelope case.
On the other hand,
for the single-bubble contribution (red),
$x$ and $y$ must be spacelike because otherwise 
the nucleation region $\delta V_{xy}$ does not exist.
See also Fig.~\ref{fig:3DLightConeCept} for 
how the left panel looks in $1 + 2$ dimensions.
}
\label{fig:LightConeCept}
\end{center}
\end{figure}
%%%%%%%%%%%%%%%%

%%%%%%%%%%%%%%%%
\begin{figure}
\begin{center}
\includegraphics[width=0.6\columnwidth]{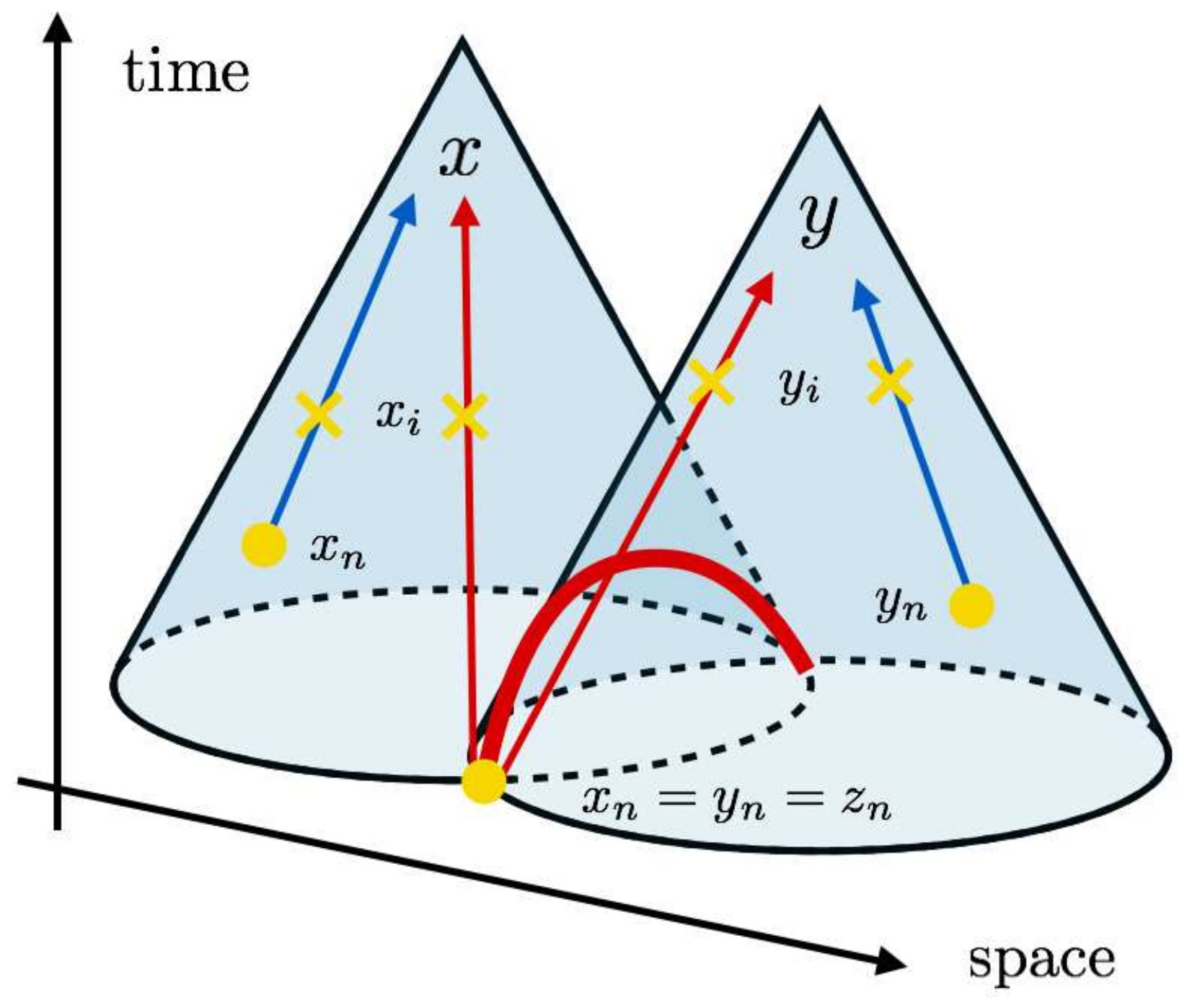}
\caption{\small
How the left panel of Fig.~\ref{fig:LightConeCept} looks in $1 + 2$ dimensions.
}
\label{fig:3DLightConeCept}
\end{center}
\end{figure}
%%%%%%%%%%%%%%%%

%%%%%%%%%%%%%%%%%%%%%%%%%%%%%%%%%%%%%%%%%%%%%%%%%%
\subsection{Single-bubble spectrum}
\label{subapp:BeyondEnvSingle}
%%%%%%%%%%%%%%%%%%%%%%%%%%%%%%%%%%%%%%%%%%%%%%%%%%

%%%%%%%%%%%%%%%%%%%%%%%%%%%%%%%%%%%%%%%%%%%%%%%%%%
\subsubsection*{Classification}
%%%%%%%%%%%%%%%%%%%%%%%%%%%%%%%%%%%%%%%%%%%%%%%%%%

Now we start the calculation of the single-bubble contribution.
See Figs.~\ref{fig:LightConeCept} and \ref{fig:3DLightConeCept}.
Note that $x$- and $y$-fragments come from the same nucleation point, 
and therefore,
we can safely limit ourselves to the spacelike combinations of $(x,y)$.
The single-bubble contribution can be classified into four classes,
depending on whether the fragments have already been intercepted or not
at the evaluation points:
\begin{itemize}
\item
$CC$ : Both $x$ and $y$-fragments have already collided with other bubbles
at the evaluation points $x$ and $y$.
\item
$CU$ : Only $x$-fragment has already collided with other bubbles and $y$-fragment remains uncollided
at the evaluation points $x$ and $y$.
\item
$UC$ : Only $y$-fragment has already collided with other bubbles and $x$-fragment remains uncollided
at the evaluation points $x$ and $y$.
\item
$UU$ : Both $x$ and $y$-fragments remain uncollided at the evaluation points $x$ and $y$.
\end{itemize}
In the following, we see that the single-bubble spectrum reduces to a simple expression 
after summing up all these classes.

%%%%%%%%%%%%%%%%%%%%%%%%%%%%%%%%%%%%%%%%%%%%%%%%%%
\subsubsection*{Simplification formula}
%%%%%%%%%%%%%%%%%%%%%%%%%%%%%%%%%%%%%%%%%%%%%%%%%%

In this subsection we introduce a ``simplification formula",
which makes the sum of the four contributions much simpler.

Let us suppose that the $UU$ contribution to the correlator 
$\left< T(x)T(y) \right>$ is written as
\begin{align}
\left< T_{ij}(x) T_{kl}(y) \right>^{(s,UU)}
&= 
\int_{-\infty}^{t_{\rm max}}dt_n 
\int_0^{2\pi} d\phi_n \;
e^{-I(x,y)} 
{\mathcal F}^{(s)}(x,y;x_i = x,y_i = y;z_n),
\label{eq:BeyondTTSUU}
\end{align}
with some function ${\mathcal F}^{(s)}$.
We denote the common nucleation point for $x$  and $y$-fragments by $z_n = (t_n,\vec{z}_n)$,
which means $x_n = y_n = z_n$.
Also, $\phi_n$ is the azimuthal angle for the nucleation point $\vec{z}_n$ 
with respect to $\vec{r} \equiv \vec{x} - \vec{y}$.
Note that $t_n$ and $\phi_n$ completely specify the nucleation point $z_n$
for given $x$ and $y$.
The explicit expression for ${\mathcal F}^{(s)}$ is found 
in the same way as Appendix~\ref{app:Envelope},
except that the energy-momentum tensor is now given by Eq.~(\ref{eq:TB})
with $\rho_B$ being Eq.~(\ref{eq:rhoBBeyond}):
\begin{align}
&{\mathcal F}^{(s)}(x,y;x_i,y_i;z_n)
\nonumber \\
&\;\;\;\;
= 
\frac{r_\perp l_B^2}{\sin(\theta_{xn\times} - \theta_{yn\times})} \Gamma(t_n)
\left( \frac{4\pi}{3}r_{xn}^{(s)3} \kappa \rho_0\frac{1}{4\pi r_{xn}^{(s)2}l_B} \right)
\left( \frac{4\pi}{3}r_{yn}^{(s)3} \kappa \rho_0\frac{1}{4\pi r_{yn}^{(s)2}l_B} \right)
N_{n\times,ijkl}
\nonumber \\
&\;\;\;\;\;\;\;\;
\times
\frac{r_B(t_{xi},t_n)^3}{r_{xn}^{(s)3}}
\frac{r_B(t_{yi},t_n)^3}{r_{yn}^{(s)3}}
D(t_x,t_{xi},t_n)
D(t_y,t_{yi},t_n).
\label{eq:BeyondCalFS}
\end{align}
Here $r_{xn}^{(s)} = r_B(t_x,t_n)$ and $r_{yn}^{(s)} = r_B(t_y,t_n)$, 
and $r_\perp$ and $N_{n\times,ijkl}$ 
are defined in the same way as Appendix~\ref{subsec:EnvSingle}.
Note that substituting $x_i = x$ and $y_i = y$ reproduces the result with the envelope approximation
(\ref{eq:EnvSingledPTT}).
Then, interestingly, the sum of the four contributions reduces to
\begin{align}
&\left< T(x) T(y) \right>^{(s,UU + UC + CU + CC)}
\nonumber \\
&\;\;\;\;\;\;\;\;
= 
\int_{-\infty}^{t_{\rm max}}dt_n 
\int_0^{2\pi} d\phi_n
\int_{t_{xn}}^{t_x}dt_{xi} 
\int_{t_{yn}}^{t_y}dt_{yi} \;
e^{-I(x_i,y_i)} 
\left[
{\mathcal F}^{(s)}(x,y;x_i,y_i,z_n)
\right]_{txi,tyi}.
\label{eq:BeyondSingleSimple}
\end{align}
We prove this formula in Appendix~\ref{app:Simple}.
Here, the expression $\left[ {\mathcal F} \right]_{txi,tyi}$ 
denotes taking the derivatives with respect to $t_{xi}$ and $t_{yi}$ 
along the propagation of the bubble wall fragments,
as explained in the previous subsection.
In the present case, ${\mathcal F}^{(s)}$ in Eq.~(\ref{eq:BeyondSingleSimple})
contains the argument $x_i = (t_{xi},\vec{x}_i)$ and $y_i = (t_{yi},\vec{y}_i)$.
When taking derivatives with respect to $t_{xi}$ and $t_{yi}$,
we regard $\vec{x}_i$ and $\vec{y}_i$ as functions of them.
This is possible once the propagation directions of $x$  and $y$-fragments are specified,
and in fact they are specified by the variables $t_n$ and $\phi_n$.

%%%%%%%%%%%%%%%%%%%%%%%%%%%%%%%%%%%%%%%%%%%%%%%%%%
\subsubsection*{Expressions for GW spectrum}
%%%%%%%%%%%%%%%%%%%%%%%%%%%%%%%%%%%%%%%%%%%%%%%%%%

Let us derive concrete expressions for $\Pi^{(s)}$ and $\Delta^{(s)}$.
First, for a general nucleation rate $\Gamma$ 
and the damping function $D$, 
we obtain the following for $\Pi^{(s)}$
by using the simplification formula (\ref{eq:BeyondSingleSimple}):
\begin{align}
\Pi^{(s)}(t_x,t_y,k)
= 
&
\int_{v|t_{x,y}|}^\infty dr 
\int_{-\infty}^{t_{\rm max}} dt_n 
\int_{t_n}^{t_x} dt_{xi}
\int_{t_n}^{t_y} dt_{yi}
\nonumber \\[1ex]
\frac{4\pi^2}{9}\kappa^2\rho_0^2
&\left[
\begin{matrix*}[l]
\displaystyle
\;
e^{-I(x_i,y_i)} \;
\Gamma(t_n) \;
\frac{r}{r_{xn}^{(s)}r_{yn}^{(s)}}
\\
\displaystyle
\times 
\left[
j_0(kr){\mathcal K}_0(n_{xn\times},n_{yn\times})
+ \frac{j_1(kr)}{kr}{\mathcal K}_1(n_{xn\times},n_{yn\times})
+ \frac{j_2(kr)}{(kr)^2}{\mathcal K}_2(n_{xn\times},n_{yn\times})
\right]
\\[2.5ex]
\displaystyle
\times \;
\partial_{txi}
\left[
r_B(t_{xi},t_n)^3
D(t_x,t_{xi},t_n)
\right]
\partial_{tyi}
\left[
r_B(t_{yi},t_n)^3
D(t_y,t_{yi},t_n)
\right]
\end{matrix*}
\right]
\label{eq:BeyondPiSConc}
\end{align}
and the following for $\Delta^{(s)}$:
\begin{align}
\Delta^{(s)}
= 
&
\int_{-\infty}^\infty dt_x
\int_{-\infty}^\infty dt_y
\int_{v|t_{x,y}|}^\infty dr 
\int_{-\infty}^{t_{\rm max}} dt_n 
\int_{t_n}^{t_x} dt_{xi}
\int_{t_n}^{t_y} dt_{yi}
\nonumber \\[1ex]
\frac{k^3}{3}
&\left[
\begin{matrix*}[l]
\displaystyle
\;
e^{-I(x_i,y_i)} \;
\Gamma(t_n) \;
\frac{r}{r_{xn}^{(s)}r_{yn}^{(s)}}
\\
\displaystyle
\times 
\left[
j_0(kr){\mathcal K}_0(n_{xn\times},n_{yn\times})
+ \frac{j_1(kr)}{kr}{\mathcal K}_1(n_{xn\times},n_{yn\times})
+ \frac{j_2(kr)}{(kr)^2}{\mathcal K}_2(n_{xn\times},n_{yn\times})
\right]
\\[2.5ex]
\displaystyle
\times \;
\partial_{txi}
\left[
r_B(t_{xi},t_n)^3
D(t_x,t_{xi},t_n)
\right]
\partial_{tyi}
\left[
r_B(t_{yi},t_n)^3
D(t_y,t_{yi},t_n)
\right]
\cos(kt_{x,y})
\end{matrix*}
\right].
\label{eq:BeyondDeltaSConc}
\end{align}
Some comments are in order.
First, the $r_\perp/\sin (\theta_{xn\times} - \theta_{yn\times})$ term in Eq.~(\ref{eq:BeyondCalFS})
has been simplified by using the same transformations as around Eq.~(\ref{eq:EnvSingledP}).
Next, the derivatives with respect to the propagation directions 
in Eq.~(\ref{eq:BeyondSingleSimple}) are now reduced to partial derivatives.
This is because $t_{xi}$ ($t_{yi}$) dependence of $\vec{x}_i$ ($\vec{y}_i$) is 
encoded in Eq.~(\ref{eq:BeyondCalFS}) through $r_B(t_{xi},t_n)$ ($r_B(t_{yi},t_n)$).
After reducing the derivatives to partial derivatives, 
we can perform the integration with respect to $\phi_n$.
Finally, the integration regions are determined as follows.
For $r$, the lower limit is set because $\delta V_{xy}$ does not exist for 
$r < v|t_{xy}|$. See \ref{fig:LightConeCept} and \ref{fig:3DLightConeCept},
and also the explanation at the beginning of Sec.~\ref{subapp:BeyondEnvSingle}.
The upper limit for $t_n$ integration comes from the same reason.
Also, the integration regions for $t_{xi}$ and $t_{yi}$ come from the simplification formula
(\ref{eq:BeyondSingleSimple}).

Here it would be a good exercise to check that these expressions coincide with
the expressions with the envelope approximation (\ref{eq:EnvPiSGeneral}) 
and (\ref{eq:EnvDeltaSGeneral})
if we assume an instant damping of the wall energy.
Such a setup is realized by setting $D(t,t_i,t_n) = \Theta(t_i - t + \epsilon)$
with $\epsilon$ being infinitesimal and positive. 
With this form, $D$ is unity for time $t$ before $t_i + \epsilon$,
while it vanishes after that.
The $t_{xi}$ and $t_{yi}$ derivatives in 
Eqs.~(\ref{eq:BeyondPiSConc}) and (\ref{eq:BeyondDeltaSConc}) 
can act both on $r_B$ and $D$,
but if either of them acts on $r_B$ the integrand vanishes 
because $D$ is unity only for the integration range $t_x - \epsilon < t_{xi} < t_x$ 
or $t_y - \epsilon < t_{yi} < t_y$.
Therefore these derivatives have to act on $D$,
and this gives $\delta(t_{xi} - t_x + \epsilon)$ and $\delta(t_{yi} - t_y + \epsilon)$.
Performing $t_{xi}$ and $t_{yi}$ integrations,
one sees that
Eqs.~(\ref{eq:BeyondPiSConc}) and (\ref{eq:BeyondDeltaSConc}) 
coincide with Eqs.~(\ref{eq:EnvPiSGeneral}) and (\ref{eq:EnvDeltaSGeneral}), respectively.

Let us finally derive the expression for $\Delta^{(s)}$ with 
the damping function (\ref{eq:D}) and the nucleation rate (\ref{eq:Gamma}).
Given these specific forms we can perform one integration,
because we can shift the whole system in the time direction without changing the geometry of bubbles. 
In fact, if we write all the time variables in Eq.~(\ref{eq:BeyondDeltaSConc})
in terms of the difference from the nucleation time $t_n$,
all the effects of time shift in the system appears 
in the exponent $I(x_i,y_i)$ and the nucleation rate $\Gamma(t_n)$.\footnote{
In fact, in Eq.~(\ref{eq:BeyondDeltaSConc}), 
$r_{xn}^{(s)} = vt_{x,n}$ and $r_{yn}^{(s)} = vt_{y,n}$ contain time differences only, 
and ${\mathcal K}$ functions contain angular variables such as 
$s_{xn\times}$ and $c_{xn\times}$, 
which can be written in terms of $r$, $t_{x,n}$ and $t_{y,n}$.
Also, from $r_B(t_{xi},t_n) = vt_{xi,n}$, $r_B(t_{yi},t_n) = vt_{yi,n}$,
$D(t_x,t_{xi},t_n) = e^{-t_{x,xi}/\tau}$ and 
$D(t_y,t_{yi},t_n) = e^{-t_{y,yi}/\tau}$,
one sees that the last line also has time differences only.
}
Therefore we can integrate out the time shift direction and obtain
\begin{align}
\Delta^{(s)}
= 
&
\int_{-\infty}^\infty dt_{x,y}
\int_{|t_{x,y}|}^\infty dr_v
\int_{r_v/2}^\infty dt_{\left< x,y \right>,n}
\int_0^{t_{x,n}} dt_{xi,n}
\int_0^{t_{y,n}} dt_{yi,n}
\nonumber \\[1ex]
\frac{v^3k^3}{3}
&
\left[
\begin{matrix*}[l]
\displaystyle
\;
\frac{e^{-(t_{xi,n} + t_{yi,n})/2}e^{-(t_{x,xi} + t_{y,yi})/\tau}}
{{\mathcal I}(t_{xi,yi},r_{xi,yi}/v)}
\frac{(3t_{xi,n}^2 + t_{xi,n}^3/\tau)(3t_{yi,n}^2 + t_{yi,n}^3/\tau)}{t_{x,n}t_{y,n}}
\\[2ex]
\displaystyle
\times \;
r_v 
\left[
j_0(vkr_v){\mathcal K}_0(n_{xn\times},n_{yn\times})
+ \frac{j_1(vkr_v)}{vkr_v}{\mathcal K}_1(n_{xn\times},n_{yn\times})
+ \frac{j_2(vkr_v)}{(vkr_v)^2}{\mathcal K}_2(n_{xn\times},n_{yn\times})
\right]
\\[2ex]
\displaystyle
\times
\cos(k(t_{x,n} - t_{y,n}))
\end{matrix*}
\right].
\label{eq:BeyondDeltaSConcFinal}
\end{align}
Note that $t_{x,n}$, $t_{y,n}$, $t_{x,xi}$ and $t_{y,yi}$
in this expression should be understood 
in terms of the integration variables as
\begin{align}
t_{x,n} 
&= \frac{t_{x,y}}{2} + t_{\left< x,y \right>,n},
\;\;\;\;
t_{y,n} 
= -\frac{t_{x,y}}{2} + t_{\left< x,y \right>,n},
\;\;\;\;
t_{x,xi} 
= t_{x,n} - t_{xi,n},
\;\;\;\;
t_{y,yi} 
= t_{y,n} - t_{yi,n},
\end{align}
where $t_{\left< x,y \right>,n} = t_{\left< x,y \right>} - t_n$.
In addition, the arguments $t_{xi,yi} = t_{xi,n} - t_{yi,n}$ and $r_{xi,yi} = |\vec{x}_i - \vec{y}_i|$ 
in the ${\mathcal I}$ function are also functions of the integration variables.
In fact, using the relation
$\vec{x}_i - \vec{y}_i = \vec{r} + v(t_{x,xi}n_ {xn\times} - t_{y,yi}n_{yn\times})$
(with the subscript $n$ indicating that the vectors $n_{x\times}$ and $n_{y\times}$ 
are on the constant-time hypersurface $\Sigma_{t_n}$),
one can express $r_{xi,yi}$ as
\begin{align}
\frac{r_{xi,yi}}{v}
&= 
\sqrt{r_v^2 + t_{x,xi}^2 + t_{y,yi}^2
+ 2r_vt_{x,xi}c_{xn\times}
- 2r_vt_{y,yi}c_{yn\times}
- 2 t_{x,xi}t_{y,yi}c_{xyn\times}},
\label{eq:riSingle}
\end{align}
where 
$c_{xn\times}$ and $c_{yn\times}$ are the cosines introduced around Eq.~(\ref{eq:cos}).
In addition, $c_{xyn\times}$ is the cosine of the angle between 
$n_{xn\times}$ and $n_{yn\times}$ given by 
$c_{xyn\times} = c_{xn\times} c_{yn\times} + s_{xn\times} s_{yn\times}$.
Alternatively, using $\vec{x}_i - \vec{y}_i = v(-t_{xi,n}n_ {xn\times} + t_{yi,n}n_{yn\times})$,
one has 
\begin{align}
\frac{r_{xi,yi}}{v}
&= 
\sqrt{t_{xi,n}^2 + t_{yi,n}^2
- 2 t_{xi,n}t_{yi,n}c_{xyn\times}},
\label{eq:riSingle2}
\end{align}
which also gives an expression of $r_{xi,yi}$ in terms of the integration variables.

%%%%%%%%%%%%%%%%%%%%%%%%%%%%%%%%%%%%%%%%%%%%%%%%%%
\subsection{Double-bubble spectrum}
\label{subapp:BeyondEnvDouble}
%%%%%%%%%%%%%%%%%%%%%%%%%%%%%%%%%%%%%%%%%%%%%%%%%%

%%%%%%%%%%%%%%%%%%%%%%%%%%%%%%%%%%%%%%%%%%%%%%%%%%
\subsubsection*{Classification}
%%%%%%%%%%%%%%%%%%%%%%%%%%%%%%%%%%%%%%%%%%%%%%%%%%

\begin{figure}
\begin{center}
\includegraphics[width=0.45\columnwidth]{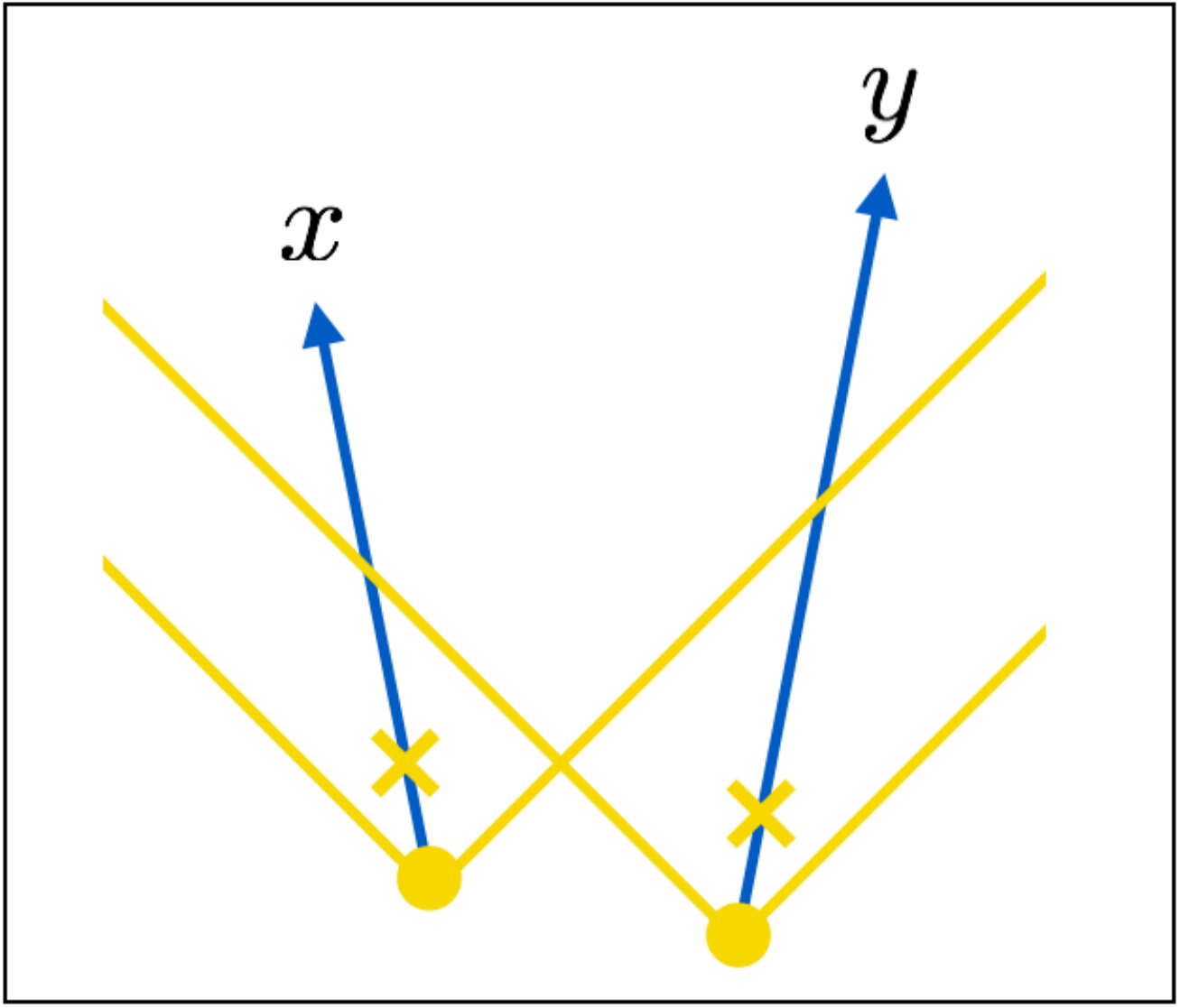}
\includegraphics[width=0.45\columnwidth]{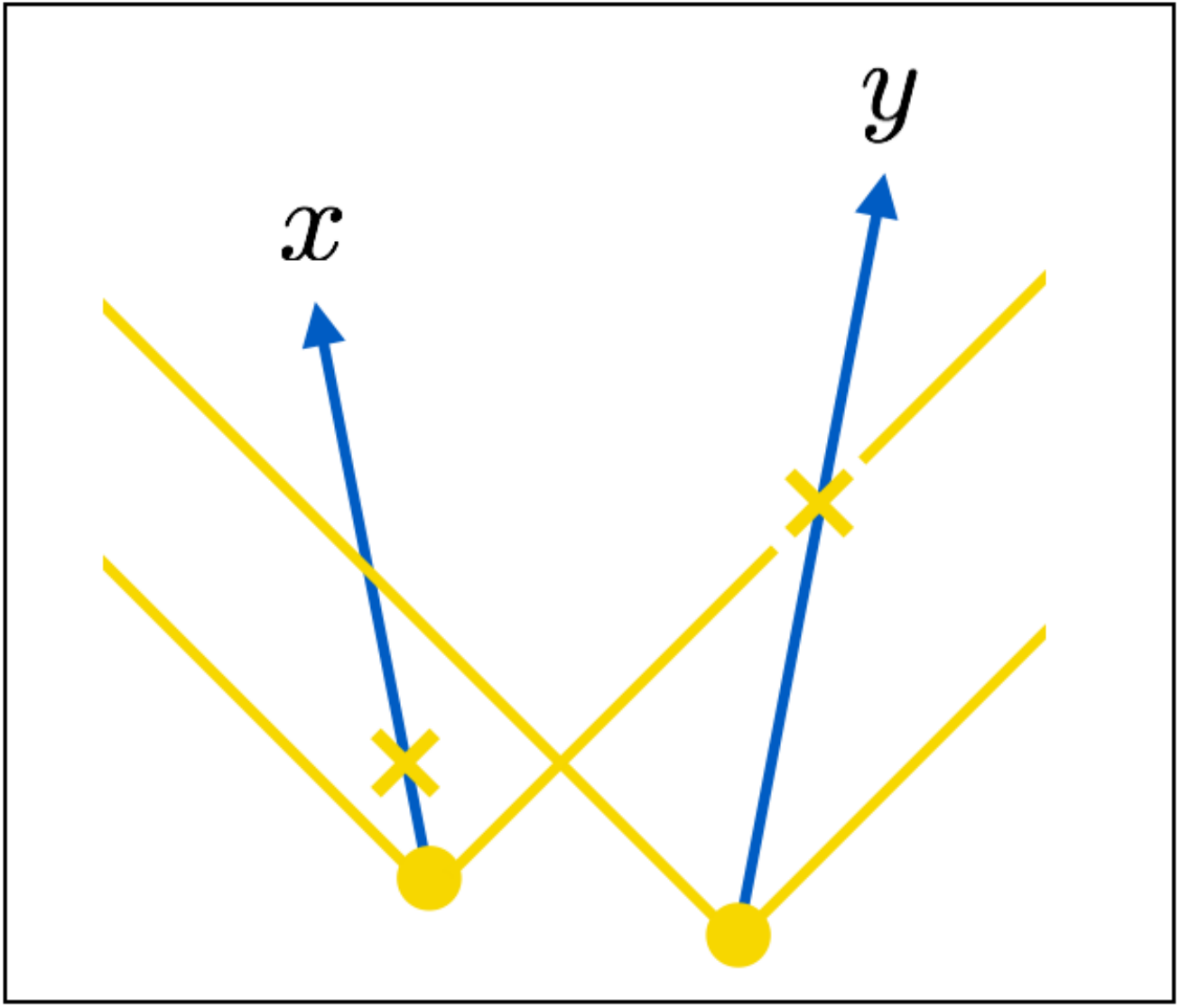}
\includegraphics[width=0.45\columnwidth]{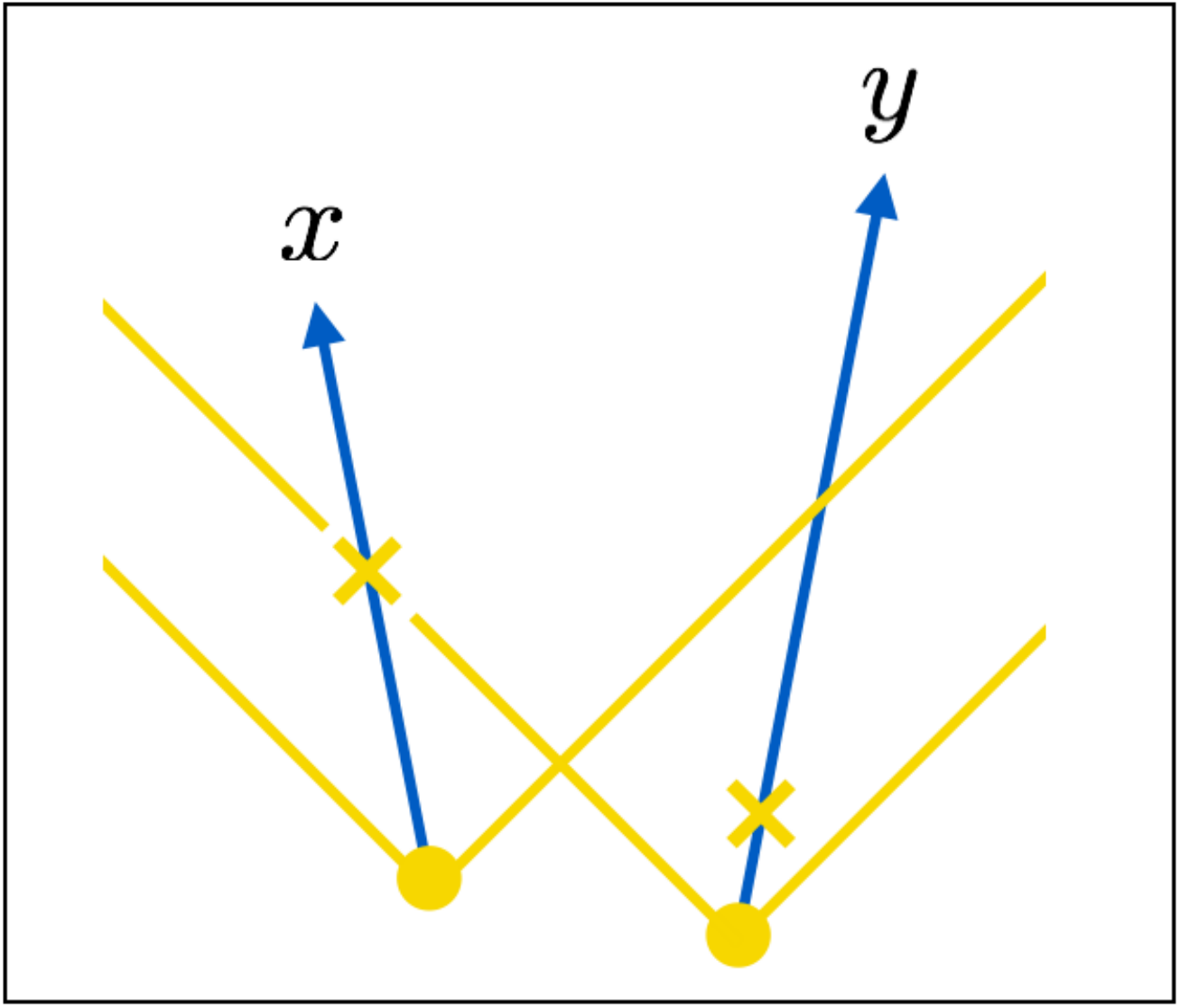}
\includegraphics[width=0.45\columnwidth]{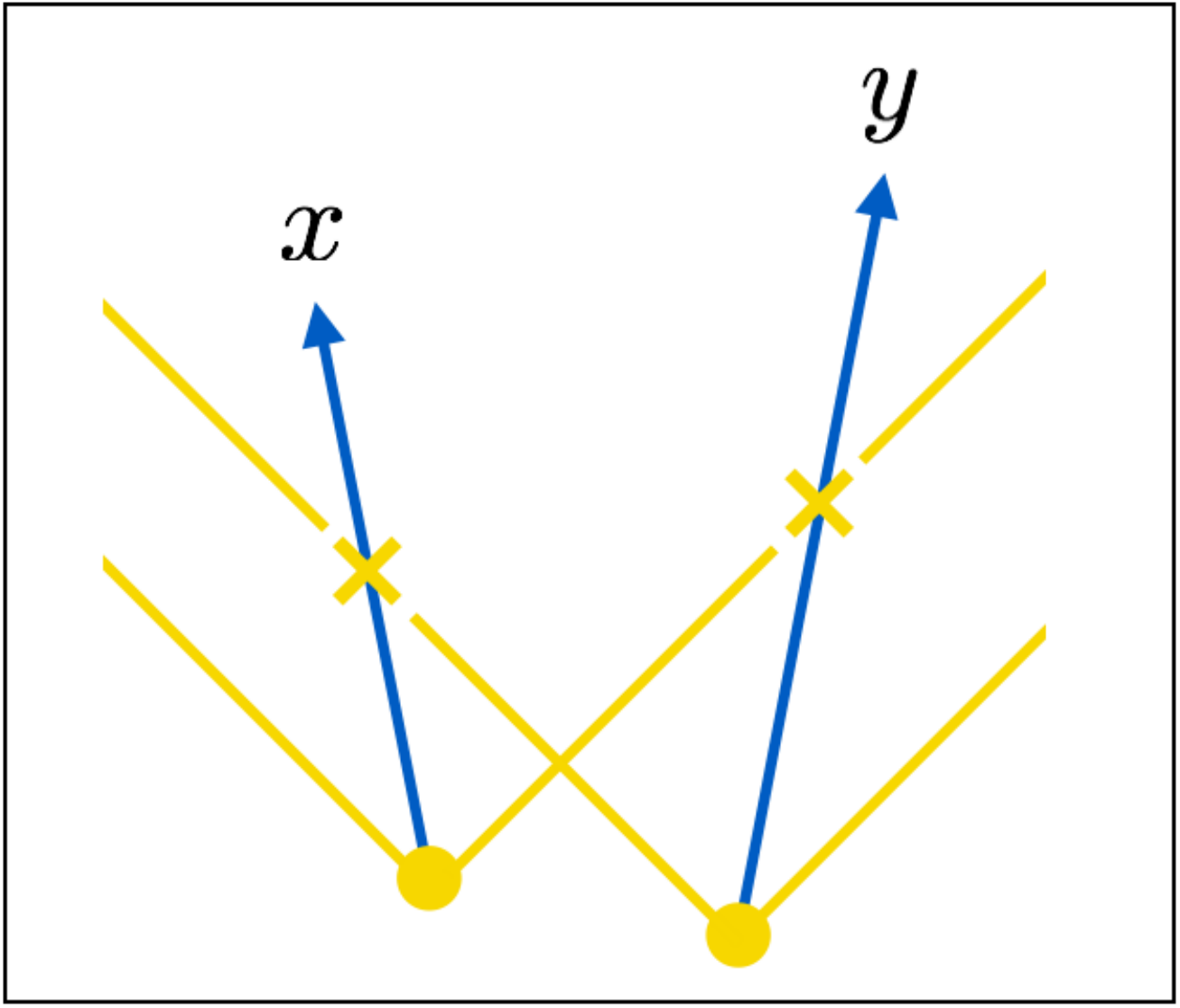}
\caption{\small
Illustration for $(C_i,C_{in}) \times (C_i,C_{in})$ in the double-bubble contribution.
The circles denote the nucleation points for the $x$- and $y$-fragments,
which are shown in blue arrows.
The crosses on these arrows denote the interception points for these fragments.
Note that some choices of $x$ and $y$ forbid
the three types including $C_{in}$ (i.e. $C_iC_{in}$, $C_{in}C_i$ and $C_{in}C_{in}$).
(Top-left) $C_iC_i$: 
the intercepting bubbles for the $x$- and $y$-fragments 
are different from the $y$- and $x$-bubbles, respectively.
(Top-right) $C_iC_{in}$:
the intercepting bubble for the $y$-fragment is the $x$-bubble.
(Bottom-left) $C_{in}C_i$: 
the intercepting bubble for the $y$-fragment is the $y$-bubble.
(Bottom-right) $C_{in}C_{in}$:
the intercepting bubbles for the $x$- and $y$-fragments are
the $y$- and $x$ bubbles, respectively.
}
\label{fig:DoubleClass}
\end{center}
\end{figure}

Next we move on to the calculation of the double-bubble contribution.
We first classify bubble wall properties as in the single-bubble case.
The bubble wall fragment reaching the evaluation point $x$ or $y$
is either already collided or still uncollided, as before.
In the double-bubble case, however, we need further classification for the former
depending on whether the bubble which intercepts the $x$-(or $y$-)fragment is 
the bubble which $y$-(or $x$-)fragment belongs to.
We illustrate this point in Fig.~\ref{fig:DoubleClass}.
The horizontal (vertical) axis is the space (time) direction,
and the blue arrows denote the propagation of the $x$- and $y$-fragments. 
Note that these blue arrows are propagating on the surface of $v$-cones
as in Fig.~\ref{fig:3DLightConeCept}.
In the upper-left panel, the interception (denoted by the crosses) occurs to 
the $x$-(or $y$-)fragment (denoted by blue arrows)
before the wall of the $y$-(or $x$-)bubble (denoted by yellow lines) catches up with 
the $x$-(or $y$-)fragment. 
In such cases the intercepting bubble is different from the one which the other fragment 
belongs to, and we label such contribution by $C_i \times C_i$.
The opposite case is shown in the lower-right panel,
where $x$-($y$-)fragment is intercepted by the bubble wall
which the other fragment belongs to.
We label such contributions by $C_{in} \times C_{in}$.
The meaning of the other two cases $C_i \times C_{in}$ and $C_{in} \times C_i$ 
is now trivial.
Note that whether contributions including $C_{in}$ exist or not
depends on the choice of the spacetime points $x$ and $y$
and on the choice of the propagation directions.

In summary, we can classify the double-bubble contribution into 
$(U,C_i,C_{in}) \times (U,C_i,C_{in})$, where
\begin{itemize}
\item
$U$ : 
$x$-(or $y$-)fragment remains uncollided at the evaluation point.

\item
$C_i$ : 
$x$-(or $y$-)fragment has already been intercepted at the evaluation point,
and the interception occurs with a collision with a bubble other than the $y$-(or $x$-)bubble.

\item
$C_{in}$ : 
$x$-(or $y$-)fragment has already been intercepted at the evaluation point,
and the interception occurs with a collision with the $y$-(or $x$-)bubble.

\end{itemize}
One may notice that $C_i \times C_i$ contribution can be classified into two classes:
the intercepting bubbles for the $x$- and $y$-fragments are identical or not.
Both of them can properly be taken into account by the simplification formula we explain below.

%%%%%%%%%%%%%%%%
%%%%%%%%%%%%%%%%

%%%%%%%%%%%%%%%%%%%%%%%%%%%%%%%%%%%%%%%%%%%%%%%%%%
\subsubsection*{Simplification formula}
%%%%%%%%%%%%%%%%%%%%%%%%%%%%%%%%%%%%%%%%%%%%%%%%%%

\begin{figure}
\begin{center}
\includegraphics[width=0.6\columnwidth]{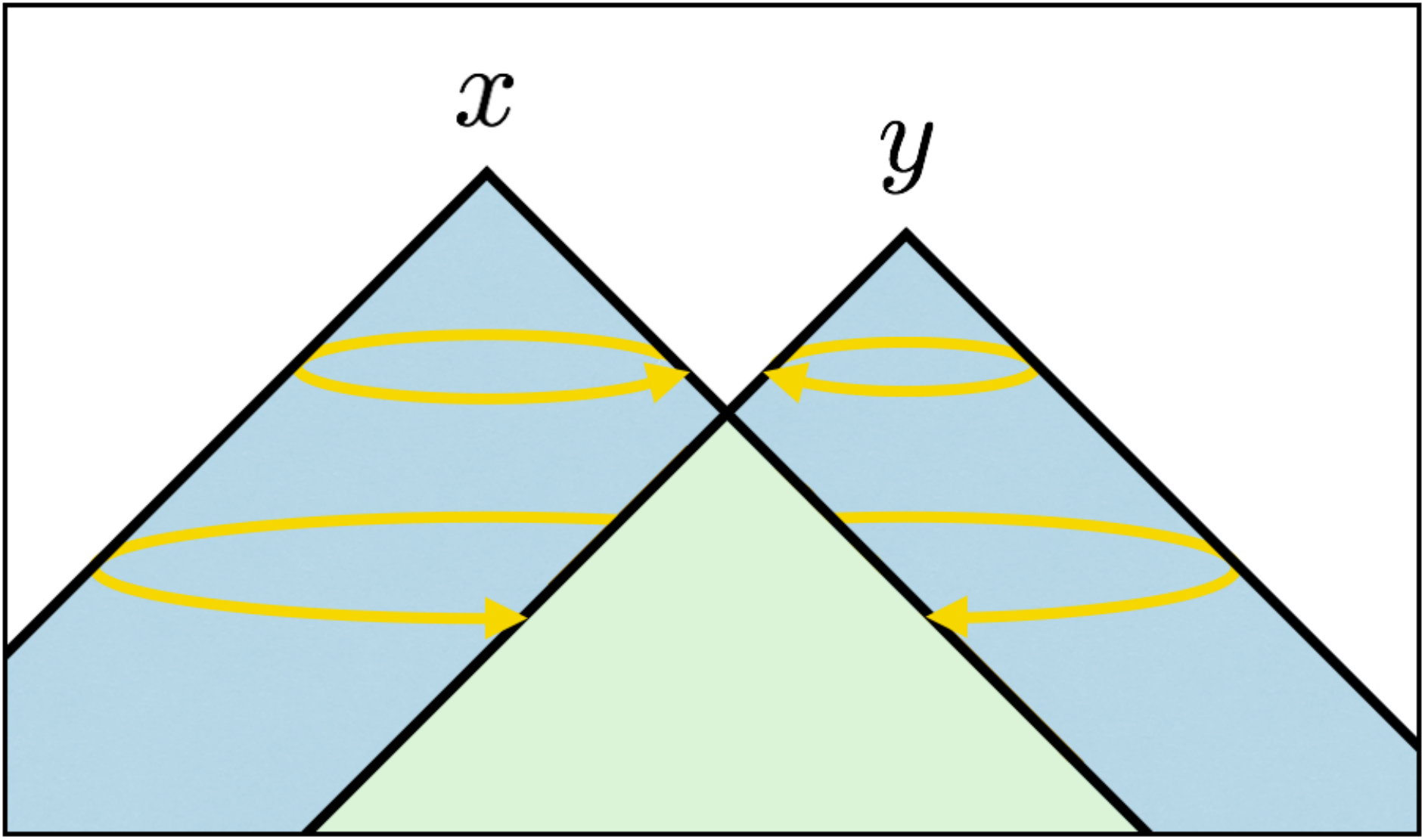}
\caption{\small
Illustration for the integration ranges for the angular variables in Eq.~(\ref{eq:TTdUU}).
The yellow arrows show how the nucleation point changes
as the propagation direction $\Omega_{xn}$ ($\Omega_{yn}$) changes
on a constant-time hypersurface.
For $t_x > t_{\rm max}$ ($t_y > t_{\rm max}$)
the spacelike theta functions in Eq.~(\ref{eq:TTdUU}) always give unity
(the upper two yellow lines),
while for $t_x < t_{\rm max}$ ($t_y < t_{\rm max}$)
they make the integrand vanish if the nucleation point $x_n$ ($y_n$)
enters the past $v$-cone of $y$ ($x$) 
(the lower two yellow lines).
Also, they always give zero for timelike combinations of $x$ and $y$.
Therefore the integration ranges are effectively constrained within 
$\delta V_x^{(y)}$ and $\delta V_y^{(x)}$ with spacelike combinations of $x$ and $y$,
and hence Eq.~(\ref{eq:TTdUU}) is consistent with the calculation in Sec.~\ref{subsec:EnvDouble}.
}
\label{fig:DoubleUU}
\end{center}
\end{figure}

A simplification formula exists in the double-bubble case as well,
which makes the sum of the nine classes $(U,C_i,C_{in}) \times (U,C_i,C_{in})$ much simpler.
First, let us write the $UU$ contribution in the following form:
\begin{align}
\left< T_{ij}(x) T_{kl}(y) \right>^{(d,UU)}
&= 
\int_{-\infty}^{t_x} dt_{xn} 
\int_{-\infty}^{t_y} dt_{yn} 
\int d\Omega_{xn} 
\int d\Omega_{yn} \;
\nonumber \\
&\;\;\;\;\;\;
\Theta_{\rm sp}(x,y_n)
\Theta_{\rm sp}(x_n,y)
e^{-I(x,y)} 
{\mathcal F}^{(d)}(x,y;x_i = x,y_i = y;x_n,y_n).
\label{eq:TTdUU}
\end{align}
Here $\Omega_{xn}$ and $\Omega_{yn}$ denote 
the propagation directions $(\theta_{xn},\phi_{xn})$ and $(\theta_{yn},\phi_{yn})$
of the wall fragments, respectively.
Note that
the constant-time hypersurfaces at the nucleation times differ for $x$- and $y$-bubbles,
since the nucleation of $x$- and $y$-bubbles occurs independently of each other.
This argument is the same as Appendix~\ref{subsec:EnvDouble}.
On the other hand, the integration ranges 
for these directions are now over the whole solid angle in Eq.~(\ref{eq:TTdUU})
in contrast to Appendix~\ref{subsec:EnvDouble}.
In Appendix~\ref{subsec:EnvDouble}
we integrate the nucleation points over 
$\delta V_x^{(y)} = \delta V_x - V_y$ and 
$\delta V_y^{(x)} = \delta V_y - V_x$,
which do not form the whole solid angle.
This might seem contradictory, but actually is not.
The spacelike theta functions in Eq.~(\ref{eq:TTdUU}) guarantee that 
the integration ranges are effectively the same.
We illustrate this point in Fig.~\ref{fig:DoubleUU}.
The nucleation point $x_n$ cannot enter the past $v$-cone of $y$
because $\Theta_{\rm sp}(x_n,y)$ gives zero in such cases,
while $\Theta_{\rm sp}(x,y_n)$ guarantees that 
$y_n$ is outside the $v$-cone of $x$.
Also note that Eq.~(\ref{eq:TTdUU}) forces $(x,y)$ to be spacelike,
because if $x$ and $y$ are timelike 
the theta functions make the integrand to vanish for any solid angle.
The explicit form of ${\mathcal F}^{(d)}$ is read off from 
the derivation in Appendix~\ref{app:Envelope}:
\begin{align}
&{\mathcal F}^{(d)}(x,y;x_i,y_i;x_n,y_n)
\nonumber \\
&\;\;\;\;
= 
r_{xn}^{(d)2}l_B\Gamma(t_{xn}) \;
r_{yn}^{(d)2}l_B\Gamma(t_{yn})
\left( \frac{4\pi}{3}r_{xn}^{(d)3} \kappa \rho_0\frac{1}{4\pi r_{xn}^{(d)2}l_B} \right)
\left( \frac{4\pi}{3}r_{yn}^{(d)3} \kappa \rho_0\frac{1}{4\pi r_{yn}^{(d)2}l_B} \right)
N_{n,ijkl}
\nonumber \\
&\;\;\;\;\;\;\;\;
\times
\frac{r_B(t_{xi},t_{xn})^3}{r_{xn}^{(d)3}}
\frac{r_B(t_{yi},t_{yn})^3}{r_{yn}^{(d)3}}
D(t_x,t_{xi},t_{xn})
D(t_y,t_{yi},t_{yn}).
\end{align}
In fact, one sees that the substitution of $x_i = x$ and $y_i = y$
reproduces Eq.~(\ref{eq:EnvDoubledPTT}).

Now, 
the sum of the nine contributions $(U,C_i,C_{in}) \times (U,C_i,C_{in})$ gives a simple formula
\begin{align}
&\left< T_{ij}(x) T_{kl}(y) \right>^{(d,(U,C_i,C_{in}) \times (U,C_i,C_{in}))}
\nonumber \\
&\;\;\;\;\;\;\;\;
= 
\int_{-\infty}^{t_x} dt_{xn} 
\int_{-\infty}^{t_y} dt_{yn} 
\int_{t_{xn}}^{t_x} dt_{xi} 
\int_{t_{yn}}^{t_y} dt_{yi} 
\int d\Omega_{xn} 
\int d\Omega_{yn} \;
\nonumber \\
&\;\;\;\;\;\;\;\;\;\;\;\;\;\;\;\;
\Theta_{\rm sp}(x_i,y_n)
\Theta_{\rm sp}(x_n,y_i)
e^{-I(x_i,y_i)} 
\left[
{\mathcal F}^{(d)}(x,y;x_i,y_i;x_n,y_n)
\right]_{txi,tyi}.
\label{eq:BeyondDoubleSimple}
\end{align}
Here the subscript on the squared parenthesis denotes 
the time derivatives along the propagation of the wall fragments
defined in Appendix~\ref{subsec:Prerequisites}.
We prove this formula in Appendix~\ref{app:Simple}.

%%%%%%%%%%%%%%%%
%%%%%%%%%%%%%%%%

%%%%%%%%%%%%%%%%%%%%%%%%%%%%%%%%%%%%%%%%%%%%%%%%%%
\subsubsection*{Expressions for GW spectrum}
%%%%%%%%%%%%%%%%%%%%%%%%%%%%%%%%%%%%%%%%%%%%%%%%%%

We now derive concrete expressions for $\Pi^{(d)}$ and $\Delta^{(d)}$.
For a general nucleation rate $\Gamma$ and the dumping function $D$, 
the simplification formula (\ref{eq:BeyondDoubleSimple}) gives
\begin{align}
\Pi^{(d)}(t_x,t_y,k)
=
&
\int_0^\infty dr 
\int_{-\infty}^{t_x} dt_{xn} 
\int_{-\infty}^{t_y} dt_{yn} 
\int_{t_{xn}}^{t_x} dt_{xi} 
\int_{t_{yn}}^{t_y} dt_{yi} 
\int_{-1}^1 dc_{xn}
\int_{-1}^1 dc_{yn}
\int_0^{2\pi} d\phi_{xn,yn}
\nonumber \\[1ex]
\frac{4\pi^2}{9} \kappa^2\rho_0^2 
&\left[
\begin{matrix*}[l]
\displaystyle
\;
\Theta_{\rm sp}(x_i,y_n)
\Theta_{\rm sp}(x_n,y_i)
e^{-I(x_i,y_i)} 
\Gamma(t_{xn})\Gamma(t_{yn})
\\[1ex]
\displaystyle
\times \;
r^2
\left[
j_0(kr){\mathcal K}_0(n_{xn},n_{yn})
+ \frac{j_1(kr)}{kr}{\mathcal K}_1(n_{xn},n_{yn})
+ \frac{j_2(kr)}{(kr)^2}{\mathcal K}_2(n_{xn},n_{yn})
\right]
\\[2ex]
\displaystyle
\times \;
\partial_{txi}
\left[
r_B(t_{xi},t_{xn})^3
D(t_x,t_{xi},t_{xn})
\right]
\partial_{tyi}
\left[
r_B(t_{yi},t_{yn})^3
D(t_y,t_{yi},t_{yn})
\right]
\end{matrix*}
\right],
\label{eq:BeyondPiDConc}
\end{align}
for the correlator, and
\begin{align}
\Delta^{(d)}
=
&
\int_{-\infty}^\infty dt_x 
\int_{-\infty}^\infty dt_y 
\nonumber \\
&
\int_0^\infty dr 
\int_{-\infty}^{t_x} dt_{xn} 
\int_{-\infty}^{t_y} dt_{yn} 
\int_{t_{xn}}^{t_x} dt_{xi} 
\int_{t_{yn}}^{t_y} dt_{yi} 
\int_{-1}^1 dc_{xn}
\int_{-1}^1 dc_{yn}
\int_0^{2\pi} d\phi_{xn,yn}
\nonumber \\[1ex]
\frac{k^3}{3}
&\left[
\begin{matrix*}[l]
\displaystyle
\;
\Theta_{\rm sp}(x_i,y_n)
\Theta_{\rm sp}(x_n,y_i)
e^{-I(x_i,y_i)} 
\Gamma(t_{xn})\Gamma(t_{yn})
\\[1ex]
\displaystyle
\times \;
r^2
\left[
j_0(kr){\mathcal K}_0(n_{xn},n_{yn})
+ \frac{j_1(kr)}{kr}{\mathcal K}_1(n_{xn},n_{yn})
+ \frac{j_2(kr)}{(kr)^2}{\mathcal K}_2(n_{xn},n_{yn})
\right]
\\[2ex]
\displaystyle
\times \;
\partial_{txi}
\left[
r_B(t_{xi},t_{xn})^3
D(t_x,t_{xi},t_{yn})
\right]
\partial_{tyi}
\left[
r_B(t_{yi},t_{yn})^3
D(t_y,t_{yi},t_{yn})
\right]
\cos(kt_{x,y})
\end{matrix*}
\right],
\label{eq:BeyondDeltaDConc}
\end{align}
for the GW spectrum.
Note that the derivatives along the propagation of the wall fragments 
have now reduced to partial derivatives with respect to the interception times $t_{xi}$ and $t_{yi}$,
because the dependence of $\vec{x}_i$ and $\vec{y}_i$ on $t_{xi}$ and $t_{yi}$, respectively,
is now encoded in the functional form of ${\mathcal F}^{(d)}$.
Also note that the integration range for $r$ is $r > 0$ in contrast to Appendix~\ref{subsec:EnvDouble}.
As explained above, this is because the spacelike theta functions 
effectively guarantee that the nucleation point for the $x$-(or $y$-)bubble 
to be spacelike to the evaluation point $y$ (or $x$).
In addition, we have integrated out the azimuthal angle $(\phi_{xn} + \phi_{yn})/2$ 
because the geometry of the system is independent of this direction.
This leaves $\phi_{xn,yn} = \phi_{xn} - \phi_{yn}$ as the only remaining azimuthal angle.

Again it would be a good exercise to check that 
Eqs.~(\ref{eq:BeyondPiDConc}) and (\ref{eq:BeyondDeltaDConc})
coincide with the expressions with the envelope approximation
(\ref{eq:EnvPiDGeneral}) and (\ref{eq:EnvDeltaDGeneral})
if we assume an instant damping of the wall energy.
As in Sec.~\ref{subapp:BeyondEnvSingle}, we take the damping function to be 
$D(t,t_i,t_n) = \Theta(t_i - t + \epsilon)$ with $\epsilon$ being infinitesimal and positive. 
The $t_{xi}$ and $t_{yi}$ derivatives have to act on the damping functions 
due to the same reason as the previous subsection, 
and we can complete $t_{xi}$ and $t_{yi}$ integrations 
using the resulting delta functions.
Now the spacelike theta functions are replaced by
$\Theta_{\rm sp}(x,y_n)$ and $\Theta_{\rm sp}(y,x_n)$, 
and hence only spacelike combinations of $(x,y_n)$ and $(y,x_n)$ are allowed.
Since the integration $t_x > t_{\rm max}$ or $t_y > t_{\rm max}$
gives vanishing contributions due to the spherical symmetry,
one can safely restrict the integration range to 
$t_x < t_{\rm max}$ and $t_y < t_{\rm max}$
to obtain Eq.~(\ref{eq:EnvPiDGeneralOnTheWay}).
The same arguments as Appendix~\ref{subsec:EnvDouble} lead to 
Eqs.~(\ref{eq:EnvPiDGeneral}) and (\ref{eq:EnvDeltaDGeneral}).

Finally we derive the GW spectrum 
with the nucleation rate (\ref{eq:Gamma}) and the damping function (\ref{eq:D}).
Given these specific forms, we can perform the integration 
with respect to the overall time shift of the system. 
In fact, one can write all the time variables in Eq.~(\ref{eq:BeyondDeltaDConc}) 
in terms of time differences except for the exponent $I(x_i,y_i)$ 
and the nucleation rate $\Gamma(t_{xn})\Gamma(t_{yn})$.
From the derivation in Appendix~\ref{subsec:EnvDouble} 
we know that such a time shift direction can be integrated out.
As a result, we have
\footnote{
In Eq.~(\ref{eq:BeyondDeltaDConc}), 
the spacelike theta functions 
as well as the second and third lines in the square parenthesis 
contain time differences only.
Therefore, overall time-shift dependence appears only in 
$I(x_i,y_i) = v^3 \Gamma_* e^{t_{\left< x,y \right>}/2}e^{-(t_{x,xi} + t_{y,yi} - r_{xi,yi}/v)/2} 
{\mathcal I}(t_{xi,yi},r_{xi,yi})$ 
and the nucleation rate 
$\Gamma(t_{xn})\Gamma(t_{yn}) = \Gamma_*^2 e^{2t_{\left< x,y \right>}} e^{-(t_{x,xn} + t_{y,yn})}$,
in the form of $e^{t_{\left< x,y \right>}}$.
After rewriting $t_x$ and $t_y$ integrations into $t_{\left< x,y \right>}$ and $t_{x,y}$ integrations,
and expressing all the other time integration variables
in terms of time differences like $t_{x,xn}$, $t_{y,yn}$, $t_{x,xi}$ and $t_{y,yi}$,
we can integrate out $t_{\left< x,y \right>}$ direction to obtain Eq.~(\ref{eq:BeyondDeltaDConcConc}).
}
\begin{align}
\Delta^{(d)}
=
&
\int_{-\infty}^\infty dt_{x,y}
\int_0^\infty dr_v
\nonumber \\
&
\int_0^\infty dt_{x,xn}
\int_0^\infty dt_{y,yn}
\int_0^{t_{x,xn}} dt_{x,xi}
\int_0^{t_{y,yn}} dt_{y,yi}
\int_{-1}^1 dc_{xn}
\int_{-1}^1 dc_{yn}
\int_0^{2\pi} d\phi_{xn,yn}
\nonumber \\[1ex]
\frac{v^3k^3}{3}
&\left[
\begin{matrix*}[l]
\;
\Theta_{\rm sp}(x_i,y_n)
\Theta_{\rm sp}(x_n,y_i)
\\[1.5ex]
\displaystyle
\times \;
\frac{e^{-(t_{xi,xn} + t_{yi,yn})} e^{-(t_{x,xi} + t_{y,yi})/\tau}}
{{\mathcal I}(t_{xi,yi},r_{xi,yi}/v)^2}
(3t_{xi,xn}^2 + t_{xi,xn}^3/\tau)
(3t_{yi,yn}^2 + t_{yi,yn}^3/\tau)
\\[2ex]
\displaystyle
\times \;
r_v^2
\left[
j_0(vkr_v){\mathcal K}_0(n_{xn},n_{yn})
+ \frac{j_1(vkr_v)}{vkr_v}{\mathcal K}_1(n_{xn},n_{yn})
+ \frac{j_2(vkr_v)}{(vkr_v)^2}{\mathcal K}_2(n_{xn},n_{yn})
\right]
\\[2ex]
\times
\cos(kt_{x,y})
\end{matrix*}
\right].
\label{eq:BeyondDeltaDConcConc}
\end{align}
Note that time differences such as $t_{xi,xn}$, $t_{yi,yn}$ and $t_{xi,yi}$ are expressed in terms of the 
integration variables as $t_{xi,xn} = t_{x,xn} - t_{x,xi}$, $t_{yi,yn} = t_{y,yn} - t_{y,yi}$,
and $t_{xi,yi} = t_{x,y} - t_{x,xi} + t_{y,yi}$.
In addition, $r_{xi,yi} = |\vec{x}_i - \vec{y}_i|$ can also be written in terms of the integration variables.
In fact, the relation $\vec{x}_i - \vec{y}_i = \vec{r} + v(t_{x,xi}n_{xn} - t_{y,yi}n_{yn})$ gives
\begin{align}
\frac{r_{xi,yi}}{v}
&= 
\sqrt{r_v^2 + t_{x,xi}^2 + t_{y,yi}^2
+ 2r_vt_{x,xi}c_{xn}
- 2r_vt_{y,yi}c_{yn}
- 2 t_{x,xi}t_{y,yi}c_{xyn}},
\label{eq:riDouble}
\end{align}
where $c_{xyn}$ is the cosine of the angle between 
$n_{xn}$ and $n_{yn}$ given by 
$c_{xyn} = c_{xn} c_{yn} + s_{xn} s_{yn}\cos(\phi_{xn} - \phi_{yn})$.

\clearpage

%%%%%%%%%%%%%%%%%%%%%%%%%%%%%%%%%%%%%%%%%%%%%%%%%%
\section{Derivation of the simplification formulas}
\label{app:Simple}
\setcounter{equation}{0}
%%%%%%%%%%%%%%%%%%%%%%%%%%%%%%%%%%%%%%%%%%%%%%%%%%

In this appendix we derive the simplification formulas (\ref{eq:BeyondSingleSimple})
and (\ref{eq:BeyondDoubleSimple}).

%%%%%%%%%%%%%%%%%%%%%%%%%%%%%%%%%%%%%%%%%%%%%%%%%%
\subsection{Single-bubble simplification formula}
%%%%%%%%%%%%%%%%%%%%%%%%%%%%%%%%%%%%%%%%%%%%%%%%%%

Let us start from Eq.~(\ref{eq:BeyondTTSUU}),
which is written here again for completeness
\begin{align}
\left< T_{ij}(x) T_{kl}(y) \right>^{(s,UU)}
&= 
\int_{-\infty}^{t_{\rm max}}dt_n 
\int d\phi_n \;
e^{-I(x,y)} 
{\mathcal F}^{(s)}(x,y;x_i = x,y_i = y;z_n),
\label{eq:AppBeyondTTSUU}
\end{align}
and show that other contributions are written as
\begin{align}
\left< T(x) T(y) \right>^{(s,UC)}
&=
\int_{-\infty}^{t_{\rm max}}dt_n 
\int d\phi_n
\int_{t_{yn}}^{t_y}dt_{yi} \;
\left[ I(x,y_i) \right]_{tyi}
e^{-I(x,y_i)} {\mathcal F}^{(s)}(x,y;x_i = x,y_i;z_n),
\nonumber \\
\left< T(x) T(y) \right>^{(s,CU)}
&=
\int_{-\infty}^{t_{\rm max}}dt_n 
\int d\phi_n
\int_{t_{xn}}^{t_x}dt_{xi} \;
\left[ I(x_i,y) \right]_{txi}
e^{-I(x_i,y)} {\mathcal F}^{(s)}(x,y;x_i,y_i = y;z_n),
\nonumber \\
\left< T(x) T(y) \right>^{(s,CC)}
&= 
\int_{-\infty}^{t_{\rm max}}dt_n 
\int d\phi_n
\int_{t_{xn}}^{t_x}dt_{xi} 
\int_{t_{yn}}^{t_y}dt_{yi} \;
\nonumber \\
&\;\;\;\;\;\;
\left[
\left[ I(x_i,y_i) \right]_{txi} \left[ I(x_i,y_i) \right]_{tyi} - \left[ I(x_i,y_i) \right]_{txi,tyi}
\right] 
e^{-I(x_i,y_i)} {\mathcal F}^{(s)}(x,y;x_i,y_i,z_n),
\label{eq:AppTTsOthers}
\end{align}
where remember that 
the subscripts on the square bracket mean
taking the derivative with respect to that quantity
along the propagation of the bubble wall fragments.

In deriving the simplification formula, 
it would be instructive to see the structure of Eq.~(\ref{eq:AppBeyondTTSUU}).
It consists of two parts.
One is $P(x,y) = e^{-I(x,y)}$, which guarantees that no bubbles nucleate in $V_x \cup V_y$.
This is necessary because otherwise at least one of the $x$- and $y$-fragments 
collides before the evaluation time with bubble(s) which nucleate in $V_x \cup V_y$.
The other is ${\mathcal F}^{(s)}$,
which takes into account the probability for nucleation of the $xy$-bubble
(with the infinitesimal time intervals $dt_n$ and angle $d\phi_n$)
and the resulting values of the energy-momentum tensor at the evaluation points.

\begin{figure}[h]
\begin{center}
\includegraphics[width=0.4\columnwidth]{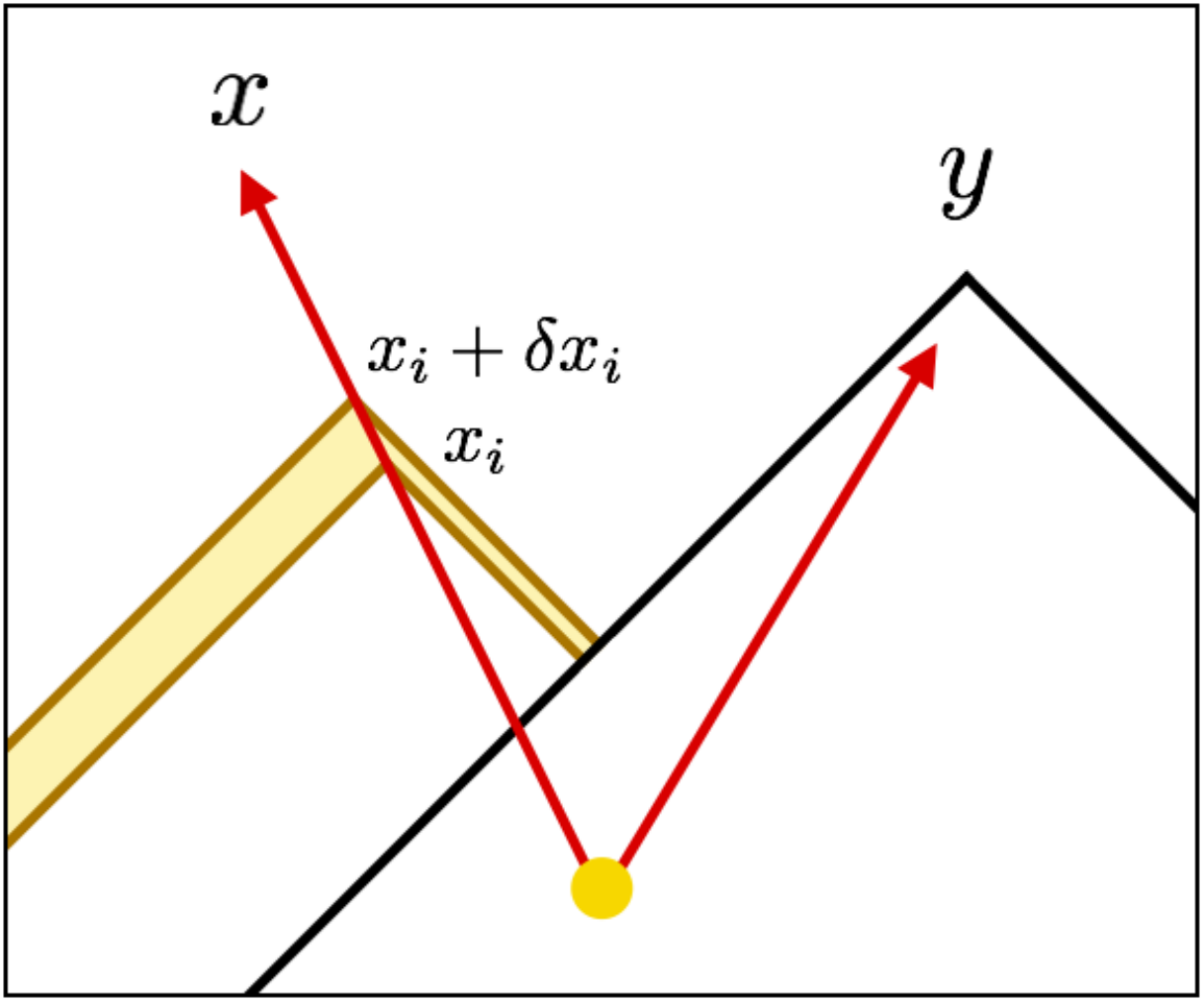}
\includegraphics[width=0.4\columnwidth]{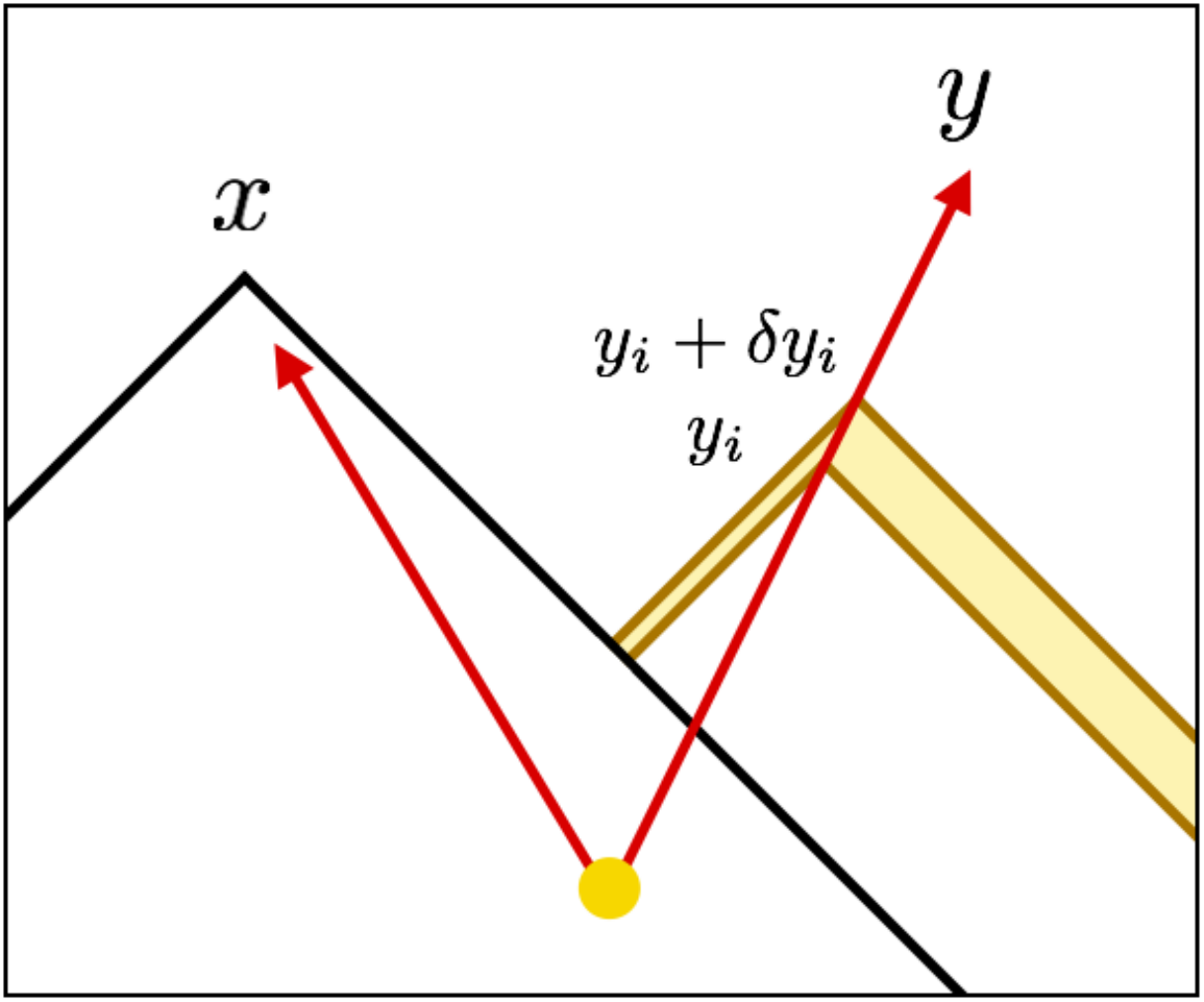}
\includegraphics[width=0.4\columnwidth]{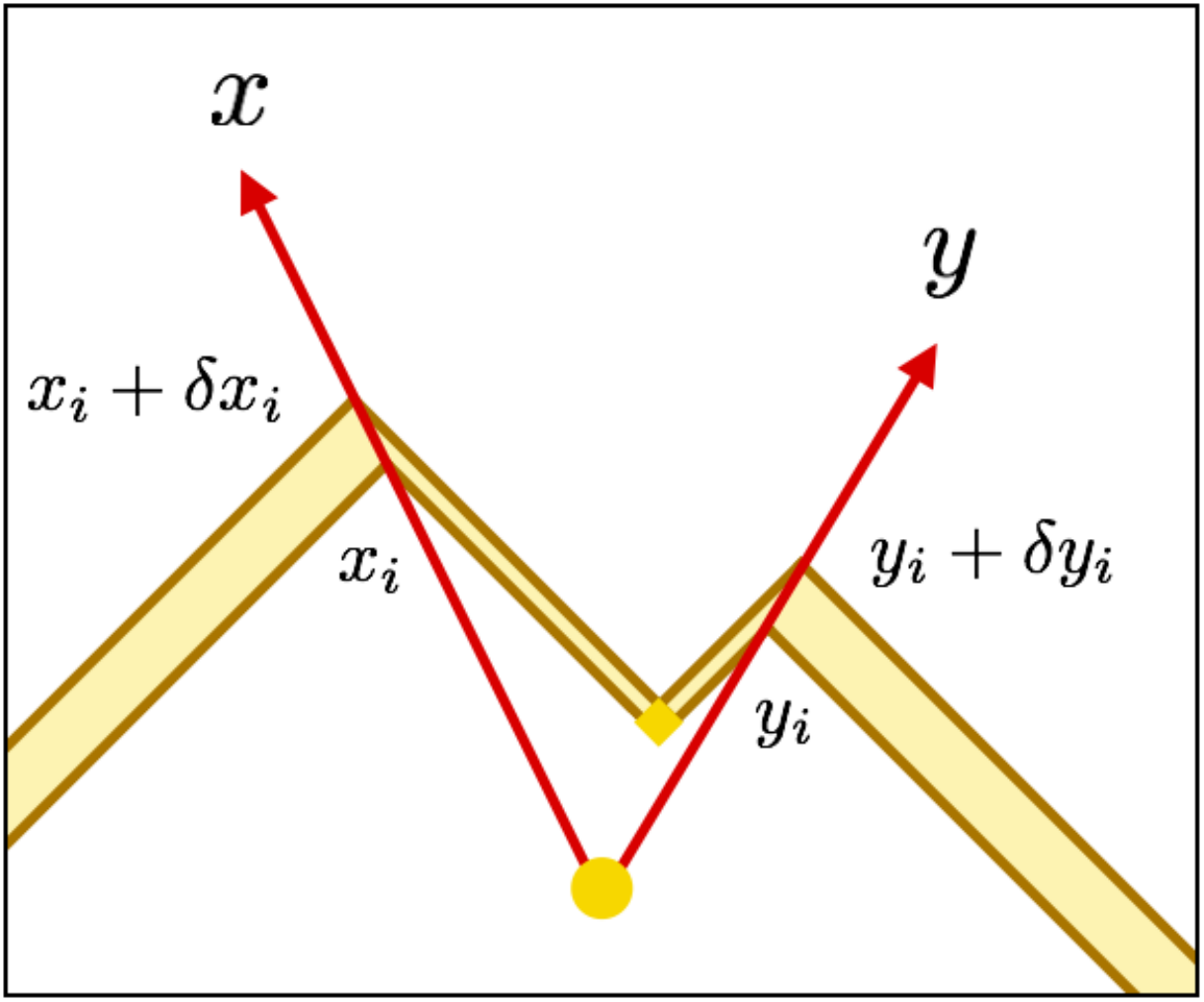}
\caption{\small
(Top-left) $CU$ contribution.
The $x$-fragment is intercepted in the time interval $[t_x,t_x + dt_x]$ 
by some bubble which does not intercept the $y$-fragment before $t_y$.
Such a bubble nucleates in the thin yellow band.
Note that the red arrows are on the past $v$-cones of $x$ and $y$, 
and their inclinations do not mean 
that bubble walls propagate slower than $v$: see Fig.~\ref{fig:3DLightCone}. 
(Top-right) 
$UC$ contribution.
Same as the top-left, except that the roles of $x$ and $y$ are interchanged.
(Bottom)
$CC$ contribution.
This contains two cases: 
one is that one bubble nucleates in each of the left and right thin yellow bands,
and the other is that only one bubble nucleates in the small yellow diamond in the center.
}
\label{fig:SingleCept}
\end{center}
\end{figure}

Let us see how these arguments change in other contributions,
taking $CU$ as an example.
See the left panel of Fig.~\ref{fig:SingleCept}.
We first consider the change in the false vacuum probability.
Since now the $x$-fragment is allowed to collide with other bubbles after it is intercepted at $x_i$
(i.e., in the left panel of Fig.~\ref{fig:SingleCept}, 
bubble walls are allowed to cross the left red arrow at anywhere from $x_i$ to $x$),
we do not require that no bubbles nucleate in $V_x \cup V_y$.
Instead, we require that the $x$-fragment remain uncollided until it reaches $x_i$:
\begin{itemize}
\item
No bubbles nucleate in the union of past $v$-cones of $x_i$ and $y$.
\end{itemize}
However, this condition does not guarantee that the $x$-fragment is intercepted at $x_i$.
Let us assume that the $x$-fragment is intercepted between the infinitesimal time interval 
$[t_{xi},t_{xi} + dt_{xi}]$,
and let us consider where such an intercepting bubble can nucleate.
Since we are now focusing on $C$``$U$" contribution, 
we require the following for such a bubble:
\begin{itemize}
\item
The bubble which intercepts the $x$-fragment nucleates somewhere
on the past $v$-cone of $x_i$, 
and it does not intercept the $y$-fragment before the evaluation time $t_y$.
\end{itemize}
The nucleation region satisfying this requirement is shown 
as the yellow band in the left panel of Fig.~\ref{fig:SingleCept}.
Here $x_i + \delta x_i$ is the spacetime point of the $x$-fragment at time $t_x + dt_x$.

The probability for both of the two events above to occur is given by
$P(x_i + \delta x_i,y) - P(x_i,y) = \left[ I(x_i,y) \right]_{txi}e^{-I(x_i,y)}dt_{xi}$.
The reason is simple: if there is any difference between 
the false vacuum probabilities $P(x_i + \delta x_i,y)$ and $P(x_i,y)$,
the difference corresponds to the probability for 
a transition from the false to the true vacuum occurs somewhere 
between $x_i$ and $x_i + \delta x_i$.
This transition means that 
a fragment propagating from $x_i$ and $x_i + \delta x_i$ 
is intercepted by a bubble which nucleates 
somewhere in the yellow band in the left panel of Fig.~\ref{fig:SingleCept}
(i.e. $V_{x_i + \delta x_i} - (V_{x_i} \cup V_y)$).
Note that we do not have to care about how many intercepting bubbles nucleate,
because in the thin-wall limit the number of such bubbles reduces to unity.
Another way to understand this expression is to note that 
$\left[ I(x_i,y) \right]_{txi}dt_{xi} = I(x_i + \delta x_i,y) - I(x_i,y)$
gives the integration of the nucleation probability $\Gamma$
over the yellow band,
which gives the probability for a single bubble nucleation in this region in the thin-wall limit,
while $e^{-I(x_i,y)}$ guarantees that no bubbles nucleate in $V_{x_i} \cup V_y$.

Now, the replacement for ${\mathcal F}^{(s)}$ is easy to find.
The only change is in the value of the energy-momentum tensor
due to the fact that the $x$-fragment cannot be sourced after the interception at $x_i$.
Therefore we use ${\mathcal F}^{(s)}(x,y;x_i,y_i = y;z_n)$ 
instead of ${\mathcal F}^{(s)}(x,y;x_i = x,y_if = y;z_n)$.
Taking all the above into account, and integrating with the interception time,
one has the $CU$ contribution in Eq.~(\ref{eq:AppTTsOthers}).
The $UC$ contribution is given by the same argument with the roles of $x_\bullet$ and $y_\bullet$ interchanged.

Lastly, let us consider the $CC$ contribution.
See the lower panel of Fig.~\ref{fig:SingleCept}.
Suppose that the interceptions occur in the time intervals $[t_x,t_x + dt_x]$ and $[t_y,t_y + dt_y]$ 
for the $x$- and $y$-fragments, respectively.
There are two cases for this:
\begin{itemize}
\item
One bubble nucleates in $V_{x_i + \delta x_i} - (V_{x_i} \cup V_{y_i + \delta y_i})$
(the left yellow band in the lower panel of Fig.~\ref{fig:SingleCept}),
and another nucleates in $V_{y_i + \delta y_i} - (V_{x_i + \delta x_i} \cup V_{y_i})$
(the right yellow band in the same figure).
\item
One bubble nucleates in the small spacetime region 
$(V_{x_i + \delta x_i} \cup V_{y_i + \delta y_i} - V_{x_i} \cup V_{y_i}) 
- (V_{x_i + \delta x_i} - V_{x_i}) 
- (V_{y_i + \delta y_i} - V_{y_i})$
(the small yellow diamond in the same figure).
\end{itemize}
The former means that the two intercepting bubbles are different,
while in the latter a single bubble intercepts both the $x$- and $y$-fragments.
Under the condition that no bubbles nucleate in $V_{x_i} \cup V_{y_i}$
(which occurs with probability $e^{-I(x_i,y_i)}$),
the probability for the former is given by $\left[ I (x_i,y_i)\right]_{txi} \left[ I(x_i,y_i) \right]_{tyi} dt_{xi}dt_{yi}$,
while that for the latter is given by $- \left[ I(x_i,y_i) \right]_{txi,tyi} dt_{xi}dt_{yi}$.
For the ${\mathcal F}^{(s)}$ function, we have only to use ${\mathcal F}^{(s)}(x,y;x_i,y_i;z_n)$.
Therefore we obtain the expression in Eq.~(\ref{eq:AppTTsOthers}).

Once we have Eqs.~(\ref{eq:AppBeyondTTSUU}) and (\ref{eq:AppTTsOthers}), 
their sum becomes much simpler.
Noting that ${\mathcal F}^{(s)}(x,y;x_i = x_n,y_i;z_n)$
and ${\mathcal F}^{(s)}(x,y;x_i,y_i = y_n;z_n)$ vanish
because those bubbles intercepted just after nucleation have vanishing energy and momentum,
we obtain 
\begin{align}
&\left< T(x) T(y) \right>^{(s,UU + UC + CU + CC)}
\nonumber \\
&\;\;\;\;\;\;\;\;
= 
\int_{-\infty}^{t_{\rm max}}dt_n 
\int d\phi_n
\int_{t_{xn}}^{t_x}dt_{xi} 
\int_{t_{yn}}^{t_y}dt_{yi} \;
e^{-I(x_i,y_i)} 
\left[
{\mathcal F}^{(s)}(x,y;x_i,y_i,z_n)
\right]_{txi,tyi},
\label{eq:AppBeyondSingleSimple}
\end{align}
after integration by parts.
This is what we call the ``simplification formula" for the single-bubble.

%%%%%%%%%%%%%%%%
%%%%%%%%%%%%%%%%

%%%%%%%%%%%%%%%%%%%%%%%%%%%%%%%%%%%%%%%%%%%%%%%%%%
\subsection{Double-bubble simplification formula}
%%%%%%%%%%%%%%%%%%%%%%%%%%%%%%%%%%%%%%%%%%%%%%%%%%

%%%%%%%%%%%%%%%%%%%%%%%%%%%%%%%%%%%%%%%%%%%%%%%%%%
\subsubsection*{$(U,C_i) \times (U,C_i)$ contributions}
%%%%%%%%%%%%%%%%%%%%%%%%%%%%%%%%%%%%%%%%%%%%%%%%%%

Let us start from Eq.~(\ref{eq:BeyondTTSUU}), 
which is written here again for completeness
\begin{align}
\left< T_{ij}(x) T_{kl}(y) \right>^{(d,UU)}
&= 
\int_{-\infty}^{t_x} dt_{xn} 
\int_{-\infty}^{t_y} dt_{yn} 
\int d\Omega_{xn} 
\int d\Omega_{yn} \;
\nonumber \\
&\;\;\;\;
\Theta_{\rm sp}(x,y_n)
\Theta_{\rm sp}(x_n,y)
e^{-I(x,y)} 
{\mathcal F}^{(d)}(x,y;x_i = x,y_i = y;x_n,y_n).
\label{eq:AppTTdUU}
\end{align}
Below we first show that
the other contributions in $(U,C_i) \times (U,C_i)$ are expressed as
\begin{align}
\left< T_{ij}(x) T_{kl}(y) \right>^{(d,UC_i)}
&= 
\int_{-\infty}^{t_x} dt_{xn} 
\int_{-\infty}^{t_y} dt_{yn} 
\int_{t_{yn}}^{t_y} dt_{yi} 
\int d\Omega_{xn} 
\int d\Omega_{yn} \;
\nonumber \\
&\;\;\;\;\;\;
\Theta_{\rm sp}(x,y_n)
\Theta_{\rm sp}(x_n,y_i)
\left[ I(x,y_i) \right]_{tyi}
e^{-I(x,y_i)} 
{\mathcal F}^{(d)}(x,y;x_i = x,y_i;x_n,y_n),
\nonumber \\
\left< T_{ij}(x) T_{kl}(y) \right>^{(d,C_iU)}
&= 
\int_{-\infty}^{t_x} dt_{xn} 
\int_{-\infty}^{t_y} dt_{yn} 
\int_{t_{xn}}^{t_x} dt_{xi} 
\int d\Omega_{xn} 
\int d\Omega_{yn} \;
\nonumber \\
&\;\;\;\;\;\;
\Theta_{\rm sp}(x_i,y_n)
\Theta_{\rm sp}(x_n,y)
\left[ I(x_i,y) \right]_{txi}
e^{-I(x_i,y)} 
{\mathcal F}^{(d)}(x,y;x_i,y_i = y;x_n,y_n),
\nonumber \\
\left< T_{ij}(x) T_{kl}(y) \right>^{(d,C_iC_i)}
&= 
\int_{-\infty}^{t_x} dt_{xn} 
\int_{-\infty}^{t_y} dt_{yn} 
\int_{t_{xn}}^{t_x} dt_{xi} 
\int_{t_{yn}}^{t_y} dt_{yi} 
\int d\Omega_{xn} 
\int d\Omega_{yn} \;
\nonumber \\
&\;\;\;\;\;\;
\Theta_{\rm sp}(x_i,y_n)
\Theta_{\rm sp}(x_n,y_i)
\left[ 
\left[ I(x_i,y_i) \right]_{txi}\left[ I(x_i,y_i) \right]_{tyi}
- \left[ I(x_i,y_i) \right]_{txi,tyi}
\right]
\nonumber \\
&\;\;\;\;\;\;
e^{-I(x_i,y_i)} 
{\mathcal F}^{(d)}(x,y;x_i,y_i;x_n,y_n).
\label{eq:AppTTdOthers}
\end{align}
Let us remember the arguments in Appendix~\ref{app:BeyondEnv} 
when we derive Eq.~(\ref{eq:AppTTdUU}).
The spacelike theta functions require $(x,y_n)$ and $(x_n,y)$ to be spacelike.
This is because, for timelike $(x,y_n)$ for example, either of the following two occurs:
\begin{itemize}
\item
For $t_x < t_{yn}$, $y_n$ is in the future $v$-cone of $x$
and therefore $\vec{y}_n$ is already inside the $x$-bubble at the nucleation time $t_{yn}$.
This makes the nucleation of the $y$-bubble impossible.
\item
For $t_x > t_{yn}$, $x$ is in the future $v$-cone of $y_n$ 
and therefore $\vec{x}$ is already inside the $y$-bubble at the evaluation time $t_x$.
This is not ``$U$" type contribution.
\end{itemize}
Therefore, the spacelike theta functions select out proper combinations of the nucleation points $x_n$ and $y_n$
for given $x$ and $y$.
Also, the false vacuum probability $P(x,y) = e^{-I(x,y)}$ guarantees that no bubbles nucleate in $V_x \cup V_y$,
while the probability for bubble nucleation at proper spacetime points and 
the value of the energy-momentum tensor at the evaluation points are encoded in ${\mathcal F}^{(d)}$.

\begin{figure}
\begin{center}
\includegraphics[width=0.45\columnwidth]{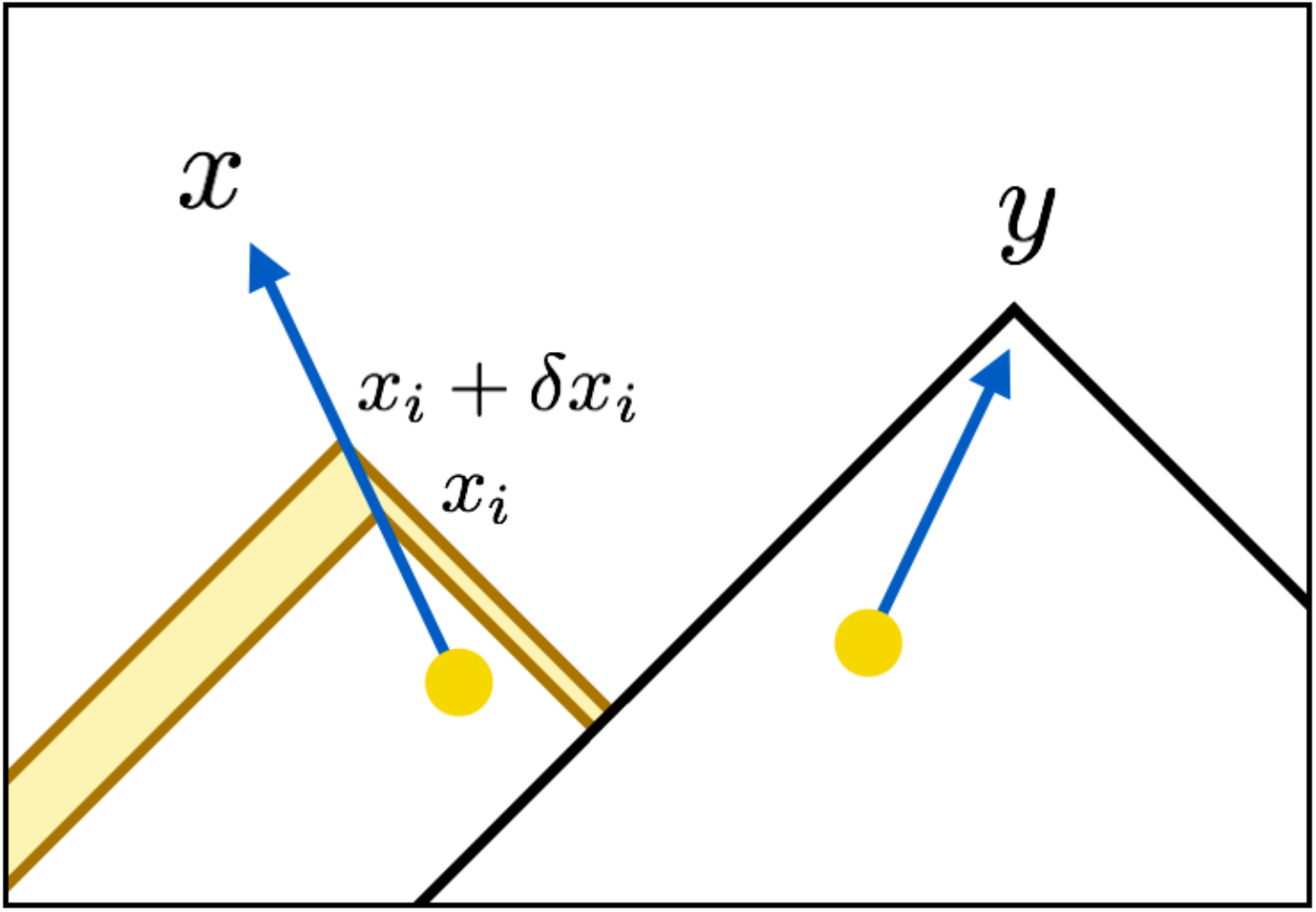}
\includegraphics[width=0.45\columnwidth]{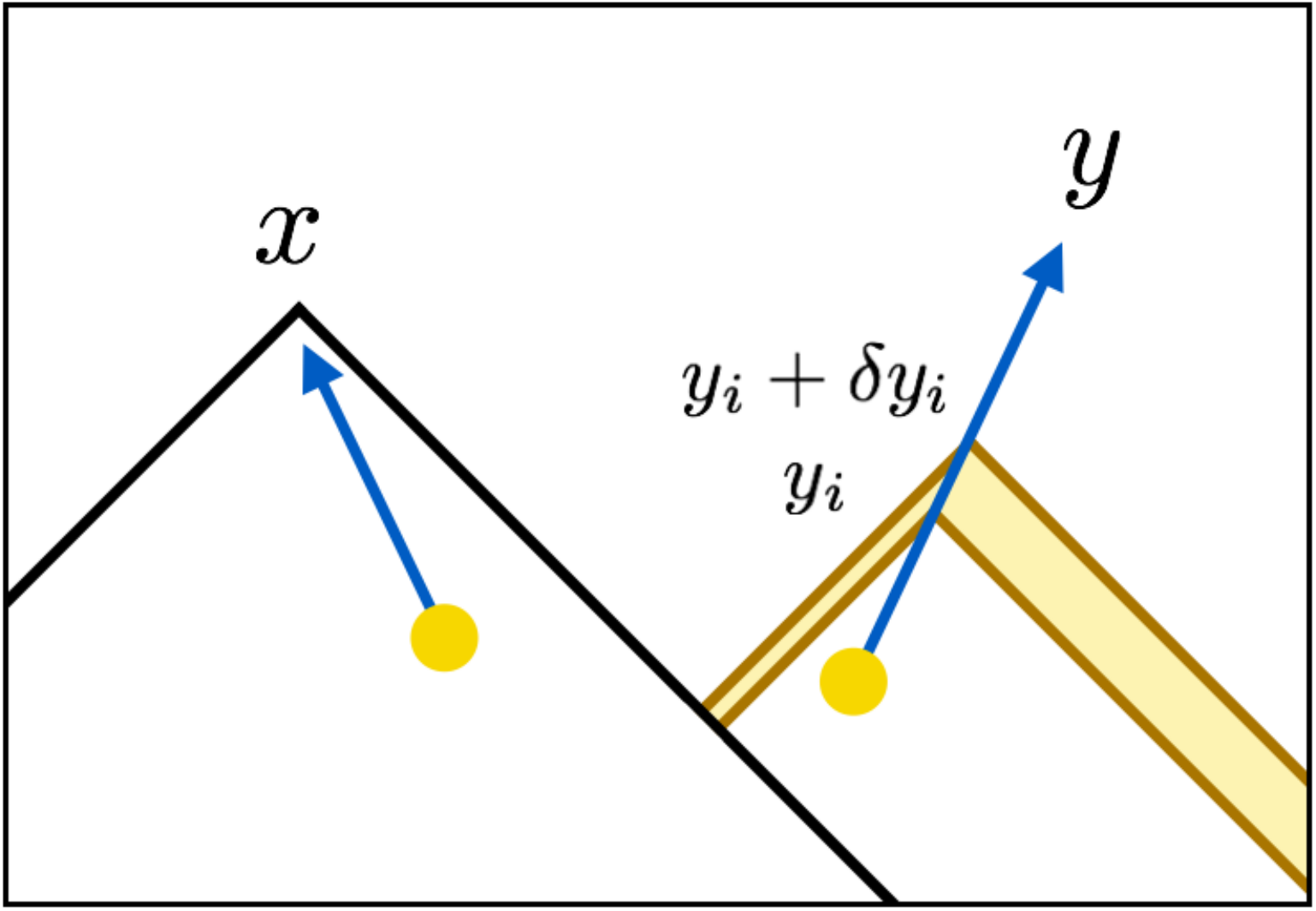}
\includegraphics[width=0.5\columnwidth]{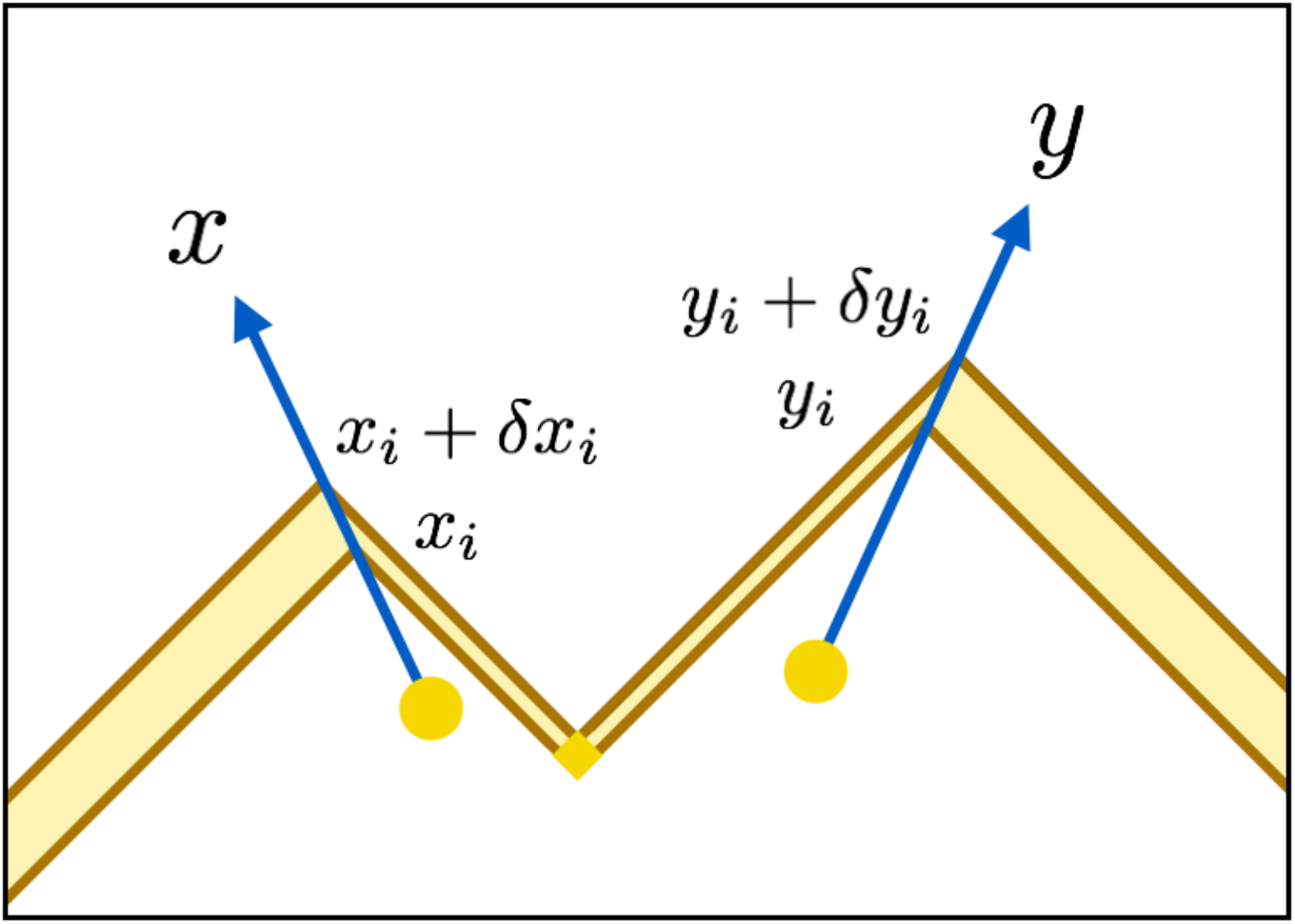}
\caption{\small
(Top-left) 
$C_iU$ contribution.
The $x$-fragment is intercepted in the time interval $[t_x,t_x + dt_x]$ 
by some bubble which do not intercept the $y$-fragment before $t_y$.
(Top-right) 
$UC_i$ contribution.
Same as the left panel, except that the role of $x$ and $y$ are interchanged.
(Bottom) 
$C_iC_i$ contribution.
One possibility is that 
one bubble nucleates in each of the left and right thin yellow region,
and the other possibility is that only one bubble nucleates in the small 
diamond-shaped region.
}
\label{fig:DoubleCept}
\end{center}
\end{figure}

These arguments are modified in the other contributions in $(U,C_i) \times (U,C_i)$.
Let us take $C_iU$ as an example. See the left panel of Fig.~\ref{fig:DoubleCept}.
Now the $x$-fragment has already been intercepted at the evaluation point $t_x$,
and therefore, in contrast to $UU$ contribution, aaa
we do not have to guarantee the ``$U$"(uncollided)-ness of the $x$-fragment by $\Theta_{\rm sp}(x,y_n)$.
Instead, $\Theta_{\rm sp}(x,y_n)$ must be replaced with $\Theta_{\rm sp}(x_i,y_n)$.
The reason is as follows. 
Even though the $x$-fragment is allowed to collide with other bubbles before the evaluation time $t_x$,
it is not allowed before $t_{xi}$
because $x_i$ is the interception point, or the first collision point, by definition.
Therefore we have to guarantee the ``$U$"-ness of the $x$-fragment until the interception point $x_i$,
and that is why we replace $\Theta_{\rm sp}(x,y_n)$ with $\Theta_{\rm sp}(x_i,y_n)$.\footnote{
After this replacement, 
$x$ and $y_n$ are no longer spacelike.
One might worry that, 
if $x$ and $y_n$ are timelike with $t_x < t_{yn}$,
nucleation of the $y$-bubble is not guaranteed 
because $y_n$ is inside the future $v$-cone of $x$,
which is inside the future $v$-cone of $x_n$
(that is, $\vec{y}_n$ has already been past by the $x$-bubble 
at the nucleation time $t_{yn}$).
However, this never occurs.
If $y_n$ is inside the future $v$-cone of $x$,
it means that $y_n$ is also inside the future $v$-cone of $x_i$
because the latter $v$-cone contains the former.
Such a combination of $x_i$ and $y_n$ is forbidden by $\Theta_{\rm sp}(x_i,y_n)$.
}
On the other hand, the other spacelike theta function $\Theta_{\rm sp}(x_n,y)$ remains the same,
because otherwise the ``$U$"-ness of the $y$-fragment is not guaranteed.
The derivation of the remaining part (i.e. $\left[ I(x,y_i) \right]_{tyi} e^{-I(x,y_i)}$ 
and ${\mathcal F}^{(d)}$) is basically the same as the single-bubble simplification formula.
Since the spacelike theta functions themselves 
do not guarantee that proper bubble configurations are realized,
we have to impose the following in order to realize such bubble configurations:
\begin{itemize}
\item
No bubbles nucleate in the union of the past $v$-cones of $x_i$ and $y$.
\item
The bubble which intercepts the $x$-fragment nucleates somewhere 
on the past $v$-cone of $x_i$, 
and it does not intercept the $y$-fragment before the evaluation time $t_y$.
\end{itemize}
Just as the single-bubble simplification formula,
these are simultaneously taken into account by considering
$P(x_i + \delta x_i,y) - P(x_i,y) = \left[ I(x_i,y) \right]_{txi} e^{-I(x_i,y)}$
for the interception of the $x$-fragment in the time interval $[t_{xi},t_{xi} + dt_{xi}]$.
For the ${\mathcal F}^{(d)}$ function, we have only to use 
${\mathcal F}^{(d)}(x,y;x_i = x,y_i;x_n,y_n)$ to replace the value of the energy-momentum tensor 
at the evaluation points.
These considerations give the expression for $C_iU$ in Eq.~(\ref{eq:AppTTdOthers}).
For the $UC_i$ contribution, 
the same argument holds with the roles of $x_{\bullet}$ and $y_{\bullet}$ interchanged.

For the $C_iC_i$ contributions there are two possibilities, 
just like $CC$ in the single-bubble contribution (see the bottom panel of Fig.~\ref{fig:DoubleCept}):
\begin{itemize}
\item
One bubble nucleates in $V_{x_i + \delta x_i} - (V_{x_i} \cup V_{y_i + \delta y_i})$,
and another nucleates in $V_{y_i + \delta y_i} - (V_{x_i + x_i + \delta x_i} \cup V_{y_i})$.
\item
One bubble nucleates in the small spacetime region 
$(V_{x_i + \delta x_i} \cup V_{y_i + \delta y_i} - V_{x_i} \cup V_{y_i}) 
- (V_{x_i + \delta x_i} - V_{x_i}) 
- (V_{y_i + \delta y_i} - V_{y_i})$.
\end{itemize}
The sum of these two probabilities gives 
$\left[ \left[ I(x_i,y_i) \right]_{txi} \left[ I(x_i,y_i) \right]_{tyi} - \left[ I(x_i,y_i) \right]_{txi,tyi} \right] dt_{xi}dt_{yi}$.
For the ${\mathcal F}^{(d)}$ function, we use ${\mathcal F}^{(d)}(x,y;x_i,y_i;x_n,y_n)$
to take both interceptions into account.
This completes the proof of Eq.~(\ref{eq:AppTTdOthers}).
Now, using integration by parts as in deriving Eq.~(\ref{eq:AppBeyondSingleSimple}), we have
\begin{align}
&\left< T_{ij}(x) T_{kl}(y) \right>^{(d,UU + UC_i + C_iU + C_iC_i)}
\nonumber \\
&\;\;\;\;\;\;
= 
\int_{-\infty}^{t_x} dt_{xn} 
\int_{-\infty}^{t_y} dt_{yn} 
\int_{t_{xn}}^{t_x} dt_{xi} 
\int_{t_{yn}}^{t_y} dt_{yi} 
\int d\Omega_{xn} 
\int d\Omega_{yn} \;
\nonumber \\
&\;\;\;\;\;\;\;\;\;\;\;\;
e^{-I(x_i,y_i)} 
\left[
\Theta_{\rm sp}(x_i,y_n)
\Theta_{\rm sp}(x_n,y_i)
{\mathcal F}^{(d)}(x,y;x_i,y_i;x_n,y_n)
\right]_{txi,tyi},
\label{eq:AppBeyondDoubleSimple1}
\end{align}
as the sum of the four contributions.

\begin{figure}
\begin{center}
\includegraphics[width=0.45\columnwidth]{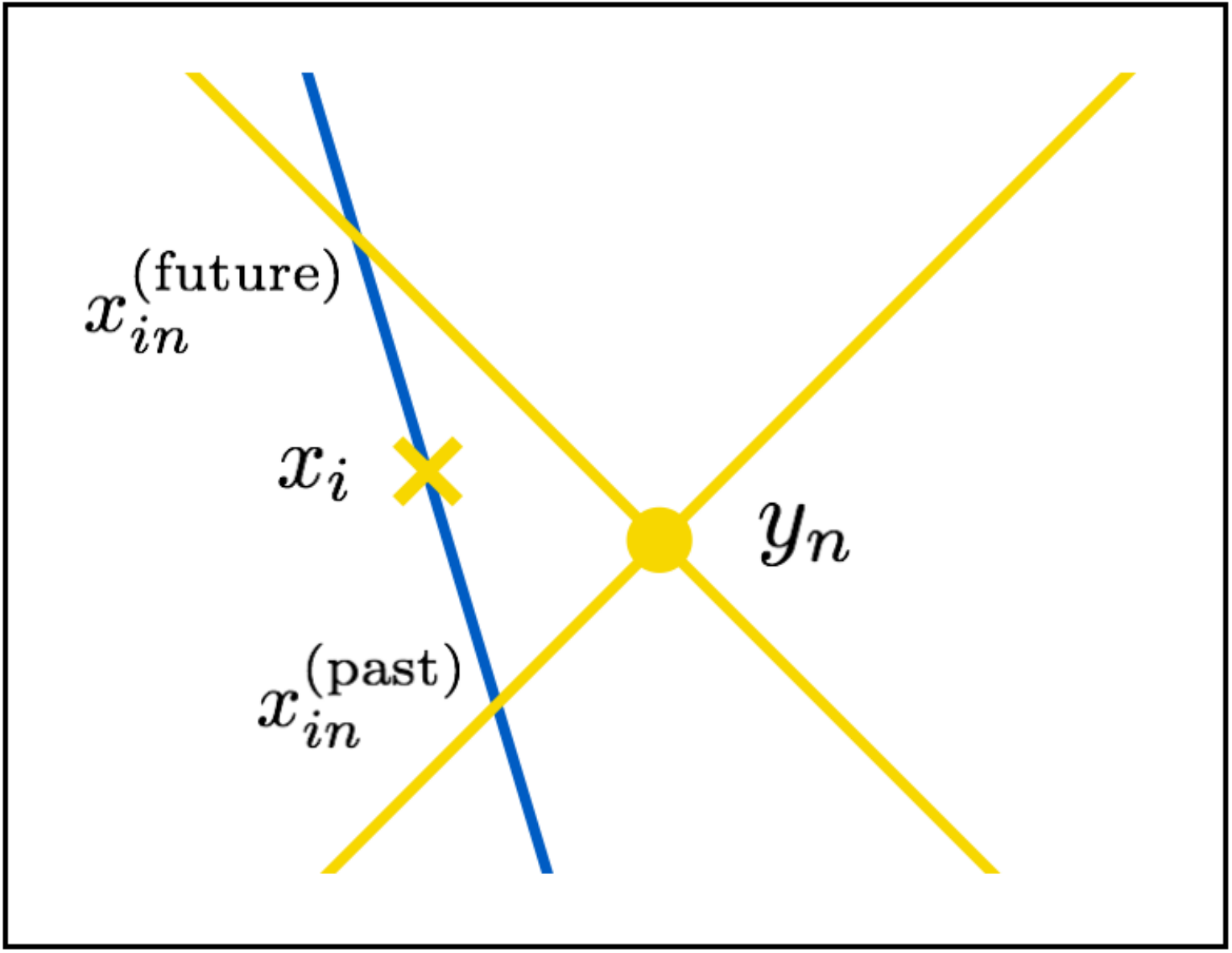}
\includegraphics[width=0.45\columnwidth]{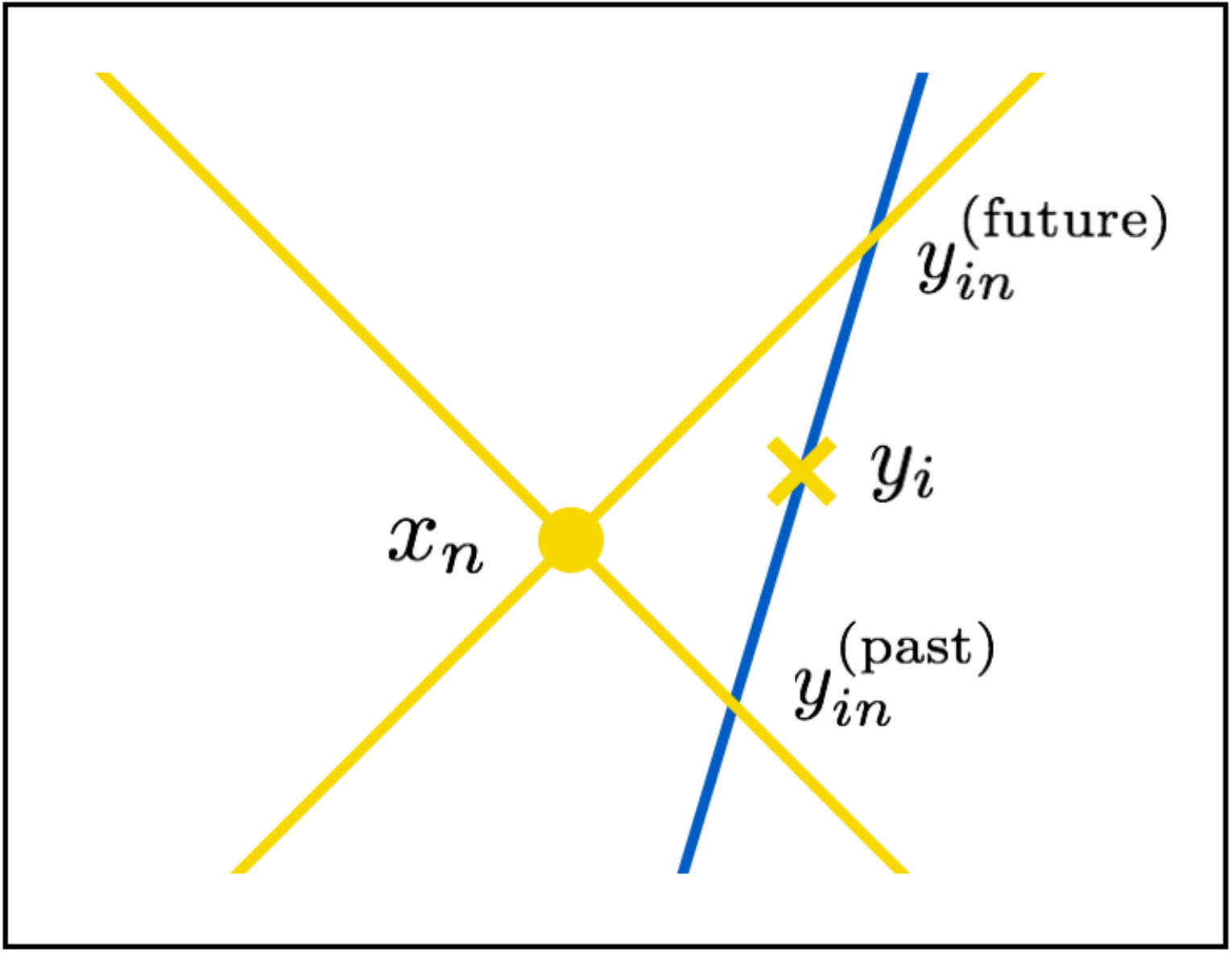}
\caption{\small
Definition of $x_{in}^{\rm (past)}$, $x_{in}^{\rm (future)}$ (left) 
and $y_{in}^{\rm (past)}$, $y_{in}^{\rm (future)}$ (right).
They are the spacetime crossing points of 
the propagation line of the $x$-($y$-)fragment and the $v$-cones of $y_n$ ($x_n$).
If the crossing point is on the past (future) $v$-cone, we label it with past (future).
The derivative of the spacelike theta function 
$\left[ \Theta_{\rm sp}(x_i,y_n) \right]_{txi}$ or $\left[ \Theta_{\rm sp}(x_n,y_i) \right]_{tyi}$
is nonzero only at these points.
Note that there is only either of $x_{in}^{\rm (past)}$ or $x_{in}^{\rm (future)}$
($y_{in}^{\rm (past)}$ or $y_{in}^{\rm (future)}$).
See the main text on this point.
}
\label{fig:xinyin}
\end{center}
\end{figure}

Below we see the behavior of the time derivatives acting on the spacelike theta functions
in Eq.~(\ref{eq:AppBeyondDoubleSimple1}),
because such terms partly cancel out with the other five contributions in $(U,C_i,C_{in}) \times (U,C_i,C_{in})$.
In the following, for given $x$ and $y$, 
we regard Eq.~(\ref{eq:AppBeyondDoubleSimple1}) as 
first fixing $(t_{xn},t_{yn})$ and $(\Omega_{xn},\Omega_{yn})$
and then scanning $(t_{xi},t_{yi})$, 
which means that we scan $(x_i,y_i)$ along the propagation lines of $x$- and $y$-fragments.
Now, let us consider $\left[ \Theta_{\rm sp}(x_i,y_n) \right]_{txi}$ for example.
From the definition of the subscript on the square bracket,
we take the derivative by changing $t_{xi}$ in $\Theta_{\rm sp}(x_i,y_n)$ 
along the propagation of the $x$-fragment.
This derivative is nonzero only when the spacetime relation between $(x_i,y_n)$ 
changes from timelike to spacelike ($\Theta_{\rm sp}$ changes from $0$ to $1$)
or spacelike to timelike ($\Theta_{\rm sp}$ changes from $1$ to $0$).
Such a spacetime point is either
\begin{itemize}
\item
On the past $v$-cone of $y_n$.
\item
On the future $v$-cone of $y_n$.
\end{itemize}
We denote these points by $x_{in}^{\rm (past)}$ and $x_{in}^{\rm (future)}$, respectively.
See Fig.~\ref{fig:xinyin} for illustration.
Though in this figure both $x_{in}^{\rm (past)}$ and $x_{in}^{\rm (future)}$ seem to exist,
this is an artifact arising from plotting in two-dimensions.
In fact, after some consideration, one sees that there is only one such point
along the propagation line of the $x$-fragment 
for given evaluation points $(x,y)$ and propagation directions $(\Omega_{xn},\Omega_{yn}).$\footnote{
In fact, the spacetime crossing point between the propagation line of the $x$-fragment 
and both the past and future $v$-cones of $y_n$ can be calculated 
by squaring the identity $\vec{x}_{in} - \vec{y}_n = (\vec{x}_{in} - \vec{x}) - (\vec{y}_n - \vec{x})$
with $|\vec{x}_{in} - \vec{y}_n| = v|t_{xin,yn}|$, $\vec{x}_{in} - \vec{x} = -vt_{x,xin}n_x$
and $\vec{y}_n - \vec{x} = -\vec{r} + vt_{y,yn}n_y$.
Also, the crossing point between the propagation line of the $y$-fragment 
and the past and future $v$-cones of $x_n$ is obtained by 
squaring $\vec{y}_{in} - \vec{x}_n = (\vec{y}_{in} - \vec{y}) - (\vec{x}_n - \vec{y})$
with $|\vec{y}_{in} - \vec{x}_n| = v|t_{yin,xn}|$, $\vec{y}_{in} - \vec{y} = -vt_{y,yin}n_y$
and $\vec{x}_n - \vec{y} = \vec{r} + vt_{xn,x}n_x$.
The resulting expressions are
\begin{align}
t_{xin,x}
&= \frac{-t_{x,yn}^2 + t_{y,yn}^2 + r^2 - 2t_{y,yn}rc_y}{2(t_{x,yn} + rc_x - t_{y,yn}c_{xy})},
\;\;\;\;
t_{yin,y}
= \frac{-t_{y,xn}^2 + t_{x,xn}^2 + r^2 + 2t_{x,xn}rc_x}{2(t_{y,xn} - rc_y - t_{x,xn}c_{xy})}.
\label{eq:AppBeyondtxintyin}
\end{align}
Here $c_{xy} = n_x \cdot n_y$ is the cosine between the propagation angles of $x$- and $y$-fragments,
and also we have presented these crossing points in terms of time differences $t_{xin,x} = t_{xin} - t_x$ 
and $t_{yin,y} = t_{yin} - t_y$.
This means that there is a unique spacetime crossing point between the propagation line and the $v$-cones.
\label{fn:AppCross}
}\footnote{
For special propagation directions with which the denominator of $t_{xin,x}$ or $t_{yin,y}$ 
in Eq.~(\ref{eq:AppBeyondtxintyin}) vanishes, 
there is no spacetime crossing between the propagation line and the $v$-cone.
Also, if the numerator vanishes in addition, there are infinite number of the crossing points.
However, since such contributions occupy a vanishing measure in the integration,
we can safely neglect them.
}
After further consideration, one sees that only the future crossing point can appear in
the integration region for $t_{xi}$,
and the past crossing point never appears in the integration.\footnote{
The past crossing point does not appear in the integration region for the following reason.
Suppose that it appears. Then, denoting the past crossing point by $x_{in}$ for brevity, one finds that 
\begin{itemize}
\item[]
\begin{center}
$(x_n,x_{in})$ is null with $t_{xn} < t_{xin}$.
\end{center}
\end{itemize}
Also one finds that
\begin{itemize}
\item[]
\begin{center}
$(x_{in},y_n)$ is null with $t_{xin} < t_{yn}$,
\;\;\;\;
$(y_n,y_i)$ is null with $t_{yn} < t_{yi}$.
\end{center}
\end{itemize}
Spacetime points $x_n$ and $y_i$ satisfying these three conditions are timelike 
(except for limited cases with vanishing integration measure).
Therefore the other spacelike theta function $\Theta_{\rm sp}(x_n,y_i)$ 
in Eq.~(\ref{eq:AppBeyondDoubleSimple1}) vanishes.
}
Therefore we write $x_{in}^{\rm (future)}$ simply as $x_{in}$, 
and take only this into account.
The future crossing point between the propagation line of the $y$-fragment and the $v$-cones of $x_n$
is defined in the same way, and we call it $y_{in}$.
Now the time derivatives give
\begin{align}
\left[ \Theta_{\rm sp}(x_i,y_n) \right]_{txi}
&= 
-\delta(t_{xi} - t_{xin}),
\;\;\;\;
\left[ \Theta_{\rm sp}(x_n,y_i) \right]_{tyi}
= 
-\delta(t_{yi} - t_{yin}),
\end{align}
where the minus signs arise because $\Theta_{\rm sp}$ changes from $1$ to $0$ 
at the future crossing points.
Therefore, we can rewrite Eq.~(\ref{eq:AppBeyondDoubleSimple1}) as 
\begin{align}
&\left< T_{ij}(x) T_{kl}(y) \right>^{(d,UU + UC_i + C_iU + C_iC_i)}
\nonumber \\
&\;\;\;\;\;\;
= 
\int_{-\infty}^{t_x} dt_{xn} 
\int_{-\infty}^{t_y} dt_{yn} 
\int_{t_{xn}}^{t_x} dt_{xi} 
\int_{t_{yn}}^{t_y} dt_{yi} 
\int d\Omega_{xn} 
\int d\Omega_{yn} \;
\nonumber \\
&\;\;\;\;\;\;\;\;\;\;\;\;
e^{-I(x_i,y_i)} 
\left[
\begin{matrix*}[l]
\;
\delta(t_{xi} - t_{xin}) \delta(t_{yi} - t_{yin}){\mathcal F}^{(d)}(x,y;x_i,y_i,x_n,y_n)
\\[1.5ex]
- \; \delta(t_{xi} - t_{xin}) \Theta_{\rm sp}(x_n,y_i) \left[ {\mathcal F}^{(d)}(x,y;x_i,y_i,x_n,y_n) \right]_{tyi}
\\[1.5ex]
- \; \Theta_{\rm sp}(x_i,y_n) \delta(t_{yi} - t_{yin}) \left[ {\mathcal F}^{(d)}(x,y;x_i,y_i,x_n,y_n) \right]_{txi}
\\[1.5ex]
+ \; \Theta_{\rm sp}(x_i,y_n) \Theta_{\rm sp}(x_n,y_i) \left[ {\mathcal F}^{(d)}(x,y;x_i,y_i,x_n,y_n) \right]_{txi,tyi}
\end{matrix*}
\right].
\label{eq:AppBeyondDoubleSimple2}
\end{align}
Below we see some of these terms cancel out with the other five contributions to give a simple formula.

%%%%%%%%%%%%%%%%
%%%%%%%%%%%%%%%%

%%%%%%%%%%%%%%%%
%%%%%%%%%%%%%%%%

%%%%%%%%%%%%%%%%%%%%%%%%%%%%%%%%%%%%%%%%%%%%%%%%%%
\subsubsection*{$(U,C_i,C_{in}) \times (U,C_i,C_{in})$ contributions}
%%%%%%%%%%%%%%%%%%%%%%%%%%%%%%%%%%%%%%%%%%%%%%%%%%

Let us see the other five contributions, 
$UC_{in}$, $C_{in}U$, $C_iC_{in}$, $C_{in}C_i$ and $C_{in}C_{in}$.
Remember that $C_{in}$ refers to those contributions where the $x$-(or $y$-)fragment is intercepted by
the $y$-(or $x$-)bubble.
Therefore, whether such a contribution exists or not 
depends on whether the future crossing point $t_{xin}$ or $t_{yin}$ 
lies in the integration region $[t_{xn},t_x]$ or $[t_{yn},t_y]$, respectively.
However, in any cases, the five contributions can be written as follows:
\begin{align}
\left< T_{ij}(x) T_{kl}(y) \right>^{(d,UC_{in})}
= 
&
\int_{-\infty}^{t_x} dt_{xn} 
\int_{-\infty}^{t_y} dt_{yn}
\int_{t_{yn}}^{t_y} dt_{yi} 
\int d\Omega_{xn} 
\int d\Omega_{yn} \;
\nonumber \\
&
\Theta_{\rm sp}(x,y_n) \delta(t_{yi} - t_{yin}) 
e^{-I(x,y_i)} 
{\mathcal F}^{(d)}(x,y;x,y_i;x_n,y_n),
\label{eq:AppBeyondDoubleSimpleUCin}
\end{align}
\begin{align}
\left< T_{ij}(x) T_{kl}(y) \right>^{(d,C_{in}U)}
= 
&
\int_{-\infty}^{t_x} dt_{xn} 
\int_{-\infty}^{t_y} dt_{yn}
\int_{t_{xn}}^{t_x} dt_{xi} 
\int d\Omega_{xn} 
\int d\Omega_{yn} \;
\nonumber \\
&
\delta(t_{xi} - t_{xin}) \Theta_{\rm sp}(x_n,y) 
e^{-I(x_i,y)} 
{\mathcal F}^{(d)}(x,y;x_i,y;x_n,y_n),
\label{eq:AppBeyondDoubleSimpleCinU}
\end{align}
\begin{align}
\left< T_{ij}(x) T_{kl}(y) \right>^{(d,C_iC_{in})}
= 
&
\int_{-\infty}^{t_x} dt_{xn} 
\int_{-\infty}^{t_y} dt_{yn}
\int_{t_{xn}}^{t_x} dt_{xi} 
\int_{t_{yn}}^{t_y} dt_{yi} 
\int d\Omega_{xn} 
\int d\Omega_{yn} \;
\nonumber \\
&
\Theta_{\rm sp}(x_i,y_n) \delta(t_{yi} - t_{yin}) 
\left[ I(x_i,y_i) \right]_{txi}
e^{-I(x_i,y_i)} 
{\mathcal F}^{(d)}(x,y;x_i,y_i;x_n,y_n),
\label{eq:AppBeyondDoubleSimpleCiCin}
\end{align}
\begin{align}
\left< T_{ij}(x) T_{kl}(y) \right>^{(d,C_{in}C_i)}
= 
&
\int_{-\infty}^{t_x} dt_{xn} 
\int_{-\infty}^{t_y} dt_{yn}
\int_{t_{xn}}^{t_x} dt_{xi} 
\int_{t_{yn}}^{t_y} dt_{yi} 
\int d\Omega_{xn} 
\int d\Omega_{yn} \;
\nonumber \\
&
\delta(t_{xi} - t_{xin}) \Theta_{\rm sp}(x_n,y_i) 
\left[ I(x_i,y_i) \right]_{tyi}
e^{-I(x_i,y_i)} 
{\mathcal F}^{(d)}(x,y;x_i,y_i;x_n,y_n),
\label{eq:AppBeyondDoubleSimpleCinCi}
\end{align}
\begin{align}
\left< T_{ij}(x) T_{kl}(y) \right>^{(d,C_{in}C_{in})}
= 
&
\int_{-\infty}^{t_x} dt_{xn} 
\int_{-\infty}^{t_y} dt_{yn}
\int_{t_{xn}}^{t_x} dt_{xi} 
\int_{t_{yn}}^{t_y} dt_{yi} 
\int d\Omega_{xn} 
\int d\Omega_{yn} \;
\nonumber \\
&
\delta(t_{xi} - t_{xin}) \delta(t_{yi} - t_{yin}) 
e^{-I(x_i,y_i)} 
{\mathcal F}^{(d)}(x,y;x_i,y_i;x_n,y_n).
\label{eq:AppBeyondDoubleSimpleCinCin}
\end{align}
First let us see the $C_{in}U$ contribution (\ref{eq:AppBeyondDoubleSimpleCinU}).
When the future crossing point $t_{xin}$ does not lie in the integration region $[t_{xn},t_x]$,
this expression trivially holds since it vanishes.
On the other hand, when $t_{xin}$ lies in the integration region $[t_{xn},t_x]$,
one can perform the $t_{xi}$ integration. This replaces all the $x_i$ in the integrand with $x_{in}$.
Now the meanings of the factors in the integrand are clear:
$\Theta_{\rm sp}(x_n,y)$ guarantees the ``$U$"-ness of the $y$-fragment.
$P(x_{in},y) = e^{-I(x_{in},y)}$ guarantees that no bubbles nucleate in the union of the past $v$-cones 
of $x_{in}$ and $y$, which is necessary for $x_{in}$ to be the interception (first collision) point 
and for $y$ to remain uncollided until the evaluation time.
${\mathcal F}^{(d)}$ with $x_i$ replaced by $x_{in}$ represents 
the proper value of the energy-momentum tensor.
Note that no spacelike theta function appears for $x_i$ and $y_n$ 
because they are null with each other in the present case.
The same arguments apply to $UC_{in}$ contribution (\ref{eq:AppBeyondDoubleSimpleUCin}).

Next let us see the $C_{in}C_i$ contribution (\ref{eq:AppBeyondDoubleSimpleCinCi}).
This trivially holds when $t_{xin}$ does not lie in the integration region $[t_{xn},t_x]$.
In the opposite case, one can perform $t_{xi}$ integration to replace
all the $x_i$ in the integrand with $x_{in}$.
Now the meanings of the factors in the integrand are as follows:
$\Theta_{\rm sp}(x_n,y_i)$ guarantees that $y_i$ is the interception (first collision) point.
$\left[ I(x_{in},y_i) \right]_{tyi} e^{-I(x_{in},y_i)}$ guarantees 
that no bubbles nucleate in the union of the past $v$-cones of $x_{in}$ and $y_i$, 
and also that the intercepting bubble for the $y$-fragment nucleate somewhere 
on the past $v$-cone of $y_i$ so that the interception occurs in the time interval $[t_{yi},t_{yi} + dt_{yi}]$.
${\mathcal F}^{(d)}$ with $x_i$ replaced by $x_{in}$ represents 
the proper value of the energy-momentum tensor.
The same arguments apply to $C_iC_{in}$ contribution (\ref{eq:AppBeyondDoubleSimpleCiCin}).

The last is the $C_{in}C_{in}$ contribution (\ref{eq:AppBeyondDoubleSimpleCinCin}).
We consider those cases where both $t_{xin}$ and $t_{yin}$ are in the integration regions 
since otherwise trivial.
We can perform $t_{xi}$ and $t_{yi}$ integrations,
and the resulting false vacuum probability $P(x_{in},y_{in}) = e^{-I(x_{in},y_{in})}$
guarantees that no bubbles nucleate in the union of the past $v$-cones of $x_{in}$ and $y_{in}$,
while ${\mathcal F}^{(d)}$ with $(x_i,y_i)$ replaced by $(x_{in},y_{in})$ properly represents 
the value of the energy-momentum tensor in the $C_{in}C_{in}$ contribution.

Finally, we sum up all the contributions 
(\ref{eq:AppBeyondDoubleSimple2})--(\ref{eq:AppBeyondDoubleSimpleCinCin}).
All the terms except for the last one in Eq.~(\ref{eq:AppBeyondDoubleSimple2}) cancel out to give\footnote{
The procedure is as follows.
Eq.~(\ref{eq:AppBeyondDoubleSimpleCiCin}) and the third line in the square bracket in
Eq.~(\ref{eq:AppBeyondDoubleSimple2}) give a term like $\left[ -e^{-I(x_i,y_i)}{\mathcal F}^{(d)} \right]_{txi}$.
Integration by parts with respect to $t_{xi}$ 
gives a surface term at $t_{xi} = t_x$ and a term with two $\delta$-functions.
The former cancels out with Eq.~(\ref{eq:AppBeyondDoubleSimpleUCin}).
Similar procedure on
Eq.~(\ref{eq:AppBeyondDoubleSimpleCinCi}) and the second line in
Eq.~(\ref{eq:AppBeyondDoubleSimple2})
gives a surface term at $t_{yi} = t_y$ and a term with two $\delta$-functions.
The former cancels out with Eq.~(\ref{eq:AppBeyondDoubleSimpleCinU}).
The remaining term cancels out with 
the first line in Eq.~(\ref{eq:AppBeyondDoubleSimple2})
and Eq.~(\ref{eq:AppBeyondDoubleSimpleCinCin}).
This leaves only the last line in Eq.~(\ref{eq:AppBeyondDoubleSimple2}).
}
\begin{align}
&\left< T_{ij}(x) T_{kl}(y) \right>^{(d,(U,C_i,C_{in}) \times (U,C_i,C_{in}))}
\nonumber \\
&\;\;\;\;\;\;\;\;
= 
\int_{-\infty}^{t_x} dt_{xn} 
\int_{-\infty}^{t_y} dt_{yn} 
\int_{t_{xn}}^{t_x} dt_{xi} 
\int_{t_{yn}}^{t_y} dt_{yi} 
\int d\Omega_{xn} 
\int d\Omega_{yn} \;
\nonumber \\
&\;\;\;\;\;\;\;\;\;\;\;\;\;\;\;\;
\Theta_{\rm sp}(x_i,y_n)
\Theta_{\rm sp}(x_n,y_i)
e^{-I(x_i,y_i)} 
\left[
{\mathcal F}^{(d)}(x,y;x_i,y_i;x_n,y_n)
\right]_{txi,tyi}.
\label{eq:AppBeyondDoubleSimple}
\end{align}
This is what we call the ``simplification formula" for the double-bubble.

\clearpage

%%%%%%%%%%%%%%%%%%%%%%%%%%%%%%%%%%%%%%%%%%%%%%%%%%
\section{Techniques for numerical evaluation}
\label{app:tech}
\setcounter{equation}{0}
%%%%%%%%%%%%%%%%%%%%%%%%%%%%%%%%%%%%%%%%%%%%%%%%%%

In this appendix we explain several techniques used for numerical evaluation of the GW spectrum.

%%%%%%%%%%%%%%%%%%%%%%%%%%%%%%%%%%%%%%%%%%%%%%%%%%
\subsection{Equivalent ways to impose the spacelike conditions}
\label{subapp:space}
%%%%%%%%%%%%%%%%%%%%%%%%%%%%%%%%%%%%%%%%%%%%%%%%%%

In the double-bubble contribution (\ref{eq:BeyondDeltaDConcConc}),
we encounter two spacelike theta functions.
In this subsection, we show that they can be replaced by one spacelike theta function.
This replacement makes it possible to integrate out $t_{x,xn}$ and $t_{y,yn}$ directions,
and as a result leaves seven-dimensional integration.

\begin{figure}
\begin{center}
\includegraphics[width=0.45\columnwidth]{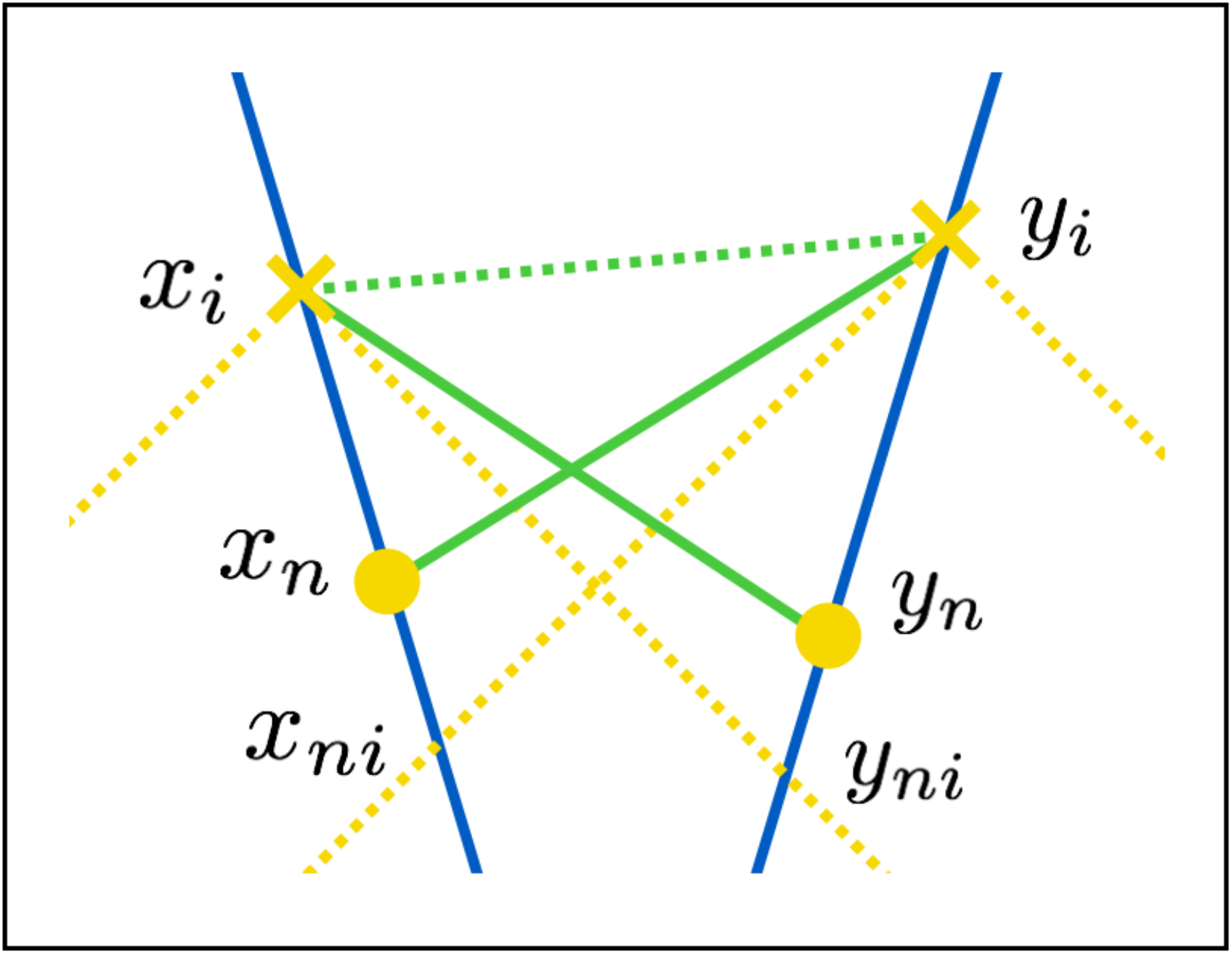}
\includegraphics[width=0.45\columnwidth]{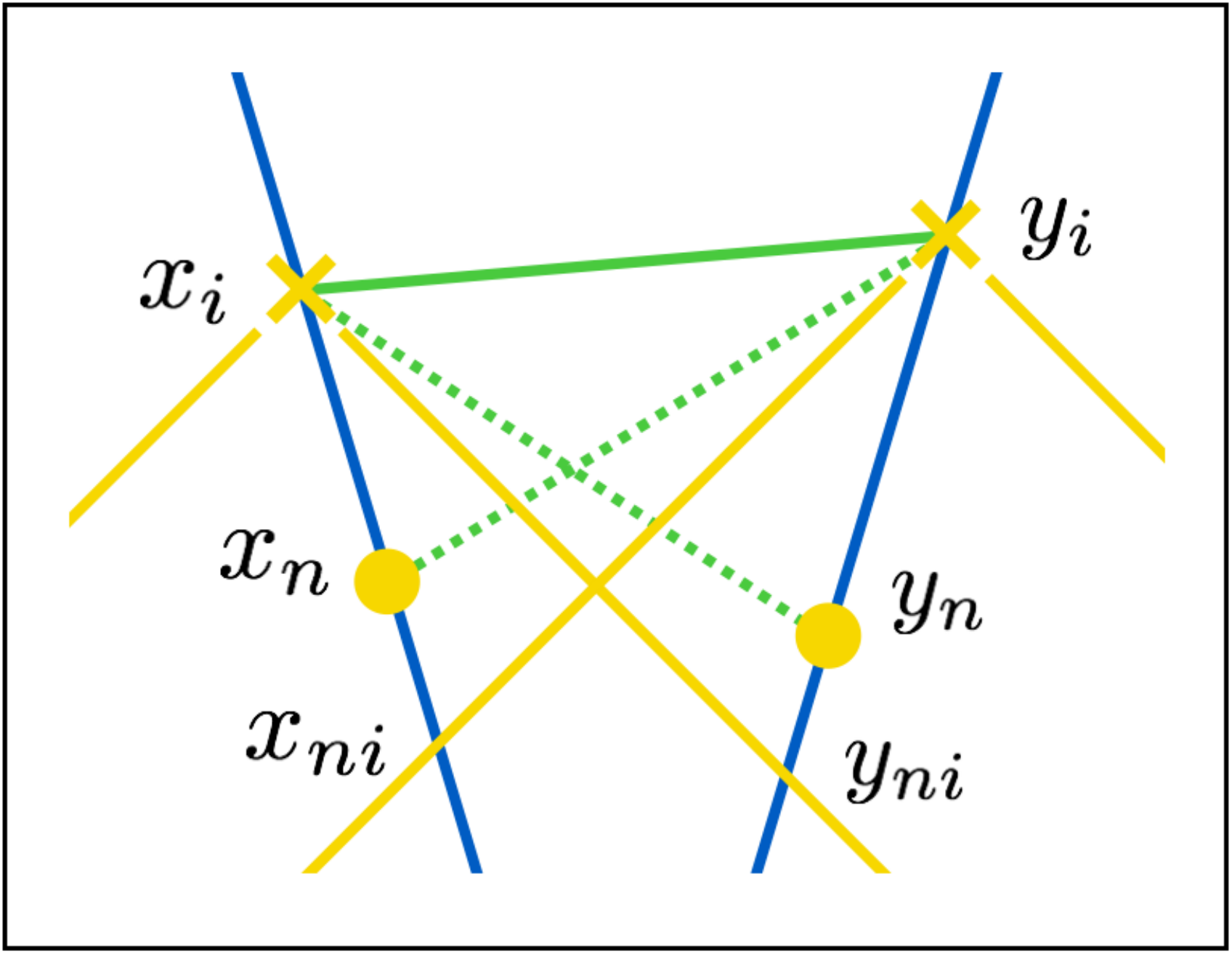}
\caption{\small
Two equivalent ways of imposing the spacelike conditions.
Spacelike points are connected with green lines.
(Left) Fix the four points so that $(x_i,y_n)$ and $(x_n,y_i)$ are spacelike.
(Right) Fix $x_i$ and $y_i$ so that they are spacelike,
and then choose $x_n$ and $y_n$ so that 
they do not enter the past $v$-cones of $y_i$ and $x_i$, respectively.
By adopting the way shown in the right panel,
$t_{xn}$ and $t_{yn}$ can be integrated 
from $t_{xni}$ to $t_{xi}$ and from $t_{yni}$ to $t_{yi}$, respectively.
}
\label{fig:Spacelike}
\end{center}
\end{figure}

First, we write Eq.~(\ref{eq:BeyondDeltaDConcConc}) here again for completeness:
\begin{align}
\Delta^{(d)}
=
&
\int_{-\infty}^\infty dt_{x,y}
\int_0^\infty dr_v
\nonumber \\
&
\int_0^\infty dt_{x,xn}
\int_0^\infty dt_{y,yn}
\int_0^{t_{x,xn}} dt_{x,xi}
\int_0^{t_{y,yn}} dt_{y,yi}
\int_{-1}^1 dc_{xn}
\int_{-1}^1 dc_{yn}
\int_0^{2\pi} d\phi_{xn,yn}
\nonumber \\[1ex]
\frac{v^3k^3}{3}
&\left[
\begin{matrix*}[l]
\;
\Theta_{\rm sp}(x_i,y_n)
\Theta_{\rm sp}(x_n,y_i)
\\[1.5ex]
\displaystyle
\times \;
\frac{e^{-(t_{xi,xn} + t_{yi,yn})} e^{-(t_{x,xi} + t_{y,yi})/\tau}}
{{\mathcal I}(t_{xi,yi},r_{xi,yi}/v)^2}
(3t_{xi,xn}^2 + t_{xi,xn}^3/\tau)
(3t_{yi,yn}^2 + t_{yi,yn}^3/\tau)
\\[2ex]
\displaystyle
\times \;
r_v^2
\left[
j_0(vkr_v){\mathcal K}_0(n_{xn},n_{yn})
+ \frac{j_1(vkr_v)}{vkr_v}{\mathcal K}_1(n_{xn},n_{yn})
+ \frac{j_2(vkr_v)}{(vkr_v)^2}{\mathcal K}_2(n_{xn},n_{yn})
\right]
\\[2ex]
\times
\cos(kt_{x,y})
\end{matrix*}
\right].
\label{eq:AppBeyondDeltaDConcConc}
\end{align}
Suppose that we first fix the evaluation points $(x,y)$ and the propagation directions
of the wall fragments $(\Omega_{xn},\Omega_{yn})$.
Then, the spacelike theta functions in this expression require 
two nucleation points $(x_n,y_n)$ and two interception points $(x_i,y_i)$ to satisfy
\begin{itemize}
\item
$(x_i,y_n)$ are spacelike.
\item
$(x_n,y_i)$ are spacelike.
\end{itemize}
We illustrate this procedure in the left panel of Fig.~\ref{fig:Spacelike}.
Fixing $(x,y)$ and $(\Omega_{xn},\Omega_{yn})$ corresponds to 
fixing the separation and directions of the two blue lines in this figure.
Then finding $(x_n,y_n)$ and $(x_i,y_i)$ satisfying the above conditions means
finding four points on these lines so that $(x_i,y_n)$ and $(x_n,y_i)$ are spacelike 
(denoted by the green solid lines).
Now we argue that this procedure is equivalent to the following 
\begin{itemize}
\item
Find spacelike $(x_i,y_i)$.
\item
Fix $(x_n,y_n)$ so that 
$x_n$ lies between $x_{ni}$ and $x_i$
and
$y_n$ lies between $y_{ni}$ and $y_i$.
\end{itemize}
Here $x_{ni}$ and $y_{ni}$ are defined as 
the spacetime crossing points between the propagation lines of the $x$- and $y$-fragments and 
the past $v$-cone of $y_i$ and $x_i$, respectively.
We illustrate this procedure in the right panel of Fig.~\ref{fig:Spacelike}.

The equivalence is proven as follows.
First, suppose that we have spacelike combinations of $(x_i,y_n)$ and $(x_n,y_i)$.
Then $(x_i,y_i)$ are spacelike 
because otherwise either of $(x_i,y_n)$ or $(x_n,y_i)$ becomes timelike
(note that $t_{xn} < t_{xi}$ and $t_{yn} < t_{yi}$).
Also $x_n$ ($y_n$) lies between $x_{ni}$ and $x_i$ ($y_{ni}$ and $y_i$) 
because otherwise it enters the past $v$-cone of $y_i$ ($x_i$) and becomes timelike to $y_i$ ($x_i$).
Second, suppose that $(x_i,y_i)$ are spacelike and
that $x_n$ ($y_n$) is chosen so that 
it is between $x_{ni}$ and $x_n$ ($y_{ni}$ and $y_n$).
Then $(x_i,y_n)$ and $(x_n,y_i)$ are spacelike combinations because $x_n$ and $y_n$ are 
outside the past $v$-cone of $y_i$ and $x_i$, respectively.
This proves the equivalence.

Here one may notice that $x_{ni}$ and $y_{ni}$ do not necessarily exist,
just as $x_{in}$ and $y_{in}$ do not necessarily exist in Appendix~\ref{subapp:BeyondEnvDouble}.
In fact, there is only one spacetime crossing point between the propagation line of 
the $x$-(or $y$-)fragment and the past and future $v$-cones of $y_i$ (or $x_i$).
They are given by\footnote{
These are obtained by squaring the identity
$\vec{x}_{ni} - \vec{y}_i = (\vec{x}_{ni} - \vec{x}) - (\vec{y}_i - \vec{x})$
with $|\vec{x}_{ni} - \vec{y}_i| = v|t_{xni,yi}|$, $\vec{x}_{ni} - \vec{x} = vt_{x,xni}n_x$
and $\vec{y}_i - \vec{x} = -\vec{r} + vt_{y,yi}n_y$,
and the identity
$\vec{y}_{ni} - \vec{x}_i = (\vec{y}_{ni} - \vec{y}) - (\vec{x}_i - \vec{y})$
with $|\vec{y}_{ni} - \vec{x}_i| = v|t_{yni,xi}|$, $\vec{y}_{ni} - \vec{y} = vt_{y,yni}n_y$
and $\vec{x}_i - \vec{y} = \vec{r} + vt_{x,xi}n_x$, respectively.
}
\begin{align}
t_{xni,x}
&= \frac{-t_{x,yi}^2 + t_{y,yi}^2 + r^2 - 2t_{y,yi}rc_y}{2(t_{x,yi} + rc_x - t_{y,yi}c_{xy})},
\;\;\;\;
t_{yni,y}
= \frac{-t_{y,xi}^2 + t_{x,xi}^2 + r^2 + 2t_{x,xi}rc_x}{2(t_{y,xi} - rc_y - t_{x,xi}c_{xy})}.
\label{eq:txnityni}
\end{align}
Therefore the past crossing point $x_{ni}$ or $y_{ni}$ does not exist 
if the solution (\ref{eq:txnityni}) indicates the crossing point on the future $v$-cone.
Even in such cases, the equivalence holds if one uses 
$t_{xni} = -\infty$ or $t_{yni} = -\infty$,
because the absence of $x_{ni}$ or $y_{ni}$ just means that 
the nucleation point $x_n$ or $y_n$ can be chosen to be infinitely past
without violating the spacelikeness of $(x_i,y_n)$ or $(x_n,y_i)$.

%%%%%%%%%%%%%%%%
%%%%%%%%%%%%%%%%

\begin{figure}
\begin{center}
\includegraphics[width=0.6\columnwidth]{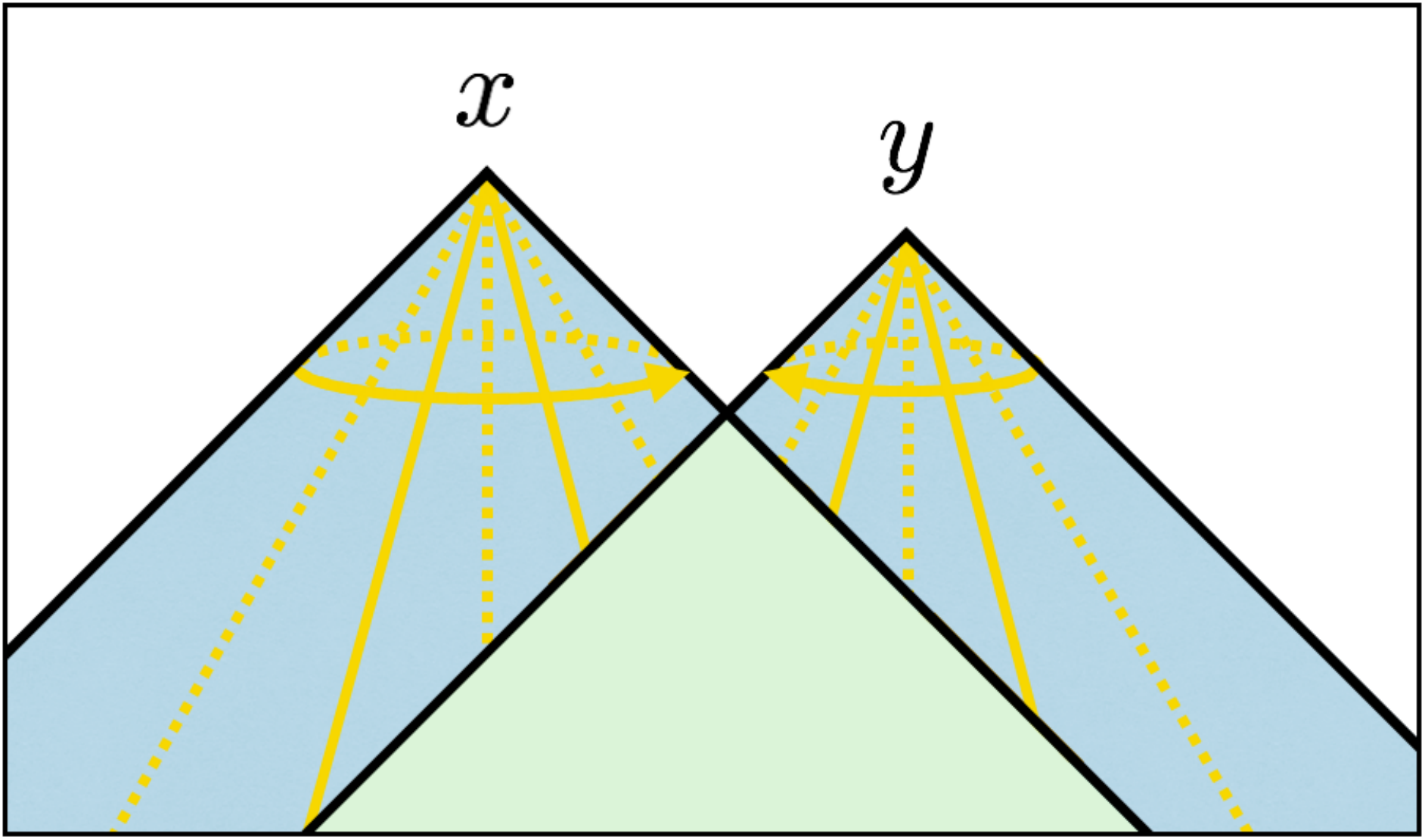}
\caption{\small
Illustration for the envelope limit $\tau \to 0$ in 
Eq.~(\ref{eq:AppBeyondDeltaDConcConcFinal}).
For each yellow line 
(which corresponds to the propagation line of a wall fragment),
the time integration from the upper endpoint to the lower endpoint
is performed in the first place.
The dependence on the lower endpoint appear in 
Eq.~(\ref{eq:AppBeyondDeltaDConcConcFinal})
as $t_{xni}$ or $t_{yni}$ through the ${\mathcal E}$ function.
}
\label{fig:DoubleUU2}
\end{center}
\end{figure}

Using this equivalence, one can integrate $t_{x,xn}$ and $t_{y,yn}$ 
in Eq.~(\ref{eq:AppBeyondDeltaDConcConc}).
Rewriting the integration with $t_{x,xn}$, $t_{y,yn}$, $t_{x,xi}$ and $t_{y,yi}$ as
\begin{align}
\int_0^\infty dt_{x,xn}
\int_0^\infty dt_{y,yn}
\int_0^{t_{x,xn}} dt_{x,xi}
\int_0^{t_{y,yn}} dt_{y,yi}
\;
\to
\;
\int_0^\infty dt_{x,xi}
\int_0^\infty dt_{y,yi}
\int_0^\infty dt_{xi,xn}
\int_0^\infty dt_{yi,yn},
\end{align}
one sees that the integrand vanishes for $t_{xi,xn}$ and $t_{yi,yn}$ 
larger than $t_{xi,xni}$ and $t_{yi,yni}$, respectively, because of the above arguments.
Then one can perform $t_{xi,xn}$ and $t_{yi,yn}$ integration 
on $[0,t_{xi,xni}]$ and $[0,t_{yi,yni}]$, respectively.
Note that $t_{xi,xni}$ or $t_{yi,yni}$ must be taken to be $+\infty$ if $x_{ni}$ or $y_{ni}$ does not exist.
As a result, one obtains
\begin{align}
\Delta^{(d)}
=
&
\int_{-\infty}^\infty dt_{x,y}
\int_0^\infty dr_v
\int_0^\infty dt_{x,xi}
\int_0^\infty dt_{y,yi}
\int_{-1}^1 dc_{xn}
\int_{-1}^1 dc_{yn}
\int_0^{2\pi} d\phi_{xn,yn}
\nonumber \\[1ex]
12v^3k^3
&\left[
\begin{matrix*}[l]
\displaystyle
\;
\Theta_{\rm sp}(x_i,y_i)
\\[1.5ex]
\displaystyle
\times
\left(1 + \frac{1}{\tau} \right)^2
\left[ 1 - {\mathcal E}(\tau;t_{xi,xni}) \right]
\left[ 1 - {\mathcal E}(\tau;t_{yi,yni}) \right]
\frac{e^{-(t_{x,xi} + t_{y,yi})/\tau}}
{{\mathcal I}(t_{xi,yi},r_{xi,yi}/v)^2}
\\[2ex]
\displaystyle
\times \;
r_v^2
\left[
j_0(vkr_v){\mathcal K}_0(n_{xn},n_{yn})
+ \frac{j_1(vkr_v)}{vkr_v}{\mathcal K}_1(n_{xn},n_{yn})
+ \frac{j_2(vkr_v)}{(vkr_v)^2}{\mathcal K}_2(n_{xn},n_{yn})
\right]
\\[2ex]
\times
\cos(kt_{x,y})
\end{matrix*}
\right],
\label{eq:AppBeyondDeltaDConcConcFinal}
\end{align}
where the ${\mathcal E}$ function is defined as
\begin{align}
{\mathcal E}(\tau;t_{i,ni})
&\equiv
e^{-t_{i,ni}}
\left[ \frac{1}{6(1 + \tau)}t_{i,ni}^3 + \frac{1}{2}t_{i,ni}^2 + t_{i,ni} + 1 \right].
\end{align}
Note that the two ${\mathcal E}$'s in the second line
vanish for $t_{xi,xni} = \infty$ and $t_{yi,yni} = \infty$, respectively.
Therefore, one sees that nonzero ${\mathcal E}$'s arise 
when we cannot take the nucleation points 
infinitely past for given evaluation points and propagation directions.
Also note that the behavior in $\tau \to 0$ limit (envelope limit) is safely taken,
because $(1 + 1/\tau)^2$ and $e^{-(t_{x,xi} + t_{y,yi})/\tau}$ 
in the second line in the square parenthesis work as $\delta$-functions in this limit:
$e^{-t_{x,xi}/\tau}/\tau \to \delta(t_{x,xi} - \epsilon)$ 
and $e^{-t_{y,yi}/\tau}/\tau \to \delta(t_{y,yi} - \epsilon)$
with $\epsilon$ being infinitesimal and positive.
Therefore $t_{x,xi}$ and $t_{y,yi}$ integrations can be completed in that limit,
and all $t_{xi}$ and $t_{yi}$ in the integrand are substituted by $t_x$ and $t_y$, respectively.
Numerically, it is confirmed in Sec.~\ref{sec:Numerical} 
that Eq.~(\ref{eq:AppBeyondDeltaDConcConcFinal}) gives the same value as the envelope case
for small values of $\tau$.\footnote{
Though Eq.~(\ref{eq:AppBeyondDeltaDConcConcFinal}) numerically gives the same value as 
the envelope case in $\tau \to 0$ limit,
it is difficult to directly compare this expression with 
the one we obtain in Appendix~\ref{subsec:EnvDouble}.
This is because parametrization of the integration is different in these expressions.
The difference is understood from Figs.~\ref{fig:EnvDouble}, \ref{fig:DoubleUU} and \ref{fig:DoubleUU2}.
After taking the envelope limit in Eq.~(\ref{eq:AppBeyondDeltaDConcConcFinal}), 
the resulting expression parametrizes the integral as shown in Fig.~\ref{fig:DoubleUU2}.
Fixing the propagation directions, i.e. fixing $c_{xn}$, $c_{yn}$ and $\phi_{xn,yn}$, 
determines the yellow lines for $x$- and $y$-fragments in this figure.
In Eq.~(\ref{eq:AppBeyondDeltaDConcConcFinal}), 
the time integrations along these yellow lines are already performed.
The lower endpoints of these yellow lines are encoded as 
the direction-dependent $t_{xni}$ and $t_{yni}$ in this equation.
Therefore, the parametrization of the integral is different 
from the envelope calculation (\ref{eq:EnvDeltaDGeneral}) or Fig.~\ref{fig:EnvDouble}, 
and also from the original parametrization of the $UU$ contribution 
(\ref{eq:TTdUU}) or Fig.~\ref{fig:DoubleUU}.
However, these are just difference in parametrization and the physics is the same.
}

%%%%%%%%%%%%%%%%
%%%%%%%%%%%%%%%%

%%%%%%%%%%%%%%%%%%%%%%%%%%%%%%%%%%%%%%%%%%%%%%%%%%
\subsection{$\left< T(x) \right> \left< T(y) \right>$ subtraction}
%%%%%%%%%%%%%%%%%%%%%%%%%%%%%%%%%%%%%%%%%%%%%%%%%%

Numerically, the double-bubble spectrum (\ref{eq:AppBeyondDeltaDConcConcFinal}) 
shows a poor convergence especially when $\tau$ is much larger than $1/\beta$.
There is a physical reason for this behavior.
Below we first explain the reason, and then provide a remedy for it.

\begin{figure}
\begin{center}
\includegraphics[width=0.6\columnwidth]{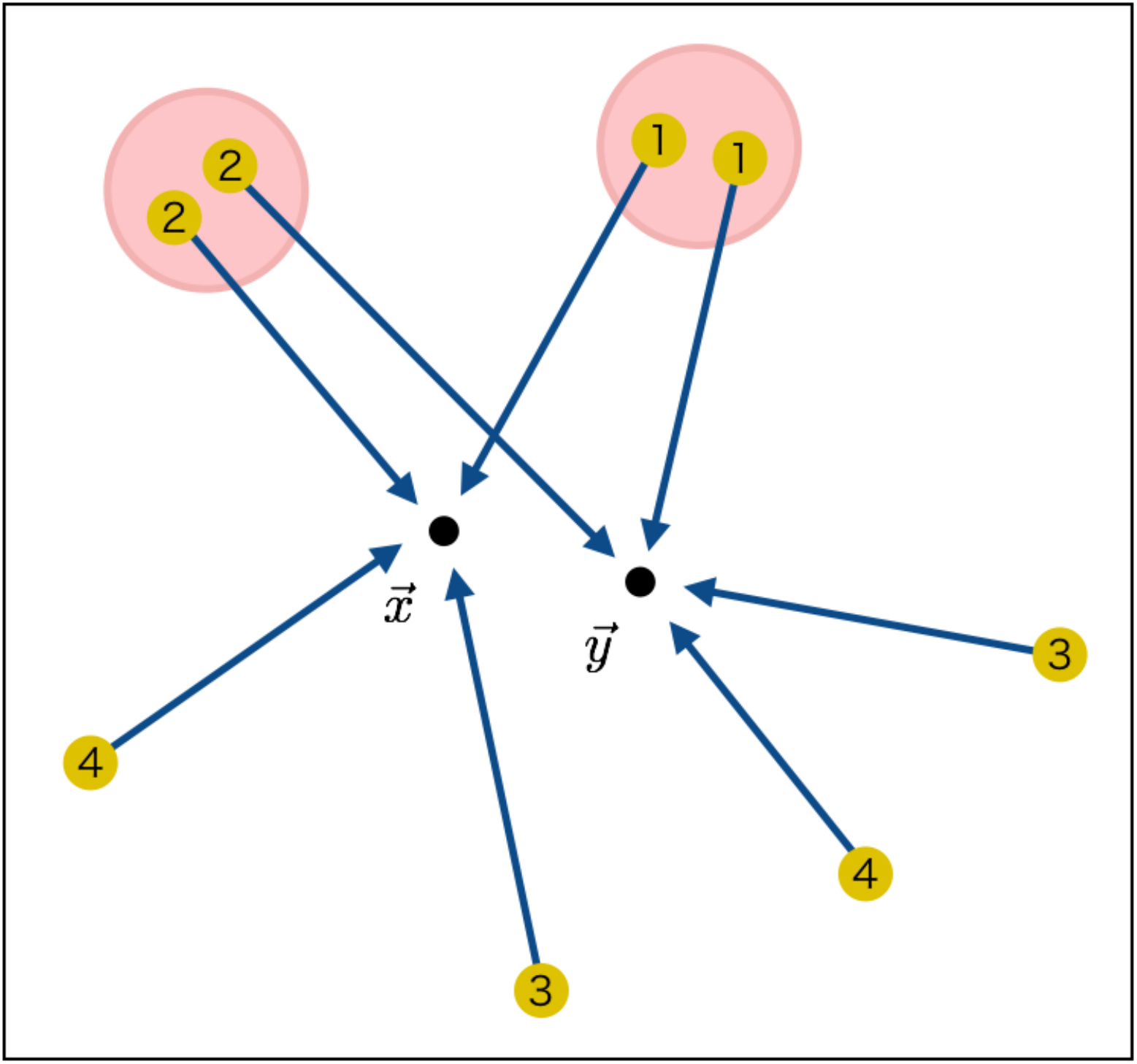}
\caption{\small
Illustration for how correlation between the $x$- and $y$-bubbles arises.
If the $y$-bubble nucleates at point $1$, 
the $x$-bubble tend not to nucleate around the $y$-bubble nucleation point,
or tend to have a weak bubble wall even if it nucleates around there, 
due to the existence of the $y$-bubble.
Such a realization pattern is denoted by label $1$.
The pink circles denote the correlation region, which typically have size $\sim 1/\beta$.
The same applies to the nucleation pattern $2$.
For the nucleation patterns $3$ or $4$, the $x$- and $y$-bubbles nucleate 
and experience subsequent collisions with other bubbles 
almost independently of each other.
For long-lasting walls (i.e. $\tau \gg 1/\beta$),
there are much more cases like $3$ or $4$ than like $1$ or $2$.
}
\label{fig:Cherry}
\end{center}
\end{figure}

First remember that $1/\beta$ determines the typical timescale from bubble nucleation to collisions
(see the arguments around Eq.~(\ref{eq:Gamma})).
Then suppose that $\tau$ is much larger than $1/\beta$.
The energy and momentum of the bubble walls decrease only with power-law 
after collisions for time $\sim \tau$, and then vanishes rapidly.
What makes the poor convergence is such long-lasting walls.
In order to explain how such walls cause a problem, 
we fix the evaluation points $x$ and $y$
so that both time and spacial separations of them are much smaller than $\tau$
(but not necessarily smaller than $1/\beta$).
These evaluation points can be either timelike or spacelike,
because both contributions exist in the double-bubble beyond the envelope.   
Then, as shown in Fig.~\ref{fig:Cherry} with the yellow circles numbered from $1$ to $4$, 
the energy-momentum tensor at these evaluation points is contributed by bubble walls 
from every direction which nucleated far from the evaluation points.

Let us consider the evaluation point $x$ for the moment,
and see how these wall fragments contribute to the energy-momentum tensor at $x$.
If they come from every direction with exactly the same condition,
the resulting ensemble average of the energy-momentum tensor at $x$ 
is isotropic and it does not contribute to the GW spectrum.
We used similar arguments in Appendix~\ref{subsec:EnvDouble}:
we restricted the integration regions for $t_{xn}$ and $t_{yn}$ before $t_{\rm max}$,
because those bubbles which nucleate after $t_{\rm max}$ give an isotropic contribution
after taking the ensemble average.
However, in the present case, 
the wall fragments do not necessarily come from every direction in the same way.
Given that the $y$-bubble has nucleated somewhere, 
$x$-fragment tend not to come from the direction where $y$-bubble has nucleated.
Fig.~\ref{fig:Cherry} illustrates this.
If the $y$-bubble has already nucleated at point $1$,
the $x$-bubble cannot nucleate around that point
(because spacial regions around the $y$-bubble soon experience the transition from the false to true vacuum) 
or tends to be weaker even if it nucleates
(because such a bubble soon collides with the $y$-bubble).
If the $y$-bubble has nucleated at point $2$,
the $x$-bubble tends not to nucleate or tends to be weaker around that point.
In this sense, 
what breaks the isotropy is the correlation between the two bubbles.

The above arguments give us two different viewpoints on the double-bubble contribution:
\begin{itemize}
\item
All the nucleation patterns of the $x$- and $y$-bubbles 
such as pattern $1$ to $4$ in Fig.~\ref{fig:Cherry}
sum up to give the GW spectrum.
\item
Most of the nucleation patterns, like pattern $3$ or $4$ in Fig.~\ref{fig:Cherry},
do not contribute to the GW spectrum because they give only isotropic contributions.
Only when the two nucleation points have correlation,
like pattern $1$ or $2$,
is the isotropy broken and contributions to the GW spectrum arise.
\end{itemize}
These two viewpoints give the same GW spectrum,
but require different numerics.
The first one gives the spectrum only after 
summing up the contributions from all directions.
Eq.~(\ref{eq:AppBeyondDeltaDConcConcFinal}) is based on this viewpoint,
and therefore shows a poor convergence.
Hence, it would be natural to look for an expression based on the second viewpoint.
It is provided by subtracting $\left< T(x) \right> \left< T(y) \right>$.
We explain this below.

The system we consider in this paper is isotropic, 
and therefore Eq.~(\ref{eq:Pi}) with $\left< T(x) T(y) \right>$ replaced by 
$\left< T(x) \right> \left< T(y) \right>$ vanishes due to the projection operator.
We denote the unequal-time correlator $\Pi$ with such a replacement by
$\Pi_\infty$, where the meaning of the subscript will be made clear soon.
Correspondingly, we obtain $\Delta_\infty$ 
by substituting $\Pi_\infty$ instead of $\Pi$ into Eq.~(\ref{eq:DeltaPi}).
These $\Pi_\infty$ and $\Delta_\infty$ are strictly zero, but we calculate them anyway.
They can be calculated by following a similar procedure to Appendix~\ref{subapp:BeyondEnvDouble}.
However, there is a straightforward way to derive the resulting expression:
taking $r_{xi,yi} \to \infty$ in Eq.~(\ref{eq:AppBeyondDeltaDConcConcFinal}).
This is because considering $\left< T(x) \right> \left< T(y) \right>$
corresponds to considering two copies of the system which are unrelated with each other,
and calculate $\left< T(x) \right>$ and $\left< T(y) \right>$ in each of the two, respectively.
Such two uncorrelated systems are effectively realized by taking $r_{xi,yi} \to \infty$ limit.
In this limit, the spacelike theta function always gives unity and can be eliminated from the expression.
Also, ${\mathcal E}$ functions vanish because $t_{xni}$ and $t_{yni}$ go to $-\infty$.
In addition, the ${\mathcal I}$ function is replaced by ${\mathcal I}_\infty$, where
\begin{align}
{\mathcal I}_\infty (t_{x,y})
&= 8\pi \left[ e^{t_{x,y}/2} + e^{-t_{x,y}/2} \right].
\end{align}
As a result, we obtain the following expression for the double-bubble spectrum
\begin{align}
\Delta^{(d)} 
=
&\;
\Delta^{(d)} 
- \Delta_\infty
\nonumber \\
=
&
\int_{-\infty}^\infty dt_{x,y}
\int_0^\infty dr_v
\int_0^\infty dt_{x,xi}
\int_0^\infty dt_{y,yi}
\int_{-1}^1 dc_{xn}
\int_{-1}^1 dc_{yn}
\int_0^{2\pi} d\phi_{xn,yn}
\nonumber \\[1ex]
12v^3k^3
&\left[
\begin{matrix*}[l]
\displaystyle
\;
\left(1 + \frac{1}{\tau} \right)^2
e^{-(t_{x,xi} + t_{y,yi})/\tau}
\\[1.5ex]
\displaystyle
\times
\left[
\Theta_{\rm sp}(x_i,y_i)
\frac{
\left[ 1 - {\mathcal E}(\tau;t_{xi,xni}) \right]
\left[ 1 - {\mathcal E}(\tau;t_{yi,yni}) \right]
}{{\mathcal I}(t_{xi,yi},r_{xi,yi}/v)^2}
-
\frac{1}{{\mathcal I}_\infty(t_{xi,yi})^2}
\right]
\\[2ex]
\displaystyle
\times \;
r_v^2
\left[
j_0(vkr_v){\mathcal K}_0(n_{xn},n_{yn})
+ \frac{j_1(vkr_v)}{vkr_v}{\mathcal K}_1(n_{xn},n_{yn})
+ \frac{j_2(vkr_v)}{(vkr_v)^2}{\mathcal K}_2(n_{xn},n_{yn})
\right]
\\[2ex]
\times
\cos(kt_{x,y})
\end{matrix*}
\right].
\label{eq:AppBeyondDeltaDConcConcFinalSubtracted}
\end{align}
This expression properly realizes the second viewpoint explained above:
for bubble nucleation patterns like $3$ or $4$ in Fig.~\ref{fig:Cherry},
the second line in the square parenthesis almost vanishes
(note that for a large separation $\gg 1/\beta$ between the nucleation points like pattern $3$ or $4$,
the separation between the interception points tend to be also large).
Also, one can check that $\Delta_\infty$ vanishes:
the dependence on the propagation directions appears only in the ${\mathcal K}$ functions,
which vanish after $c_{xn}$, $c_{yn}$ and $\phi_{xn,yn}$ integrations.
We use Eq.~(\ref{eq:AppBeyondDeltaDConcConcFinalSubtracted})
in the numerical evaluation in Sec.~\ref{sec:Numerical}.

%%%%%%%%%%%%%%%%
%%%%%%%%%%%%%%%%

In the following, we derive the expression for $\Delta_\infty$ one by one for completeness.
The procedure is almost the same as Sec.~\ref{subapp:BeyondEnvDouble}.
First note that there are two types of contributions to $\left< T(x) \right>$:
\begin{itemize}
\item
$U$ : $x$-fragment remains uncollided at the evaluation point.
\item
$C$ : $x$-fragment has already been intercepted at the evaluation point.
\end{itemize}
They are written as
\begin{align}
\left< T(x) \right>^{(U)}
&= 
\int_{-\infty}^{t_x} dt_{xn}
\int d\Omega_{xn} \;
e^{-I(x)}
{\mathcal F}_\infty(x;x_i = x;x_n),
\\
\left< T(x) \right>^{(C)}
&= 
\int_{-\infty}^{t_x} dt_{xn}
\int_{t_{xn}}^{t_x} dt_{xi}
\int d\Omega_{xn}
\left[ I(x_i) e^{-I(x_i)} \right]_{txi}
{\mathcal F}_\infty(x;x_i;x_n),
\end{align}
where ${\mathcal F}_\infty$ is given by
\begin{align}
{\mathcal F}_\infty (x,x_i,x_n)
&= 
r_{xn}^{(d)2}l_B\Gamma(t_{xn}) \;
\left( \frac{4\pi}{3}r_{xn}^{(d)3} \kappa \rho_0\frac{1}{4\pi r_{xn}^{(d)2}l_B} \right)
\frac{r_B(t_{xi},t_{xn})^3}{r_{xn}^{(d)3}}
D(t_x,t_{xi},t_{xn}).
\end{align}
Note that now the false vacuum probability has only one argument.
These two are combined to give
\begin{align}
\left< T(x) \right>^{(U + C)}
&= 
\int_{-\infty}^{t_x} dt_{xn}
\int_{t_{xn}}^{t_x} dt_{xi}
\int d\Omega_{xn} \;
e^{-I(x_i)}
\left[ {\mathcal F}_\infty(x;x_i;x_n) \right]_{txi},
\end{align}
where we used ${\mathcal F}_\infty (x;x_i = x_n;x_n) = 0$.
A similar expression is obtained for $\left< T(y) \right>^{(U + C)}$, and 
we obtain the following expression for $\Delta_\infty$ for a general nucleation rate and 
the damping function
\begin{align}
\Delta_\infty
=
&
\int_{-\infty}^\infty dt_x 
\int_{-\infty}^\infty dt_y 
\nonumber \\
&
\int_0^\infty dr 
\int_{-\infty}^{t_x} dt_{xn} 
\int_{-\infty}^{t_y} dt_{yn} 
\int_{t_{xn}}^{t_x} dt_{xi} 
\int_{t_{yn}}^{t_y} dt_{yi} 
\int_{-1}^1 dc_{xn}
\int_{-1}^1 dc_{yn}
\int_0^{2\pi} d\phi_{xn,yn}
\nonumber \\[1ex]
\frac{k^3}{3}
&\left[
\begin{matrix*}[l]
\displaystyle
\;
e^{-I(x_i)}e^{-I(y_i)} 
\Gamma(t_{xn})\Gamma(t_{yn})
\\[1ex]
\displaystyle
\times \;
r^2
\left[
j_0(kr){\mathcal K}_0(n_{xn},n_{yn})
+ \frac{j_1(kr)}{kr}{\mathcal K}_1(n_{xn},n_{yn})
+ \frac{j_2(kr)}{(kr)^2}{\mathcal K}_2(n_{xn},n_{yn})
\right]
\\[2ex]
\displaystyle
\times \;
\partial_{txi}
\left[
r_B(t_{xi},t_{xn})^3
D(t_x,t_{xi},t_{yn})
\right]
\partial_{tyi}
\left[
r_B(t_{yi},t_{yn})^3
D(t_y,t_{yi},t_{yn})
\right]
\cos(kt_{x,y})
\end{matrix*}
\right],
\label{eq:AppBeyondDeltaInfConc}
\end{align}
Following the same procedure as Appendix~\ref{subapp:space}, we obtain
the following expression for the nucleation rate (\ref{eq:Gamma}) 
and the damping function (\ref{eq:D}):
\begin{align}
\Delta_\infty
=
&
\int_{-\infty}^\infty dt_{x,y}
\int_0^\infty dr_v
\int_0^\infty dt_{x,xi}
\int_0^\infty dt_{y,yi}
\int_{-1}^1 dc_{xn}
\int_{-1}^1 dc_{yn}
\int_0^{2\pi} d\phi_{xn,yn}
\nonumber \\[1ex]
12v^3k^3
&\left[
\begin{matrix*}[l]
\displaystyle
\;
\left(1 + \frac{1}{\tau} \right)^2
\frac{e^{-(t_{x,xi} + t_{y,yi})/\tau}}{{\mathcal I}_\infty(t_{xi,yi})^2}
\\[2ex]
\displaystyle
\times \;
r_v^2
\left[
j_0(vkr_v){\mathcal K}_0(n_{xn},n_{yn})
+ \frac{j_1(vkr_v)}{vkr_v}{\mathcal K}_1(n_{xn},n_{yn})
+ \frac{j_2(vkr_v)}{(vkr_v)^2}{\mathcal K}_2(n_{xn},n_{yn})
\right]
\\[2ex]
\times
\cos(kt_{x,y})
\end{matrix*}
\right].
\label{eq:AppBeyondDeltaInfConcConc}
\end{align}

\clearpage

%%%%%%%%%%%%%%%%%%%%%%%%%%%%%%%%%%%%%%%%%%%%%%%%%%
\section{Analytic proof on the spectrum behavior}
\label{app:Analytic}
\setcounter{equation}{0}
%%%%%%%%%%%%%%%%%%%%%%%%%%%%%%%%%%%%%%%%%%%%%%%%%%

In this appendix we give proofs on the behavior 
of the GW spectrum for small $k$ region in $\tau \to \infty$ limit.
In Sec.~\ref{sec:Numerical} and Appendix~\ref{app:Other}
we have numerically seen that the spectrum behaves linear in $k$
in $\tau \to \infty$ limit.
The aim of this appendix is to justify this behavior by analytic arguments.
In the following, note that we adopt $\beta = 1$ unit.

%%%%%%%%%%%%%%%%%%%%%%%%%%%%%%%%%%%%%%%%%%%%%%%%%%
\subsection{Single-bubble spectrum}
%%%%%%%%%%%%%%%%%%%%%%%%%%%%%%%%%%%%%%%%%%%%%%%%%%

Let us start with Eq.~(\ref{eq:BeyondDeltaSConc}) with 
$D = 1$ (long-lasting limit $\tau \to \infty$) and $r_{xi,yi}$ given by Eq.~(\ref{eq:riSingle2}).
The key to observe the behavior of the spectrum for small $k$ is to classify the integration variables
depending on the values they typically take:
\begin{itemize}
\item
Variable $r$ takes $\sim 1/k$, because GWs with wavenumber $k$ are mostly sourced by
objects with distance $\sim 1/k$ away from each other.
\item
Variables $t_{x,n}$ and $t_{y,n}$ take $\sim 1/vk$, because $x$- and $y$-fragments 
propagate distances $\sim r \sim 1/k$ after nucleation to the evaluation times.
\item
Variables $t_{xi,n}$ and $t_{yi,n}$ take $\sim 1$, because the whole space is 
covered by bubbles typically with timescale $\sim 1$.
\end{itemize}
Based on these observations, 
we first rewrite the integration variable in Eq.~(\ref{eq:BeyondDeltaSConc}) as
\begin{align}
&
\int_{-\infty}^\infty dt_x
\int_{-\infty}^\infty dt_y
\int_{v|t_{x,y}|}^\infty dr 
\int_{-\infty}^{t_{\rm max}} dt_n 
\int_{t_n}^{t_x} dt_{xi}
\int_{t_n}^{t_y} dt_{yi}
\nonumber \\
&
\;\;\;\;
=
\;\;\;\;
\int_0^\infty dt_{x,n}
\int_0^\infty dt_{y,n}
\int_{v|t_{x,n} - t_{y,n}|}^{v(t_{x,n} + t_{y,n})} dr 
\int_{-\infty}^\infty dt_n 
\int_0^{t_{x,n}} dt_{xi,n}
\int_0^{t_{y,n}} dt_{yi,n}.
\end{align}
Note that time variables $t_{xn}$, $t_{yn}$, $t_{xi}$ and $t_{yi}$ are
now expressed in terms of differences from $t_n$.
Then we rescale those variables which typically take $\sim 1/k$ or $\sim 1/vk$ by multiplying $k$:
\begin{align}
\Delta^{(s)}
= 
&
\int_0^\infty dt'_{x,n}
\int_0^\infty dt'_{y,n}
\int_{v|t'_{x,n} - t'_{y,n}|}^{v(t'_{x,n} + t'_{y,n})} dr'
\int_{-\infty}^\infty dt_n 
\int_0^{t'_{x,n}/k} dt_{xi,n}
\int_0^{t'_{y,n}/k} dt_{yi,n}
\nonumber \\[1ex]
\frac{k}{3}
&\left[
\begin{matrix*}[l]
\displaystyle
\;
e^{-I(x_i,y_i)} \;
\Gamma(t_n) \;
\frac{r'}{r_{xn}^{'(s)}r_{yn}^{'(s)}}
\\
\displaystyle
\times 
\left[
j_0(r'){\mathcal K}_0(n_{xn\times},n_{yn\times})
+ \frac{j_1(r')}{r'}{\mathcal K}_1(n_{xn\times},n_{yn\times})
+ \frac{j_2(r')}{r^{'2}}{\mathcal K}_2(n_{xn\times},n_{yn\times})
\right]
\\[2.5ex]
\displaystyle
\times \;
\partial_{txi}
\left[
r_B(t_{xi},t_n)^3
\right]
\partial_{tyi}
\left[
r_B(t_{yi},t_n)^3
\right]
\cos(k(t'_{x,n} - t'_{y,n}))
\end{matrix*}
\right],
\label{eq:BeyondDeltaSAnalytic}
\end{align}
where variables with primes mean $\bullet' \equiv k\bullet$.
Now we argue that this expression does not depend on $k$ except for the overall $k$
in the small $k$ limit:
\begin{itemize}
\item
The upper limits for $t_{xi,n}$ and $t_{yi,n}$ go to infinity and thus become independent of $k$.
\item
The variables $r^{'(s)}_{xn} = vt'_{x,n}$ and $r^{'(s)}_{yn} = vt'_{y,n}$ do not depend on $k$.
\item
$I(x_i,y_i) = v^3 \Gamma_* e^{t_n + (t_{xi,n} + t_{yi,n})/2} {\mathcal I}(t_{xi,n} - t_{yi,n},r_{xi,yi})$
does not depend on $k$. 
Note that $r_{xi,yi}$ is independent of $k$.
This is because, in Eq.~(\ref{eq:riSingle2}), 
$c_{xyn\times}$ is written in terms of 
$c_{xn\times}$, $c_{yn\times}$, $s_{xn\times}$ and $s_{yn\times}$,
which can be expressed without $k$ dependence:
\begin{align}
c_{xn\times}
&= 
-\frac{r^{'2} + r_{xn}^{'(s)2} - r_{yn}^{'(s)2}}{2r'r_{xn}^{'(s)}},
\;\;\;\;
c_{yn\times}
= 
\frac{r^{'2} + r_{yn}^{'(s)2} - r_{xn}^{'(s)2}}{2r'r_{yn}^{'(s)}}.
\end{align}
\item
The angular variables $n_{xn\times}$ and $n_{yn\times}$ do not depend on $k$ 
because of the same reason as above.
\end{itemize}
Therefore $\Delta^{(s)}$ behaves linearly in $k$ for small $k$ 
in the long-lasting limit of the bubble walls.

%%%%%%%%%%%%%%%%%%%%%%%%%%%%%%%%%%%%%%%%%%%%%%%%%%
\subsection{Double-bubble spectrum}
%%%%%%%%%%%%%%%%%%%%%%%%%%%%%%%%%%%%%%%%%%%%%%%%%%

In the double-bubble case the procedure is basically the same as the single-bubble case.
However, in contrast to the single-bubble case,
it might be nontrivial that $r_{xi,yi}$ takes $\sim 1$, not $\sim 1/k$, at first sight
(see Eq.~(\ref{eq:riDouble})). 
First let us check by a rough argument that $r_{xi,yi}$ typically does not take $\sim 1/k$.
We consider Eq.~(\ref{eq:BeyondDeltaDConc}) with the subtraction of 
Eq.~(\ref{eq:AppBeyondDeltaInfConc}). It contains a factor 
\begin{align}
\Theta_{\rm sp}(x_i,y_n) \Theta_{\rm sp}(x_n,y_i) e^{-I(x_i,y_i)} 
- e^{-I(x_i)} e^{-I(y_i)}.
\end{align}
Suppose that $r_{xi,yi} \sim 1/k$. 
We can safely assume that $t_{xi}$ and $t_{yi}$ are different by only $\sim 1$ or smaller,
because otherwise at least either of the nucleation times $t_{xn}$ and $t_{yn}$
(which are typically only $\sim 1$ away from $t_{xi}$ and $t_{yi}$)
is away from the typical nucleation times by more than $\sim 1$, 
and the probability for such a bubble configuration to occur is exponentially suppressed.
Then we can set
$\Theta_{\rm sp}(x_i,y_n) \Theta_{\rm sp}(x_n,y_i)$ to unity.
Now we see that $e^{-I(x_i,y_i)}$ and $e^{-I(x_i)} e^{-I(y_i)}$ almost cancels out 
($e^{-I(x_i,y_i)} - e^{-I(x_i)} e^{-I(y_i)} \sim e^{-{\mathcal O}(1)/k} \times e^{-I(x_i,y_i)}$)
for such a large separation $r_{xi,yi}$ between the interception points, 
and therefore such configurations give only suppressed contributions to the GW spectrum.
Hence we can take $r_{xi,yi} \sim v$, not $\sim 1/k$, in $\Delta^{(d)} - \Delta_\infty$,
and by the same procedure as the single-bubble case, we obtain $\propto k$ behavior.
Below we see these arguments more rigorously.

We start with 
Eq.~(\ref{eq:BeyondDeltaDConc}) subtracted by Eq.~(\ref{eq:AppBeyondDeltaInfConc}):
\begin{align}
\Delta^{(d)}
=
&\;
\Delta^{(d)}
- \Delta_\infty
\nonumber \\
=
&
\int_{-\infty}^\infty dt_x 
\int_{-\infty}^\infty dt_y 
\nonumber \\
&
\int_0^\infty dr 
\int_{-\infty}^{t_x} dt_{xn} 
\int_{-\infty}^{t_y} dt_{yn} 
\int_{t_{xn}}^{t_x} dt_{xi} 
\int_{t_{yn}}^{t_y} dt_{yi} 
\int_{-1}^1 dc_{xn}
\int_{-1}^1 dc_{yn}
\int_0^{2\pi} d\phi_{xn,yn}
\nonumber \\[1ex]
\frac{k^3}{3}
&\left[
\begin{matrix*}[l]
\displaystyle
\;
\left[
\Theta_{\rm sp}(x_i,y_n)
\Theta_{\rm sp}(x_n,y_i)
e^{-I(x_i,y_i)} 
-
e^{-I(x_i)} e^{-I(y_i)}
\right]
\Gamma(t_{xn})\Gamma(t_{yn})
\\[1ex]
\displaystyle
\times \;
r^2
\left[
j_0(kr){\mathcal K}_0(n_{xn},n_{yn})
+ \frac{j_1(kr)}{kr}{\mathcal K}_1(n_{xn},n_{yn})
+ \frac{j_2(kr)}{(kr)^2}{\mathcal K}_2(n_{xn},n_{yn})
\right]
\\[2ex]
\displaystyle
\times \;
\partial_{txi}
\left[
r_B(t_{xi},t_{xn})^3
\right]
\partial_{tyi}
\left[
r_B(t_{yi},t_{yn})^3
\right]
\cos(kt_{x,y})
\end{matrix*}
\right],
\end{align}
where we have taken $D = 1$ (long-lasting limit $\tau \to \infty$).
We rewrite the integration variable $r$ in terms of $r_i \equiv r_{xi,yi}$ through the relation (\ref{eq:riDouble}).
However, before that procedure, we replace the integration range for $r$ to $(-\infty,\infty)$
and compensate it by multiplying an overall factor $1/2$:
\begin{align}
\Delta^{(d)}
=
&
\int_{-\infty}^\infty dt_x 
\int_{-\infty}^\infty dt_y 
\nonumber \\
&
\int_{-\infty}^\infty dr 
\int_{-\infty}^{t_x} dt_{xn} 
\int_{-\infty}^{t_y} dt_{yn} 
\int_{t_{xn}}^{t_x} dt_{xi} 
\int_{t_{yn}}^{t_y} dt_{yi} 
\int_{-1}^1 dc_{xn}
\int_{-1}^1 dc_{yn}
\int_0^{2\pi} d\phi_{xn,yn}
\nonumber \\[1ex]
\frac{k^3}{6}
&\left[
\begin{matrix*}[l]
\displaystyle
\;
\left[
\Theta_{\rm sp}(x_i,y_n)
\Theta_{\rm sp}(x_n,y_i)
e^{-I(x_i,y_i)} 
-
e^{-I(x_i)} e^{-I(y_i)}
\right]
\Gamma(t_{xn})\Gamma(t_{yn})
\\[1ex]
\displaystyle
\times \;
|r|^2
\left[
j_0(k|r|){\mathcal K}_0(n_{xn},n_{yn})
+ \frac{j_1(k|r|)}{k|r|}{\mathcal K}_1(n_{xn},n_{yn})
+ \frac{j_2(k|r|)}{(k|r|)^2}{\mathcal K}_2(n_{xn},n_{yn})
\right]
\\[2ex]
\displaystyle
\times \;
\partial_{txi}
\left[
r_B(t_{xi},t_{xn})^3
\right]
\partial_{tyi}
\left[
r_B(t_{yi},t_{yn})^3
\right]
\cos(kt_{x,y})
\end{matrix*}
\right],
\end{align}
where we give the distance between the interception points $r_{xi,yi}$ as a function of $r$
by (\ref{eq:riDouble}) for both positive and negative $r$.
It can be shown that $r<0$ gives an identical contribution to $r>0$.
Then we rewrite $r$ in terms of $r_i$ through Eq.~(\ref{eq:riDouble}),
and also write the time variables in terms of differences:
\begin{align}
\Delta^{(d)}
=
&
\int_0^\infty dt_{x,xi} 
\int_0^\infty dt_{y,yi} 
\nonumber \\
&
\int_{r_{i,{\rm min}}}^\infty dr_i
\int_{-\infty}^0 dt_{xn,xi} 
\int_{-\infty}^0 dt_{yn,yi} 
\int_{-\infty}^\infty dt_{xi} 
\int_{-\infty}^\infty dt_{yi} 
\int_{-1}^1 dc_{xn}
\int_{-1}^1 dc_{yn}
\int_0^{2\pi} d\phi_{xn,yn}
\nonumber \\[1ex]
\frac{k^3}{6}
&\left[
\begin{matrix*}[l]
\displaystyle
\;
\left[
\Theta_{\rm sp}(x_i,y_n)
\Theta_{\rm sp}(x_n,y_i)
e^{-I(x_i,y_i)} 
-
e^{-I(x_i)} e^{-I(y_i)}
\right]
\Gamma(t_{xn})\Gamma(t_{yn})
\\[1ex]
\displaystyle
\times \;
\frac{r_i}{\sqrt{r_i^2 - r_{i,{\rm min}}^2}}
\left[
|r_+|^2
j{\mathcal K} (k|r_+|,n_{xn},n_{yn})
+
|r_-|^2
j{\mathcal K} (k|r_-|,n_{xn},n_{yn})
\right]
\\[2ex]
\displaystyle
\times \;
\partial_{txi}
\left[
r_B(t_{xi},t_{xn})^3
\right]
\partial_{tyi}
\left[
r_B(t_{yi},t_{yn})^3
\right]
\cos(kt_{x,y})
\end{matrix*}
\right],
\end{align}
with 
\begin{align}
\frac{r_{i,{\rm min}}}{v}
&\equiv
\sqrt{(t_{x,xi}s_{xn} - t_{y,yi}s_{xn})^2 + 2t_{x,xi}t_{y,yi}(1 - c_{\phi_{xn,yn}})},
\end{align}
and
\begin{align}
j{\mathcal K} (kr,n_{xn},n_{yn})
&\equiv
j_0(kr){\mathcal K}_0(n_{xn},n_{yn})
+ \frac{j_1(kr)}{kr}{\mathcal K}_1(n_{xn},n_{yn})
+ \frac{j_2(kr)}{(kr)^2}{\mathcal K}_2(n_{xn},n_{yn}).
\end{align}
Also, the function $r_\pm$ is given by
\begin{align}
\frac{r_\pm}{v}
&\equiv
- t_{x,xi}c_{xn} 
+ t_{y,yi}c_{yn}
\pm  
\sqrt{\frac{r_i^2}{v^2} - \frac{r_{i,{\rm min}}^2}{v^2}}.
\end{align}
Now we rescale those variables which typically take $\propto 1/k$
as in the single-bubble case:
\begin{align}
\Delta^{(d)}
=
&
\int_0^\infty dt'_{x,xi} 
\int_0^\infty dt'_{y,yi} 
\nonumber \\
&
\int_{r_{i,{\rm min}}}^\infty dr_i
\int_{-\infty}^0 dt_{xn,xi} 
\int_{-\infty}^0 dt_{yn,yi} 
\int_{-\infty}^\infty dt_{xi} 
\int_{-\infty}^\infty dt_{yi} 
\int_{-1}^1 dc_{xn}
\int_{-1}^1 dc_{yn}
\int_0^{2\pi} d\phi_{xn,yn}
\nonumber \\[1ex]
\frac{1}{6k}
&\left[
\begin{matrix*}[l]
\displaystyle
\;
\left[
\Theta_{\rm sp}(x_i,y_n)
\Theta_{\rm sp}(x_n,y_i)
e^{-I(x_i,y_i)} 
-
e^{-I(x_i)} e^{-I(y_i)}
\right]
\Gamma(t_{xn})\Gamma(t_{yn})
\\[1ex]
\displaystyle
\times \;
\frac{r_i}{\sqrt{r_i^2 - r_{i,{\rm min}}^2}}
\left[
|r'_+|^2
j{\mathcal K} (|r'_+|,n_{xn},n_{yn})
+
|r'_-|^2
j{\mathcal K} (|r'_-|,n_{xn},n_{yn})
\right]
\\[2ex]
\displaystyle
\times \;
\partial_{txi}
\left[
r_B(t_{xi},t_{xn})^3
\right]
\partial_{tyi}
\left[
r_B(t_{yi},t_{yn})^3
\right]
\cos(kt_{x,y})
\end{matrix*}
\right],
\label{eq:DeltaDPropk}
\end{align}
where the prime denotes $\bullet' \equiv k\bullet$.
One sees that this expression does not depend on $k$ except for
\begin{align}
\frac{r_{i,{\rm min}}}{v}
&=
\frac{1}{k}\sqrt{(t'_{x,xi}s_{xn} - t'_{y,yi}s_{xn})^2 + 2t'_{x,xi}t'_{y,yi}(1 - c_{\phi_{xn,yn}})},
\label{eq:rimin}
\end{align}
and
\begin{align}
\frac{r'_\pm}{v}
&=
- t'_{x,xi}c_{xn} 
+ t'_{y,yi}c_{yn}
\pm  
k\sqrt{\frac{r_i^2}{v^2} - \frac{r_{i,{\rm min}}^2}{v^2}}.
\end{align}
Let us consider the implications of these $k$ dependences.
In Eq.~(\ref{eq:rimin}), the quantities in the square root are naively expected to take
values $\sim 1$.
However, if the square root takes $\sim 1$, 
the $1/k$ factor makes $r_{i,{\rm min}}$ much larger than unity.
We have seen in the beginning of this subsection 
that such a large $r_i$ $(> r_{i,{\rm min}})$ makes the exponentials of the $I$ function 
cancel out with each other.
Therefore, we see that Eq.~(\ref{eq:rimin}) with the integration region $r_i > r_{i,{\rm min}}$ 
works as $\delta$-functions for $(t'_{x,xi}s_{xn} - t'_{y,yi}s_{xn})$ and 
$2t'_{x,xi}t'_{y,yi}(1 - c_{\phi_{xn,yn}})$ in small $k$ limit.
After properly taking proportionality factors into account, 
we find that the following replacement
\begin{align}
\frac{1}{\sqrt{r_i^2 - r_{i,{\rm min}}^2}}
&\to
2\pi r_i \frac{k^2}{v^2}
\delta \left( t'_{x,xi}s_{xn} - t'_{y,yi}s_{xn} \right)
\delta \left( \sqrt{2t'_{x,xi}t'_{y,yi}(1 - c_{\phi_{xn,yn}})} \right),
\label{eq:deltaReplace}
\end{align}
is justified for small $k$ limit.\footnote{
Discussion here can be illustrated with the following example:
\begin{align}
\Delta
&=
\int_{x_{\rm min}}^\infty dx 
\int_0^\infty dy \;
\frac{f(x)}{\sqrt{x^2 - x_{\rm min}^2}},
\end{align}
with $x_{\rm min} = y/k$ and $f$ denoting some function which drops rapidly for $x \gg 1$.
Here $x$ corresponds to $r_i$, while $y$ corresponds to 
the other time and angular variables in Eq.~(\ref{eq:DeltaDPropk}).
We can explicitly calculate $y$ integration to obtain
\begin{align}
\Delta
&=
\int_0^\infty dx 
\int_0^{x > x_{\rm min}} dy \;
\frac{f(x)}{\sqrt{x^2 - x_{\rm min}^2}}
=
\int_0^\infty dx \;
\frac{\pi}{2} k f(x).
\end{align}
In the original expression, this corresponds to the replacement
\begin{align}
\frac{1}{\sqrt{x^2 - x_{\rm min}^2}}
&\to
\pi \delta (x_{\rm min}),
\end{align}
in small $k$ limit.
Similar arguments lead to Eq.~(\ref{eq:deltaReplace}).
}
Also, $r'_\pm / v \to - t'_{x,xi}c_{xn} + t'_{y,yn}c_{yn}$ holds in the same limit.
As a result, we obtain
\begin{align}
\Delta^{(d)}
\to
&
\int_0^\infty dt'_{x,xi} 
\int_0^\infty dt'_{y,yi} 
\nonumber \\
&
\int_0^\infty dr_i
\int_{-\infty}^0 dt_{xn,xi} 
\int_{-\infty}^0 dt_{yn,yi} 
\int_{-\infty}^\infty dt_{xi} 
\int_{-\infty}^\infty dt_{yi} 
\int_{-1}^1 dc_{xn}
\int_{-1}^1 dc_{yn}
\int_0^{2\pi} d\phi_{xn,yn}
\nonumber \\[1ex]
\frac{\pi}{3}k
&\left[
\begin{matrix*}[l]
\displaystyle
\;
\left[
\Theta_{\rm sp}(x_i,y_n)
\Theta_{\rm sp}(x_n,y_i)
e^{-I(x_i,y_i)} 
-
e^{-I(x_i)} e^{-I(y_i)}
\right]
\Gamma(t_{xn})\Gamma(t_{yn})
\\[1ex]
\displaystyle
\times \;
\frac{r_i^2}{v^2} \;
\delta \left( t'_{x,xi}s_{xn} - t'_{y,yi}s_{xn} \right)
\delta \left( \sqrt{2t'_{x,xi}t'_{y,yi}(1 - c_{\phi_{xn,yn}})} \right)
\\[2ex]
\times
\left[
|r'_+|^2
j{\mathcal K} (|r'_+|,n_{xn},n_{yn})
+
|r'_-|^2
j{\mathcal K} (|r'_-|,n_{xn},n_{yn})
\right]
\\[2ex]
\displaystyle
\times \;
\partial_{txi}
\left[
r_B(t_{xi},t_{xn})^3
\right]
\partial_{tyi}
\left[
r_B(t_{yi},t_{yn})^3
\right]
\cos(kt_{x,y})
\end{matrix*}
\right],
\end{align}
in small $k$ limit.
Therefore $\Delta^{(d)}$ behaves linearly in $k$ for small $k$ 
in the long-lasting limit of the bubble walls.

\clearpage

%%%%%%%%%%%%%%%%%%%%%%%%%%%%%%%%%%%%%%%%%%%%%%%%%%
\section{Useful equations}
\label{app:useful}
\setcounter{equation}{0}
%%%%%%%%%%%%%%%%%%%%%%%%%%%%%%%%%%%%%%%%%%%%%%%%%%

In this appendix we summarize useful formulas.
In performing the angular part of the Fourier transformation 
with respect to $\vec{r}$,
the following formula holds for arbitrary unit vectors $n_x$ and $n_y$,
which are defined through their angles measured from $\vec{r}$ (see Fig.~\ref{fig:Circle}):\footnote{
The formula (\ref{eq:AppKN}) can easily be derived by noting that
\begin{align}
\int d\Omega_r \int_0^{2\pi} d\phi_{\left< x,y \right>} \;
K_{ijkl}(\hat{k})e^{i \vec{k} \cdot \vec{r}}N_{ijkl}
&= 
\int d\Omega_k \int_0^{2\pi} d\phi_{\left< x,y \right>} \;
K_{ijkl}(\hat{k})e^{i \vec{k} \cdot \vec{r}}N_{ijkl}.
\end{align}
}
\begin{align}
&\int d\Omega_r \int_0^{2\pi} d\phi_{\left< x,y \right>} \;
K_{ijkl}(\hat{k})e^{i \vec{k} \cdot \vec{r}}N_{ijkl} 
\nonumber \\
&\;\;\;\;\;\;
= 
4\pi^2
\left[ 
j_0(kr) {\mathcal K}_0(n_x,n_y)
+ \frac{j_1(kr)}{kr} {\mathcal K}_1(n_x,n_y)
+ \frac{j_2(kr)}{(kr)^2} {\mathcal K}_2(n_x,n_y)
\right].
\label{eq:AppKN}
\end{align}
Here $\phi_{\left< x,y \right>} \equiv (\phi_x + \phi_y)/2$,
and ${\mathcal K}_n$ are
\begin{align}
{\mathcal K}_0(n_x,n_y)
&=
s_x^2s_y^2 \cos (2(\phi_x - \phi_y)), 
\label{eq:AppK0}
\\ 
{\mathcal K}_1(n_x,n_y)
&= 
8s_xc_xs_yc_y \cos (\phi_x - \phi_y)
- 2s_x^2s_y^2 \cos (2(\phi_x - \phi_y)), 
\label{eq:AppK1}
\\
{\mathcal K}_2(n_x,n_y)
&= 
2(3c_x^2 - 1)(3c_y^2 - 1)
- 16s_xc_xs_yc_y \cos (\phi_x - \phi_y)
+ s_x^2s_y^2 \cos (2(\phi_x - \phi_y)),
\label{eq:AppK2}
\end{align}
with $c_x$ and $s_x$ ($c_y$ and $s_y$) denoting 
$\cos \theta_x$ and $\sin \theta_x$ ($\cos \theta_y$ and $\sin \theta_y$),
Also, $j_n$ are the spherical Bessel functions
\begin{align}
j_0(x)
&= \frac{\sin x}{x},
\;\;\;\;
j_1(x)
= \frac{\sin x - x \cos x}{x^2},
\;\;\;\;
j_2(x)
= \frac{(3 - x^2)\sin x - 3x \cos x}{x^3}.
\label{eq:j}
\end{align}
Note that the vectors $n_x$ and $n_y$ 
change their directions as $\vec{r}$ changes its direction
because they are defined through their angles measured from $\vec{r}$.
The arguments $\theta_x$, $\theta_y$, $\phi_x$ and $\phi_y$ 
in Eqs.~(\ref{eq:AppK0})--(\ref{eq:AppK2}) denote such angles.

%%%%%%%%%%%%%%%%%%%%%%%%%%%%%%%%%%%%%%%%%%%%%%%%%%
\section{Other numerical results}
\label{app:Other}
\setcounter{equation}{0}
%%%%%%%%%%%%%%%%%%%%%%%%%%%%%%%%%%%%%%%%%%%%%%%%%%

In this appendix we show the results presented in Sec.~\ref{sec:Numerical}
in terms of single- and double-bubble contributions.

\begin{figure}
\begin{center}
\includegraphics[width=0.7\columnwidth]{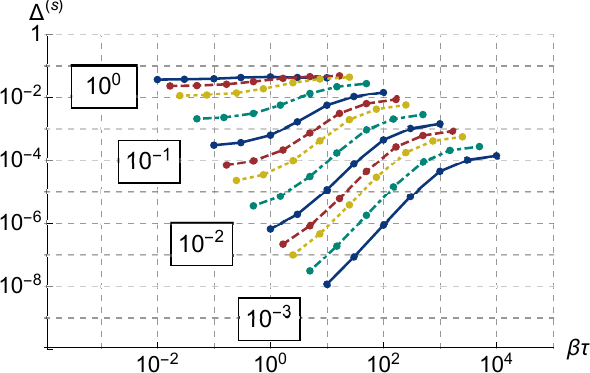}
\caption{\small
The single-bubble spectrum $\Delta^{(s)}$ as a function of the duration time $\tau$ for $v = 1$.
The blue, red, yellow and green lines correspond to 
$k = (1, 0.6, 0.4, 0.2) \times 10^{-n}$ ($n \in \mathbb{Z}$), respectively.
}
\label{fig:tauDeltaS_v=1}
\end{center}
\begin{center}
\includegraphics[width=0.7\columnwidth]{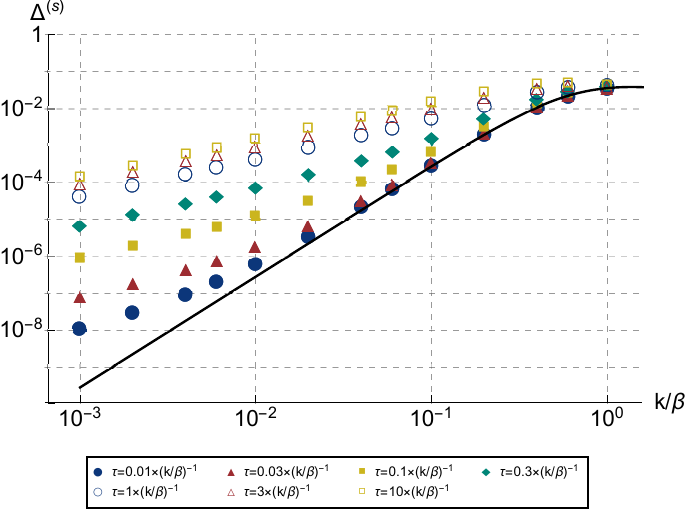}
\caption{\small
The single-bubble spectrum $\Delta^{(s)}$ as a function of wavenumber $k$ for $v = 1$.
Each data point corresponds to $\tau = (0.01, 0.03, 0.1, 0.3, 1, 3, 10) \times (1/k)$,
while the black line shows the single-bubble spectrum with the envelope approximation 
in Ref.~\cite{Jinno:2016vai}.
}
\label{fig:kDeltaS_v=1}
\end{center}
\end{figure}

First, Figs.~\ref{fig:tauDeltaS_v=1}, \ref{fig:kDeltaS_v=1} and \ref{fig:kDeltaSSlice_v=1}
are the behavior of the single-bubble spectrum for $v = 1$.
Fig.~\ref{fig:tauDeltaS_v=1} shows 
the single-bubble spectrum as a function of the duration time $\tau$
for various wavenumbers from $k = 0.001$ to $1$.
The blue, red, yellow and green lines denote 
$k = (1,0.6,0.4,0.2) \times 10^{-n}$ $(n \in \mathbb{Z})$, respectively.
Fig.~\ref{fig:kDeltaS_v=1} is essentially the same as Fig.~\ref{fig:tauDeltaS_v=1},
except that the horizontal axis is the wavenumber $k$.
Different markers correspond to different $\tau$,
and the black line is the single-bubble spectrum with the envelope approximation in Ref.~\cite{Jinno:2016vai}.
Fig.~\ref{fig:kDeltaSSlice_v=1} shows the spectrum at fixed $\tau$.
The colored lines correspond to $\tau = 1, 3, 10, 30, 100$ from bottom to top,
while the black-dashed line corresponds to $\tau = 10 \times (1/k)$.
In making this figure we have interpolated the data points shown in Figs.~\ref{fig:tauDeltaS_v=1}
and \ref{fig:kDeltaS_v=1} to make constant-$\tau$ slices.
Also, for large wavenumbers $k > 0.1$, 
we have extrapolated the value at $\tau = 10 \times (1/k)$ to larger $\tau$
by assuming that the spectrum is constant after this time.

\begin{figure}
\begin{center}
\includegraphics[width=0.7\columnwidth]{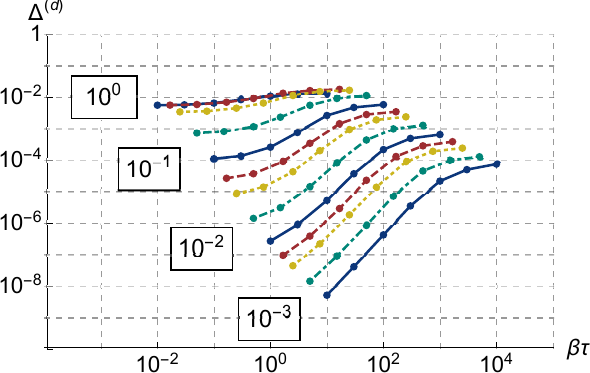}
\caption{\small
The double-bubble spectrum $\Delta^{(d)}$ as a function of the duration time $\tau$ for $v = 1$.
The blue, red, yellow and green lines correspond to 
$k = (1, 0.6, 0.4, 0.2) \times 10^{-n}$ ($n \in \mathbb{Z}$), respectively.
}
\label{fig:tauDeltaD_v=1}
\end{center}
\begin{center}
\includegraphics[width=0.7\columnwidth]{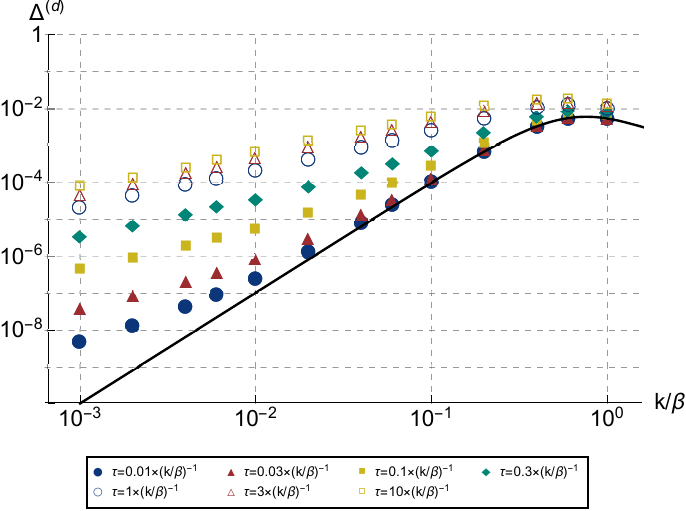}
\caption{\small
The double-bubble spectrum $\Delta^{(d)}$ as a function of wavenumber $k$ for $v = 1$.
Each data point corresponds to $\tau = (0.01, 0.03, 0.1, 0.3, 1, 3, 10) \times (1/k)$,
while the black line shows the double-bubble spectrum with the envelope approximation 
in Ref.~\cite{Jinno:2016vai}.
}
\label{fig:kDeltaD_v=1}
\end{center}
\end{figure}

\begin{figure}
\begin{center}
\includegraphics[width=0.7\columnwidth]{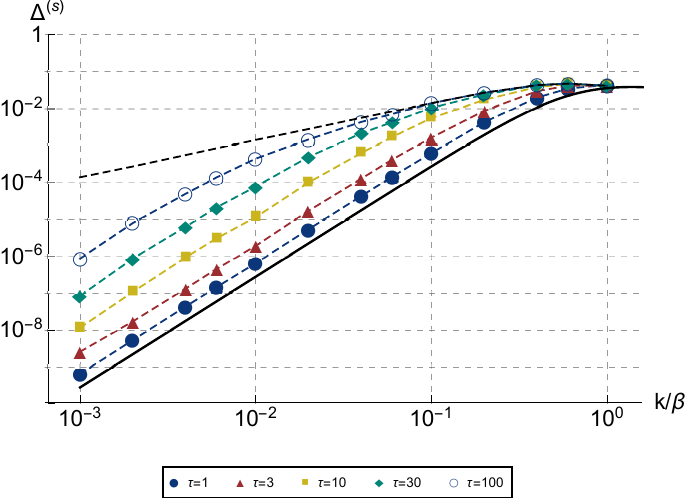}
\caption{\small
The single-bubble spectrum $\Delta^{(s)}$ as a function of wavenumber $k$ for $v = 1$.
Each colored line corresponds to $\tau = 1, 3, 10, 30, 100$ from bottom to top,
while the black-dashed line corresponds to the data points at $\tau = 10 \times (1/k)$.
The black-solid line is the same as in Fig.~\ref{fig:kDeltaS_v=1}.
}
\label{fig:kDeltaSSlice_v=1}
\end{center}
\begin{center}
\includegraphics[width=0.7\columnwidth]{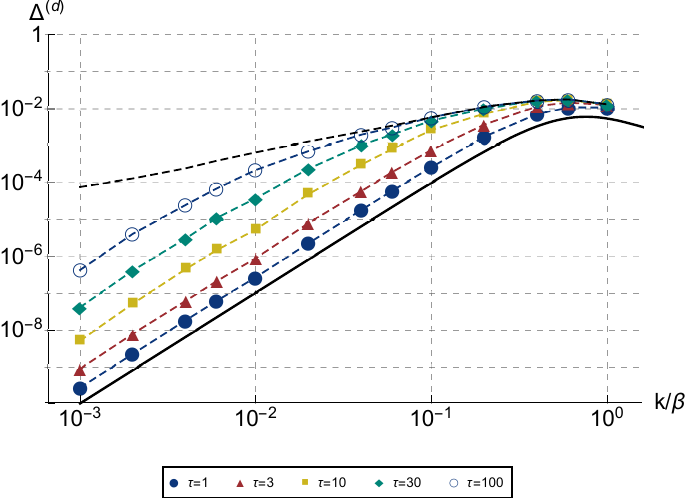}
\caption{\small
The double-bubble spectrum $\Delta^{(d)}$ as a function of wavenumber $k$ for $v = 1$.
Each colored line corresponds to $\tau = 1, 3, 10, 30, 100$ from bottom to top,
while the black-dashed line corresponds to the data points at $\tau = 10 \times (1/k)$.
The black-solid line is the same as in Fig.~\ref{fig:kDeltaD_v=1}.
}
\label{fig:kDeltaDSlice_v=1}
\end{center}
\end{figure}

Next, Figs.~\ref{fig:tauDeltaD_v=1}, \ref{fig:kDeltaD_v=1} and \ref{fig:kDeltaDSlice_v=1}
are the same as Figs.~\ref{fig:tauDeltaS_v=1}, \ref{fig:kDeltaS_v=1} and \ref{fig:kDeltaSSlice_v=1},
respectively, except that they show the double-bubble spectrum.
It is seen that the single-bubble dominates the double-bubble for the wavenumbers shown in the plot.

\begin{figure}
\begin{center}
\includegraphics[width=0.7\columnwidth]{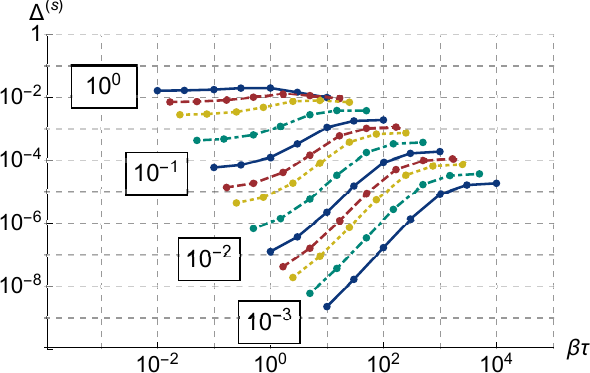}
\caption{\small
Same as Fig.~\ref{fig:tauDeltaS_v=1} except that $v = c_s$.
}
\label{fig:tauDeltaS_v=cs}
\end{center}
\begin{center}
\includegraphics[width=0.7\columnwidth]{./figs/kDeltaS_v=1.pdf}
\caption{\small
Same as Fig.~\ref{fig:kDeltaS_v=1} except that $v = c_s$.
}
\label{fig:kDeltaS_v=cs}
\end{center}
\end{figure}

\begin{figure}
\begin{center}
\includegraphics[width=0.7\columnwidth]{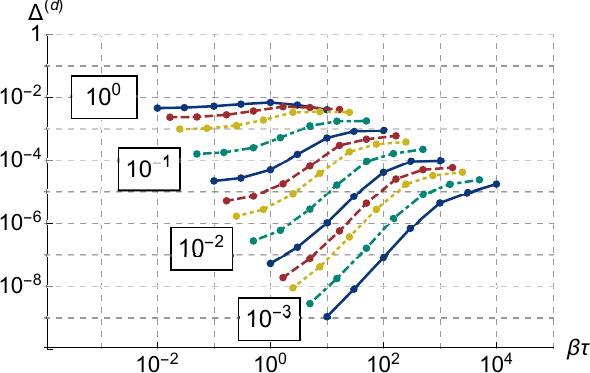}
\caption{\small
Same as Fig.~\ref{fig:tauDeltaD_v=1} except that $v = c_s$.
}
\label{fig:tauDeltaD_v=cs}
\end{center}
\begin{center}
\includegraphics[width=0.7\columnwidth]{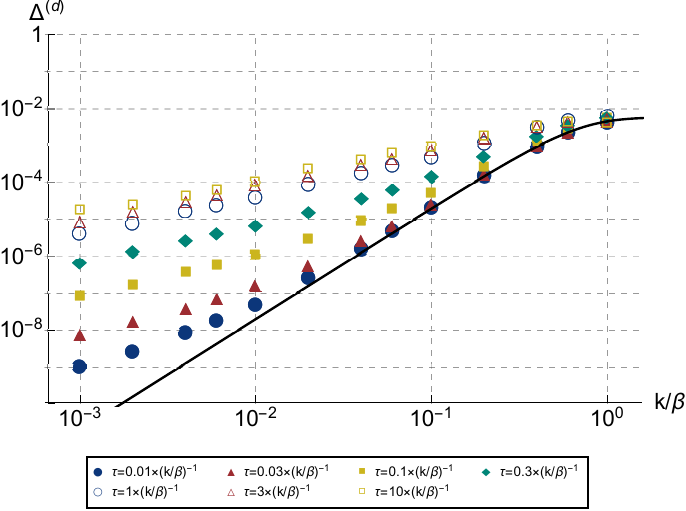}
\caption{\small
Same as Fig.~\ref{fig:kDeltaD_v=1} except that $v = c_s$.
}
\label{fig:kDeltaD_v=cs}
\end{center}
\end{figure}

\begin{figure}
\begin{center}
\includegraphics[width=0.7\columnwidth]{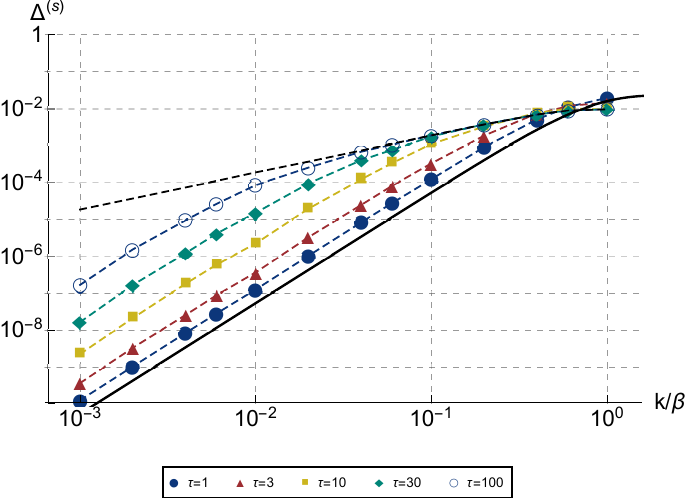}
\caption{\small
Same as Fig.~\ref{fig:kDeltaSSlice_v=1} except that $v = c_s$.
}
\label{fig:kDeltaSSlice_v=cs}
\end{center}
\begin{center}
\includegraphics[width=0.7\columnwidth]{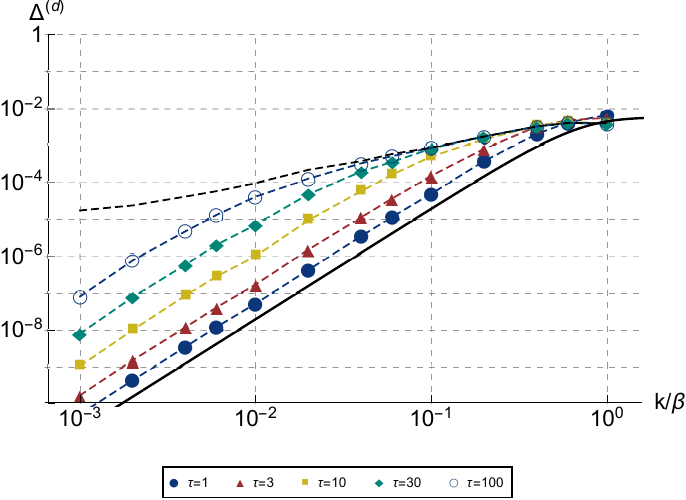}
\caption{\small
Same as Fig.~\ref{fig:kDeltaDSlice_v=1} except that $v = c_s$.
}
\label{fig:kDeltaDSlice_v=cs}
\end{center}
\end{figure}

Finally, Figs.~\ref{fig:tauDeltaS_v=cs}, \ref{fig:kDeltaS_v=cs} and \ref{fig:kDeltaSSlice_v=cs} 
are the behavior of the single-bubble spectrum for $v = c_s$,
which correspond to 
Figs.~\ref{fig:tauDeltaS_v=1}, \ref{fig:kDeltaS_v=1} and \ref{fig:kDeltaSSlice_v=1}, respectively.
The basic features are the same as $v = 1$ case.
Also, Figs.~\ref{fig:tauDeltaD_v=cs}, \ref{fig:kDeltaD_v=cs} and \ref{fig:kDeltaDSlice_v=cs} 
show the double-bubble spectrum for $v = c_s$.

%%%%%%%%%%%%%%%%
%%%%%%%%%%%%%%%%

%%%%%%%%%%%%%%%%
%%%%%%%%%%%%%%%%

%%%%%%%%%%%%%%%%
%%%%%%%%%%%%%%%%

%%%%%%%%%%%%%%%%
%%%%%%%%%%%%%%%%

%%%%%%%%%%%%%%%%
%%%%%%%%%%%%%%%%

%%%%%%%%%%%%%%%%
%%%%%%%%%%%%%%%%

%%%%%%%%%%%%%%%%%%%%%%%%%%%%%%%%%%%%%%%%%%%%%%%%%%
\section{Comment on ``spherical symmetry"}
\label{app:saru}
\setcounter{equation}{0}
%%%%%%%%%%%%%%%%%%%%%%%%%%%%%%%%%%%%%%%%%%%%%%%%%%

In this appendix we explain why the single-bubble contribution arises
in spite of the fact that spherically symmetric objects do not radiate GWs.

\begin{figure}
\begin{center}
\includegraphics[width=0.4\columnwidth]{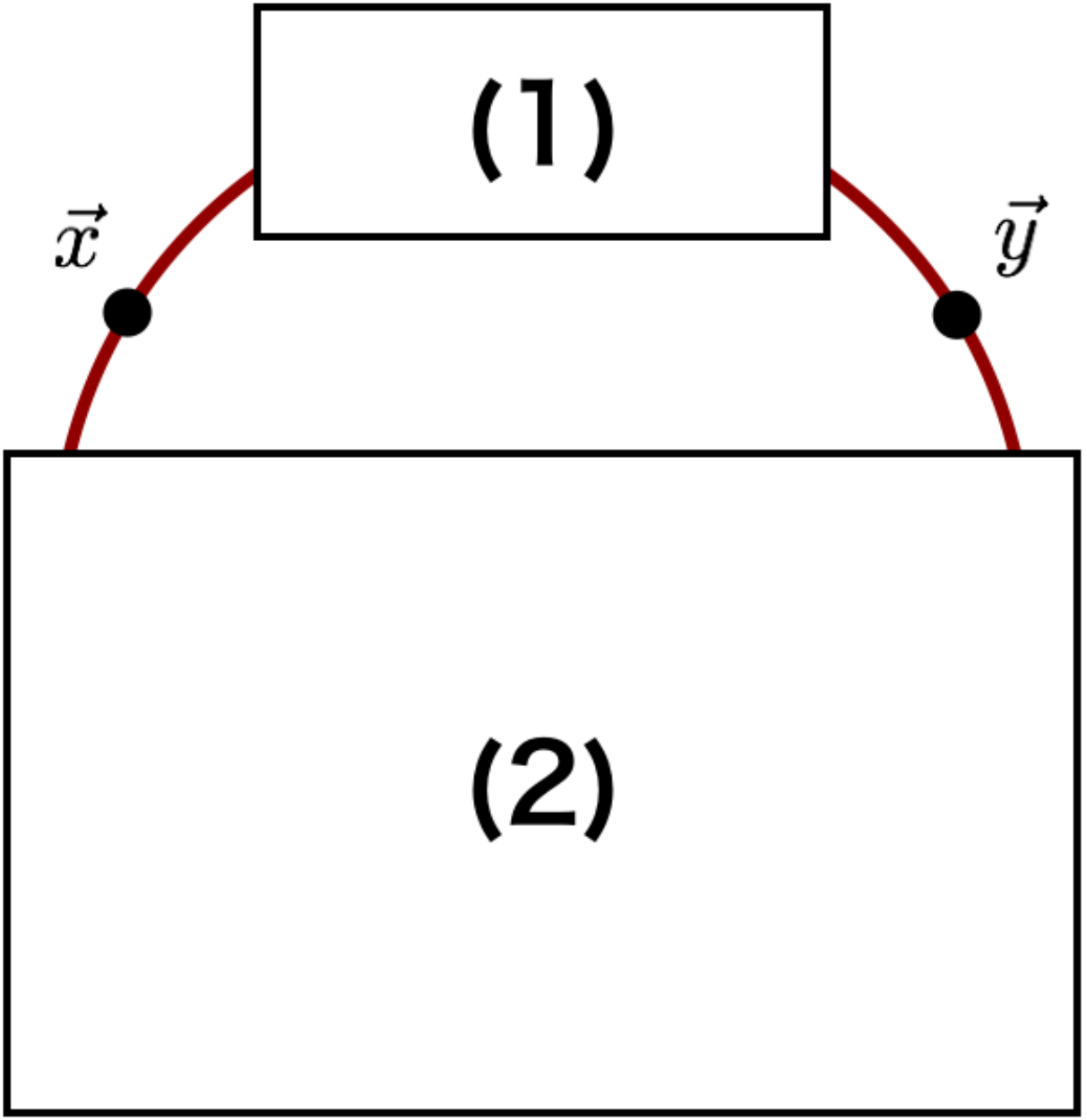} 
\\ \vspace{10mm}
\includegraphics[width=0.9\columnwidth]{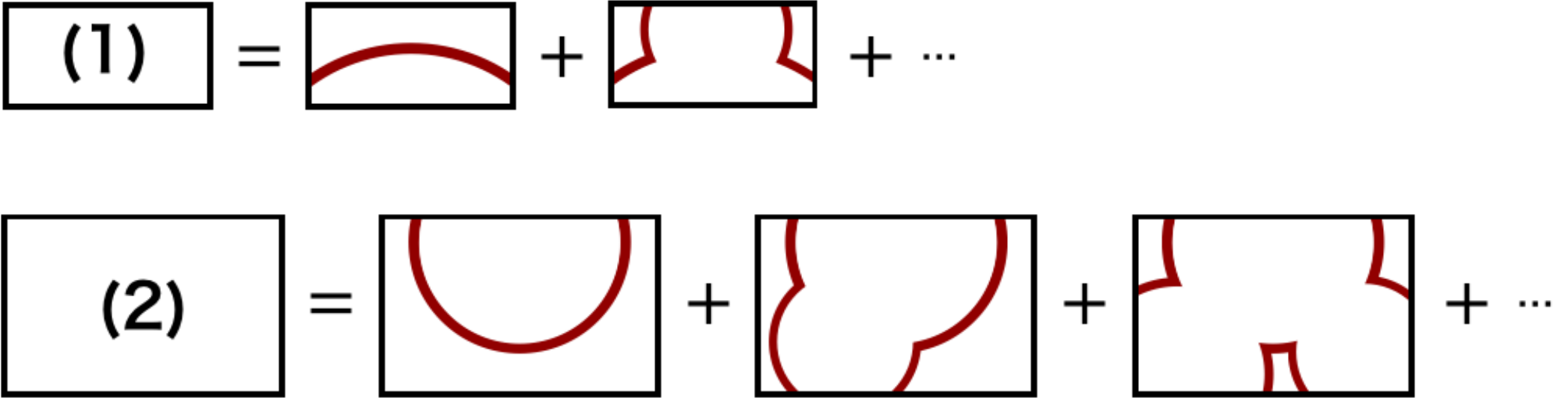}
\caption{\small
Illustration for why the single-bubble contribution arises 
in spite of the spherical symmetry of a single bubble.
In this figure we show the envelope case and set $t_x = t_y$.
Same figure as in Ref.~\cite{Jinno:2016vai}.
}
\label{fig:Saru}
\end{center}
\end{figure}

Let us take the single-bubble contribution in the envelope case as an example.
See Fig.~\ref{fig:Saru}. 
In this figure we take the evaluation times to satisfy $t_x = t_y$ for an illustrative purpose.
The point is that, 
in the formalism illustrated in Sec.~\ref{sec:Analytic} and 
used in Appendix~\ref{app:Envelope}--\ref{app:Simple},
we do not scan the spacetime points $x$ and $y$ over the surface of a single bubble
but rather first fix them and then sum up all the bubble configurations 
which gives nonzero $T(x)T(y)$.
One sees that the spherical symmetry of a single bubble is in fact broken 
by the process of fixing $x$ and $y$:
the bubble wall fragments propagating towards $\vec{x}$ and $\vec{y}$ 
cannot collide with other walls until the evaluation time 
(note that walls instantly disappear when they collide with each other within the envelope approximation),
while other parts of the wall can collide before that time, as shown in Fig.~\ref{fig:Saru}.
Therefore, our formalism automatically takes into account the breaking of 
the spherical symmetry of each bubble caused by collisions with other bubbles.

%%%%%%%%%%%%%%%%
%%%%%%%%%%%%%%%%

\clearpage

%%%%%%%%%%%%%%%%%%%%%%%%%%%%%%%%%%%%%%%%%%%%%%%%%%
\small
\bibliography{ref}
%%%%%%%%%%%%%%%%%%%%%%%%%%%%%%%%%%%%%%%%%%%%%%%%%%

%%%%%%%%%%%%%%%%%%%%%%%%%%%%%%%%%%%%%%%%%%%%%%%%%%
\end{document}